\documentclass[a4paper,fleqn,usenatbib]{mnras}
\usepackage{mathptmx}
\usepackage[T1]{fontenc}
\usepackage{ae,aecompl}
\usepackage{comment}

\usepackage{amssymb,amsmath,amsfonts,graphicx}
\usepackage{natbib,color}

\def\lesssim {<\kern-1.2em\lower1.1ex\hbox{$\sim$}~}   

\title[Spread of stellar abundance ratios of old stellar populations]{The neutron-capture and $\alpha$-elements abundance ratios  scatter in old stellar populations.  Cosmological simulations of the stellar halo}
\author[Scannapieco, Cescutti \& Chiappini]{Cecilia Scannapieco$^{1,2}$, Gabriele Cescutti$^{3,4,5,6}$ and Cristina Chiappini$^{7}$\\
$^1$ Universidad de Buenos Aires,  Facultad de Ciencias Exactas y Naturales, Departamento de F\'{\i}sica. Buenos Aires, Argentina\\
$^2$ CONICET. Buenos Aires, Argentina\\
$^3$ Dipartimento di Fisica, Sezione di Astronomia, Università di Trieste, Via G. B. Tiepolo 11, 34143 Trieste, Italy\\
$^4$ INAF, Osservatorio Astronomico di Trieste, Via Tiepolo 11,  I-34143 Trieste, Italy\\
$^5$ INFN, Sezione di Trieste, Via A. Valerio 2, I-34127 Trieste, Italy\\
$^6$ IFPU, Istitute for the Fundamental Physics of the Universe, Via Beirut, 2, I-34151 Grignano, Trieste, Italy\\
$^7$ Leibniz-Institut f\"ur Astrophysik Potsdam (AIP), An der Sternwarte 16, D-14482, Potsdam, Germany}

\begin{document}

\date{Accepted \today \ Received ...; in original form ...}

\pagerange{\pageref{firstpage}--\pageref{lastpage}} \pubyear{2014}

\maketitle

\begin{abstract}

We investigate the origin of the abundance ratios and scatter of the neutron-capture elements
Sr, Ba and Eu in the stellar halo of a Milky Way-mass
galaxy formed in a  hydrodynamical cosmological simulation,  and compare them with those of $\alpha$ elements. For
this, we implement a novel treatment for
chemical enrichment of Type II supernovae which
considers the effects of the rotation of  massive stars on the
chemical yields and  differential enrichment according to the
life-times of progenitor stars.
We find that differential enrichment has a significant impact on
the early enrichment of the interstellar medium which is translated into
broader element ratio distributions, particularly in the case of
the oldest, most metal-poor stars.
We find that the [element/Fe] ratios of the $\alpha-$elements O, Mg and Si have  systematically lower scatter compared to the neutron-capture elements ratios Sr, Ba and Eu at [Fe/H]$<-2$,
which is  $\sim 0.1-0.4$ dex for the former and  between $\sim 0.5$ and $1$ dex for the latter. The different scatter levels  found for the neutron-capture and  $\alpha$-elements is consistent with
  observations of old stars in the Milky Way.
Our model also predicts a high scatter for the [Sr/Ba]
ratio, which results from the treatment of the fast-rotating stars and
the dependence of the chemical yields on the metallicity, mass and
rotational velocities.
Such  chemical patterns
appear naturally if the different ejection times associated to stars
of different mass are properly described, without the need to invoke
for additional mixing mechanisms or a distinct treatment of the $\alpha-$ and
neutron-capture elements.

\end{abstract}


\section{Introduction}

Thanks to the Gaia satellite which provides precise measurements of parallax and proper-motion of stars (\citealt{Brown21} and references therein), as well as spectroscopic information coming from major surveys (complementing the Gaia data with radial velocities and chemistry), the complexity of the formation history of our Galaxy has become evident. In particular, the local low-metallicity regime (below [Fe/H]$\sim-$0.8) turns out to be made of a  collection of debris from past impacts with smaller galaxies (\citealt{Helmi20} and references therein) as well as early in-situ formed stars (\citealt{Miglio21}; \citealt{Montalban21} - usually more enhanced in $\alpha$-elements). 

The most important past accretion event in the early history of the Milky Way seems to have been the merger with Gaia-Enceladus \citep{Helmi18,Belokurov18}. This event could be confirmed thanks to the combination of Gaia DR2 \citep{Brown18} and APOGEE \citep{Majewski17} data. The discovery and characterization of other debris is a very active field and the several large spectroscopic surveys are playing a key role (e.g. \citealt{Koppelman19}; \citealt{Myeong19}; \citealt{Naidu20}; \citealt{ReFiorentin21}). Indeed, the information brought by detailed chemical abundances is of paramount importance in the identification of a star's origin.

The role of neutron-capture (n-capture) elements in this context 
is certainly going to be central  as suggested by recent observational results (see for instance \citealt{Aguado21}; \citealt{Limberg21}; \citealt{Gudin21}; \citealt{Gull21}). 
In particular, the results of \cite{Ryan96} and \cite{McWilliam98}, later on confirmed by \cite{Honda04} and \cite{Francois07}, showed that the [n-capture/Fe] ratios for metal-poor stars ([Fe/H]$<-2.5$) exhibit a large dispersion which can  not be attributed to observational errors, although the scatter in their [$\alpha$/Fe] and [Fe-peak/Fe] ratios as functions of [Fe/H] is very small \citep{Cayrel04}.
This striking difference between the behaviour, at low metallicity, of neutron capture elements on one side  and [$\alpha$/Fe] and [Fe-peak/Fe] ratios on the other hand,  can be explained if the production of neutron capture  elements by r-process is not homogeneous among the massive stars \citep{Cescutti08}, as the production of the other elements. Another possibility is that the sites of r-process production were diverse and rare, such as electron capture SNe (\citealt{Haynes19}, \citealt{Cescutti13}), magneto-rotationally driven SNe \citep{Cescutti14}, or neutron star mergers (\citealt{Cescutti15}, \citealt{Wehmeyer15}, \citealt{Argast04}).
Moreover, a dispersion is also present in the ratio [Sr/Ba] vs [Fe/H] at metallicity [Fe/H]$<-$2.5. This is usually attributed to the presence of a second neutron-capture process for the synthesis of the first-peak usually connected to a weak r-process \citep{Busso99,Qian00,Wanajo01} and referred to as light elements primary production (LEPP) in \cite{Travaglio04}. \cite{Montes07} showed that although weak r-process can play this role, the signature of an s-process from massive stars cannot be excluded.

The large spread in abundance ratios of n-capture elements is a consequence  of the strong dependency of the stellar yields of these elements on the stellar mass and on different production channels (e.g. massive stars or neutron-star mergers).
 Hence,  the study of the scatter among two different n-capture elements can bring important insights both on stellar nucleosynthesis and galaxy assembly (e.g. \citealt{Cescutti13}; \citealt{Cescutti14}; \citealt{Wehmeyer15}; \citealt{Wehmeyer19}; \citealt{Brauer20}; see also review by \citealt{Cowan21} and references therein). 
 For example, \cite{Limberg21} shows that the largest differences among accretion events are for n-capture elements and therefore these will be key to trace back the past accretion history of the Milky Way. 

 The fast observational progress in this area urgently calls for informative theoretical models which help the interpretation of chemo-kinematic datasets (with 6D phase space information plus chemistry and, sometimes, age - \citealt{Valentini19}; \citealt{Montalban21}; \citealt{Miglio21}).

The use of n-capture elements brings at least two important difficulties that have to be addressed before model predictions including these elements can become useful. The first is the uncertainties in the nucleosynthetic sites of production (in particular in the case of r-process elements, but also related to the early contribution of massive stars for the s-process - e.g. \citealt{Choplin18} and references therein). 
The second difficulty is related to the implementation of
the evolution of these elements in cosmological simulations,
and possible resolution/mixing effects that can affect
their predictions (e.g. \citealt{Shen15,VandeVoort15,Naiman18,Haynes19,VandeVoort20}).
Most of these previous work consider only the production
of elements by r-process events, either on-the-fly or in post-processing,
or analyse the evolution of Eu only.

In this work, we implement a detailed chemical enrichment model
which includes the production and distribution of Sr, Ba and Eu
in a cosmological simulation. In particular, we explore the dispersion
in light to heavy neutron-capture elements considering both
r-process events and the
contribution of massive stars with an s-process production
 \citep{Frischknecht16,Limongi18,Choplin18}.  As summarized in \cite{Chiappini13}, the impact of spinstars in the chemical enrichment of the earliest phases of our Galaxy is not confined to carbon and nitrogen \citep{Chiappini06,Chiappini08,Prantzos18}, but extends to s-process elements as well \citep{Pignatari08,Chiappini11}.
As a first step, in previous works we studied the scatter in n-capture abundance ratios using stochastic inhomogeneous chemical evolution models \citep{Cescutti13,Cescutti14,Cescutti15}. Such models, by construction, do not include accretion of stars formed in smaller galaxies as we know to be the case in the Milky Way. In this work we investigate the impact of accretion events and the cosmological growth of halos on these previous results, in the context of the $\Lambda$-Cold Dark Matter model.

This paper is organized as follows.  Sections~\ref{sec:simulations} and \ref{sec:isolated} describe our new model for chemical enrichment and discuss the effects of differential enrichment on the predictions for the abundance ratios. Section~\ref{sec:halo} analyses the stellar halo formed in a cosmological simulation of a Milky Way-mass galaxy, focusing on  the abundances  and scatter of neutron-capture and $\alpha-$elements, and in Section~\ref{conclu} we present our conclusions.

\section{The Simulations}\label{sec:simulations}

\subsection{Numerical Implementation}

We employ for this study
the cosmological,  hydrodynamical, Smoothed Particle
Hydrodynamics (SPH) code {\sc gadget3} \citep{Springel08},
with the  additional modules of \cite{S05,S06}  for describing
 star formation, chemical enrichment, supernova feedback,
and metal-dependent cooling. 
For this project, we have modified our standard code, to which
we refer to as the CS model, in order to properly describe
the early enrichment phases of the interstellar medium, therefore
allowing for a detailed study of the chemical abundances of the
very old stars in galaxies. 
In the following  subsections, we  describe the CS model in
terms of the physical modules relevant for this study, 
the updates we have implemented to the chemical routines, the 
chemical species and yields that we adopt, and
the setup of the simulations presented in this work.

\subsubsection{Star formation, feedback and chemical enrichment in the CS model}

In the CS model, star particles form according to the Kennicutt-Schmidt
law \citep{Kennicutt98}, in a stochastic manner \citep{Springel03}.
Gas particles are eligible for star formation if they are denser than a critical  value ($\rho_{\rm c}$) and are in a convergent flow.
For these particles, the star formation rate (SFR) per unit volume is
\begin{equation}
\rho_{\rm SFR} =  c_* {{\rho}\over{\tau_{\rm dyn}}}
\end{equation}
where $c_*$ is a star formation efficiency, $\rho$ is the gas  density and $\tau_{\rm dyn}= (4\pi G \rho)^{-1/2}$ the dynamical time of the particle.
Once formed,
star particles are treated as single stellar populations
(SSP) with a given Initial Mass Function (IMF), and 
return metals and energy to their surroundings during supernova (SN) 
explosions. As the interstellar medium (ISM) gets polluted, new stars
are produced with higher metallicities, because star particles  inherit
the element abundances of the gas from which they form.  We assume that
each gas particle can produce a maximum of two stars.

Our  model includes treatments for
chemical enrichment and energy feedback originated in Type II
(SNII) and Type Ia (SNIa) events, based on assumptions for their
rates, chemical yields and explosion times \citep{S05}, and the effects of
stars in the AGB phase \citep{Poulhazan18}.
 When a star experiences a SNII/SNIa/AGB event, the associated
chemical production is distributed into
its $N_{\rm Ngb}$ gas neighbours, in proportions given by the SPH kernel.

The CS model is based on a multiphase gas treatment which
allows coexistence of dense and diffuse phases in the same spatial region;
in this context, exploding stars will have well-defined hot and cold neighbours. 
We assume that both types of SNe eject the same amount of energy 
$E_{\rm SN}$ to the ISM,  which 
is distributed in equal proportions
to the local cold and hot gas phases of the exploding stars.
Energy feedback into hot neighbors occurs at the time of explosion;
however, for cold neighbors energy from successive explosions is
instead accumulated and deposited only after a time-delay which
depends on the local conditions of the cold and hot gas phases,
as described in \cite{S06}.  In
this way, artificial loss of SN energy in high-density regions is
prevented. 
Gas cooling is described using the metal-dependent tables of \cite{SD93}.

The CS model has been shown to be successful in producing
galaxies similar to those observed, both in the case of Milky Way (MW) mass galaxies \citep{S08,S09,S10,S11,Tissera13,Nuza14,Tissera14} and in dwarf systems
\citep{Sawala10, Sawala11, Sawala12}. In particular, our SN feedback
model is able to reproduce the formation of disks from cosmological
initial conditions \citep{S08, S09}, and to produce
galaxies that are in general terms consistent with observations
of the galaxy population \citep{S12,Guidi16}.

\subsubsection{Chemical species and  yields of SNII}

 We consider in this work the nucleosynthesis of massive stars which end up
their lives as Type II SNe, and that constitute the main drivers of the pollution of the ISM in the early phases of chemical evolution.
In particular,
 we consider the nucleosynthesis of 
the following chemical
species: H, He, C, N, O, Fe, Mg, Si, Ba, Eu, Sr and Y. 
The corresponding yields are parameterized
in terms of $5$ metallicity ranges, separated by 
$Z = 0$, $0.0001$ Z$_\odot$, $0.01$ Z$_\odot$, $0.1$ Z$_\odot$ and Z$_\odot$.

 For He, C, N, O, Mg,   Si and Fe, we have adopted the same nucleosynthesis 
 than the ones used in \cite{Chiappini06, Chiappini08} and \cite{Cescutti10}.  
   In these works, the   nucleosynthesis of He, C, N and O is metal-dependent and 
based on the work of the
Geneve group \citep{Meynet02, Hirschi07}.
The novelty of this nucleosynthesis dataset is that 
 the impact of rotation in massive stars at low metallicity is considered, 
and the rotation 
changes the final enrichment of the light elements, in particular of nitrogen. 
 Unfortunately, the yields of the Geneve group could not be implemented for Mg, Si and Fe; in fact, \cite{Meynet02} and \cite{Hirschi07} do not treat explosive nucleosynthesis, which has an important impact for these elements.
 Therefore, in this work, as well as in \cite{Chiappini06,Chiappini08} and \cite{Cescutti10}, we use the solar metallicity yields from the work of
  \cite{WW95},  slightly modified (see \citealt{Francois04}) to best match the averaged
trend of extremely metal-poor stars measured by \cite{Cayrel04}.
 For Fe,  the prescriptions are similar,
with the exception of the zero metallicity table, where 
the population III stars with masses above $20$M$_\odot$ inject negligible
amounts of iron into the ISM upon their death \citep{Cescutti10}.
 Since Mg, Si and Fe are primary elements, the use of the solar metallicity
table is a safe assumption.

For the nucleosynthesis of the neutron capture elements,
 we assume an r-process contribution 
as the one adopted in \cite{Cescutti13}:
 a strong production in a narrow mass range of $8-10$M$_\odot$
 that we call standard r-process site.  
These yields have been chosen to reproduce the mean trend 
of [Ba/Fe] versus [Fe/H] using the homogeneous chemical evolution model of \cite{Chiappini06}. 
For the remaining neutron
 capture elements, we scale the r-process contribution using the ratios 
observed in r-process rich stars \citep{Sneden08}.
We also adopt an s-process contribution coming from rotating massive stars, 
using the yields computed in \cite{Frisch16}.
Among the set of models, we have adopted 
a robust production of s-process elements, achieved by considering s-process yields of massive stars for
 $v_{\rm ini}/v_{\rm crit} = 0.5$ (fast rotators) and for a reaction rate
$^{17}$O$(\alpha,\gamma)^{21}$Ne one tenth of \cite{Caughlan88};
current uncertainties
on this rate are still large.

The stellar yields adopted here are essentially the same as the ones 
adopted in our non-cosmological stochastic model presented in \cite{Cescutti14}. 
Using similar prescriptions,  now applied to cosmological simulations, allows 
to test the impact of complex merger histories on the interpretation of the 
observed abundance scatter of different chemical elements, 
which is one of the main goals of the present work.

Throughout this paper, we have assumed the solar abundances 
 given by \cite{Grevesse10}.

\subsubsection{Implementing multiple explosion times for SNII}\label{sec:multiple}

In the  CS model, each star particle explodes as SNII  in a single event, ejecting the whole chemical production of the SSP it represents at  the time of the explosion. This time is determined by the mean age of the individual stars of the SSP.
This assumption  allows to describe the  overall evolution of chemical abundances in galaxies; however, it does not properly  describe the first stages of enrichment
of the ISM, which will be enriched gradually, starting with the most massive, most short-lived stars. Such a gradual enrichment will leave imprints in the chemical properties of the very old, metal-poor stars, as stars of different mass will contribute chemical material of different nature (e.g. end-products of r- vs s-processes vs $\alpha$-elements) at different times. It will also significantly affect the spread in chemical abundances, which we investigate in this work.

In order to better describe the enrichment of the ISM occurring
in short time-scales, we have modified our code
such that each SNII event is represented by
a series 
of different ``explosions'', which we chose to best describe
the enrichment of the different chemical elements we are interested
in. 
In particular, we assume that each star particle will
experience 5 different explosion events, which
occur at 30, 18, 9, 6 and 3 Myr from its formation. 
These correspond to stars in the mass ranges [8-9), [9,14), [14, 27.5), [27.5, 45) and [45-100] M$_\odot$,  respectively, according to the estimations of \cite{Maeder89}\footnote{Note that although small differences in the lifetimes estimations are expected for different stellar models (e.g. the Geneva ones), these will not affect our results  as the differences are smaller than the grid of explosion times considered here.}.
 The choice of  $5$  explosion events per star particle
allows to properly describe the differential enrichment produced
by the individual stars, in terms of the chemical species and yields used in this work,
avoiding large numerical overcosts.

\subsubsection{Simulation set-up}

We have tested our code in idealized simulations of the evolution
of an isolated galaxy, as well as in simulations in a cosmological context.
In order to facilitate comparison, we have
kept the same input parameters for star formation, chemical enrichment
and feedback in all runs, and performed
 simulations with the CS standard code and with the new
implementation.
We have assumed a star formation efficiency of $c_*=0.1$,
a star formation threshold of $\rho_{\rm c} = 0.1$ cm$^{-3}$, and
$E_{\rm SN} = 0.7\times 10^{51}$ erg of energy per SN. For the number
of neighbours assumed for the SPH calculations, we used $N_{\rm Ngb}=40$.
 These  choices for the input parameters produce
galaxies with realistic disk sizes \citep{S08}.
In order to facilitate comparison with the results of
the  galactic chemical
evolution  models of \cite{Cescutti13}, 
 we have assumed
a Scalo IMF.

All simulations presented in this paper have been run with the
modules for SNIa and AGB stars switched off. 
The reason for this choice is to properly identify and isolate the
effects of the fast-rotating stars and of considering differential
enrichment on the chemical abundance ratios and scatter of $\alpha-$ and n-capture-elements.
Note that we focus our analysis on the stellar halo
component, which is formed by very old stars and therefore whose
abundances are
almost exclusively determined by the very early enrichment produced
by SNII explosions.
In future papers, we will explore the impact of SNIa and 
AGBs in the  chemical enrichment histories, but 
this is beyond the scope of the present work.

\section{The effects of differential SNII enrichment}\label{sec:isolated}

In this Section we discuss the effects of implementing progressive enrichment via
SNII on the chemical abundances and scatter of the gas and stellar components of
galaxies. In particular, we compare simulations where the individual
stars that are represented in a stellar particle explode simultaneously or
progressively, better describing the first enrichment epochs. We refer to these
two  cases, correspondingly, as the {\it single
explosion} (SE) or {\it multiple explosion}  (ME) models.
In SE we assume  that each star particle formed
in the simulation 
ejects all its chemical production in a single event (i.e. as in the standard CS model). In contrast, model ME assumes
a progressive enrichment, in which the individual stars explode at different times, according to their masses and typical time-scales, as explained in Section~\ref{sec:multiple}.

In the following we compare the predictions of two simulations of the formation
of a Milky Way-mass galaxy in an idealized, isolated scenario, that assume the SE and ME models. 
We refer to them as the ``SE64'' and ``ME64'' runs, respectively\footnote{The number ``64'' encodes information on the number of particles of the simulation, see Appendix~\ref{sec:app_resolution}.}. 
 Table~\ref{table:simulations} summarizes the main properties of these simulations, as well as those of additional simulations that we run in order to test
 possible dependencies of our results on numerical
choices. These include simulations varying the number of
particles (discussed in Appendix~\ref{sec:app_resolution}), as well as the number of SPH neighbours ($N_{\rm Ngb}$, 
 which sets the smoothing region, i.e. the radius over
which the chemical elements ejected in SN explosions are distributed (discussed in Appendix~\ref{sec:app_ngb}).

\begin{table*}
\caption{Main characteristics of the simulations used in this work: reference name, type of initial conditions, number of SNII explosion events assumed ($N_{\rm SNII}$), total number of particles of the simulation ($N_{\rm tot}$), assumed number of SPH neighbours ($N_{\rm Ngb}$),
 reference Section, and  masses of the dark matter ($m_{\rm DM}$) and baryonic ($m_{\rm bar}$) particles. }

\begin{center}
\begin{tabular}{lccccccc}
\hline
\hline
 Name &  Initial Conditions & $N_{\rm SNII}$ & $N_{\rm tot}^{1}$ & $N_{\rm Ngb}$ 
& Section& $m_{\rm DM}$ [M$_\odot$]& $m_{\rm bar}$ [M$_\odot$]\\ 
\\\hline
SE64       & Isolated       &  1     & $2\times 64^3$ & 40 & \ref{sec:isolated} & $6.55\times 10^6$ & $7.28\times 10^5$\\
ME64      & Isolated       & 5       & $2\times 64^3$ & 40  & \ref{sec:isolated}& $6.55\times 10^6$ & $7.28\times 10^5$\\

SE64-N$_{\rm Ngb64}$& Isolated  &  1     & $2\times 64^3$  & 64 & \ref{sec:app_ngb}& $6.55\times 10^6$ & $7.28\times 10^5$\\
ME64-N$_{\rm Ngb64}$& Isolated & 5       & $2\times 64^3$ & 64  & \ref{sec:app_ngb}&$6.55\times 10^6$ & $7.28\times 10^5$\\

SE64-N$_{\rm Ngb128}$& Isolated  &  1     & $2\times 64^3$ & 128  & \ref{sec:app_ngb}&$6.55\times 10^6$ & $7.28\times 10^5$\\
ME64-N$_{\rm Ngb128}$& Isolated & 5       & $2\times 64^3$ & 128  & \ref{sec:app_ngb}&$6.55\times 10^6$ & $7.28\times 10^5$\\
 
\hline
SE32       & Isolated       &  1     & $2\times 32^3$   & 40  & \ref{sec:app_resolution}& $5.22\times 10^7$ & $5.80\times 10^6$\\
ME32      & Isolated       & 5       & $2\times 32^3$  & 40  & \ref{sec:app_resolution}& $5.22\times 10^7$ & $5.80\times 10^6$\\

SE32-N$_{\rm Ngb64}$& Isolated  &  1     & $2\times 32^3$  & 64 & \ref{sec:app_ngb}& $5.22\times 10^7$ & $5.80\times 10^6$\\
ME32-N$_{\rm Ngb64}$& Isolated & 5       & $2\times 32^3$  & 64 & \ref{sec:app_ngb}& $5.22\times 10^7$ & $5.80\times 10^6$\\

SE32-N$_{\rm Ngb128}$& Isolated  &  1     & $2\times 32^3$  & 128 &  \ref{sec:app_ngb}& $5.22\times 10^7$ & $5.80\times 10^6$\\
ME32-N$_{\rm Ngb128}$& Isolated & 5       & $2\times 32^3$  & 128 &\ref{sec:app_ngb}& $5.22\times 10^7$ & $5.80\times 10^6$\\

\hline
SE128       & Isolated       &  1     & $2\times 128^3$  & 40  & \ref{sec:app_resolution}& $8.19\times 10^5$ & $9.10\times 10^4$\\
ME128      & Isolated       & 5       & $2\times 128^3$ & 40 & \ref{sec:app_resolution}& $8.19\times 10^5$ & $9.10\times 10^4$\\

SE128-N$_{\rm Ngb64}$& Isolated  &  1     & $2\times 128^3$ & 64 & \ref{sec:app_ngb}& $8.19\times 10^5$ & $9.10\times 10^4$\\
ME128-N$_{\rm Ngb64}$& Isolated & 5       & $2\times 128^3$ & 64 & \ref{sec:app_ngb}& $8.19\times 10^5$ & $9.10\times 10^4$\\

SE128-N$_{\rm Ngb128}$& Isolated  &  1     & $2\times 128^3$ & 128 &  \ref{sec:app_ngb}& $8.19\times 10^5$ & $9.10\times 10^4$\\
ME128-N$_{\rm Ngb128}$& Isolated & 5       & $2\times 128^3$ & 128  &\ref{sec:app_ngb}& $8.19\times 10^5$ & $9.10\times 10^4$\\

\hline

SE       & Cosmological      & 1 &  1854223  & 40  & \ref{app:cosmo}& $2.16 \times 10^6$ &  $4.11 \times 10^5$\\
ME       & Cosmological      & 5 &  1608991 & 40  & \ref{app:cosmo}& $2.16 \times 10^6$ &  $4.11 \times 10^5$\\

\hline
\end{tabular}
\end{center}
$^1$ In the case of the cosmological simulations $N_{\rm tot}$ refers to the total number of particles within the virial radius of the galaxy at the end of the simulation ($z=0$).
\label{table:simulations}
\end{table*}

The initial conditions 
are generated by radially perturbing a spherical grid of superposed
dark matter and gas particles to produce a cloud with density profile
$\rho(r)\sim r^{-1}$,  as in \cite{Navarro93}.
The sphere is
initially in solid body rotation with an angular momentum characterized by
spin parameter $\lambda \simeq 0.1$, and the gas is cold (i.e. 
the initial thermal energy is 
only $5$ per cent of its binding energy).
The simulated system has a total mass of
$10^{12}\,{\rm M}_\odot $, 10 per cent of which is in the
form of baryons, and the initial radius is $100\,{\rm kpc}$. 
Our fiducidal tests used $N_{\rm tot} = 2 \times 64^3$ particles initially\footnote{Note that the total number of particles changes with time, as each gas particle can produce a maximum of two stars.}, yielding
particle masses of $6.55\times 10^6$ M$_\odot$ and
$7.28\times 10^5$ M$_\odot$ for dark matter and gas (Table~\ref{table:simulations}), respectively,
and we adopted a
gravitational softening length of $350$ pc for all particles.
The simulations of this Section were run for 2 Gyr,  although we focus our analysis on the early phases of the formation of the galaxies, before the first Gyr of evolution. 

We note that the idealized initial conditions of this
Section yield a  simple model for disc
formation, which is an ideal test bench for the performance and
validity of the code. However, these models are not meant to
provide a realistic scenario for the whole galaxy formation process,
in particular because there is no gas infall which, in a cosmological
context, has important effects on the growth and evolution of a galaxy.
 As explained above, we have also tested our model on cosmological simulations (see Appendix~\ref{app:cosmo}).

\subsection{Effects on the chemical enrichment of the ISM}

Our implementation of multiple
events per SN explosion is expected to affect not only the chemical
properties of the galaxies 
 but also
the star formation process, as the rate at which the gas
is enriched will impact the cooling rates, 
affecting the amount of cold/dense gas available for star formation
at different times.
This will produce an initial  change in the star formation rate due
to variations in the enrichment level, and further
changes due to the variations in the amount of available 
feedback\footnote{We note that these effects might
be enhanced/reduced in the simulations analysed in this Section
due to the idealized nature of the ICs (see Section~\ref{app:cosmo} for
simulations in a cosmological context). }.

\begin{figure*}
  \centering
  \includegraphics[width=5.5cm]{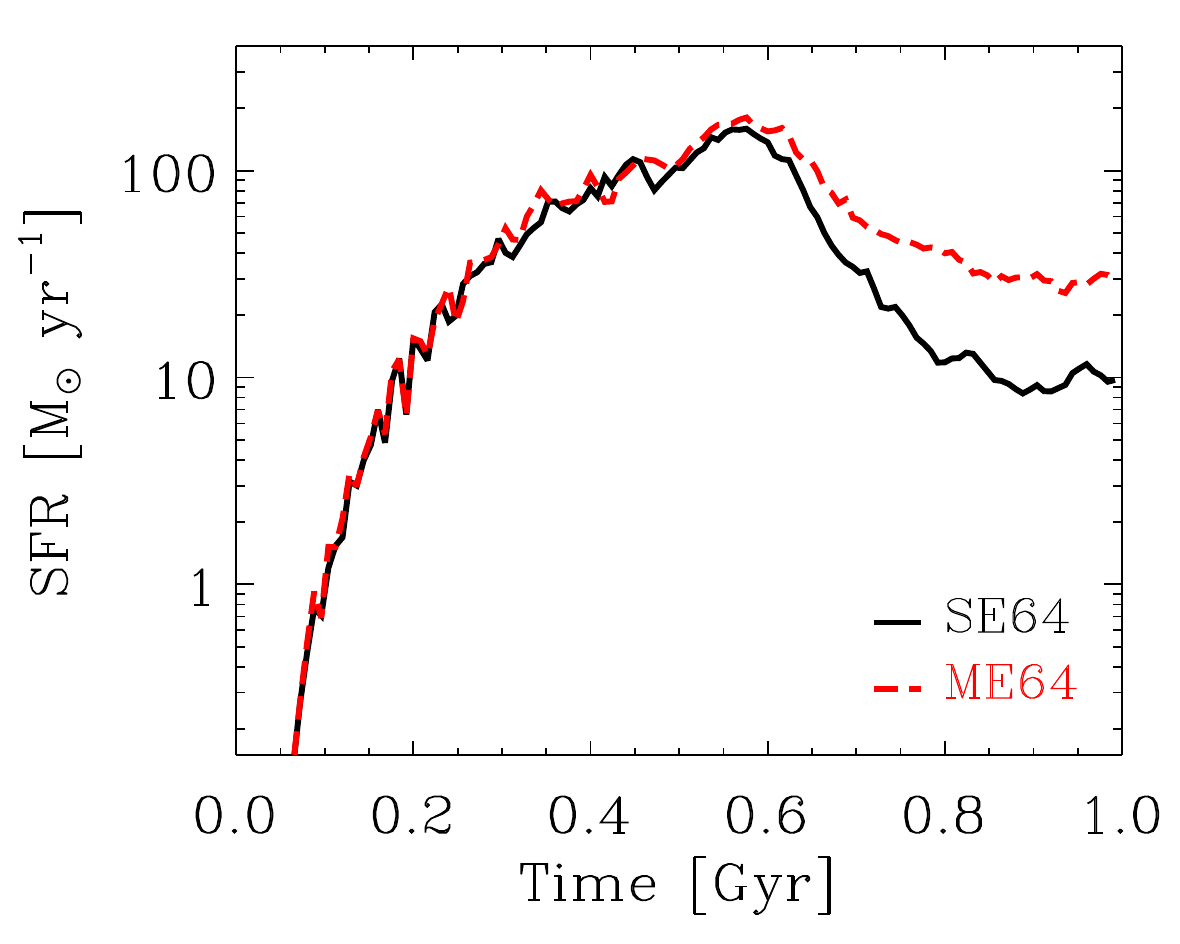}\includegraphics[width=5.5cm]{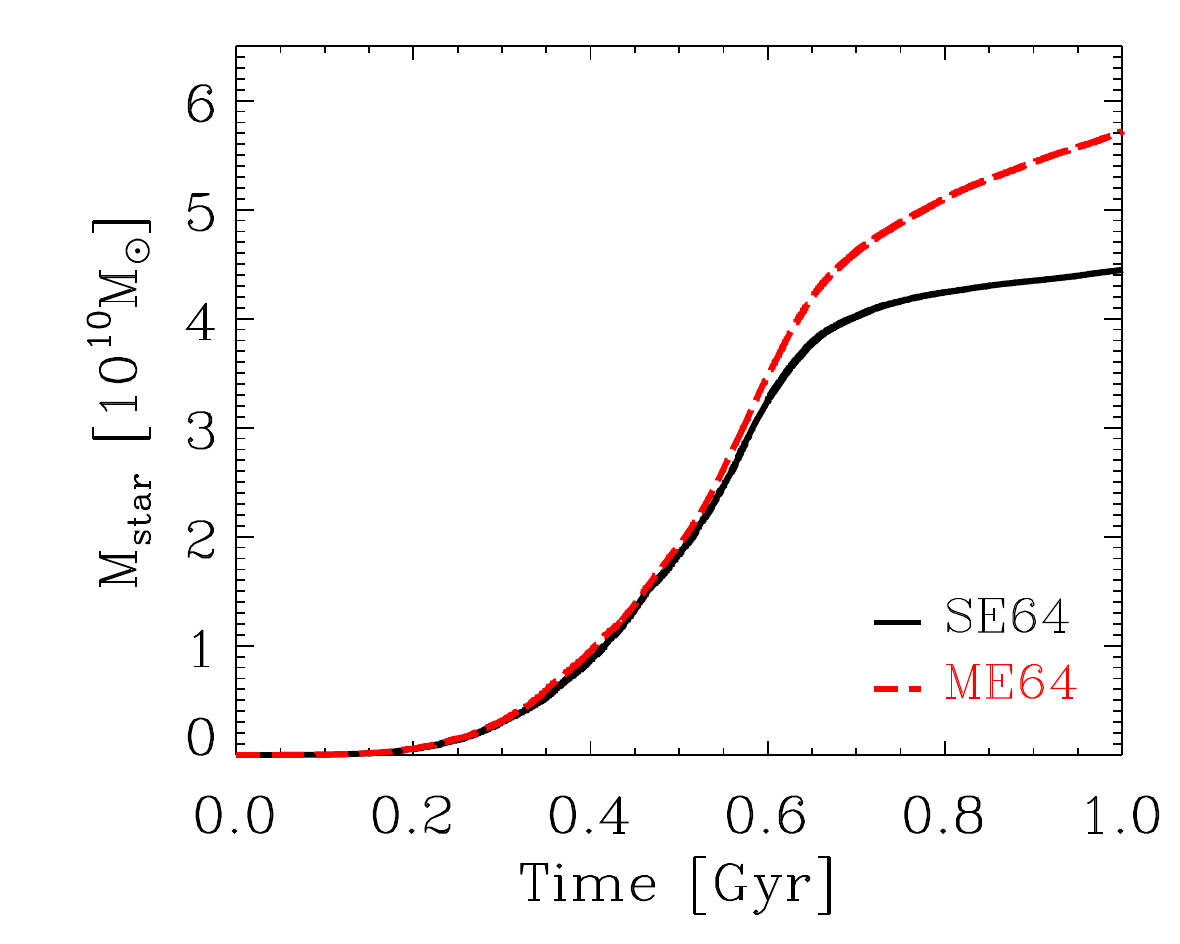}\includegraphics[width=5.5cm]{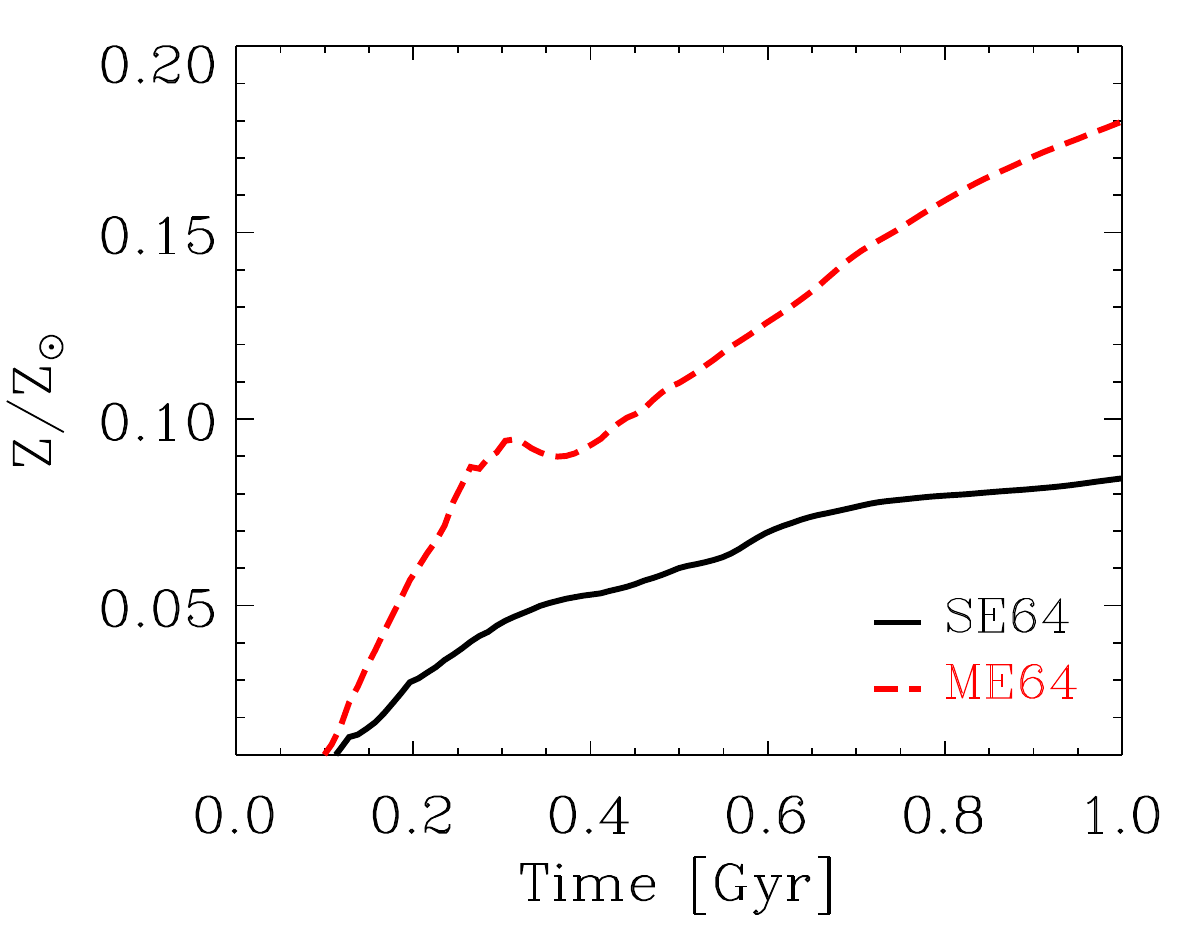}
\caption{Star formation rate, cumulative stellar mass and  stellar metallicity, as a function of time, for
our  simulations SE64 and ME64.}
\label{fig:SFR_isolated}
\end{figure*}

Fig.~\ref{fig:SFR_isolated} shows
the star formation rate (SFR), and the evolution of the
stellar mass ($M_{\rm star}$) and stellar metallicity (Z,  defined as the
 mass in metals -- all chemical elements except hydrogen and helium -- divided by the
 total mass)
in simulations SE64 and ME64. 
The two simulations have similar SFRs and integrated stellar masses during 
the first $\sim 0.5$ Gyr; however, the metallicity of the stellar population
is clearly distinct from very early times. This difference originates
in the variations of the enrichment of the ISM in the two runs: 
in ME64 the first explosion events associated to the most massive short-lived
stars occur very quickly after star formation 
(i.e. $3$ Myr, Section~\ref{sec:multiple}) 
triggering a very fast enrichment of the surrounding gas. In contrast, in SE64 
the single explosion associated to all SNII within a star particle
occurs after $\sim 15-20$ Myr from its formation\footnote{Note that the time of the explosion
of star particles associated to SNII in the CS model is the mean age of the individual stars within
a star particle that are progenitors of SNII (Section~\ref{sec:multiple}). 
The explosion time can change from particle
to particle, as it depends on the choice of the IMF and the metallicity of the star particle.}.

\begin{figure*}
  \centering
  \includegraphics[width=6cm]{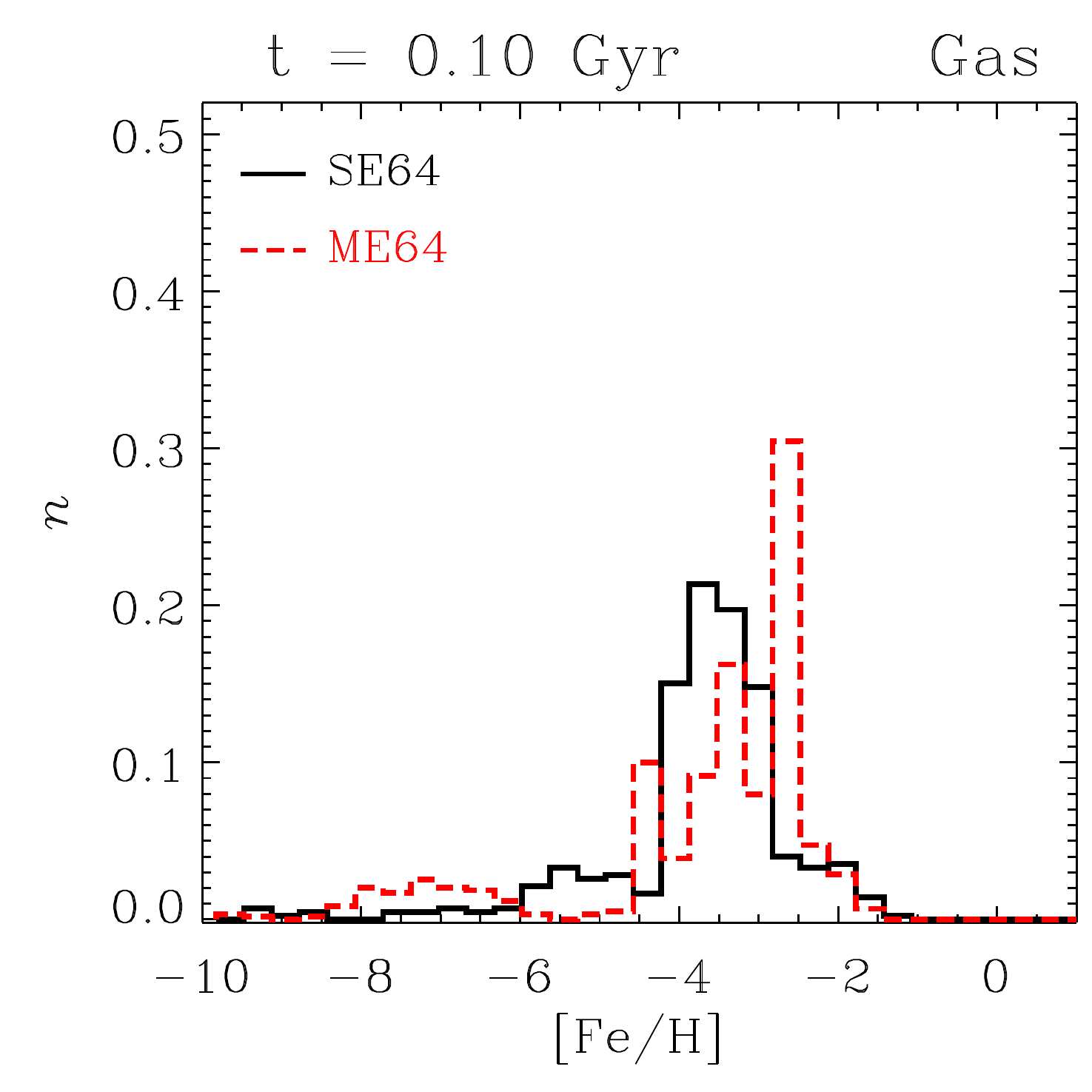}\includegraphics[width=6cm]{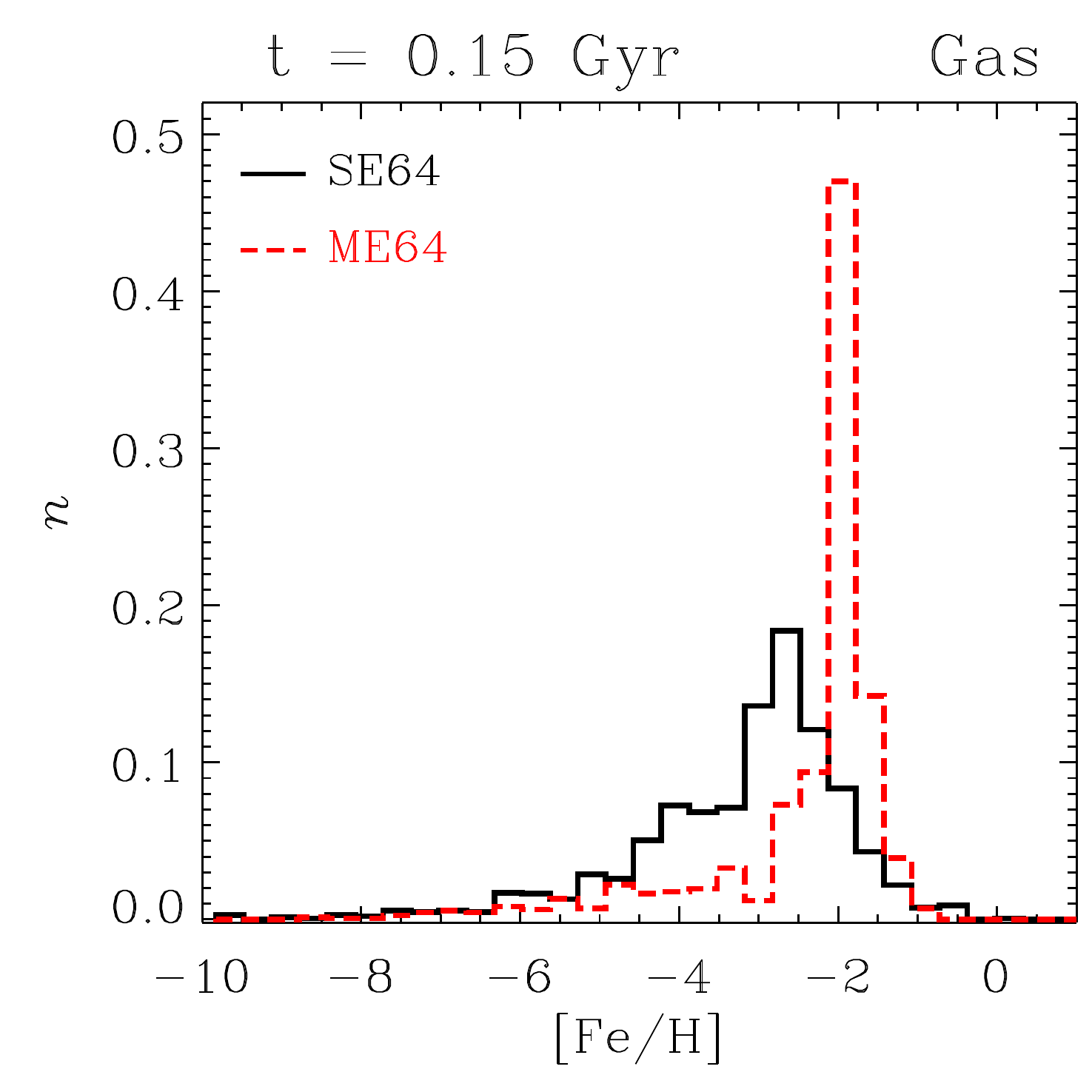}\includegraphics[width=6cm]{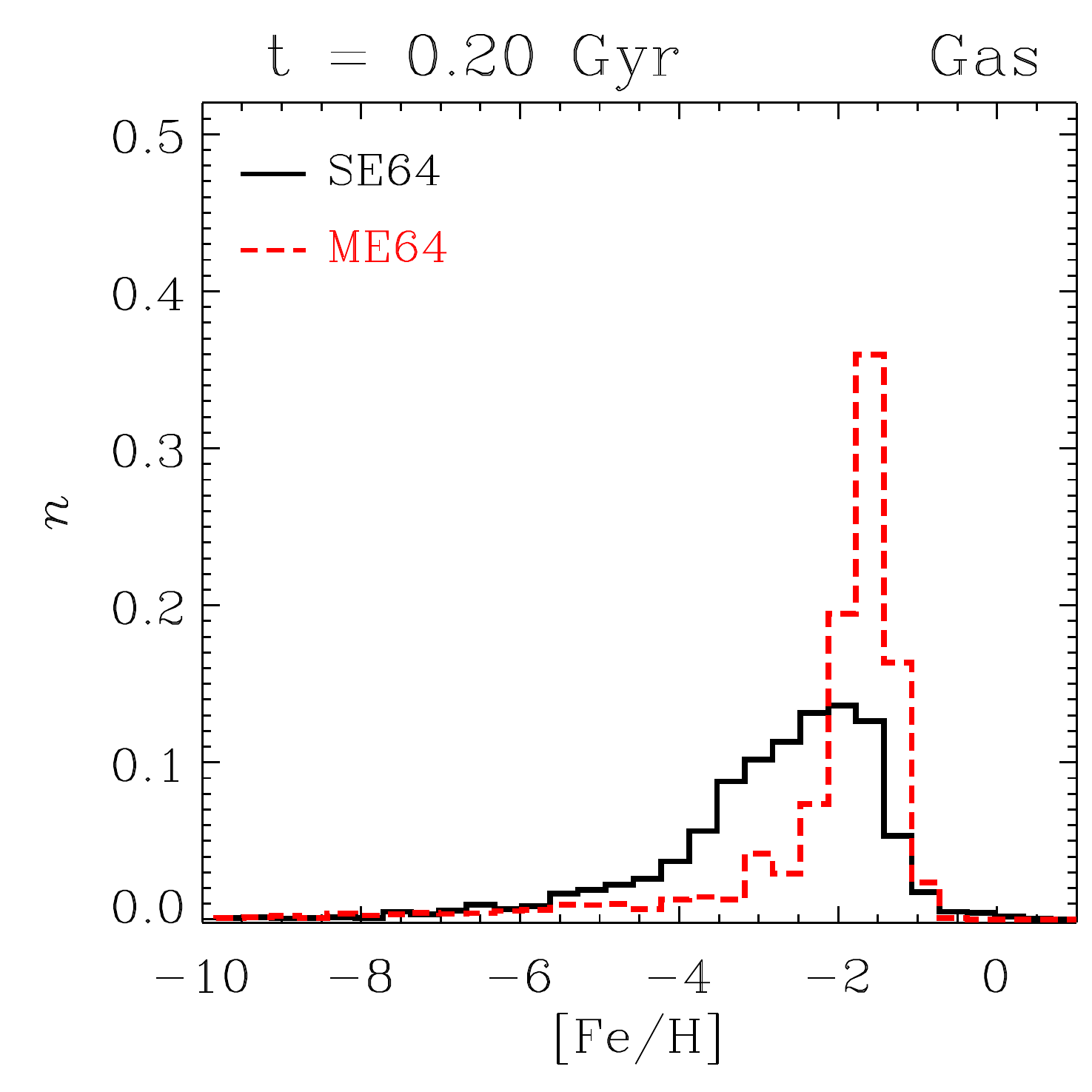}

  \vspace{0.8cm}

  \includegraphics[width=6cm]{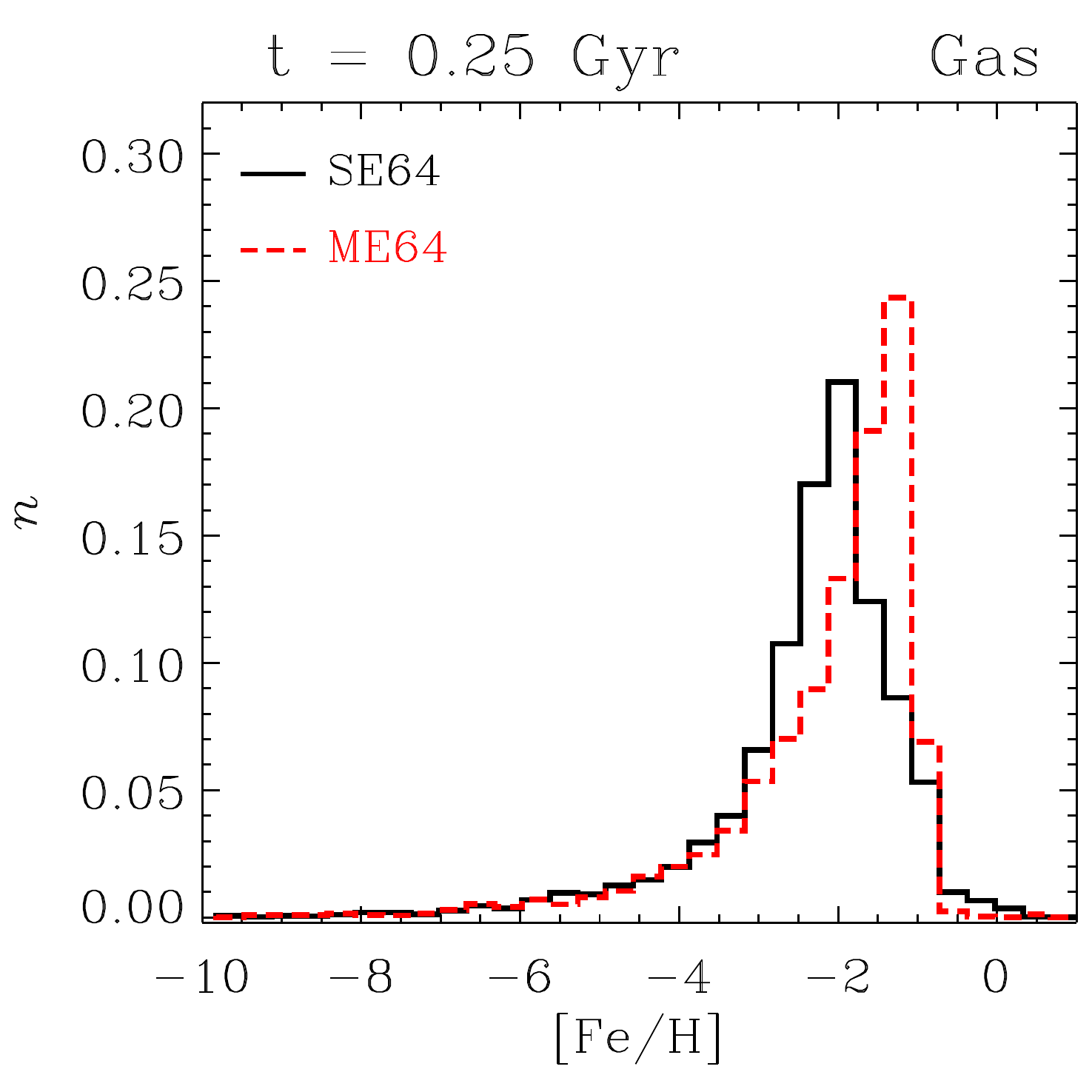}\includegraphics[width=6cm]{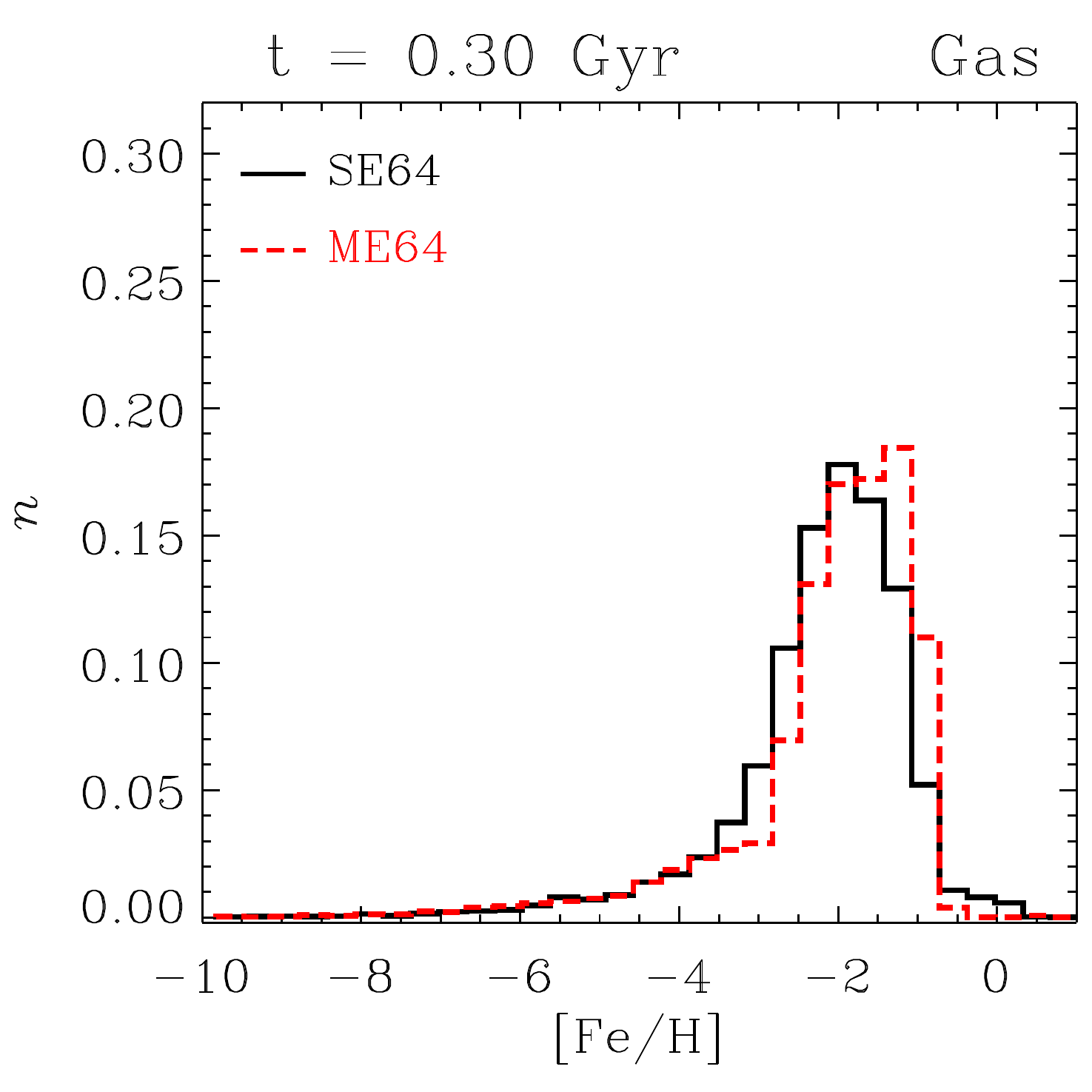}\includegraphics[width=6cm]{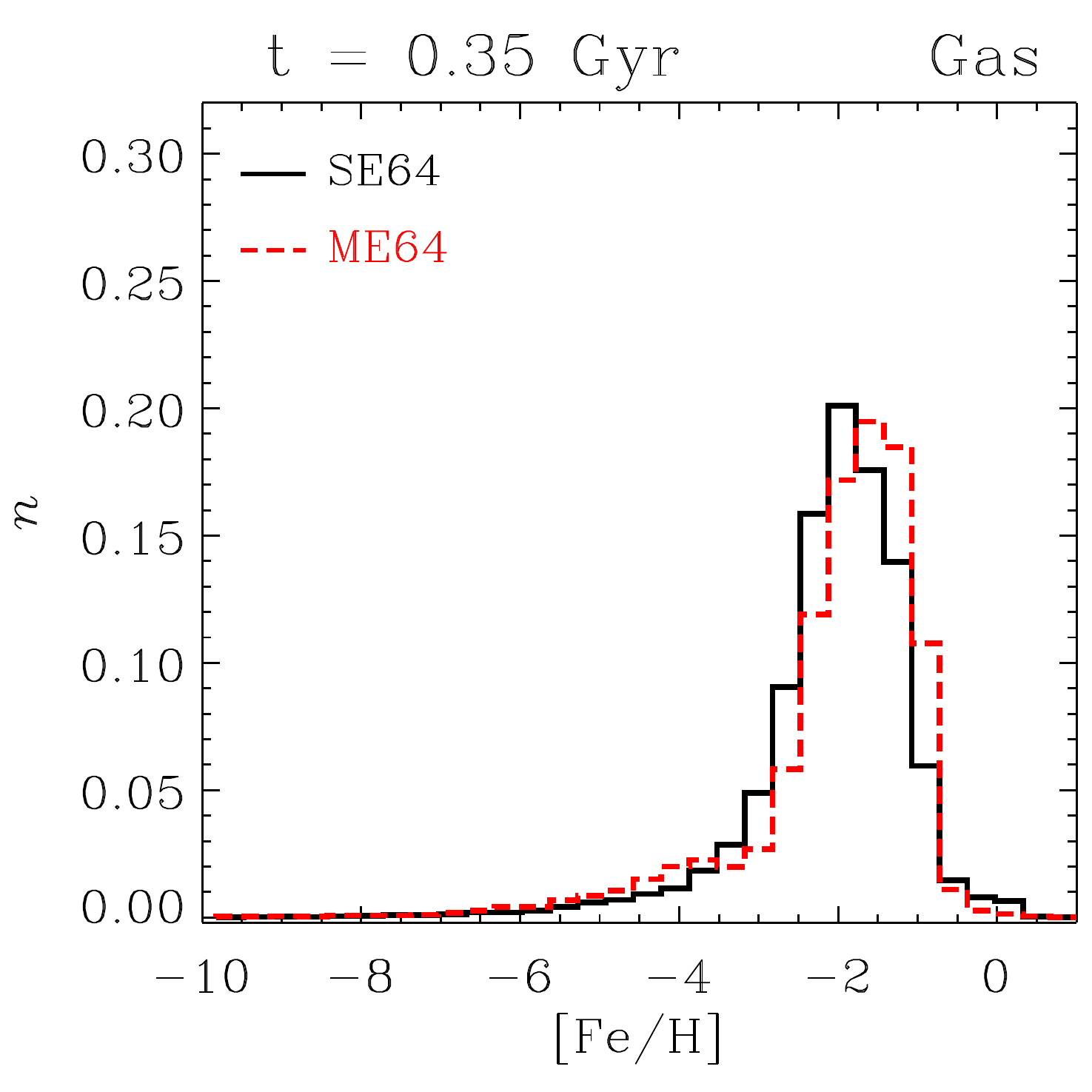}
\caption{Distribution of [Fe/H] for the gas in simulations SE64 and ME64, for various times during the
early enrichment phases.}
\label{fig:Z_gas_isolated}
\end{figure*}

The differences in the early chemical pollution of the ISM in our two simulations
can be seen in 
Fig.~\ref{fig:Z_gas_isolated}, where we compare the gas metallicities,  
in terms of the [Fe/H] abundance, 
for simulations SE64 and ME64, and for various times between $0.1$ and $0.35$ Gyr.
Even though star formation starts at the same time in both simulations, in ME64
the enrichment of the ISM is faster compared to SE64, particularly during the first
200 Myr after star formation begins. At later times, the differences 
dilute, as both the $5$ explosion events in ME64 and the single event in SE64 have
occurred, and the accumulated chemical production has been released.  As a result, the predicted number of very metal-poor stars formed within the first 200 Myr when considering the multi-explosion scenario is systematically smaller in ME64, compared to those obtained when assuming that all core-collapse SNe contributed at a single time.

The differences in the very early epochs, {\it can thus} leave clear imprints in the
galaxies, both in the enrichment levels of the stellar and gaseous components (higher
stellar metallicities in  ME64 compared to SE64) and in the cooling rates (enhanced cooling in ME64 due to the faster enrichment).
As a result, the star formation rates in ME64 are higher compared to SE64 and, 
 after 1 Gyr of evolution, 
 the stellar mass is  higher ($\sim 30 \%$) in ME compared to SE64, and
 the  stellar metallicity in ME64 is about 2 times higher (Fig.~\ref{fig:SFR_isolated}).

\begin{figure*}
  \centering
\includegraphics[width=6cm]{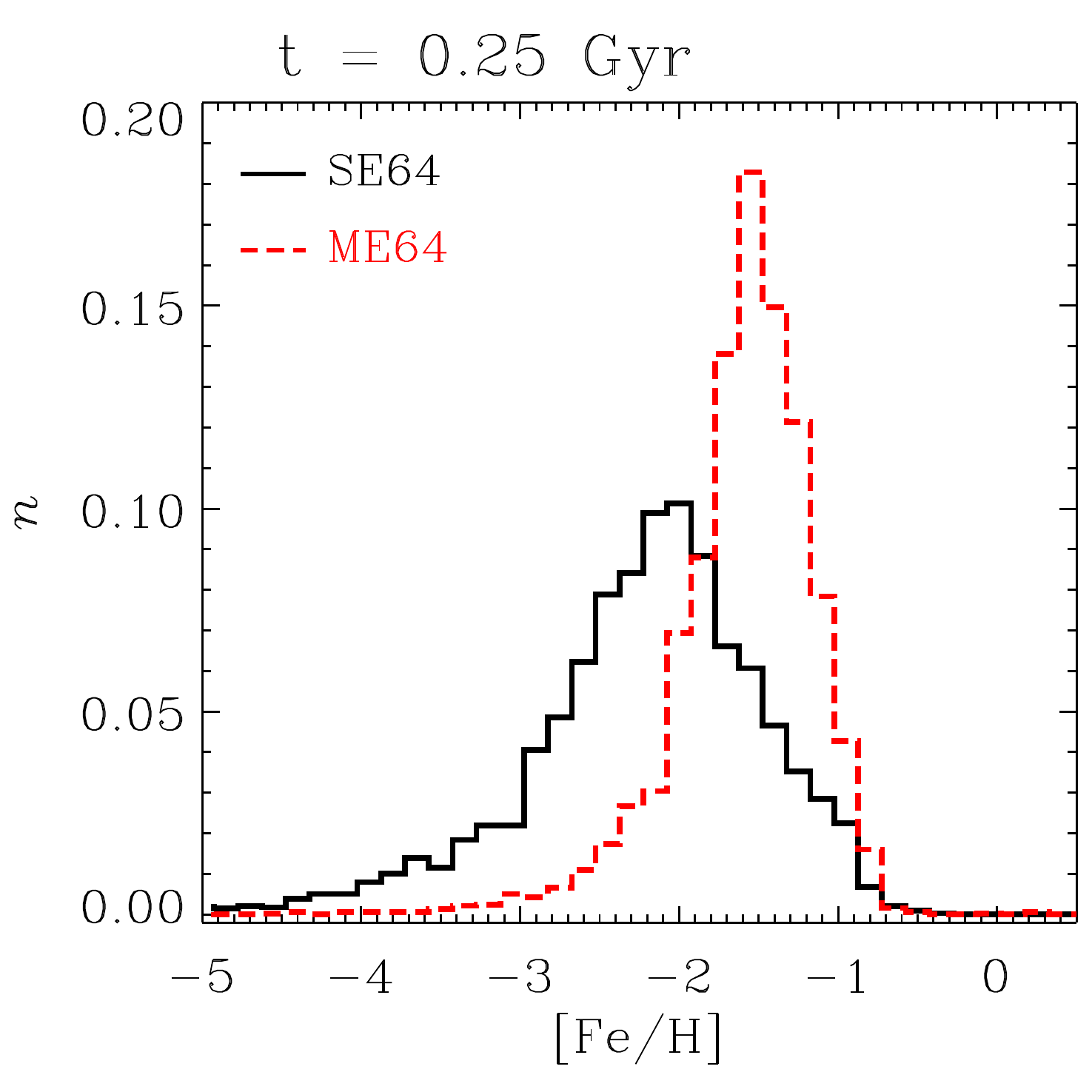}\includegraphics[width=6cm]{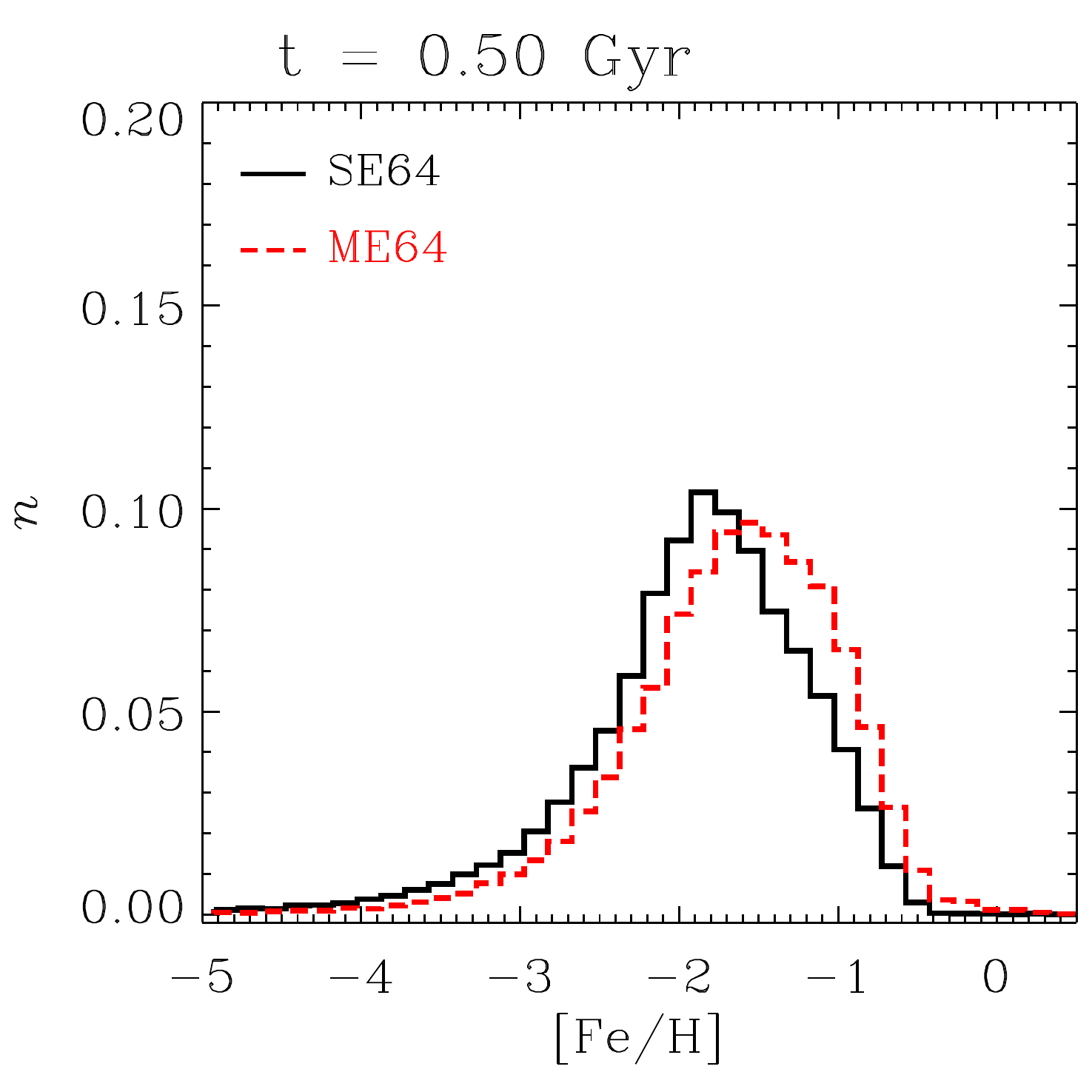}\includegraphics[width=6cm]{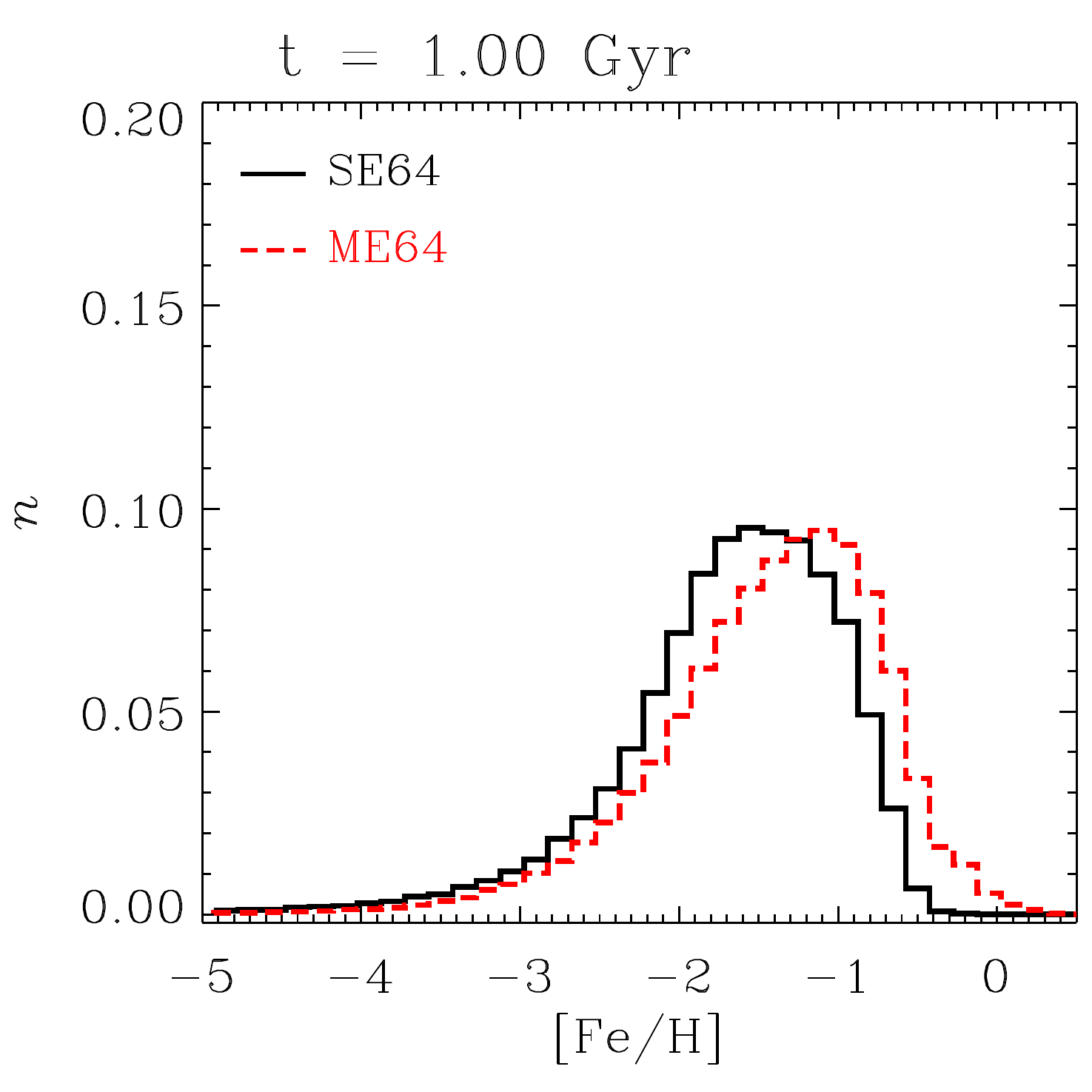}
\caption{Distribution of the stellar [Fe/H] for simulations SE64 and M64
at different times of the simulation:  0.25 Gyr, 0.5 Gyr and 1 Gyr.  
}
\label{fig:feh_hist_isolated}
\end{figure*}

Fig.~\ref{fig:feh_hist_isolated} shows the distribution functions of stellar [Fe/H] 
abundances for simulations SE64 and ME64, after
$0.25$, $0.5$ and $1$ Gyr of evolution. From these plots we can observe that, following
the characteristics of the ISM enrichment, the differences  in the [Fe/H] distributions in models SE64 and ME64 are more important at early times, both for low and high metallicity. The distributions are less dissimilar at later times; however, the differences 
originated early on on the stellar abundances can still be seen after 1 Gyr of evolution.

As explained above, imprints of the differences in the early chemical
pollution of the ISM in runs SE64 and ME64 appear in the chemical
properties of the stars. This is important given that, if properly interpreted,
 observations of the
stellar population of a galaxy could allow to reconstruct the level of enrichment of the
ISM at different times,  information that is otherwise inaccessible.

\begin{figure*}
  \centering

  \includegraphics[height=6cm]{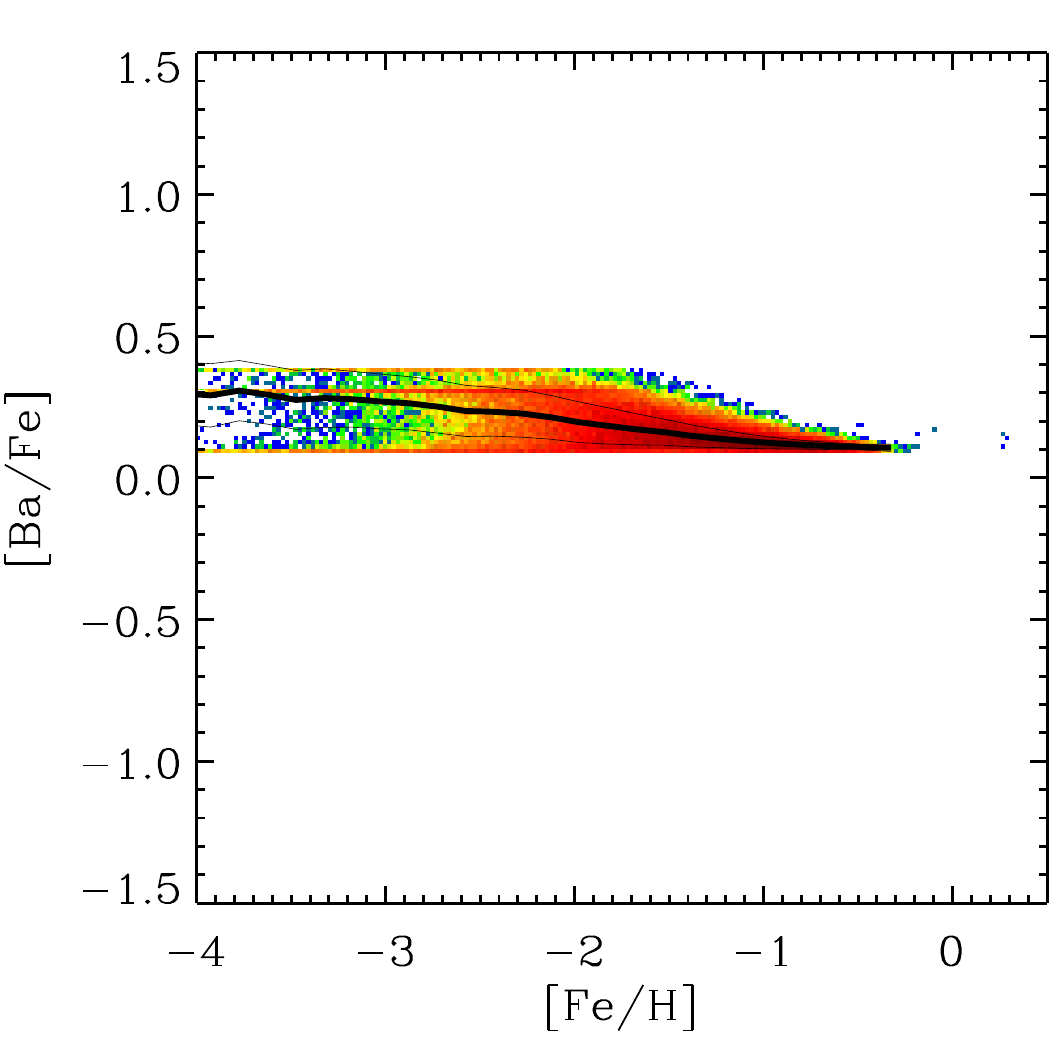}\hspace{1cm}\includegraphics[height=6cm]{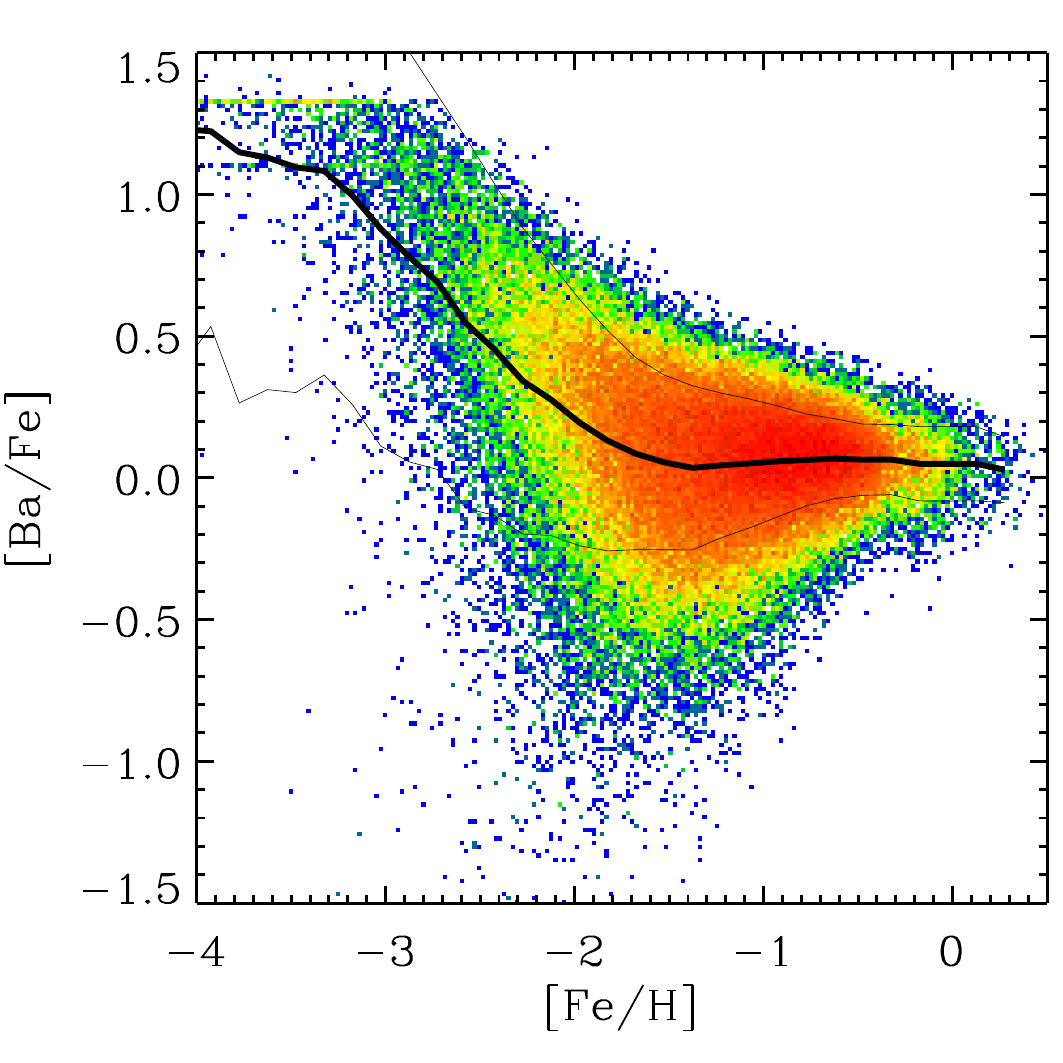}\hspace{1cm}\includegraphics[height=6cm]{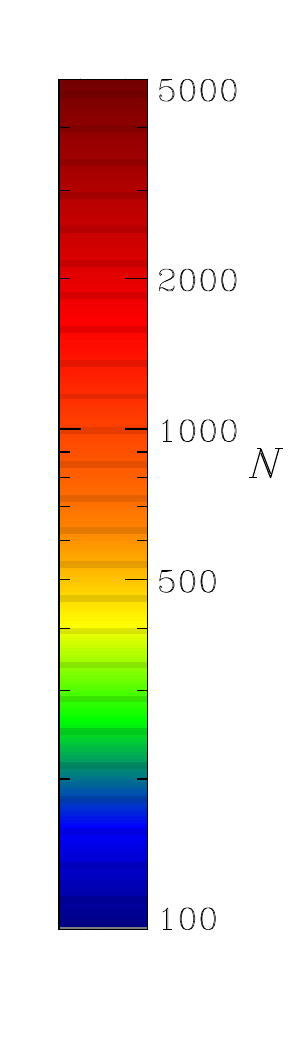}

    \includegraphics[height=6cm]{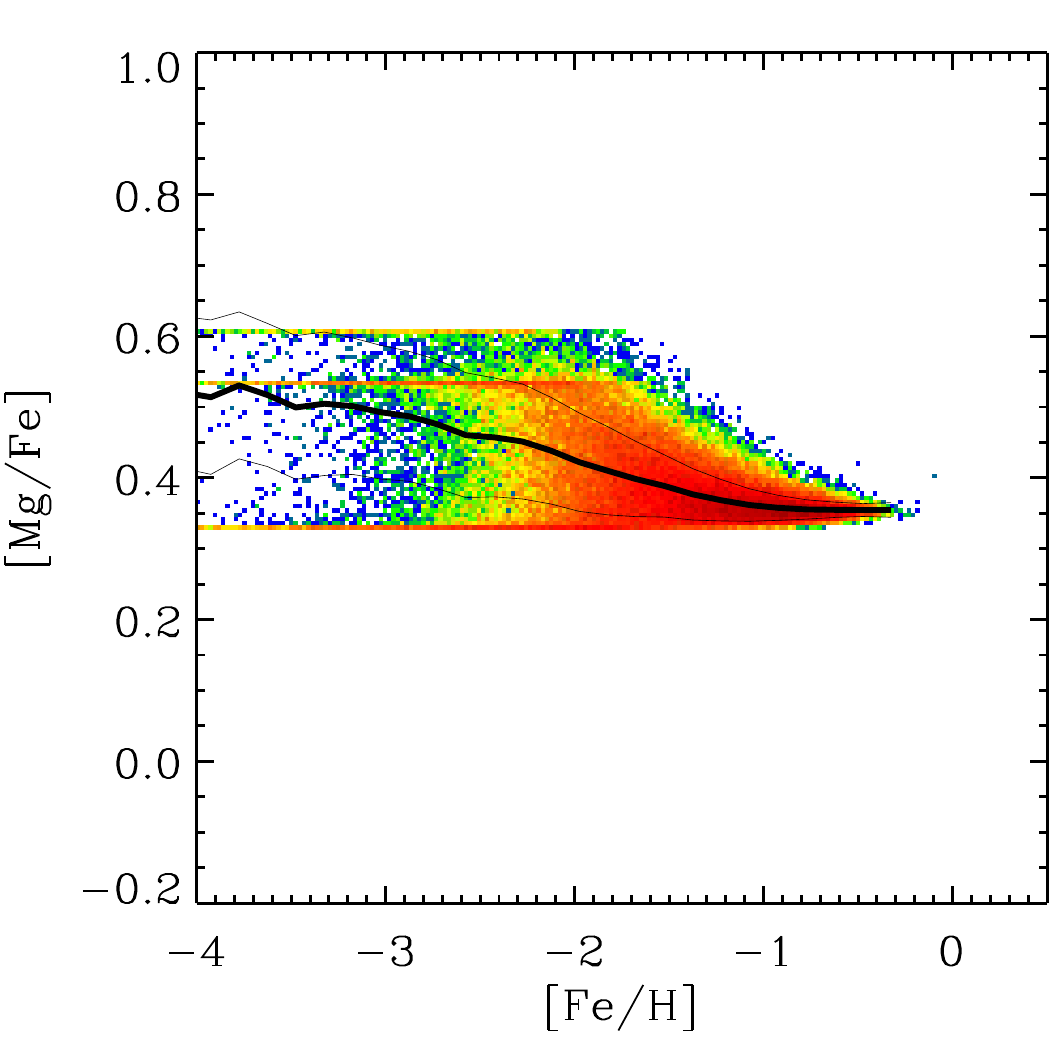}\hspace{1cm}\includegraphics[height=6cm]{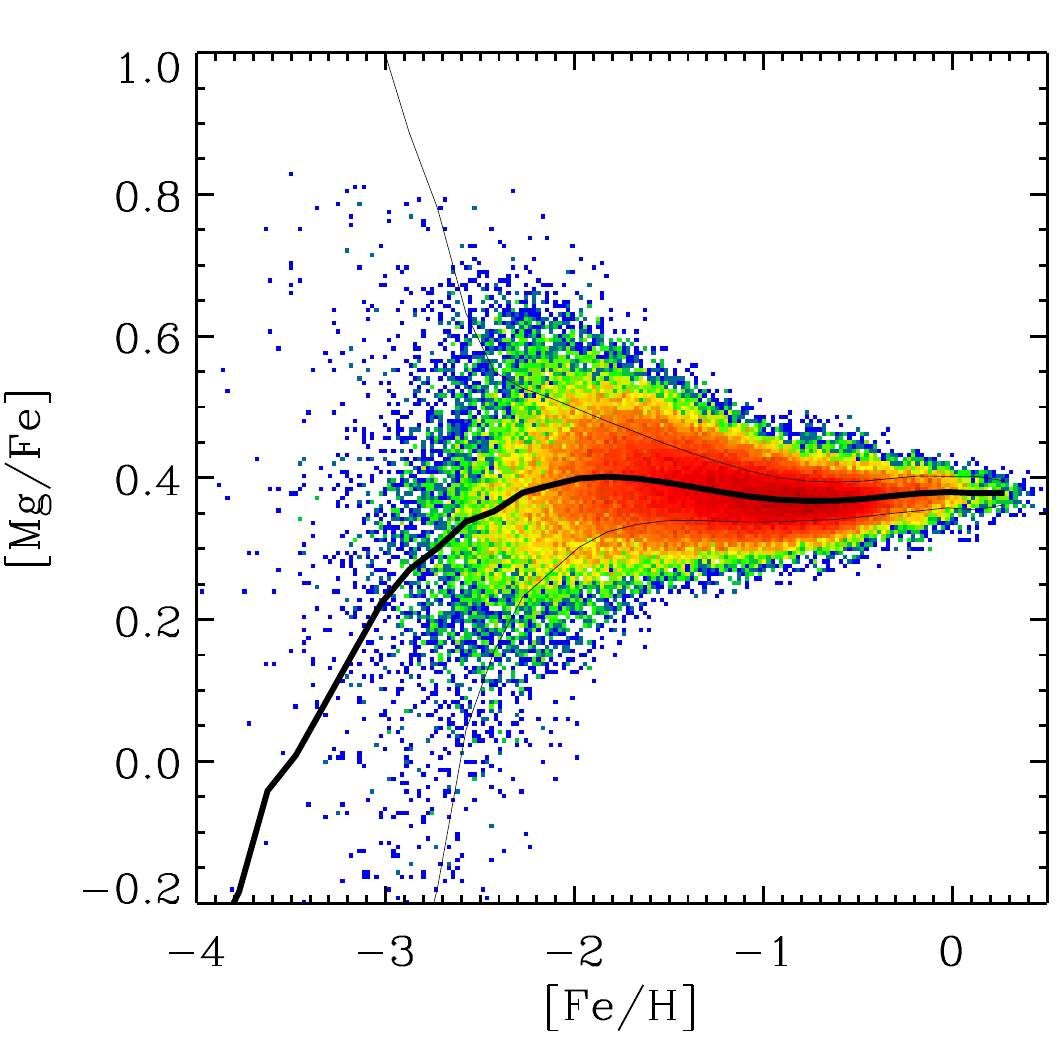}\hspace{1cm}\includegraphics[height=6cm]{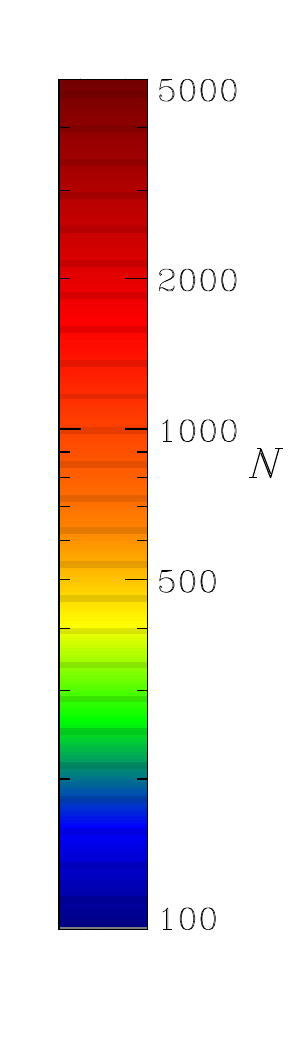}

\caption{Element ratios of the neutron-capture element Ba and the $\alpha-$element Mg, in simulations SE64 (left-hand panels) and ME64 (middle-hand panels), after 1 Gyr of evolution. The solid thick and thin lines show, correspondingly, the median and 1$-\sigma$ levels.  The colour scale is also shown. 
}
\label{fig:element_ratios_isolated}
\end{figure*}

\subsection{Effects on the abundance ratios and their scatter}

 In this Section we focus on the element ratios of various chemical species,
and the variations
originated  in simulations considering
a single or multiple explosion events per SNII.
In all cases, we analyse the distributions after 1 Gyr of evolution; as we have shown
above,  at this time
differences in simulations SE64 and ME64 have already appeared but the variations in SFRs
and total stellar masses are still moderate, allowing 
to better isolate the effects of assuming single or multiple
explosion events per SN from the feedback effects that
follow variations in the SFRs. Furthermore, this choice will allow 
a simpler comparison
with the results of our cosmological simulations of Appendix~\ref{app:cosmo}.

 We show in
Fig.~\ref{fig:element_ratios_isolated} 
the distributions of stellar 
[Ba/Fe]
as a function of [Fe/H]
in simulations SE64 and ME64, after 1 Gyr of evolution (the behaviour of the other
n-capture elements, Eu and Sr, are similar). 
We also
show the 
corresponding median values (thick lines) and
the $\pm \sigma$ levels (thin lines).  
Clearly, the implementation of multiple explosion times (ME64) has
a strong impact in the predicted abundance ratios, particularly in the case of the metal-poor
stars with [Fe/H] $\lesssim -1$. For [Fe/H] $\gtrsim -1$, the mean abundance
ratios of stars in the two runs is similar, although  the scatter in ME64 is higher.
For metal-poor stars the differences between SE64 and ME64 are dramatic,
not only in terms of 
their median values, but more importantly in terms of scatter.
 If a  single SNII explosion event is 
assumed, as in SE64, the abundance ratios are limited to
a narrow range determined by the IMF weighted mean of the stellar yields.
In this simulation, the spread in the abundance ratios  
of [Ba/Fe] 
of the order of $0.3$ at the most, for all metallicities.  
In contrast, the abundance ratio of stars in ME64
show not only a much larger scatter up to $2$ dex,
but also a clear relation between the scatter and the metallicity, such that the
scatter decreases with increasing metallicity (a behaviour that extends towards
stars with higher abundances).  
Stars in ME64
have a large variety of possible element ratios, even
for coeval stellar populations, 
which results from the dependence of the chemical yields  primarily on the stellar mass, but also on metallicity.

An important observation from  this figure is
that the lowest and highest abundance ratios that
are found in ME64  
can not  be described in SE64; in fact, in ME64 a significant fraction of the stars
is in the range of abundance ratios that are absent in SE64.

Now we turn our focus to the $\alpha$-elements, 
showing in the lower panel of Fig.~\ref{fig:element_ratios_isolated}  the distribution  of [Mg/Fe], which we use as an example for this type of element (Si
and O exhibit similar trends).
 In general terms, the behaviour of
the $\alpha-$elements 
is similar to that observed for the neutron-captures elements, in the
sense that simulation ME64 allows the abundance ratios to vary more compared to SE64,
 producing more realistic abundance patterns.  Furthermore, we can see, in ME64, 
a positive and strong correlation between  [Mg/Fe]
and [Fe/H], which is, particularly for  metal-poor stars, opposite to that found in SE64.
Similarly to our findings for the neutron-capture elements, for the $\alpha$-elements
we find that in ME64 the spread in abundance ratios decreases with increasing metallicity.
It is important to note that, in our simulation ME64, 
the abundance ratios of neutron-capture elements extend over 3 dex while a smaller spread (or around
1 dex) is predicted for the $\alpha-$ elements.

The results of this Section
show that a correct description of the different ejection-times of
release of chemical elements of different mass stars  contributing as core-collapse SNe 
is important in order to properly describe the early enrichment phases 
of the ISM in galaxies, and the abundances and scatter of the very metal-poor stars. We have shown
the effects of differential enrichment using isolated simulations, but the same results are reproduced in cosmological
runs,  in terms of the changes in abundance ratios and scatter (see Appendix~\ref{app:cosmo}) although with a higher level of complexity
due to the cosmological evolution.

\section{Abundances and scatter in the stellar halo}\label{sec:halo}

In this Section we study the chemical properties of the stellar halo
formed in a cosmological simulation of the formation of a Milky Way-mass galaxy.
This simulation, referred to as the ME run, assumes multiple explosion times per SNII\footnote{We have also run an identical simulation assuming a single explosion time for SNII; differences  between the predictions of these two simulations, as well as a comparison between the cosmological and the idealized simulations,  are discussed in Appendix~\ref{app:cosmo}.}.

The initial conditions used for our cosmological simulations 
correspond to halo named Aq-C of the Aquarius Project \citep{Springel08}. The
simulated halo has, at $z=0$, a virial mass of 
$1.6\times 10^{12}$ M$_\odot$ (see Table~\ref{table:simulations}),
a virial radius of $170$ kpc
and is, by design, mildly isolated (no
neighbour exceeding half its mass within a sphere of 2 Mpc radius). The simulations
 start
at $z=127$ and run until $z=0$.

The ICs are consistent with a $\Lambda$-Cold Dark Matter universe
with the following cosmological parameters:
$\Omega_{\rm m} = 0.25 $, $\Omega_{\Lambda} = 0.75$,  $\Omega_{\rm b} = 0.04$, 
$\sigma_8 = 0.9$ and 
$\text{H}_0 = 100 \, h \, \text{km} \text{ s}^{-1} \text{Mpc}^{-1}$,
with $h=0.73$.
The mass resolution is 
$4.1\times 10^{5}$M$_\odot$  and $2.2\times 10^{6}$M$_\odot$ 
for gas and dark matter particles, respectively, and we have adopted a gravitational
softening of 700 pc, fixed in comoving coordinates. 
The input parameters used for these simulations
are the same than those used in the idealized simulations of the previous Section,
in order to facilitate comparison.

\begin{figure*}
  \centering

\includegraphics[width=4cm]{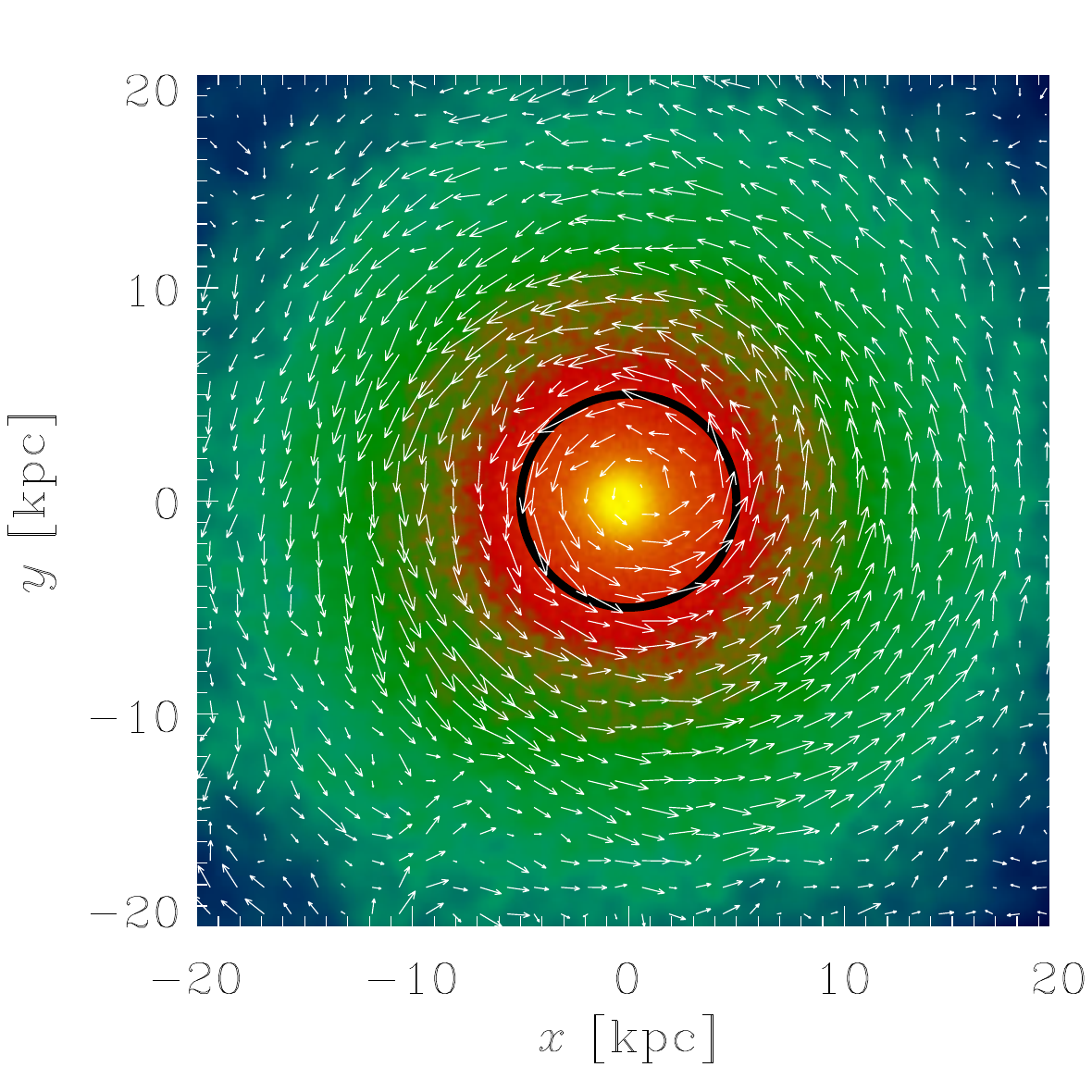}\includegraphics[width=4cm]{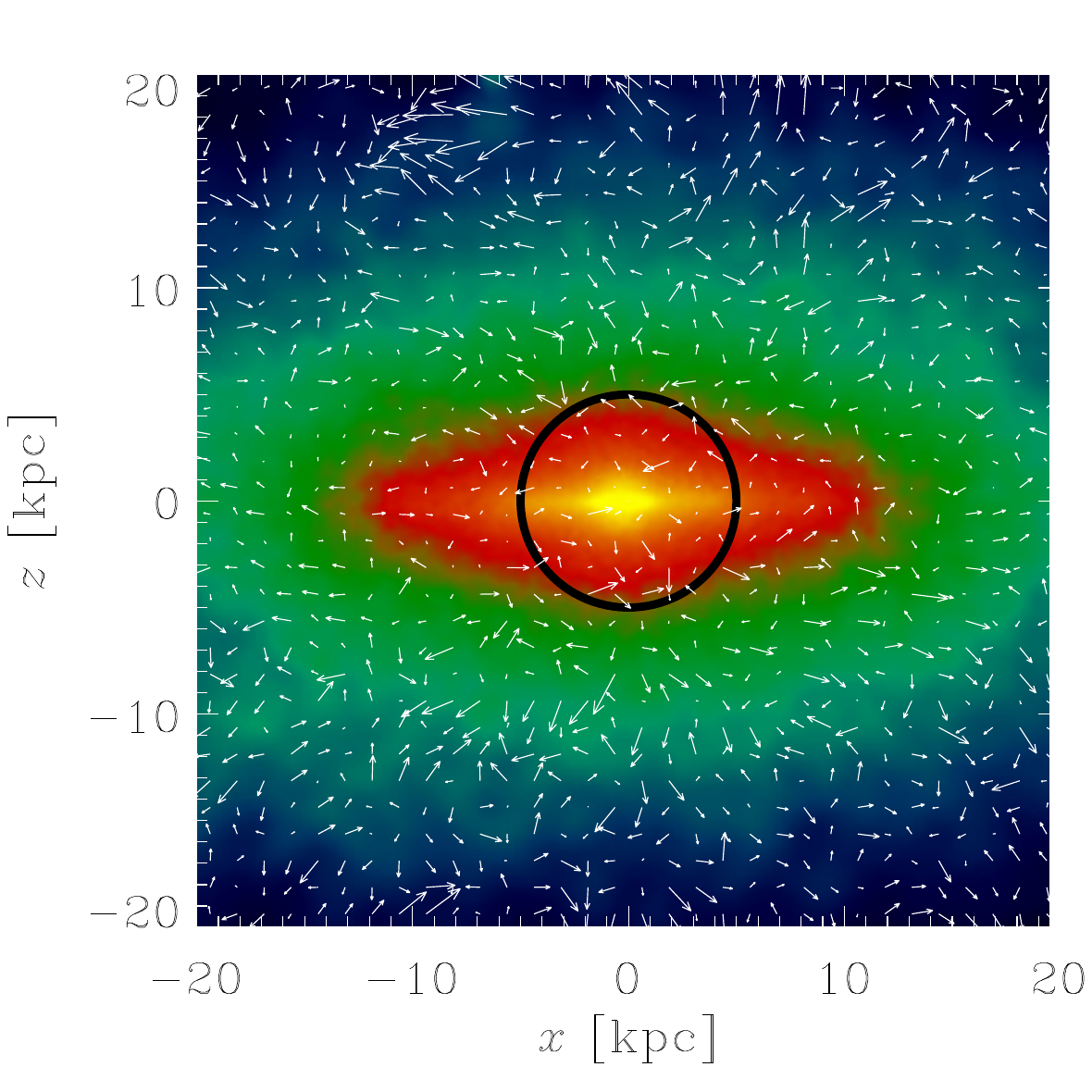} \hspace{1cm} \includegraphics[width=4cm]{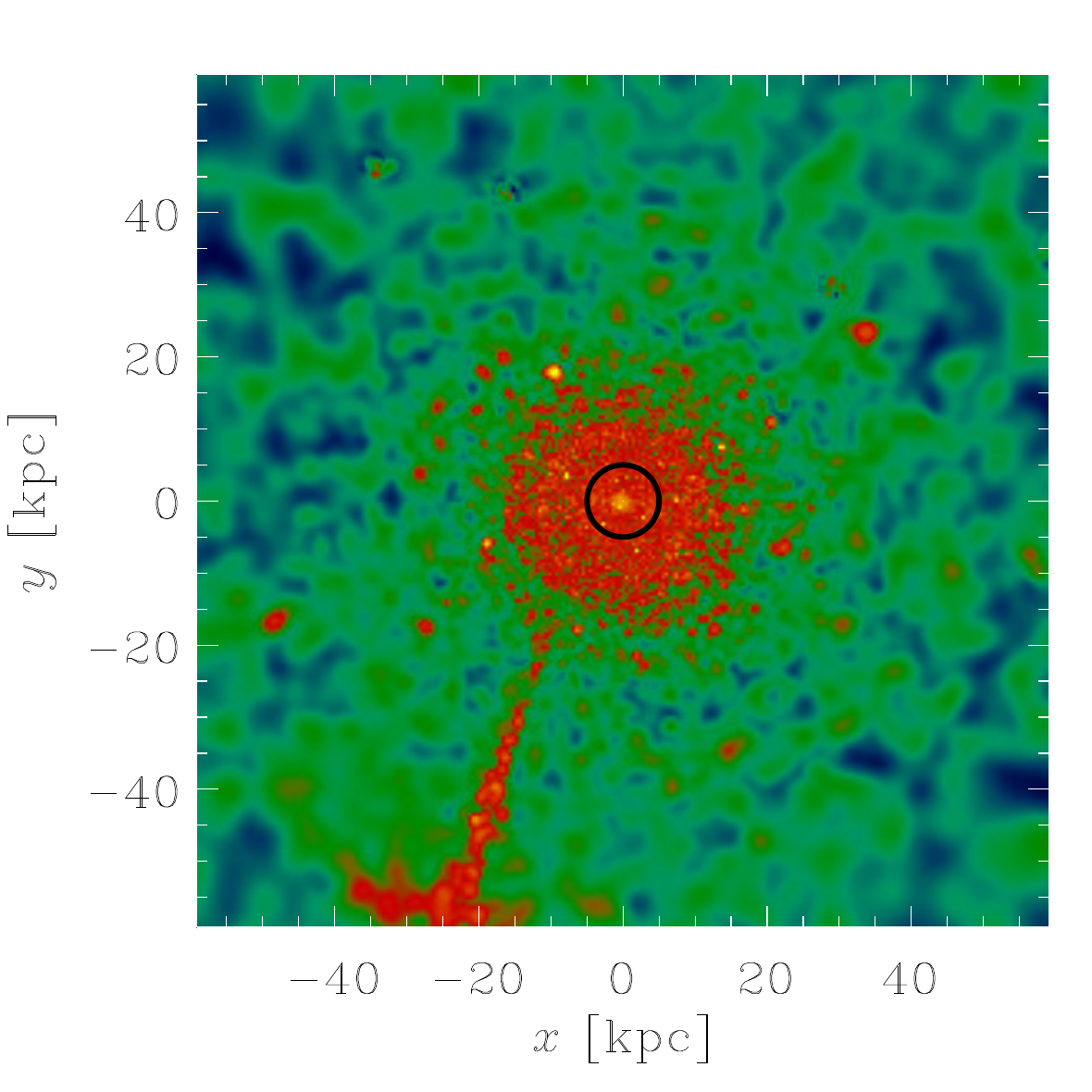}\includegraphics[width=4cm]{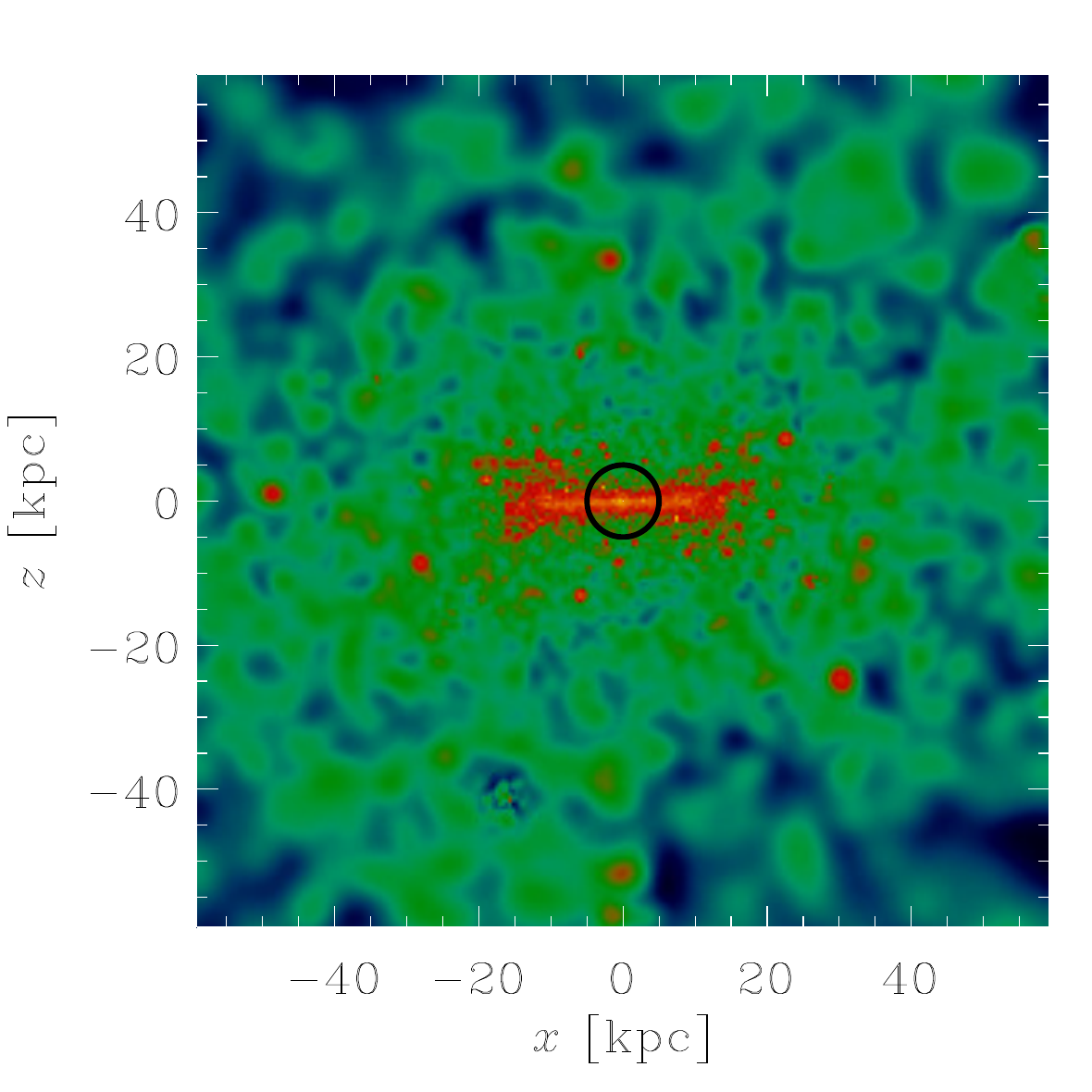}

\hspace{1cm}\includegraphics[width=6cm]{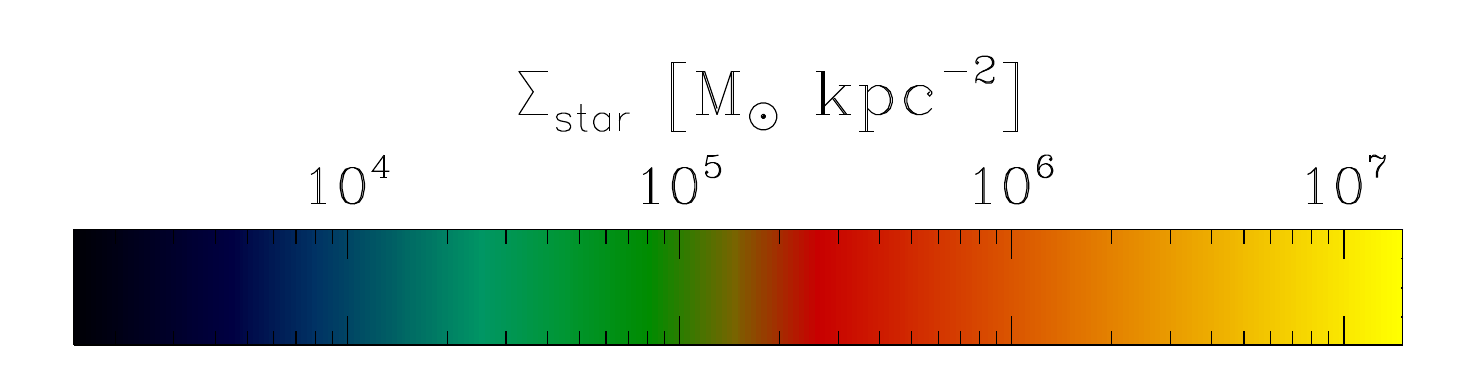}\hspace{4cm}\includegraphics[width=6cm]{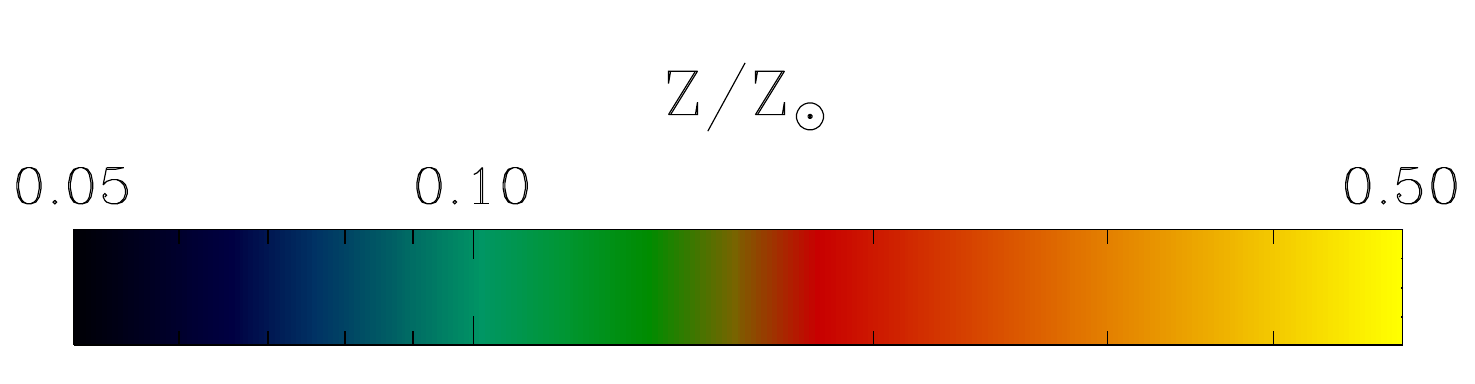}

\caption{Maps of projected stellar mass density (left-hand panels) and stellar metallicity (right-hand panels) for the simulated galaxy in our cosmological simulation ME, at $z=0$, in face-on ($xy$) and edge-on ($xz$) views. The white arrows in the mass density maps  indicate the corresponding velocity fields and reveal that, at the present time, the simulated galaxy has a rotationally-supported disk-like component, a compact bulge, and an extended stellar halo.}
\label{fig:stellar_maps_cosmo_me}
\end{figure*}

 Fig.~\ref{fig:stellar_maps_cosmo_me} shows the face-on and edge-on projections of the stellar mass density (left-hand panels) at $z=0$ for the simulated galaxy, together with the corresponding metallicity distributions (right-hand panels).
In order to make the projections, we have aligned the total angular momentum of the stars with the $z$ direction.
The simulated galaxy is, at the present time,  composed of 
 a compact, dispersion-dominated
bulge and  a rotationally-supported disk.
The rotation of the disks is clearly seen, particularly in the face-on view,  
from the velocity field included in this figure. The stellar halo is spatially
extended, and significantly fainter than the other stellar components.
The metallicity distributions reveal the different levels of enrichment in the 
different stellar components, the bulge and disc being significantly more enriched than the stellar halo region.

\begin{figure*}
  \centering

\includegraphics[width=4.2cm]{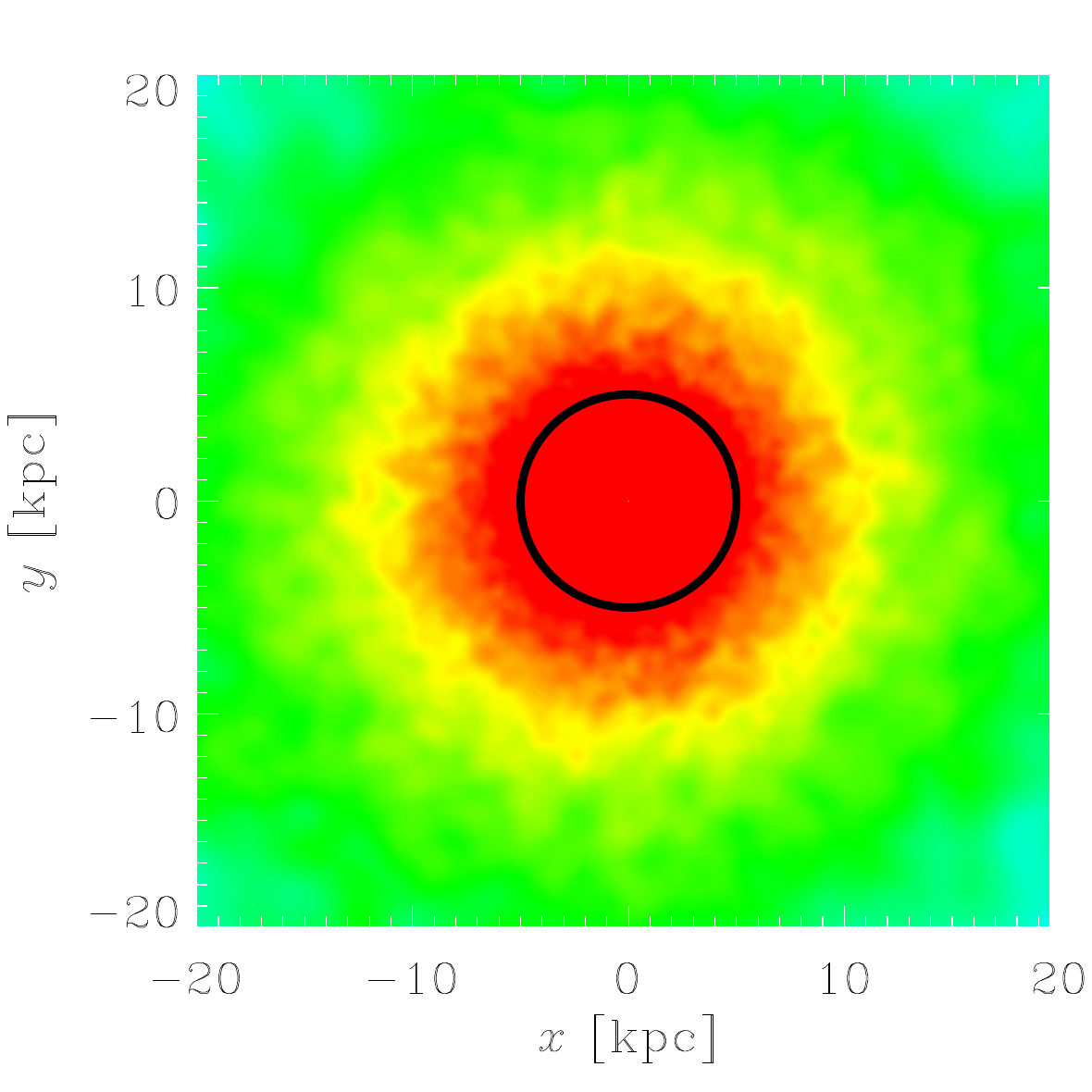}\includegraphics[width=4.2cm]{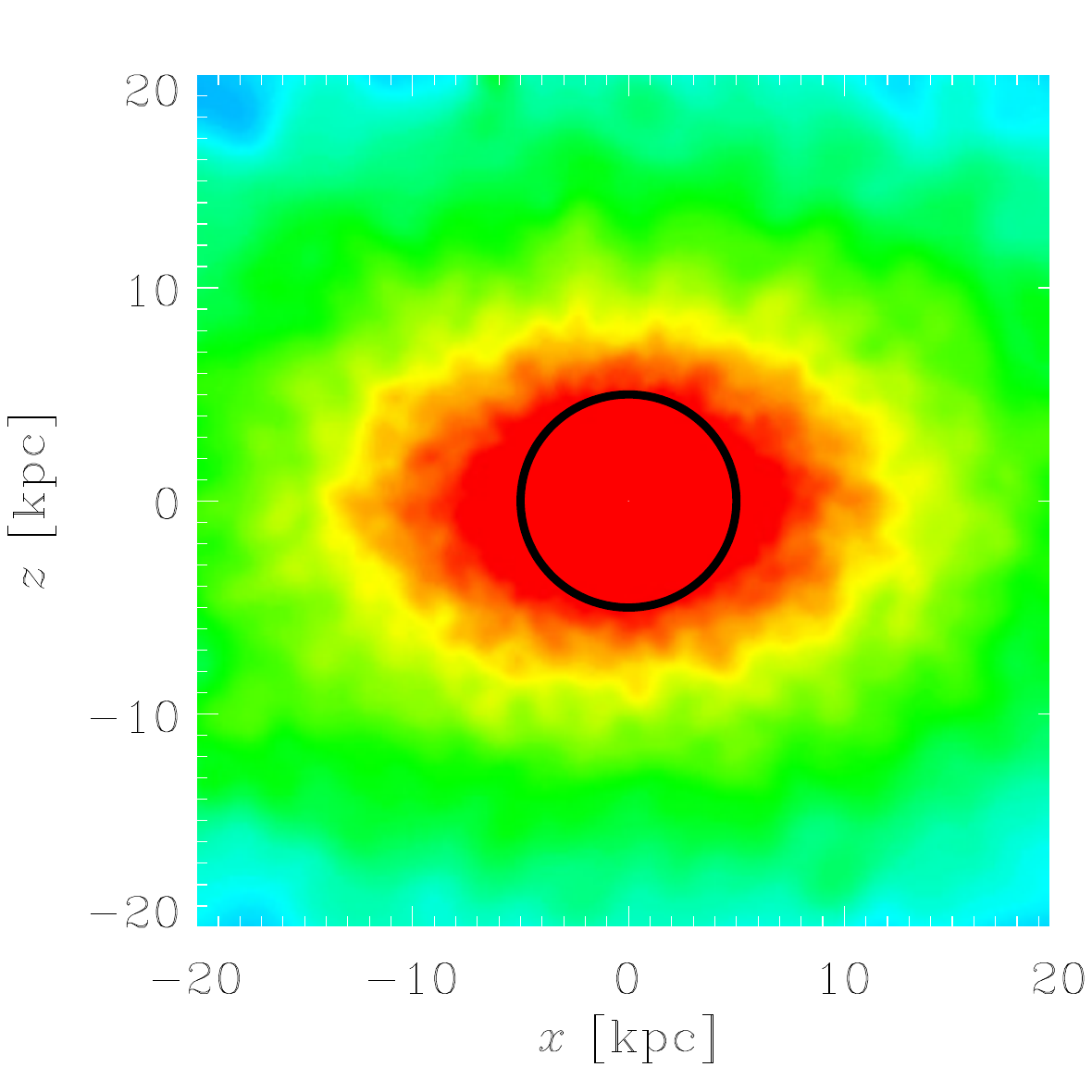}\includegraphics[width=4.2cm]{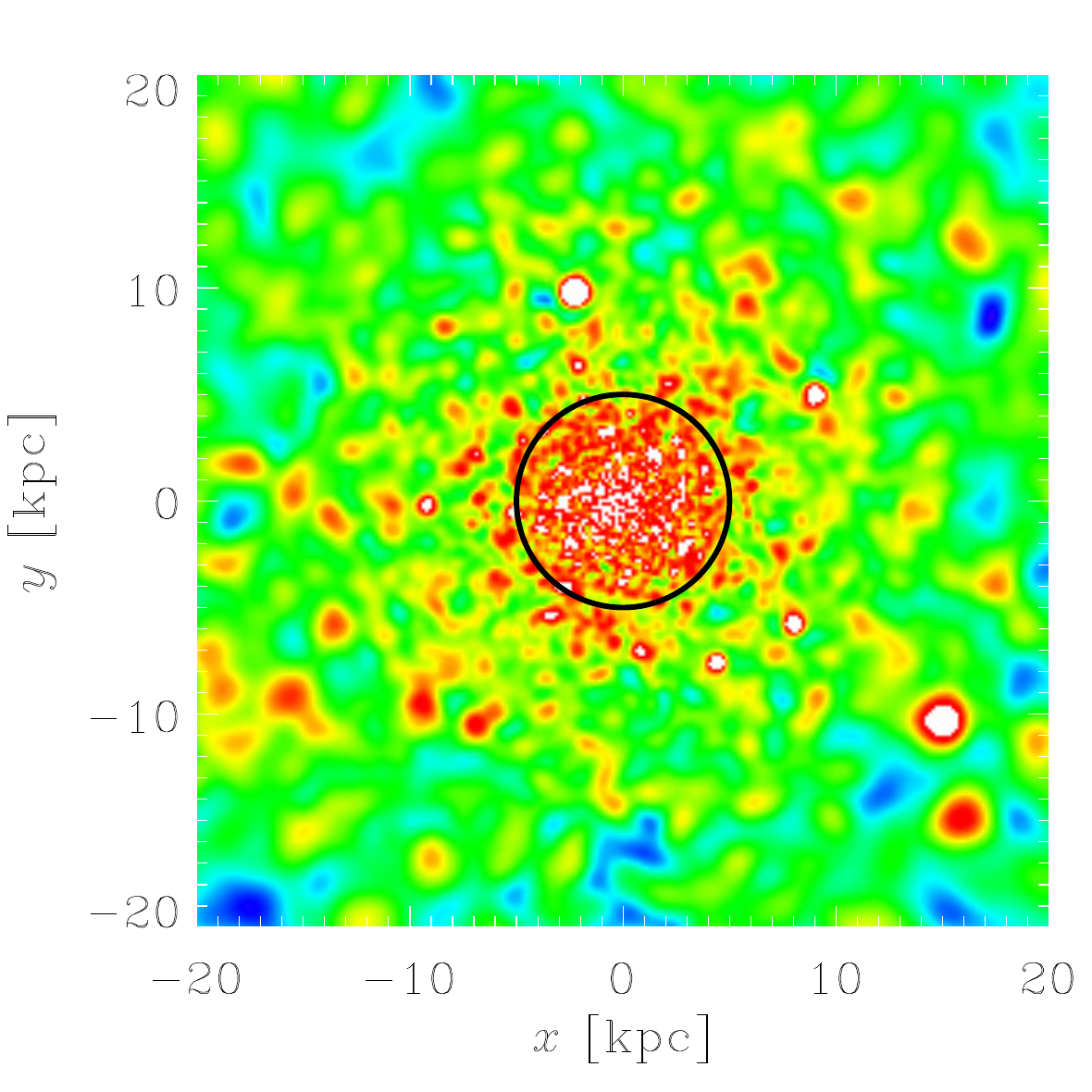}\includegraphics[width=4.2cm]{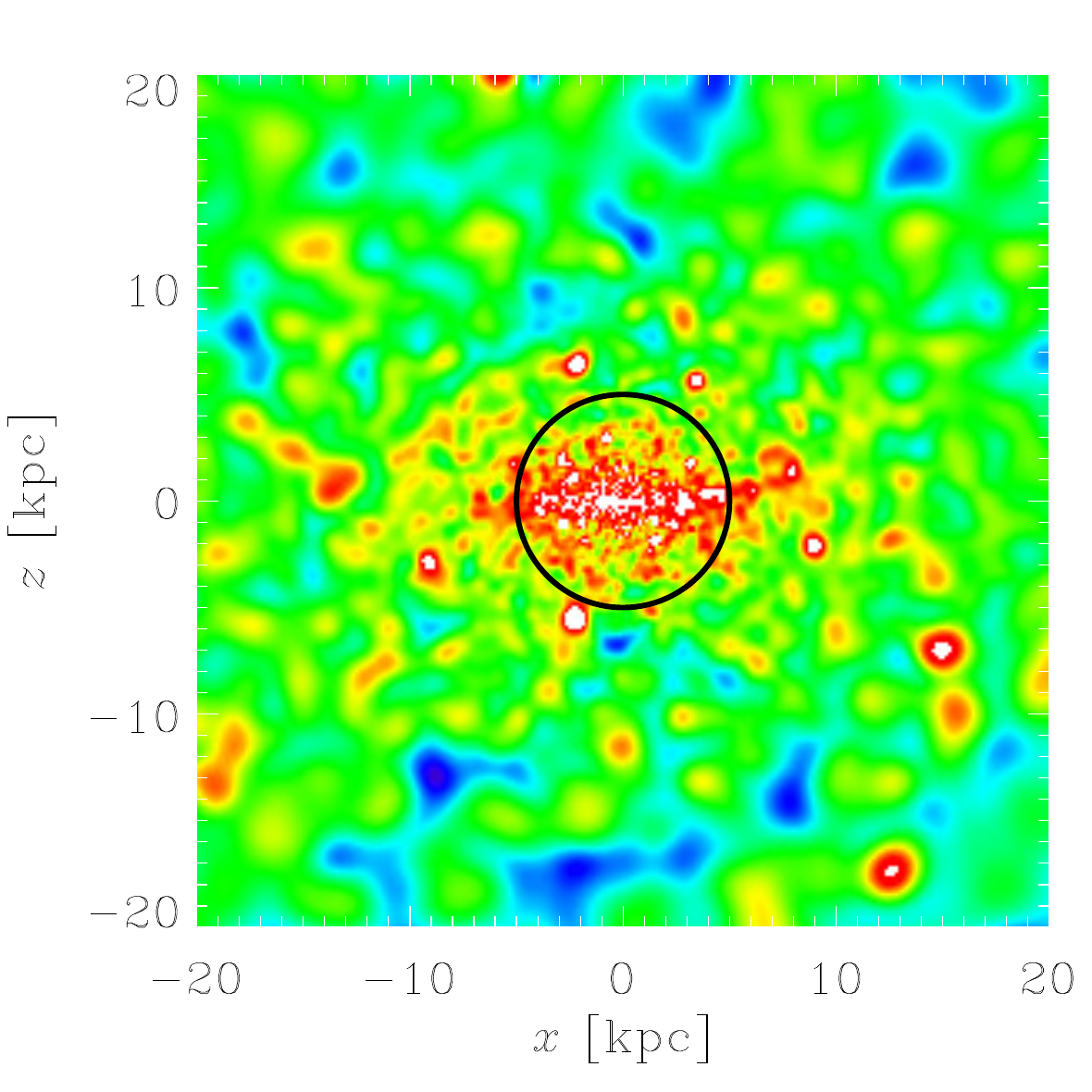}

\includegraphics[width=4.2cm]{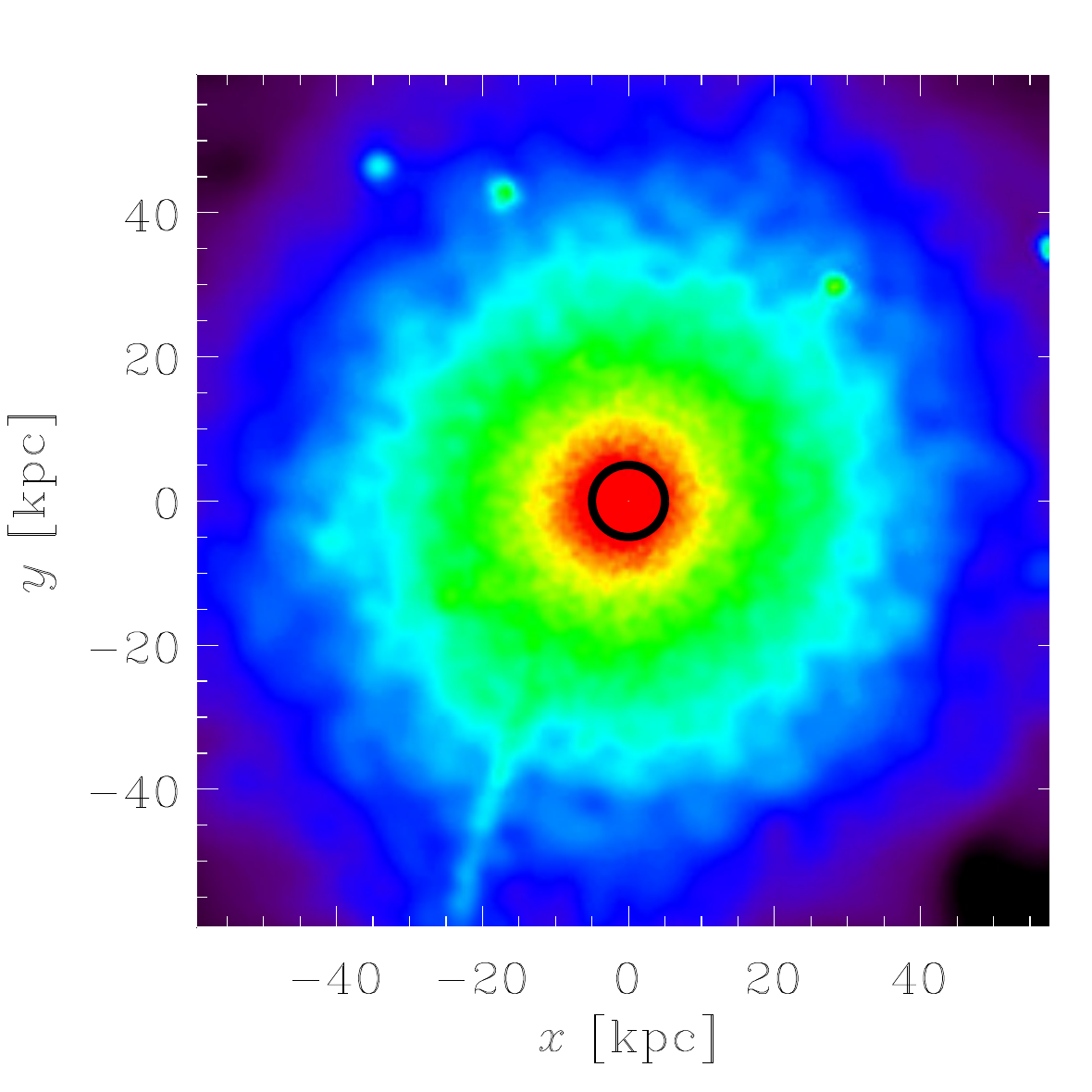}\includegraphics[width=4.2cm]{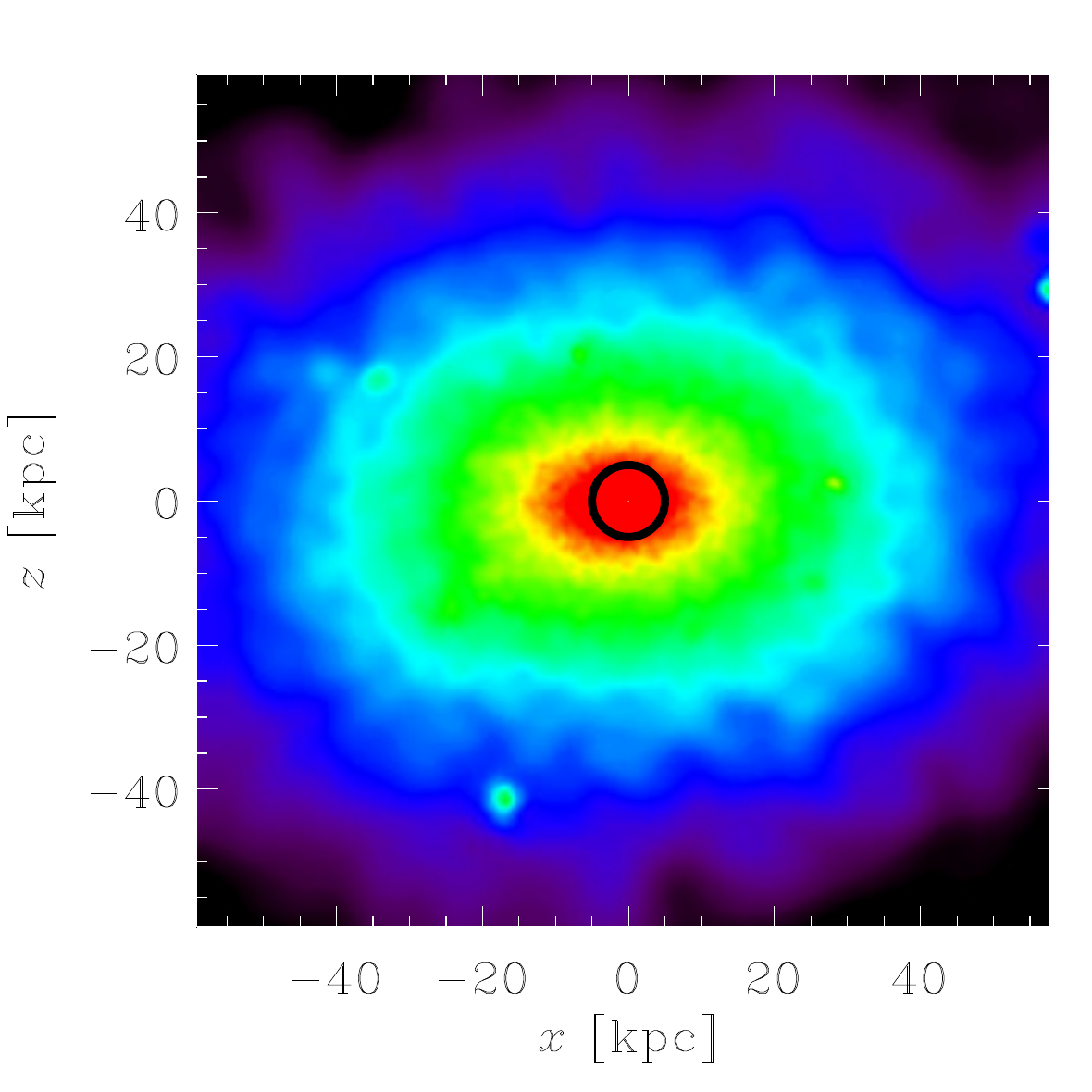}\includegraphics[width=4.2cm]{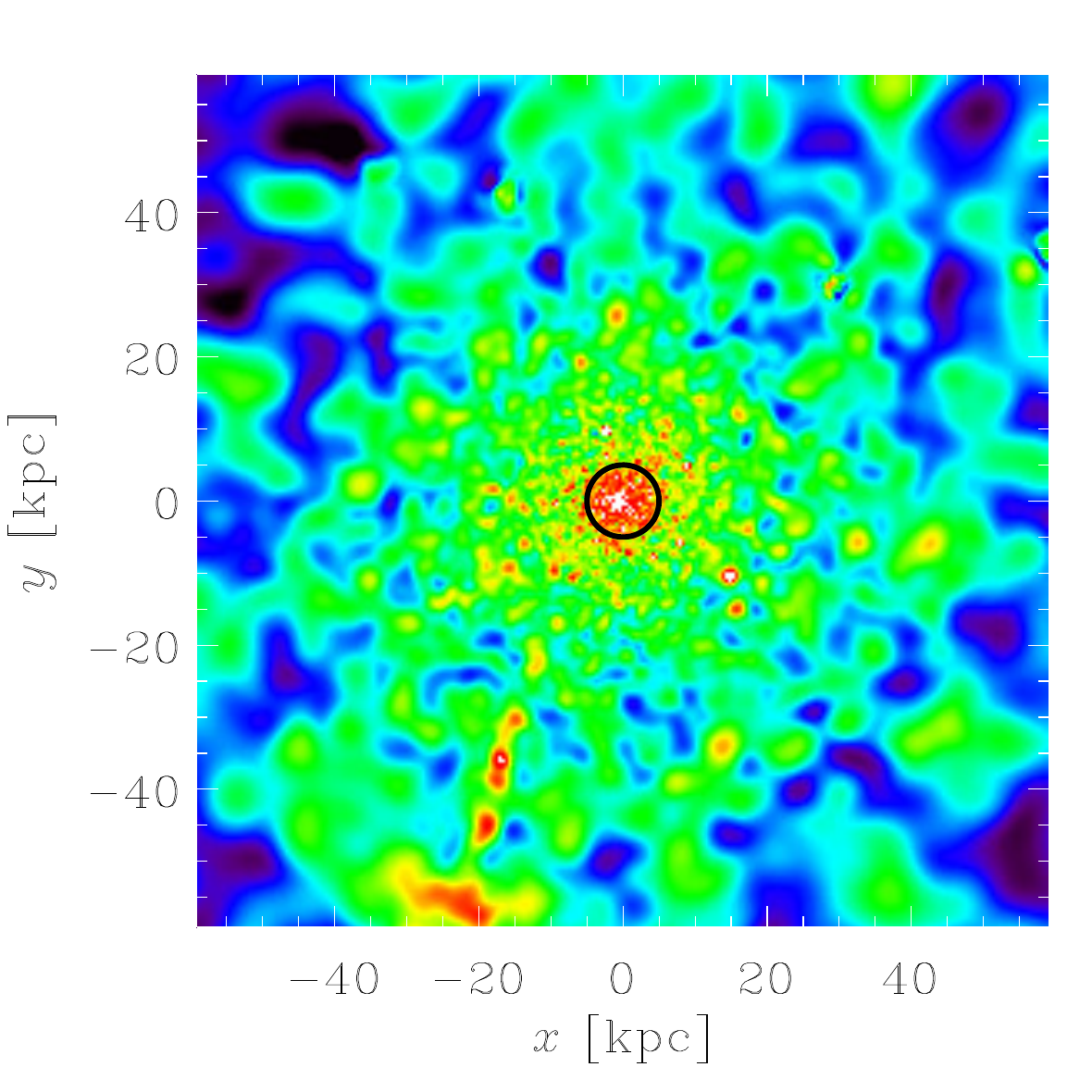}
\includegraphics[width=4.2cm]{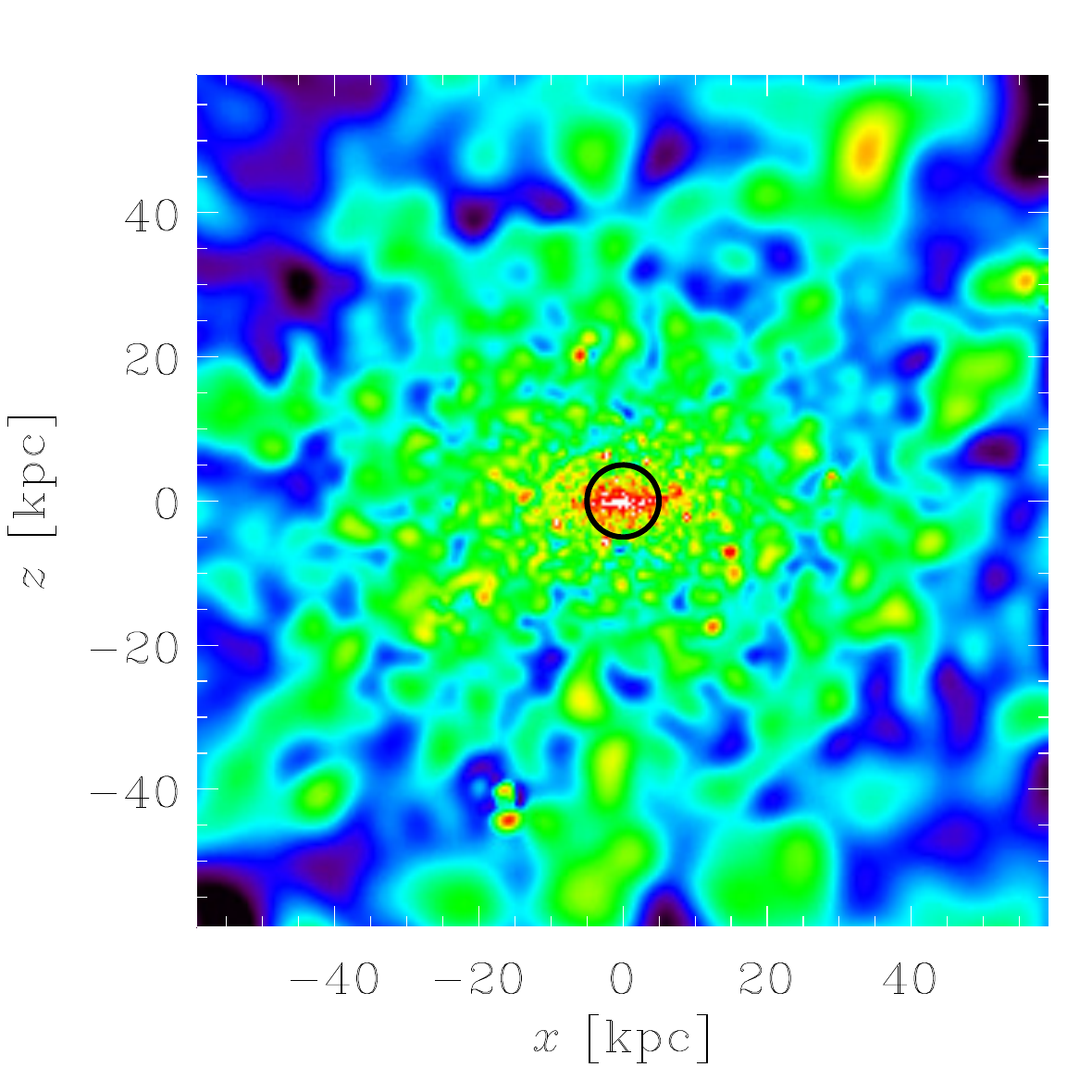}

\hspace{1cm}\includegraphics[width=6cm]{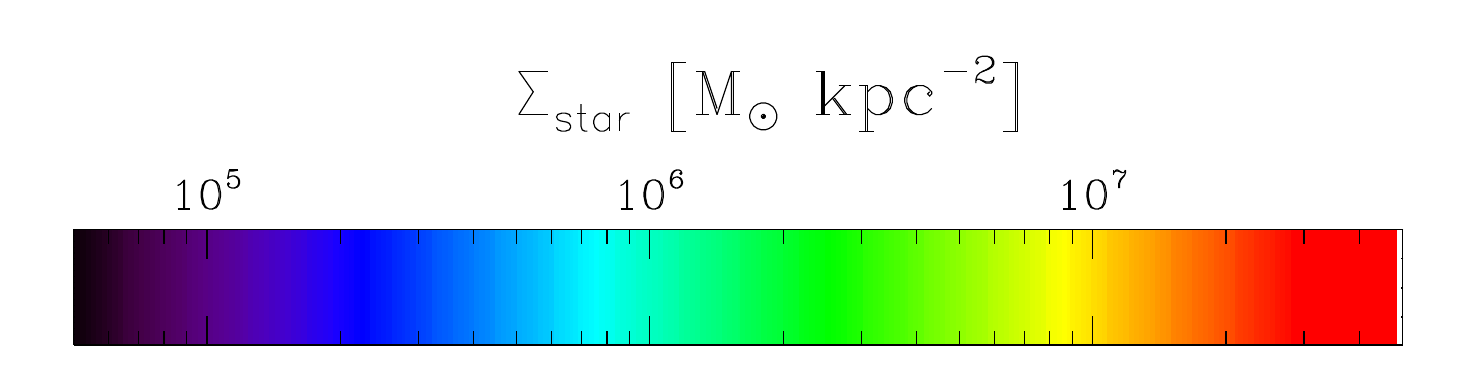}\hspace{3cm}\includegraphics[width=6cm]{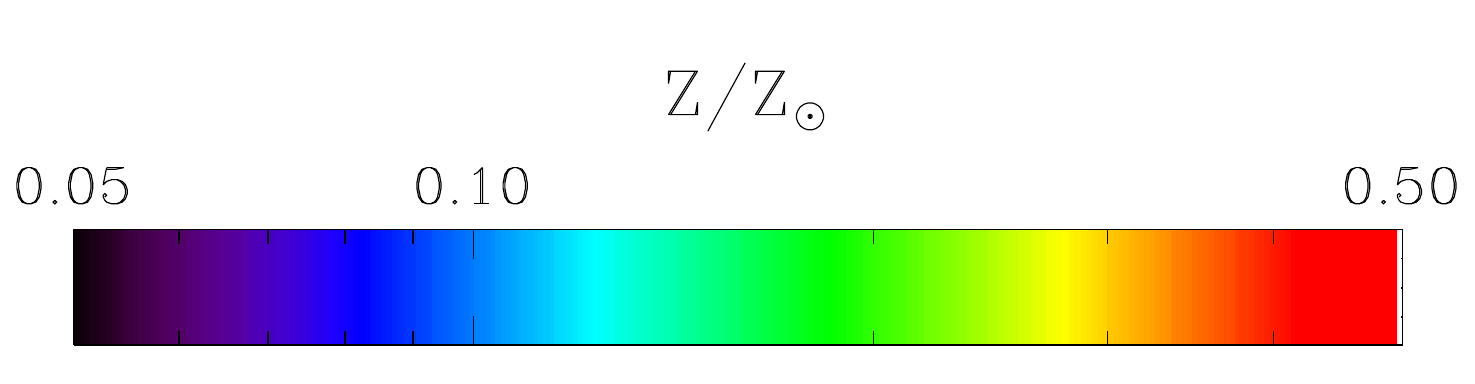}

\caption{Maps of projected stellar mass density (left-hand panels) and stellar metallicity (right-hand panels), at $z=0$, 
for our cosmological simulation ME, considering stars that formed in the first Gyr of evolution. In both cases we show two perpendicular projections, corresponding to the face-on and edge-on projections of Fig.~\ref{fig:stellar_maps_cosmo_me}. The black circle shows our threshold of 5 kpc to define halo stars. The upper- and lower panels show, respectively, the inner galaxy and a zoom-out up to 0.25 of the virial radius. 
Note that the colour code and scale have been chosen in order to highlight the halo density and metallicity structure (note that these are different from those used in Fig.~\ref{fig:stellar_maps_cosmo_me}).
}
\label{fig:stellar_maps_halo}
\end{figure*}

 In order to study the abundances in the stellar halo component,
we first need to determine how we
define the halo stars as, in the simulations,
there are many different ways to assign stars to this (or any other) component.
In this work,  we define the stellar halo using stars that formed during the 
first Gyr of evolution and that are, at the present day, outside a sphere of
5 kpc
from the center of the galaxy. As we identify the halo stars at $z=0$, the age threshold of $\tau=1$ Gyr
 assumed for the halo is included in order to exclude from the sample disc stars, which are dominant
outside the inner $5$ kpc but significantly younger. However, the exact
$\tau$ value is unimportant -- as long as it is not too low -- because disc 
and halo stars 
have very different ages.

Our choice of selecting
stellar halo particles at $z=0$ simplifies the interpretation
of results and the comparison with observational data  which integrates the full
evolutionary history of the Milky Way and provides information on the $z=0$ properties of the halo. Our choice for the simulated halo population, based only on the
age and present-day radius of stars, provides us with a simple, easy
to interpret and clean sample, without the need to make any additional assumption
on the properties of the halo (e.g. metal content or dynamics)\footnote{Note that
our sample could include thick disk stars, if this component was formed in the
first Gyr of evolution. We have verified, however, that the majority of
the simulated stellar halo stars do not have a significant degree of rotation.}. 
Furthermore,
the use of stars outside the very central region provides a sample that is uncontaminated by the old bulge population\footnote{We have tried several threshold radii to define the halo and found that
our results are not sensitive to this choice, provided $r\gtrsim 4$ kpc.}.

The left-hand panels of Fig.~\ref{fig:stellar_maps_halo} 
show maps of the present-day spatial distribution of the old stars (i.e.
formed in the first Gyr) 
in our simulation ME, up to
$20$ kpc (i.e. the inner halo) and to $1/4$ of the virial radius
(i.e. the extended halo).
Also indicated is 
the minimum radius assumed for halo stars, i.e. $5$ kpc.
 The figure shows two different projections, the face-on and edge-on views of Fig.~\ref{fig:stellar_maps_cosmo_me}, in order to highlight the morphology of
the stellar halo.
Note that, unlike the whole stellar population, 
 the old halo stars define no preferred plane, although the stellar distribution is
not fully spherical. 
The simulated stellar halo is spatially extended, has  low surface mass density, and exhibits, particularly in the outermost regions, substructure in the form of stellar clumps. A stellar stream, reminiscent of observed streams, is also detected in the face-on view.

The right-hand panels of Fig.~\ref{fig:stellar_maps_halo} show 
maps of the stellar metallicity of the old stars (formed in the first
Gyr) in simulation ME, for the same views and spatial scales shown in the left-hand panels
of this figure.
While the inner halo
($5 \le r < 20$ kpc)  is characterized by average stellar metallicities
of the order of Z/Z$_\odot \sim 0.1$ or higher, 
stars that are further away (i.e. in the outer halo)
are less enriched, with typical (spatially-averaged) metallicities between
Z/Z$_\odot\sim 0.05$ and Z/Z$_\odot\sim 0.1$. The stellar stream 
is more metal-rich compared to the
rest of the outer halo population, suggesting that these stars
were formed in an accreted satellite and were not formed {\it in-situ}.

\subsection{The formation of the stellar halo}\label{sec:formation_of_stellar_halo}

\begin{figure*}

{\hspace{0.8cm}\textbf{ All stars} \hspace{0.75cm} \textbf{$T$=1 Gyr}  \hspace{3cm} \textbf{$T$=2 Gyr}  \hspace{3cm} \textbf{$T$=3 Gyr}  \hspace{3cm} \textbf{$T$=4 Gyr} } 
\includegraphics[width=4.5cm]{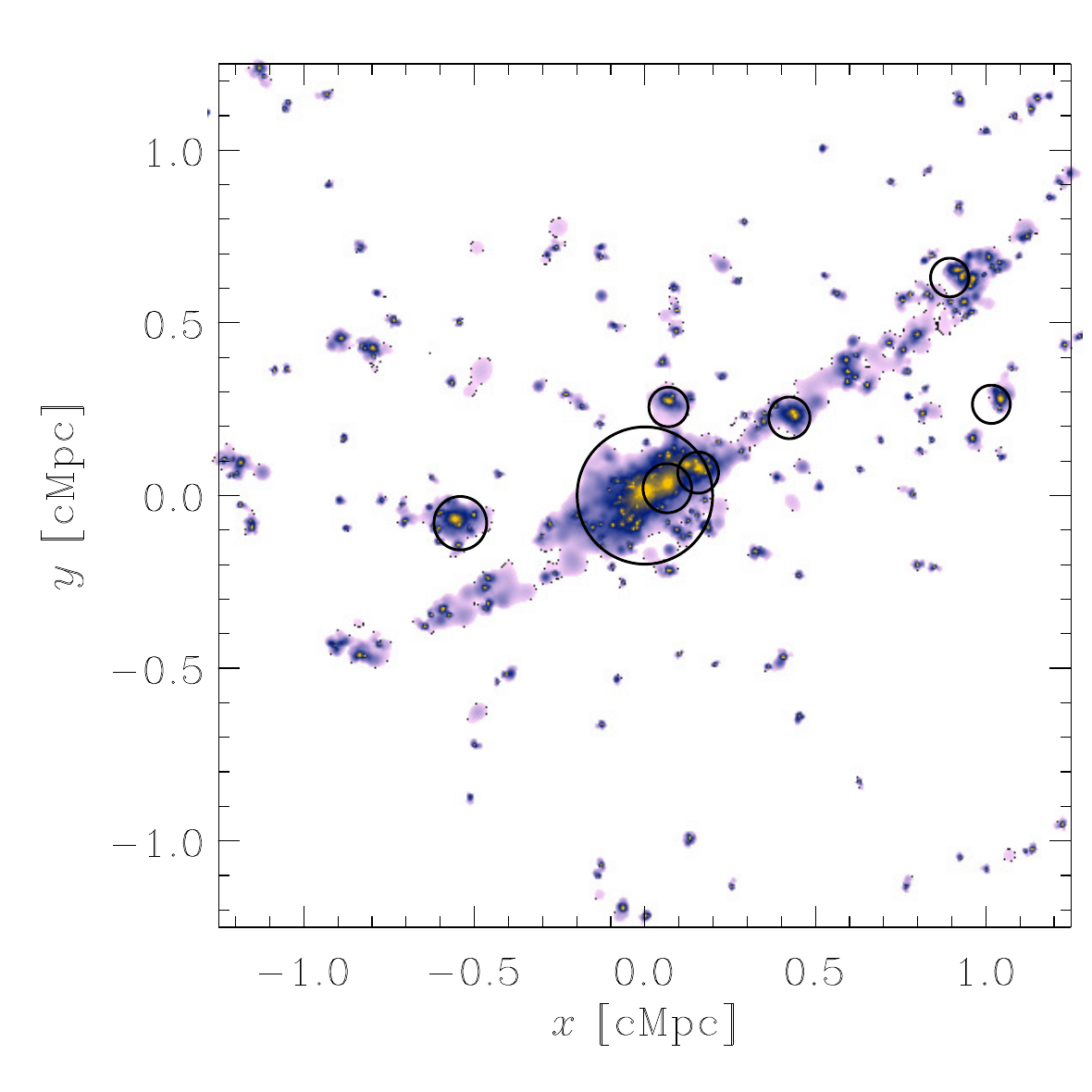}\includegraphics[width=4.5cm]{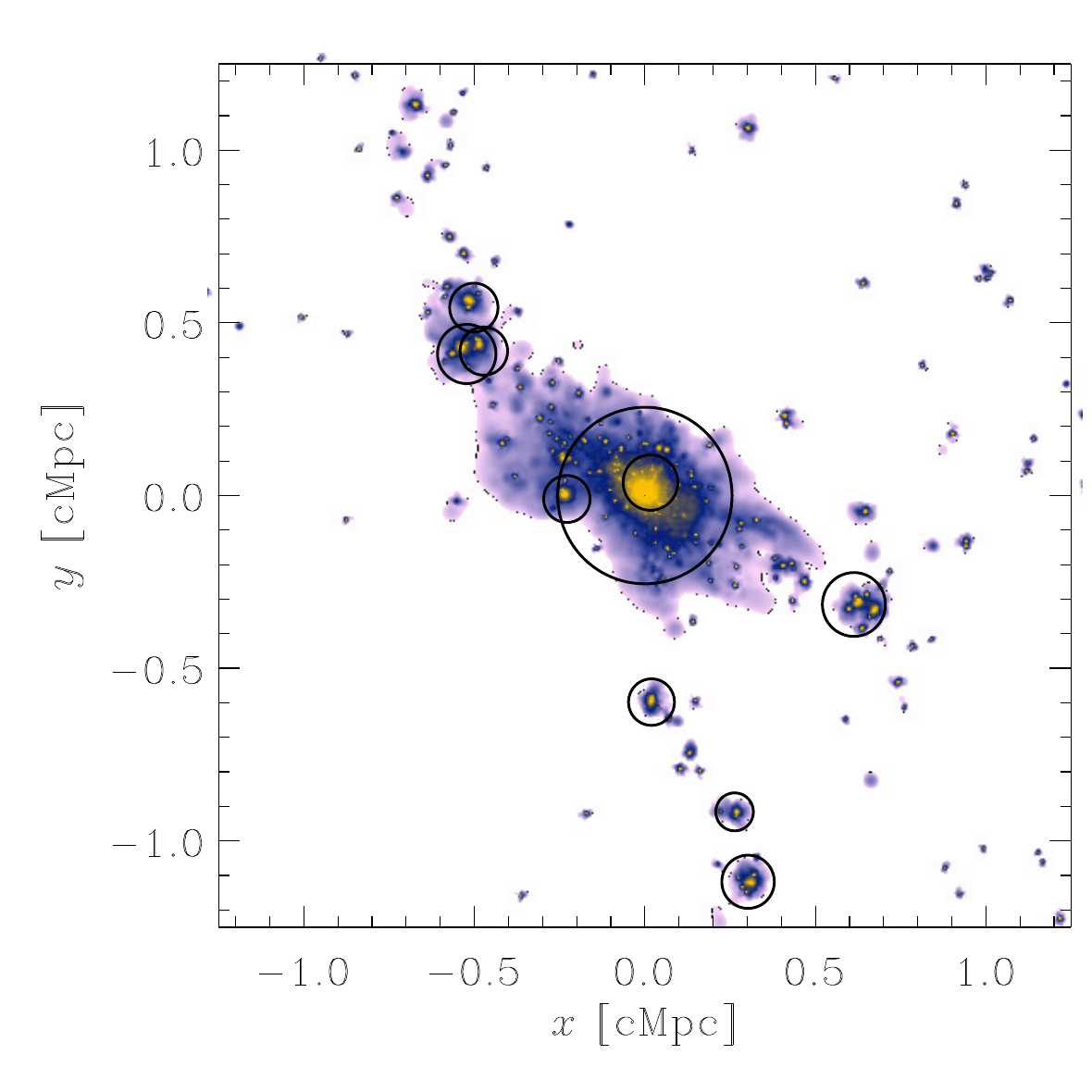}\includegraphics[width=4.5cm]{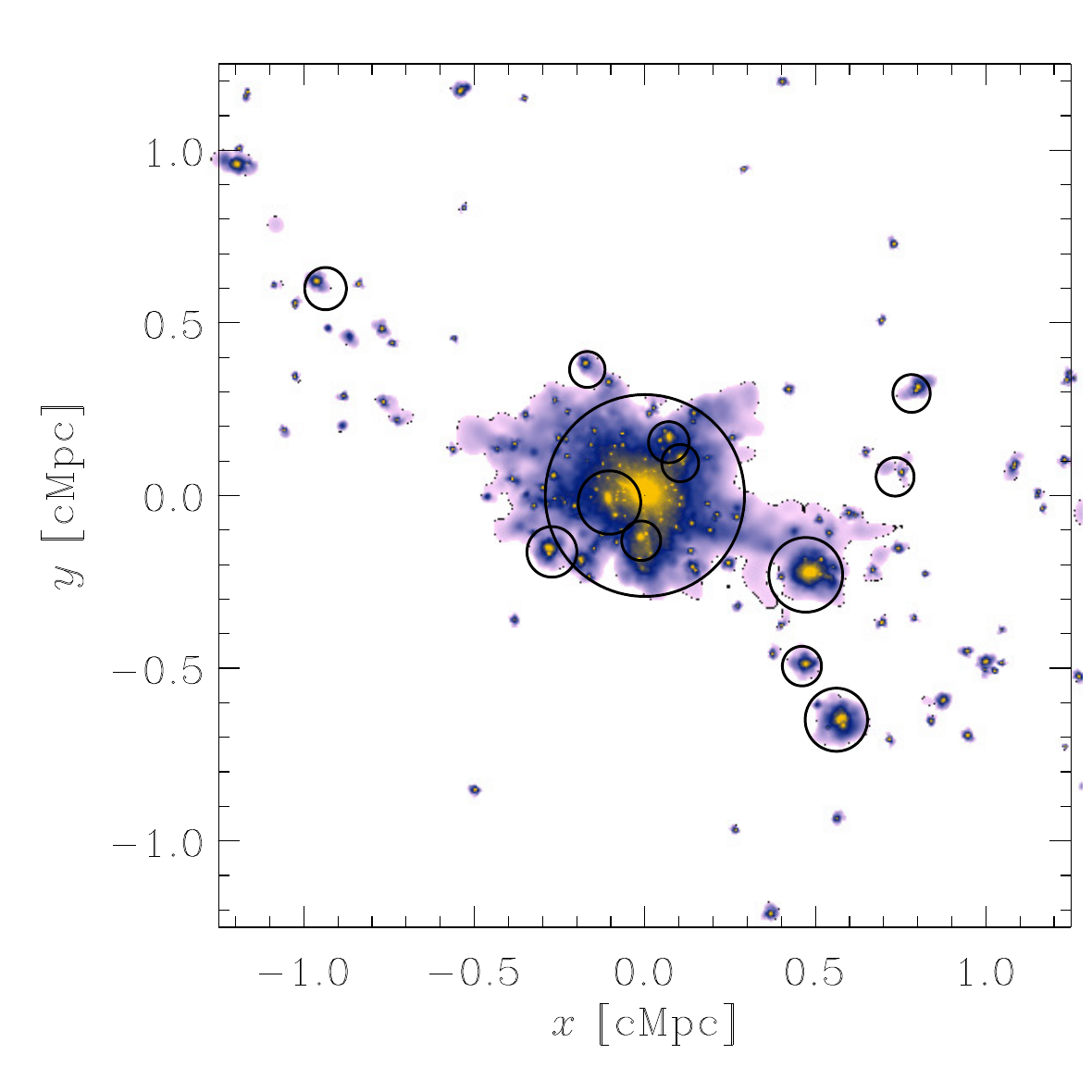}\includegraphics[width=4.5cm]{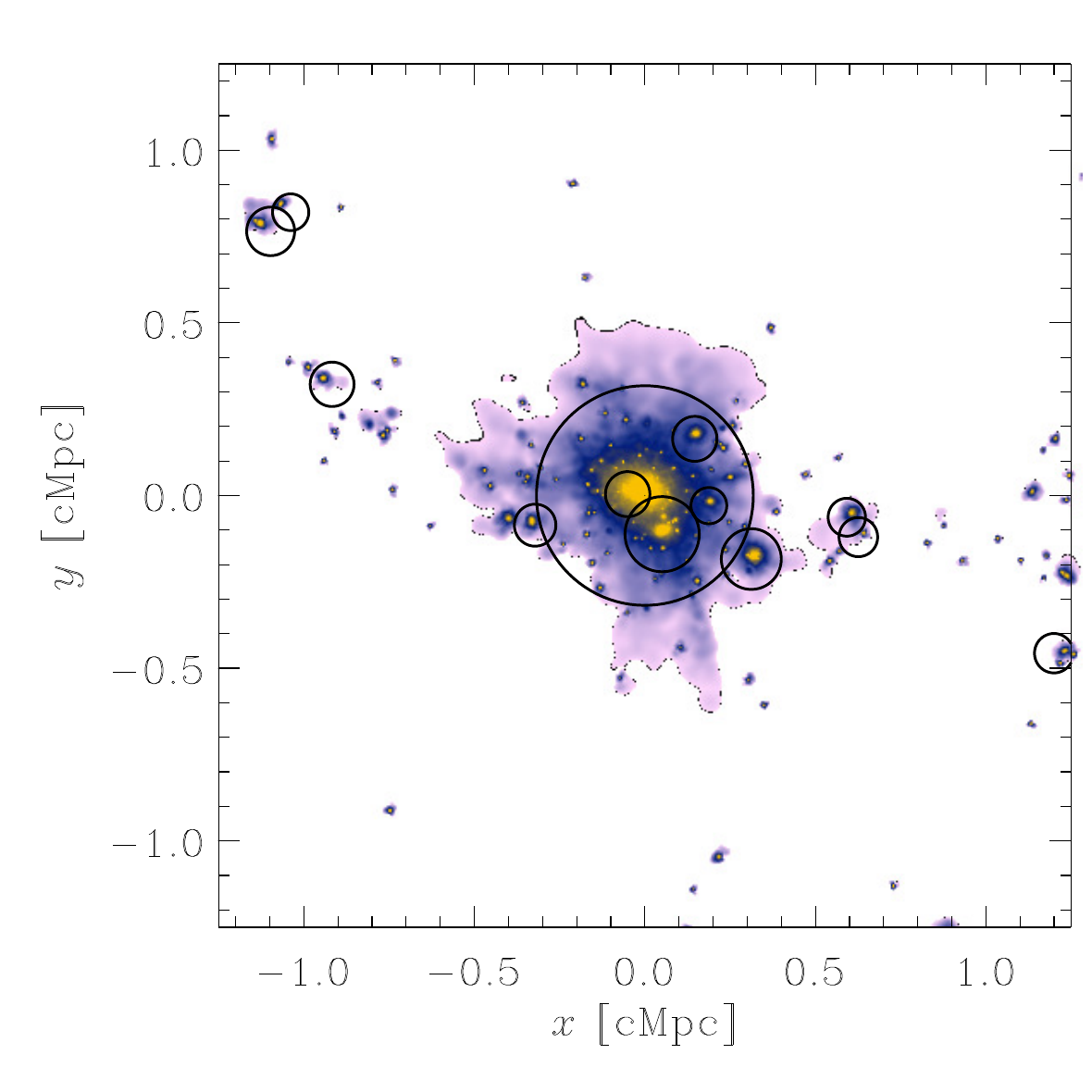}

\vspace{0.5cm}

{\hspace{0.8cm}\textbf{ Halo stars} \hspace{0.2cm} \textbf{$T$=1 Gyr}  \hspace{3cm} \textbf{$T$=2 Gyr}  \hspace{3cm} \textbf{$T$=3 Gyr}  \hspace{3cm} \textbf{$T$=4 Gyr} }\vspace{0.3cm} 
\includegraphics[width=4.5cm]{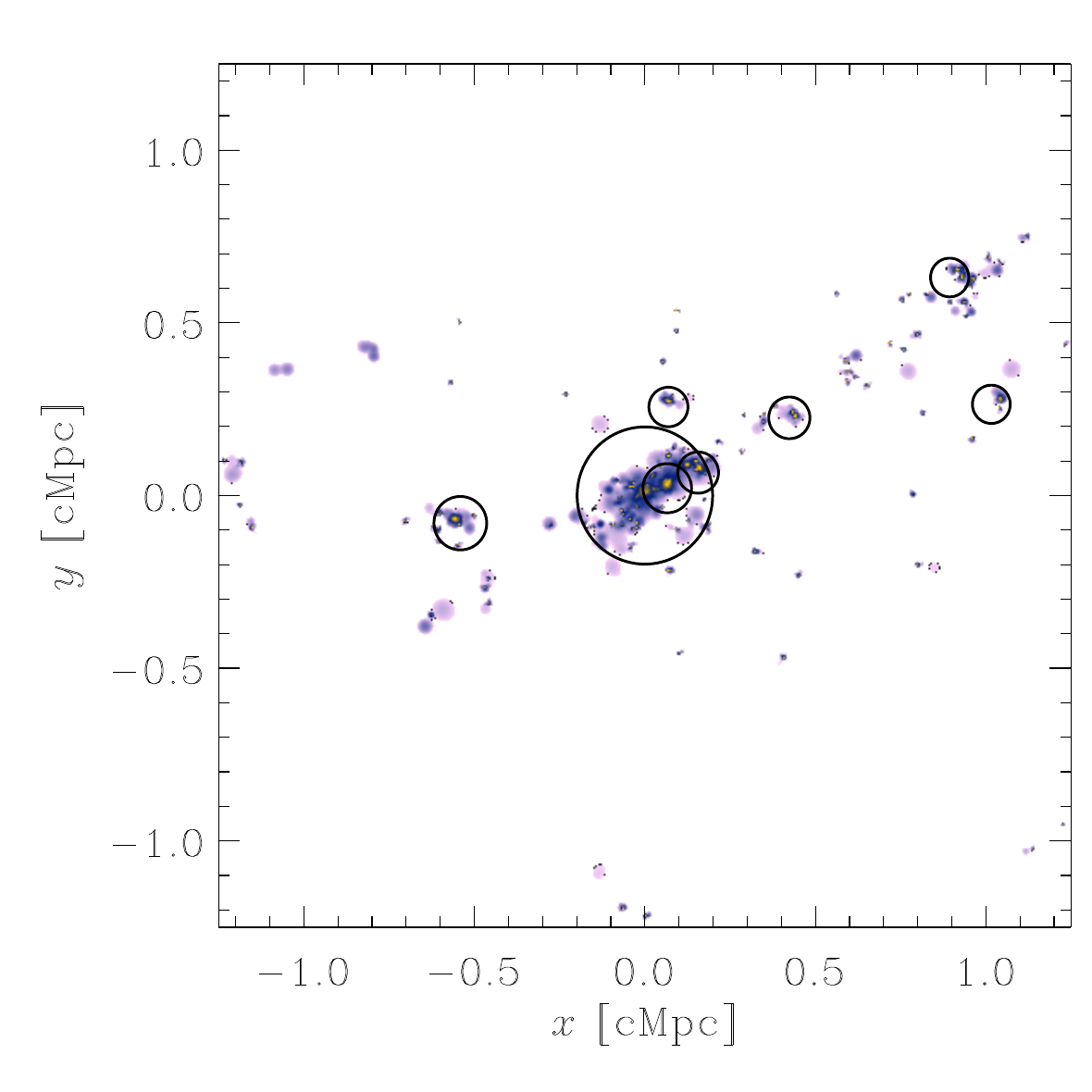}\includegraphics[width=4.5cm]{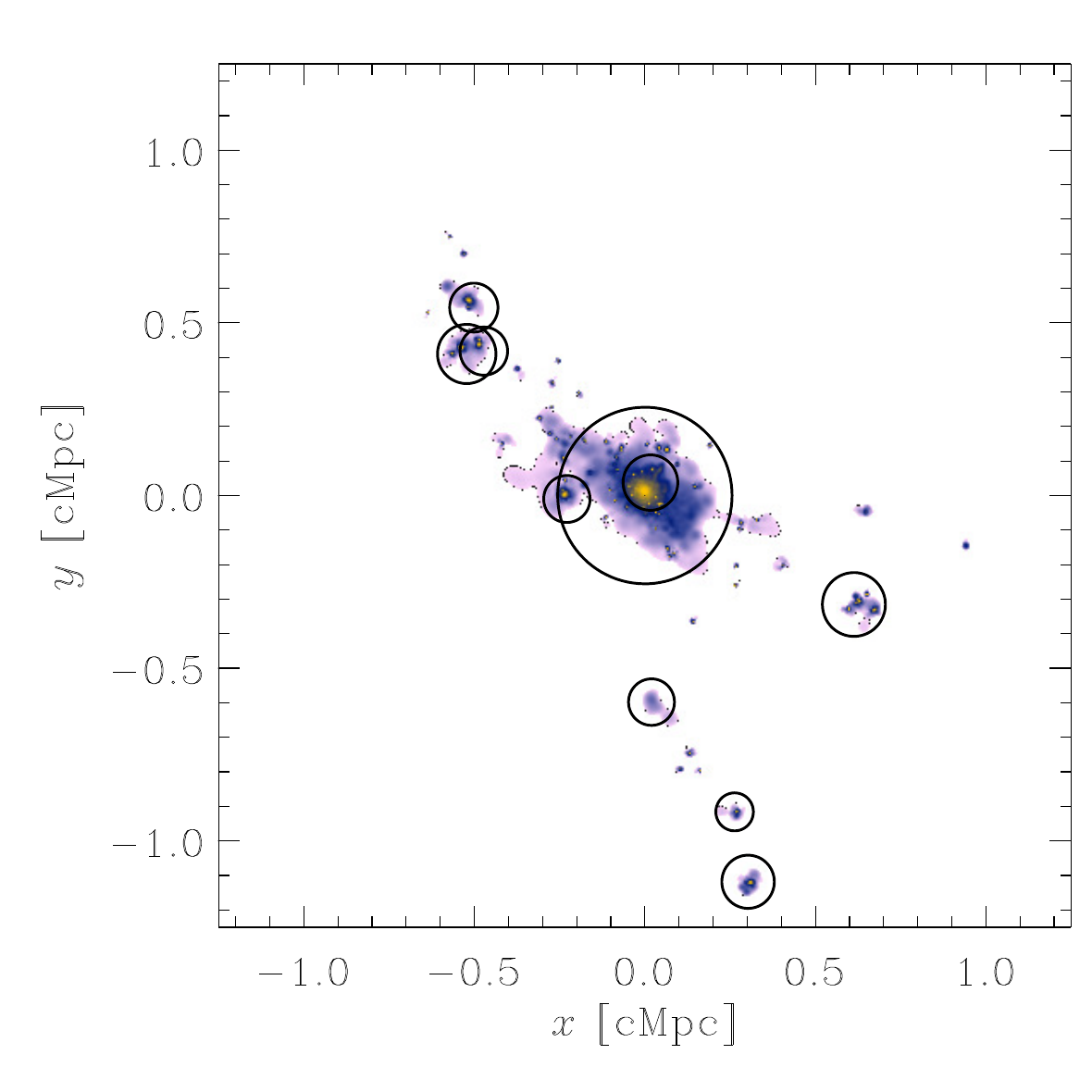}\includegraphics[width=4.5cm]{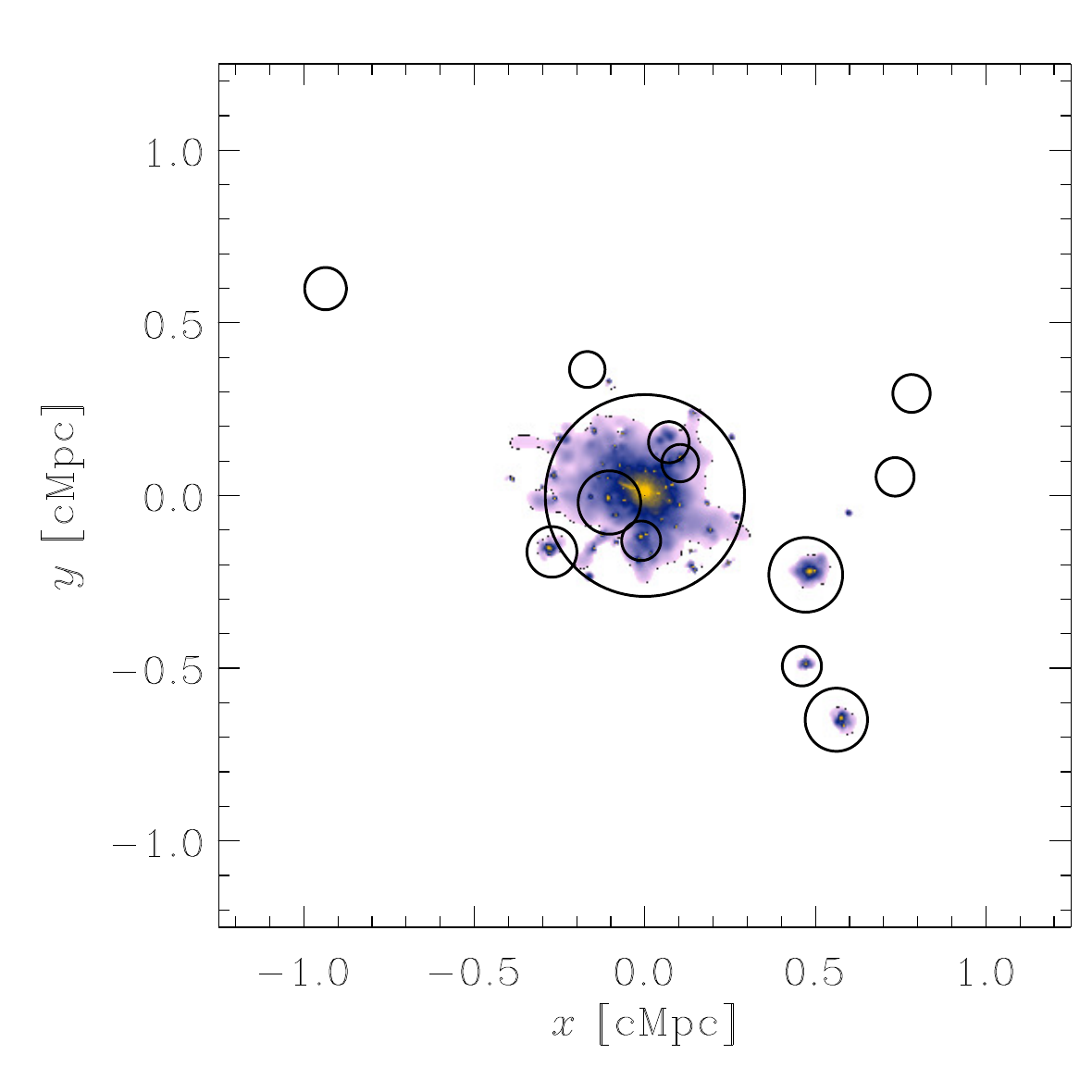}\includegraphics[width=4.5cm]{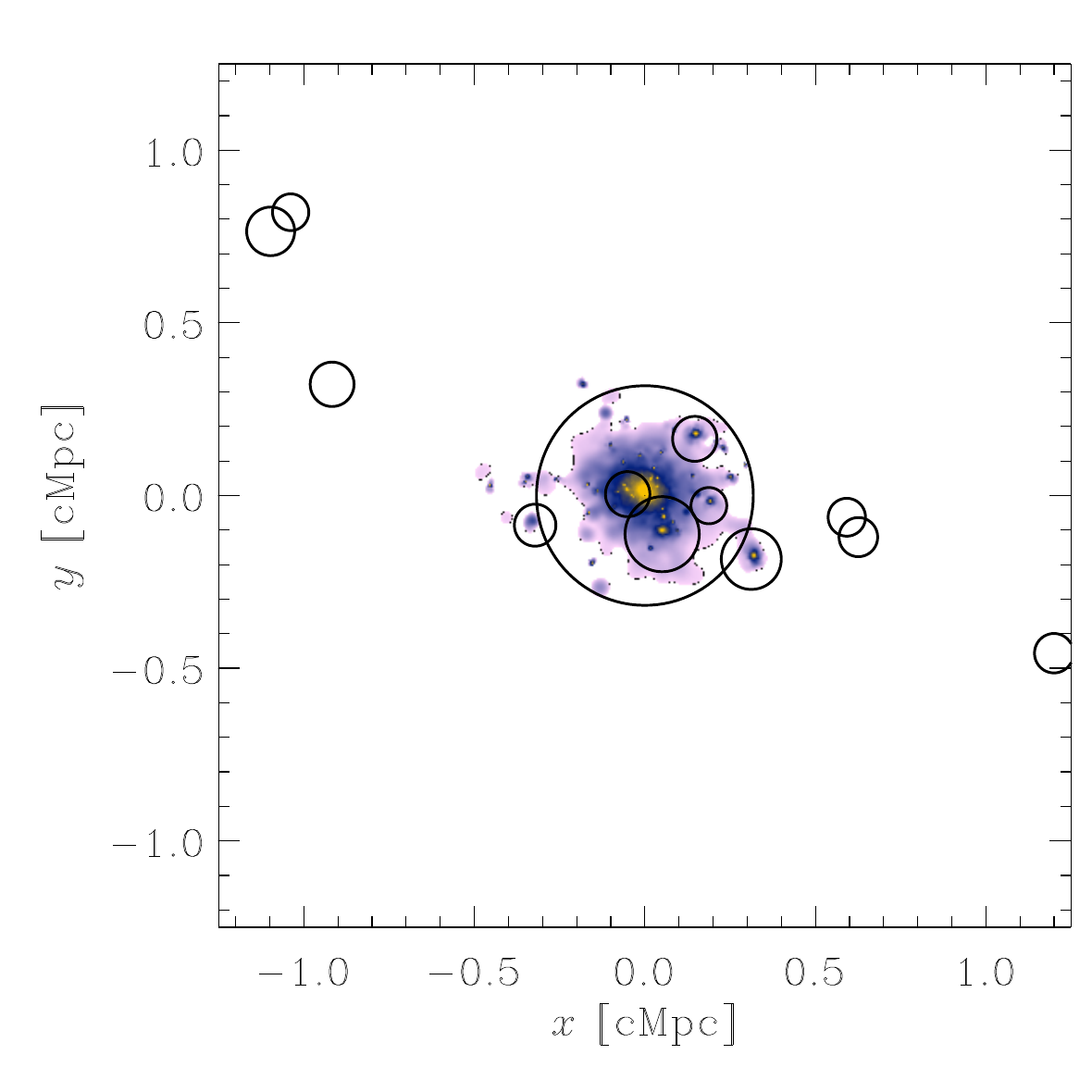}

\caption{Spatial distribution of stars that ended up in the main galaxy at $z=0$, after $1$, $2$, $3$ and $4$ Gyr of evolution. The upper panels include the whole stellar population, while the lower panels are restricted to stars that ended up in the stellar halo component. The black circles show the position of the main progenitor and the most massive satellite galaxies at the varios times; their sizes are scaled with the corresponding masses. The colour scale represents the density of stellar mass in logarithmic scale, to allow a better visualization of the halo stars and their spatial location. }
\label{fig:stellar_halo_formation}
\end{figure*}

\begin{figure*}
  \centering
\includegraphics[height=5cm]{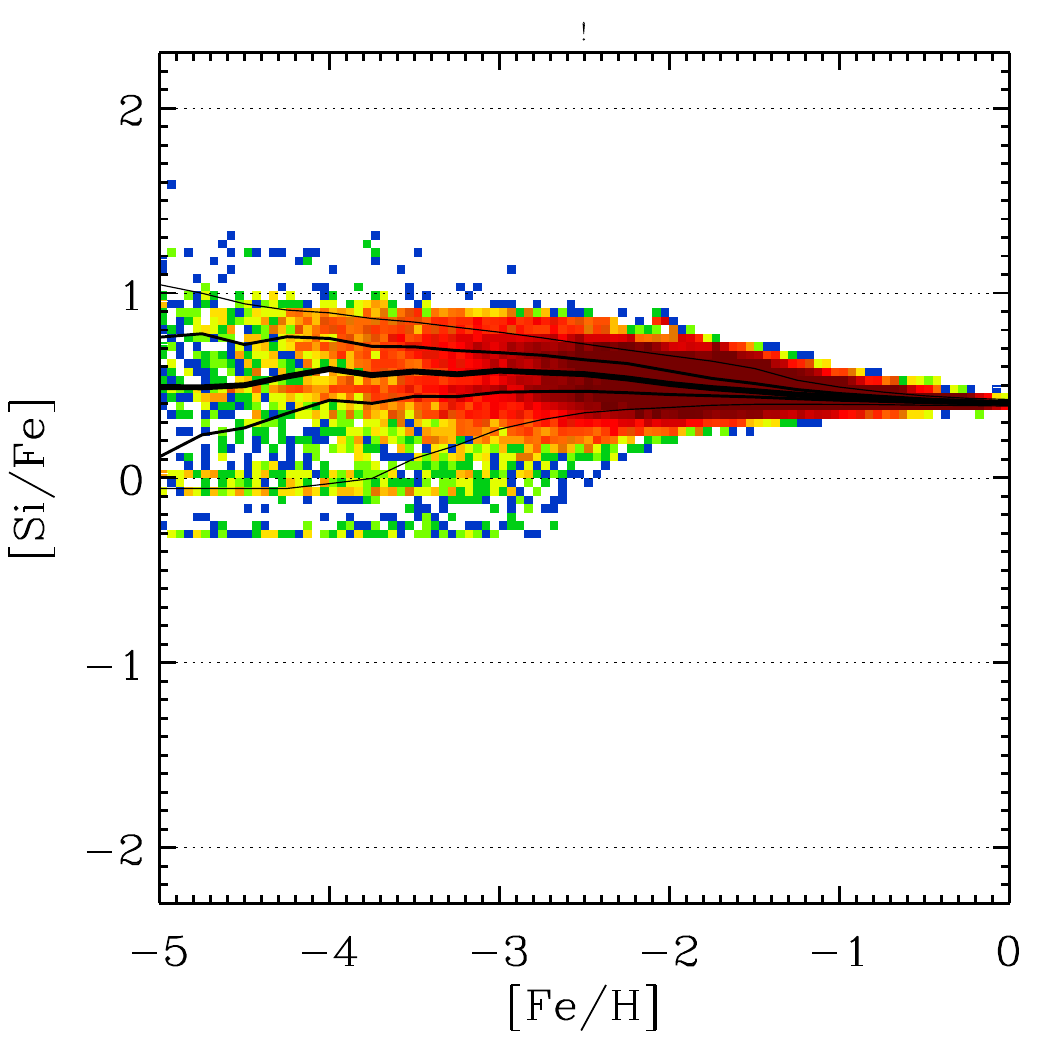}\includegraphics[height=5cm]{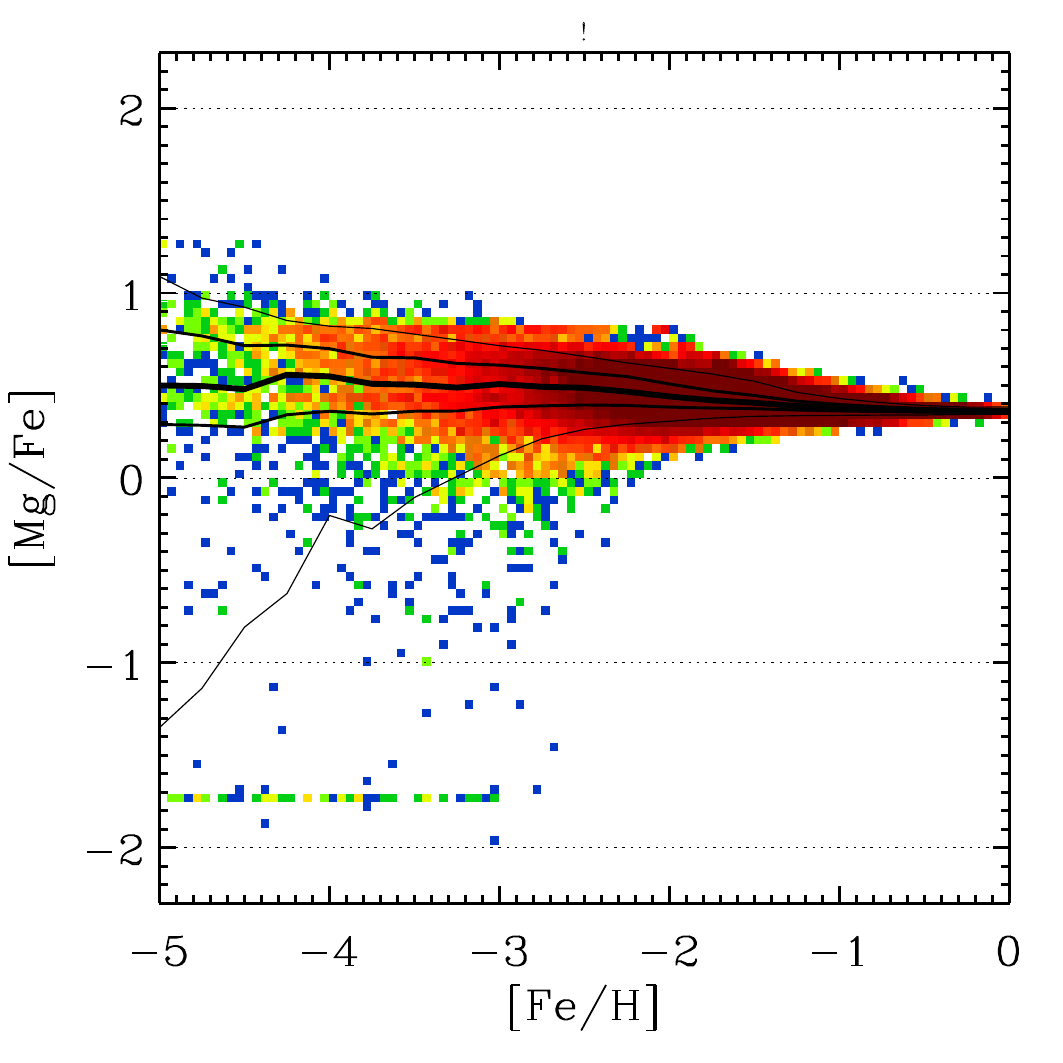}\includegraphics[height=5cm]{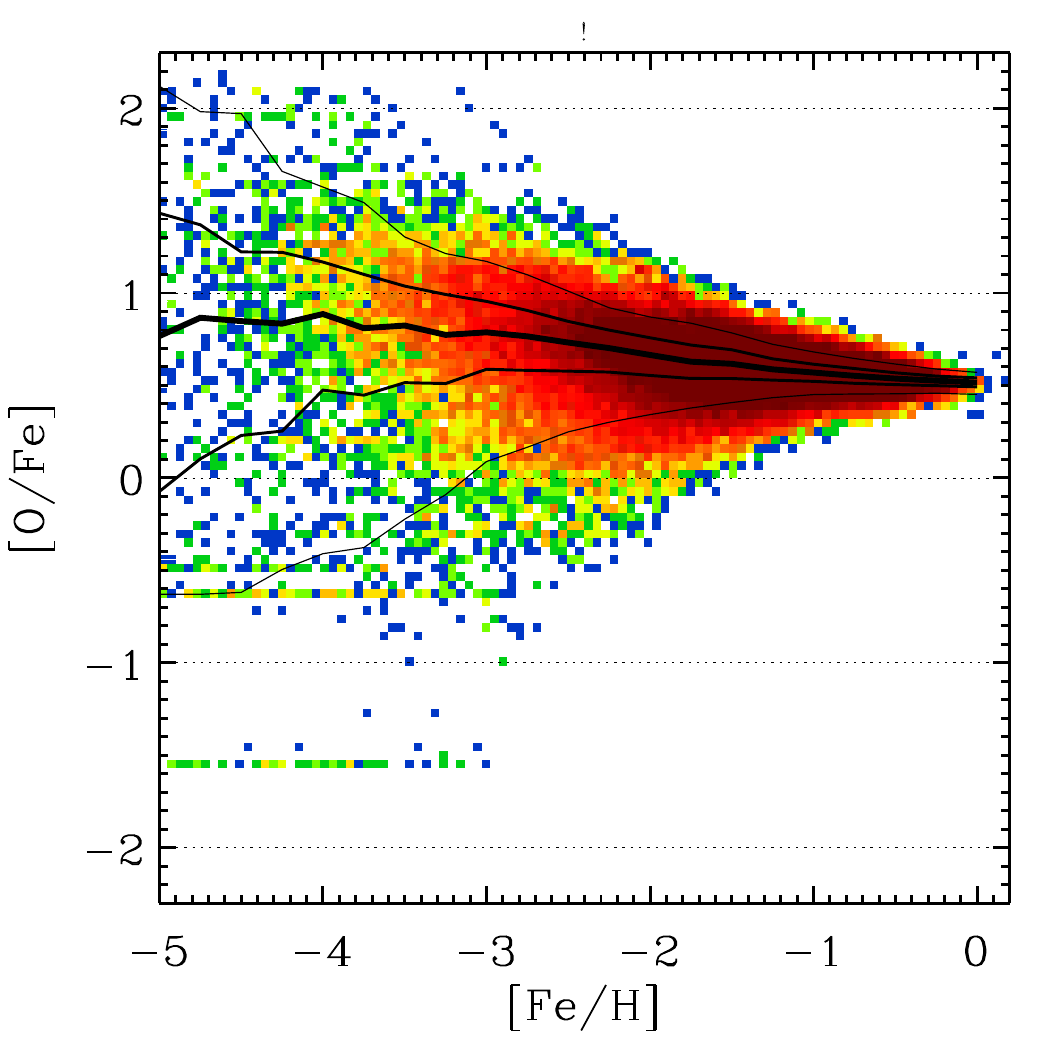}\includegraphics[height=5cm]{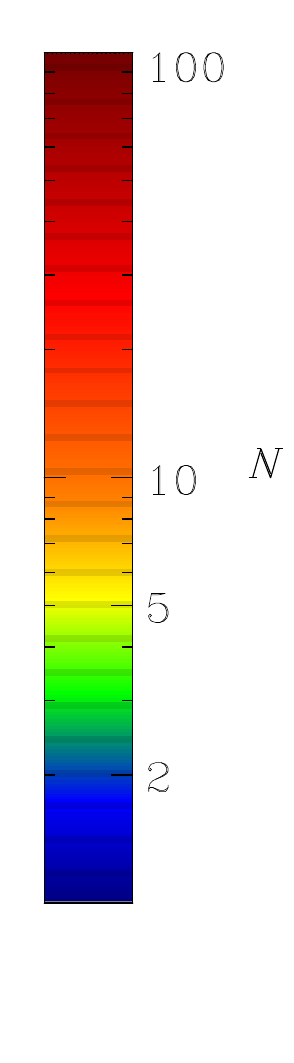}

\includegraphics[height=5cm]{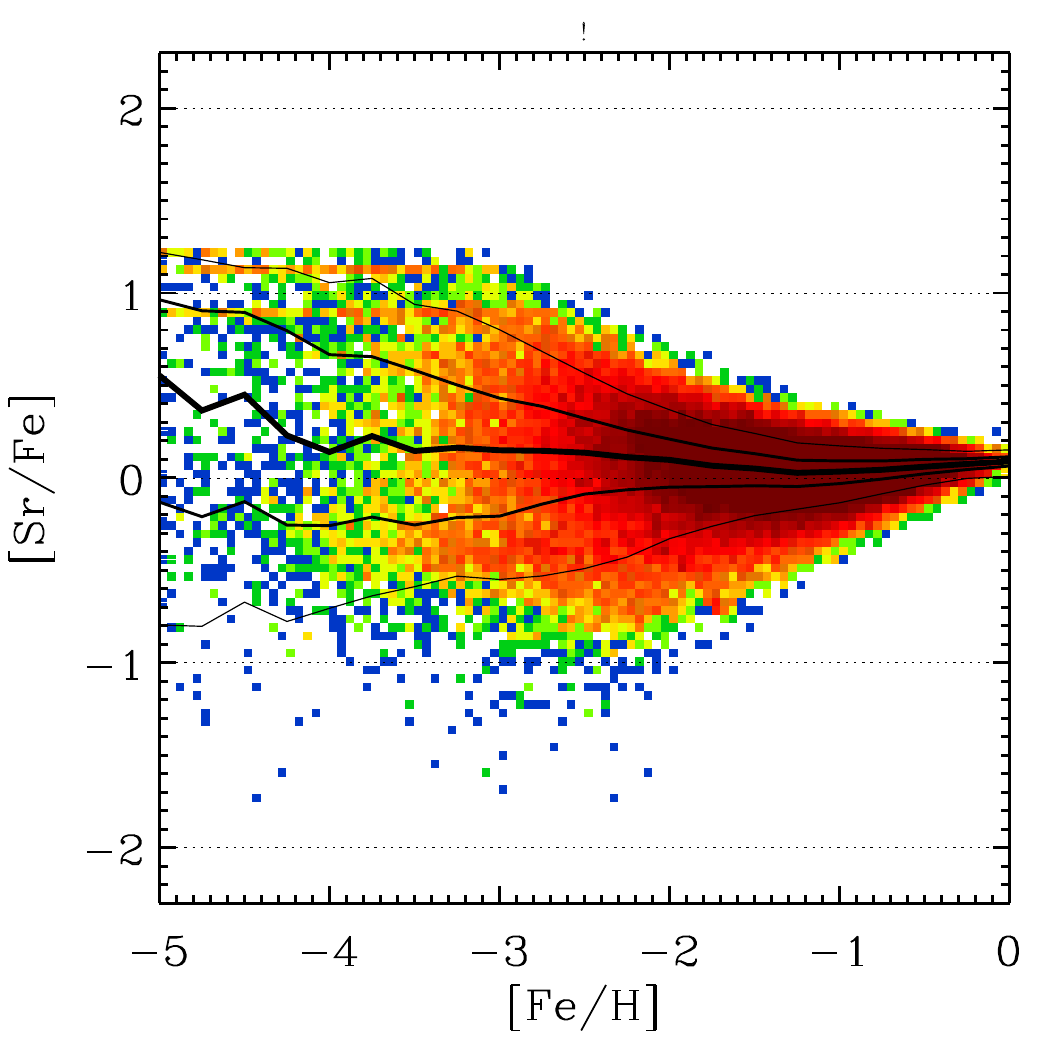}\includegraphics[height=5cm]{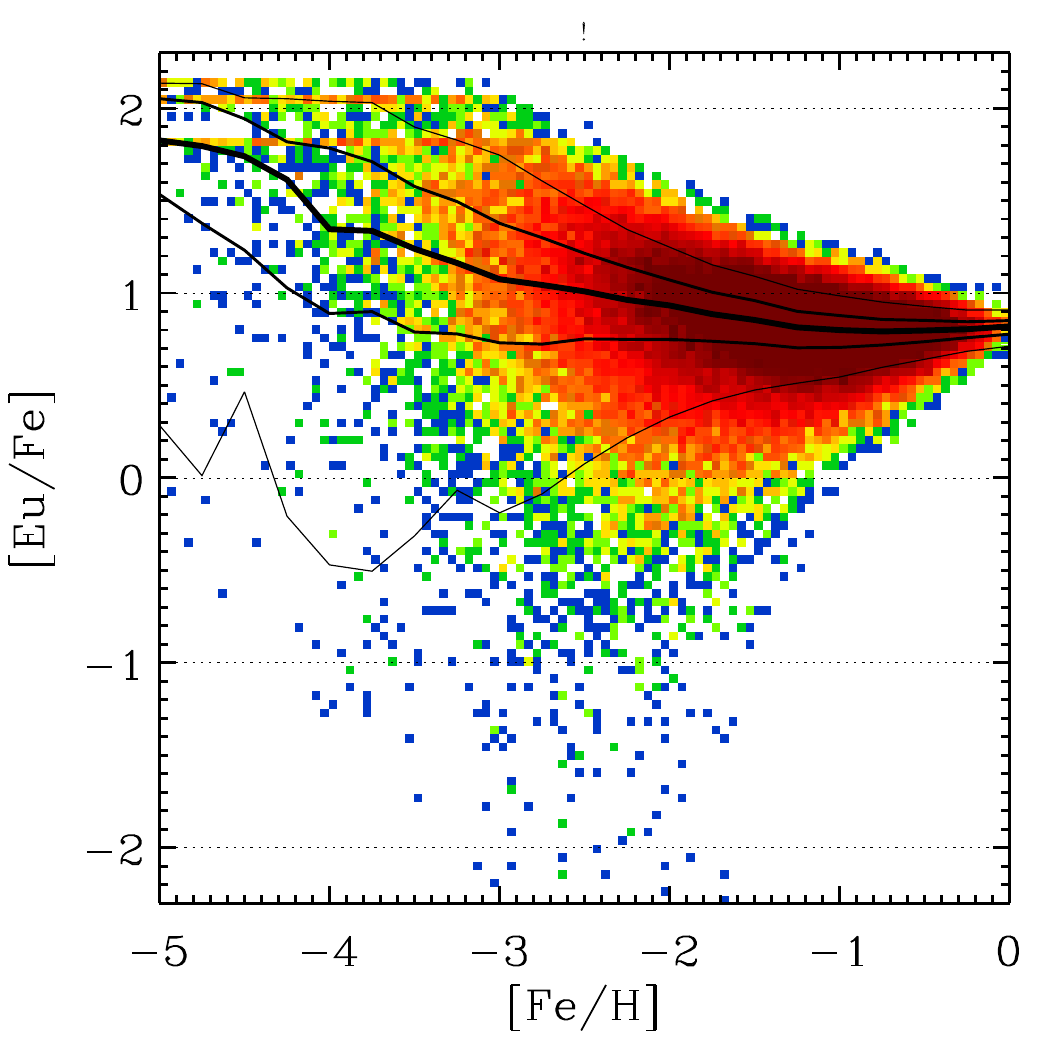}\includegraphics[height=5cm]{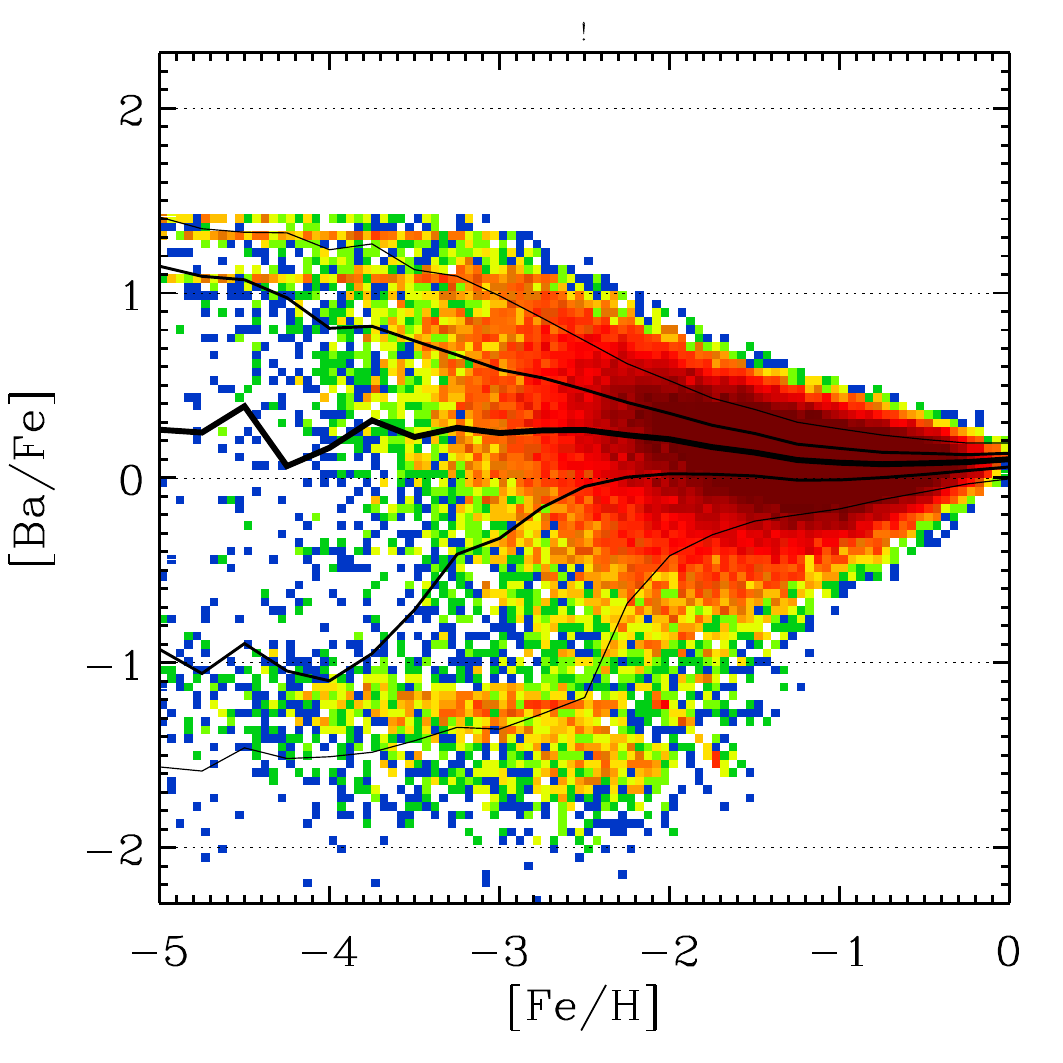}\includegraphics[height=5cm]{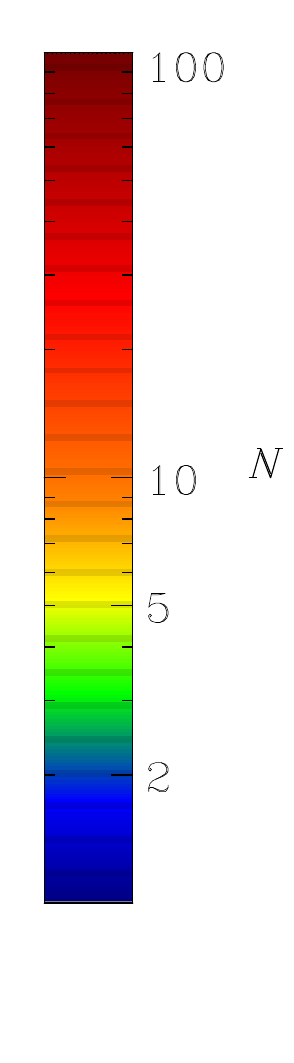}

\caption{The various abundance ratios as a function of [Fe/H] for the halo stars in our simulation ME.  The solid thick lines show the corresponding median  values and the thin lines denote the $90$ $70$, $30$ and $10$ percentiles of the distributions.
The colour scale is also shown.}
\label{fig:el_ratios_cosmo_halo}
\end{figure*}

In the cosmological context, galaxies  grow via the continuous aggregation of matter and smaller substructures which can contribute
distinct chemical patterns determined by the properties of the star formation activity within them and  the characteristics of their accretion onto a larger system.
In the simulation, we can trace back the formation of the $z=0$ system identifying,
at each time, the main progenitor -- i.e. the most massive structure at that time --
and a series of smaller systems  that later accreted onto the main progenitor. 

Figure~\ref{fig:stellar_halo_formation} shows a series of snapshots of the
distribution of all stars that ended up in the main progenitor at $z=0$ (upper panels), and in the stellar halo component (lower panels).  We focus on the early evolution, and show the distributions at 1, 2, 3 and 4 Gyr, characteristic of the formation and accretion of today's halo stars  (recall that, by definition, halo stars formed in the first 1Gyr of evolution, but they could have been accreted onto the main progenitor later on). 
The figures are centered at the mass center of the main progenitor at the 
different times, and the circles indicate the position and size of the different subhalos of the simulation, including the main progenitor. 
The upper panels of this figure reveal that, as expected within the cosmological context considered, the stellar component of the simulated galaxy has a contribution of material formed in systems other than the main progenitor.  If we consider all stars that form the $z=0$ progenitor (i.e. all stellar components), the fraction of stellar mass that has been accreted is low.

However,  the simulated stellar halo has
significant contribution of 
ex-situ stars that formed outside the main progenitor, as can be seen from the lower panels of Figure~\ref{fig:stellar_halo_formation}. The ex-situ fraction of our simulated halo is  $0.8$, i.e., only 20$\%$ of the stars in today's  halo formed in the main progenitor. 
This means that most halo stars formed in smaller, more metal-poor systems, with more episodic star formation activity, compared to the main progenitor.
 A high ex-situ fraction of the stellar halo is consistent with  observations of the Milky Way stellar halo, although in our simulation many small satellites contributed to the stellar halo, while recent discoveries suggest that a large part of the Milky Way's stellar halo could be the result of an early and massive accretion event (\citealt{Helmi18,Belokurov19} - see review by \citealt{Helmi20} and references therein).

It is worth noting that the ex-situ stars that end up in the stellar halo  at $z=0$  locate preferentially in the outskirts of their systems before accretion, as evidenced from the lower panels of the figure. The reason for this behaviour is that stars in the outer regions are the less bound stars, likely to be lost during the merger. As a result, these stars stay in the halo component and are not able to reach the bulge region, unlike the central, most bound stars in the satellites.  How this compares to the most recent observations of the Milky Way bulge will be addressed in a following paper.

As we discuss in the next subsections, the high contribution
of accreted stars to the stellar halo leaves imprints in the $z=0$ chemical properties of this component, particularly in the case of neutron-capture elements.

\subsection{Abundances ratios and scatter of halo stars}

\begin{figure}
  \centering
\includegraphics[width=6.5cm]{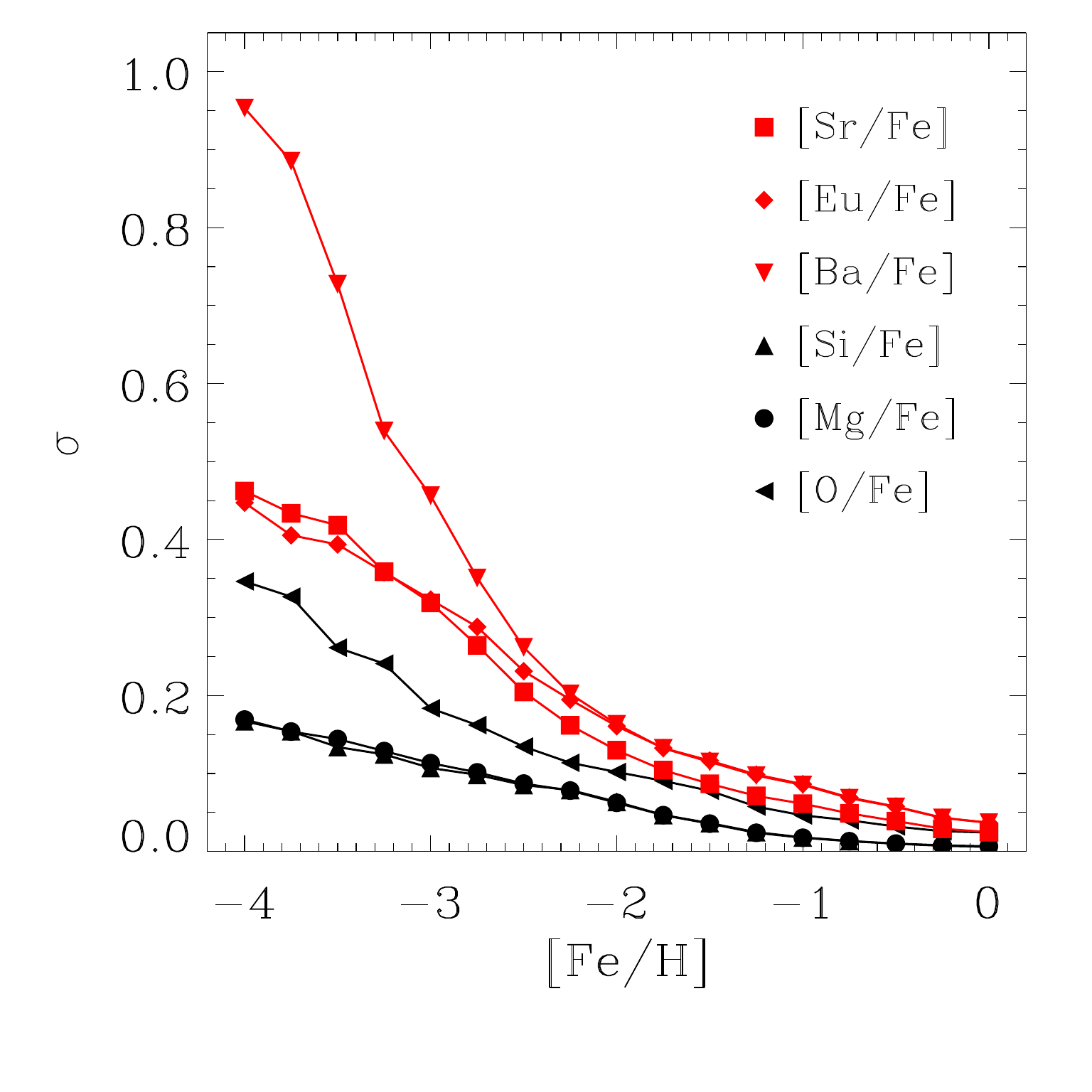}

\caption{ Estimation of the scatter $\sigma$ of the  various abundance ratios of halo stars, as a function of [Fe/H],  in our cosmological  simulation ME. The $\sigma$ value has been calculated as half the difference between the abundance ratios corresponding to the $70$ and $30$ percentiles of the distributions. }
\label{fig:sigma}
\end{figure}

We discuss in this Section the predicted abundance ratios
 and  scatter  of stars in the simulated stellar halo at $z=0$ in our simulation ME. 
Fig.~\ref{fig:el_ratios_cosmo_halo}
shows the distributions of $\alpha-$ and neutron capture-elements, relative
to Fe, as a function of the [Fe/H] abundance. 
The black thick lines indicate the median  of the distributions, while the thin lines denote the $90$, $70$, $30$ and $10$ percentile levels.
Similarly to our findings for the isolated galaxy simulations, the various elements have particular abundance levels and, more importantly, they exhibit distinct dispersions.
In particular, the $\alpha$-elements show lower scatter compared to the neutron-capture elements, and this effect  is more pronounced in the low-metallicity regime. 
The $\sigma$ values, estimated as half the difference between the values corresponding to the $70$ and $30$ percentiles of the distributions,
as a function of [Fe/H], are shown in Fig.~\ref{fig:sigma} for the various elements.
Note that the scatter varies significantly with the [Fe/H] abundance, such that the higher
the metallicity, the lower the scatter.  The typical scatter for the $\alpha-$elements  is  $\lesssim 0.35$ dex, while for the neutron capture elements the scatter levels are systematically and significantly higher,  particularly for [Fe/H]$\lesssim -2.5$.  Ba is the element with the highest $\sigma$ values, with Si having the lower dispersion levels.

\begin{figure}
  \centering
\includegraphics[width=8cm]{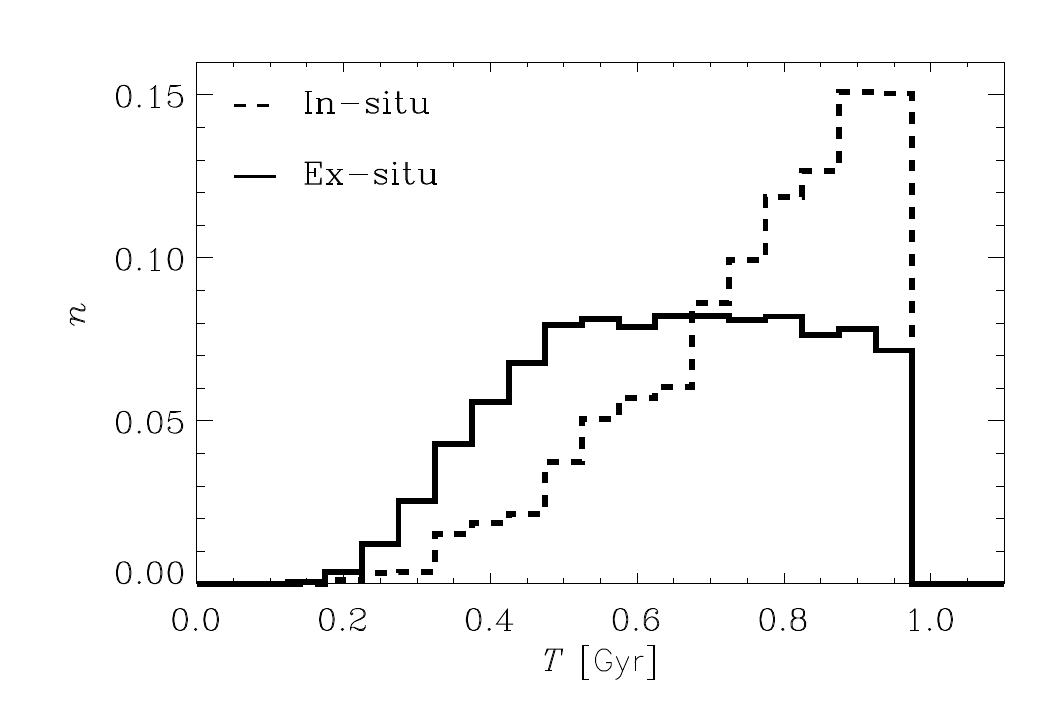}

\caption{Normalized distributions of stellar ages of halo stars, separated by ex-situ (solid lines) and in-situ (dashed lines) populations. 
}
\label{fig:hist_age_insitu_exsitu}
\end{figure}

The different scatter levels are the result of the chemical yields and their dependence with the stellar mass and metallicity, which
varies between the $\alpha-$ and neutron capture-elements, and also between elements of the same group\footnote{A similar behaviour is found in the stochastic models discussed below. Note that without considering SNIa and AGBs, the maximum scatter is due to the differences in the yields of the massive stars.}.
This is because the  predictions of the simulation for the different elements
and their scatter depend on how the yields for each element change with the stellar mass and
metallicity.
In particular, the Si yields are more tightly connected to iron and therefore the scatter is the smallest for the $\alpha-$elements. 
In the case of the neutron capture-elements, it is important to note that Eu is almost completely produced by r-processes, contrary to Ba and Sr which are also produced via s-processes, and therefore the different n-capture elements have different scatter levels.

It is worth noting that considering the cosmological evolution produces, as expected, more complex abundance ratio distributions compared to the results obtained with our idealized simulations of isolated galaxies. In particular, a small but non-negligible fraction of stars have
 [Ba/Fe]$\lesssim -1$, different from the overall distribution which has
higher [Ba/Fe] values. We discuss the origin of this group in the next Sections.

\begin{figure*}
  \centering

\includegraphics[width=5.5cm]{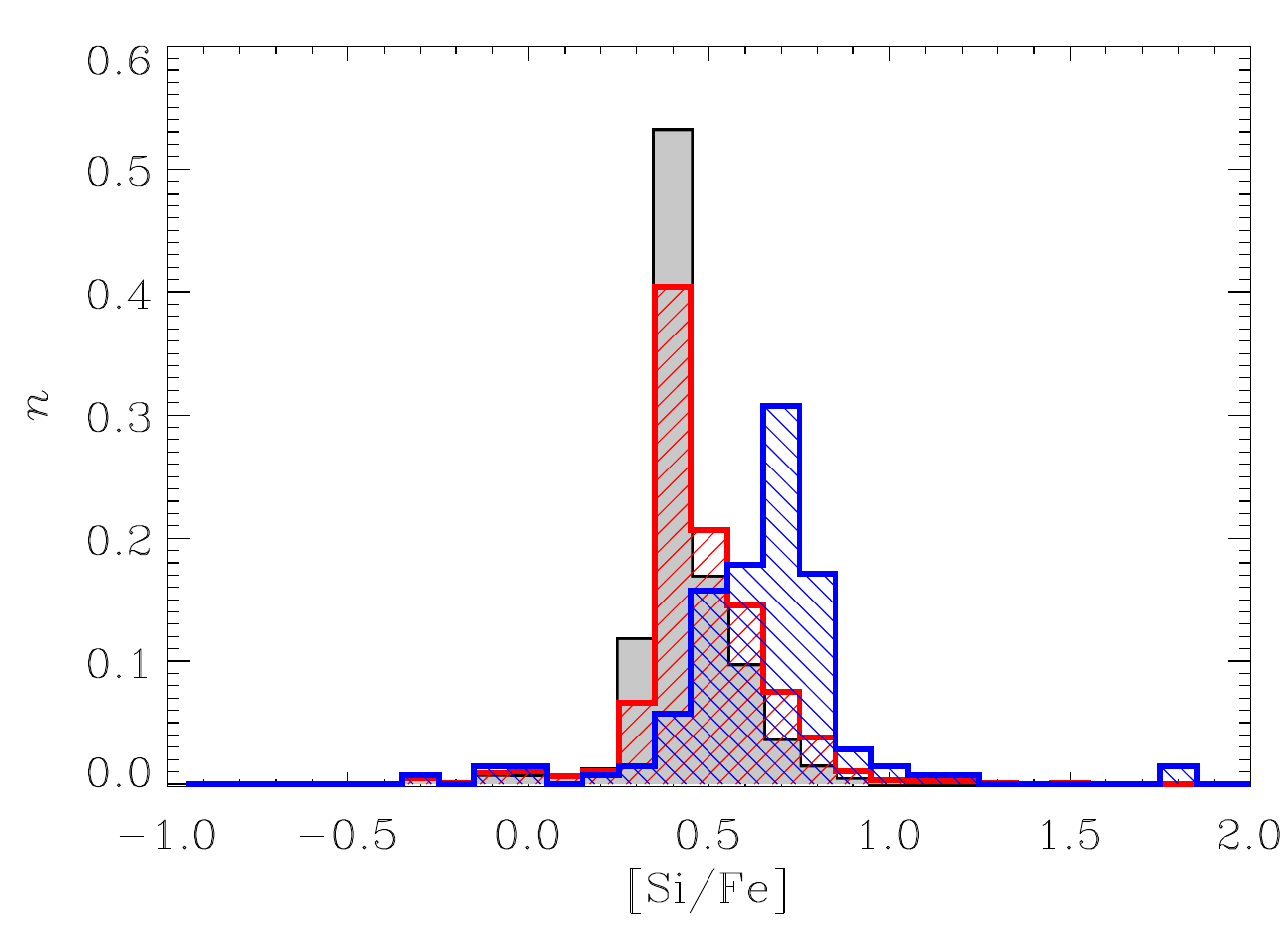}\includegraphics[width=5.5cm]{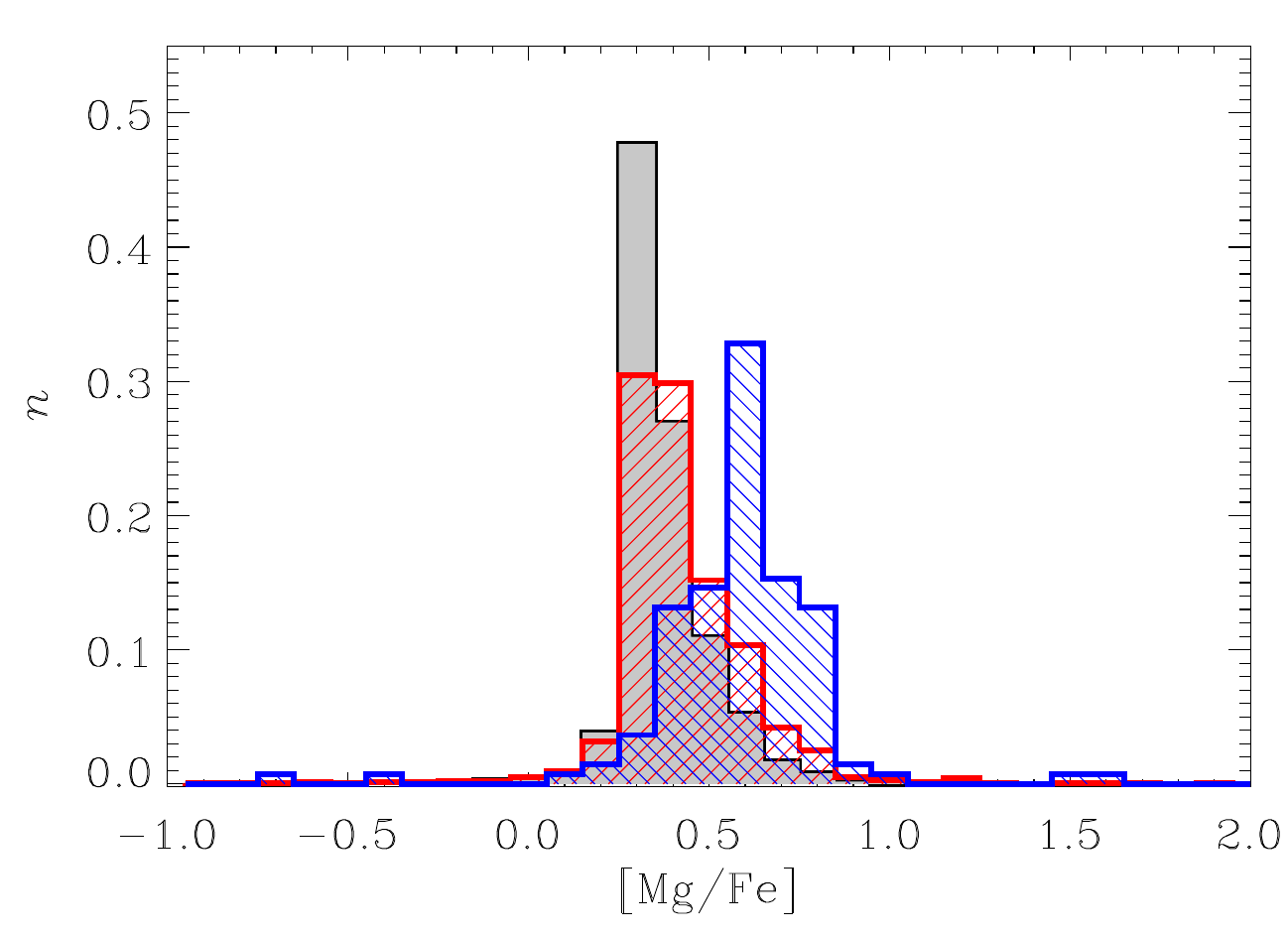}\includegraphics[width=5.5cm]{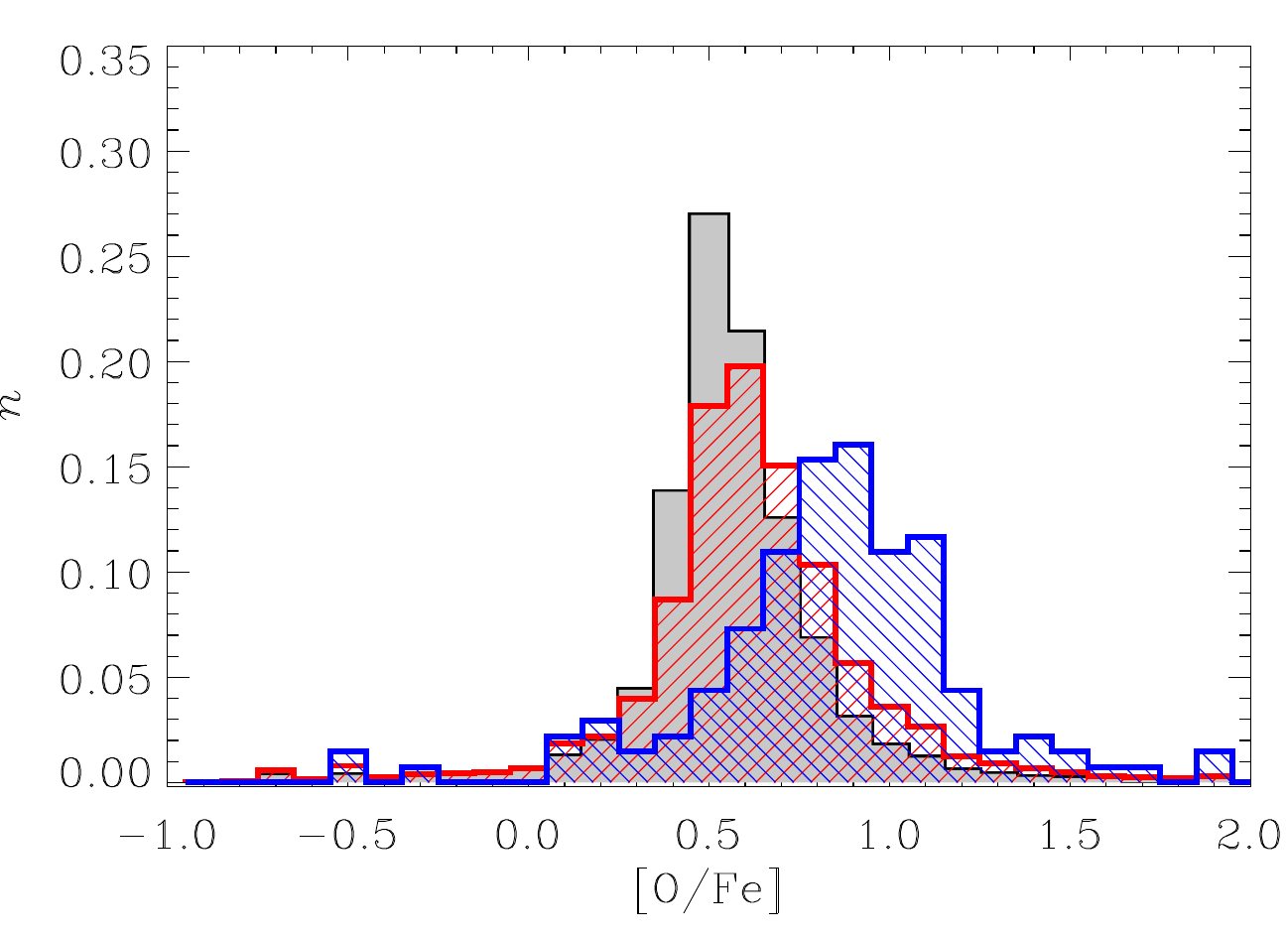}

\includegraphics[width=5.5cm]{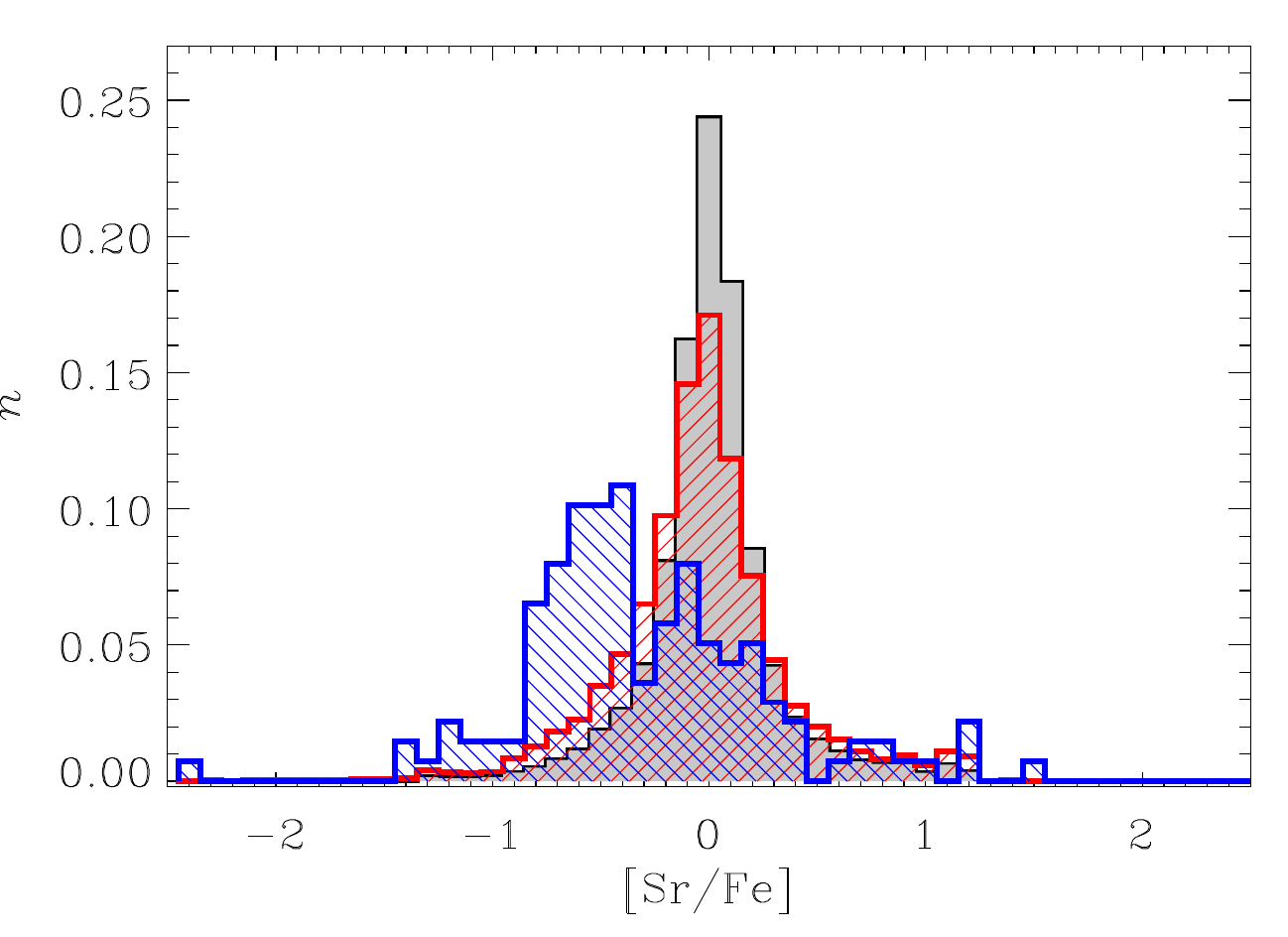}\includegraphics[width=5.5cm]{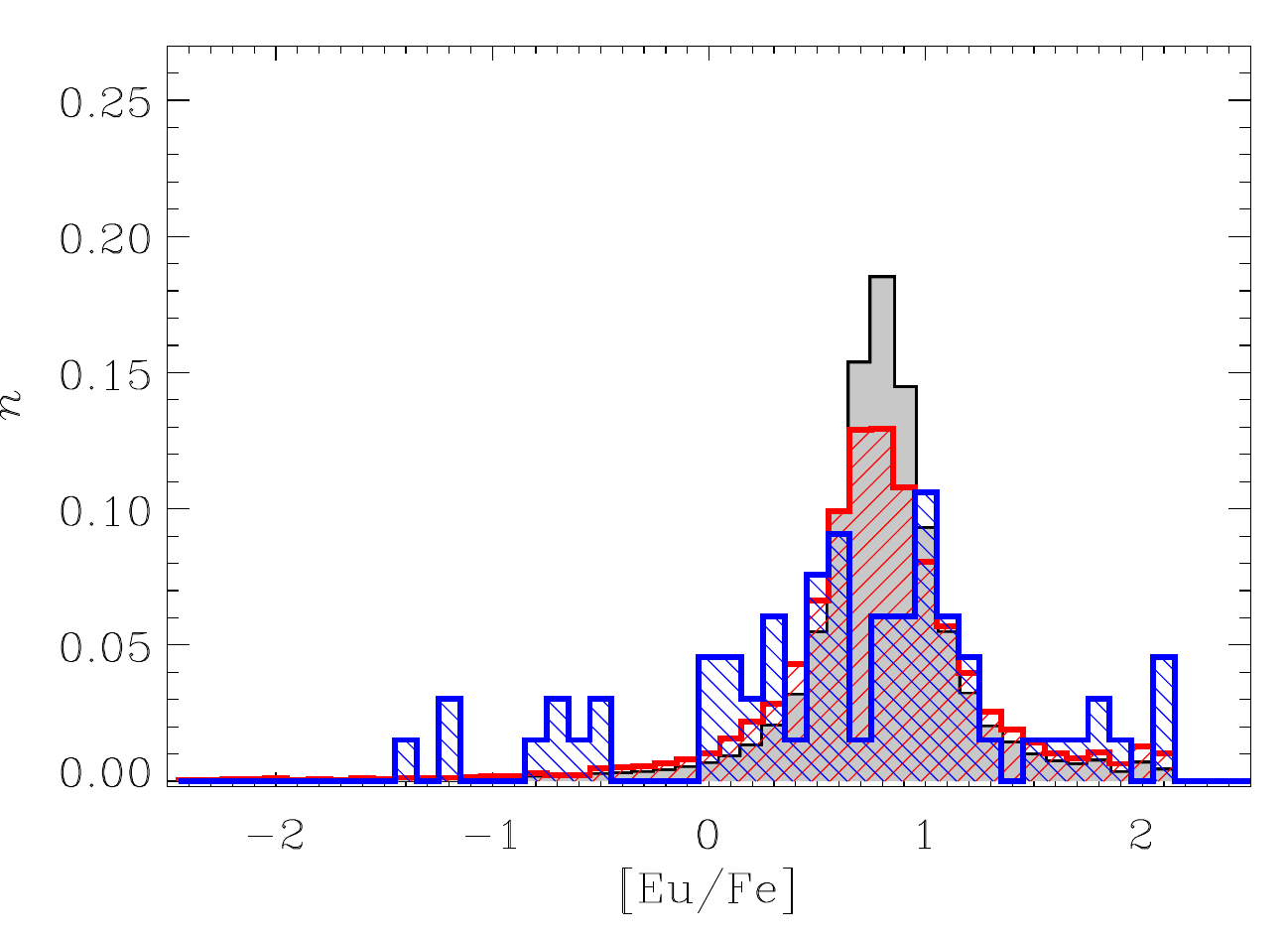}\includegraphics[width=5.5cm]{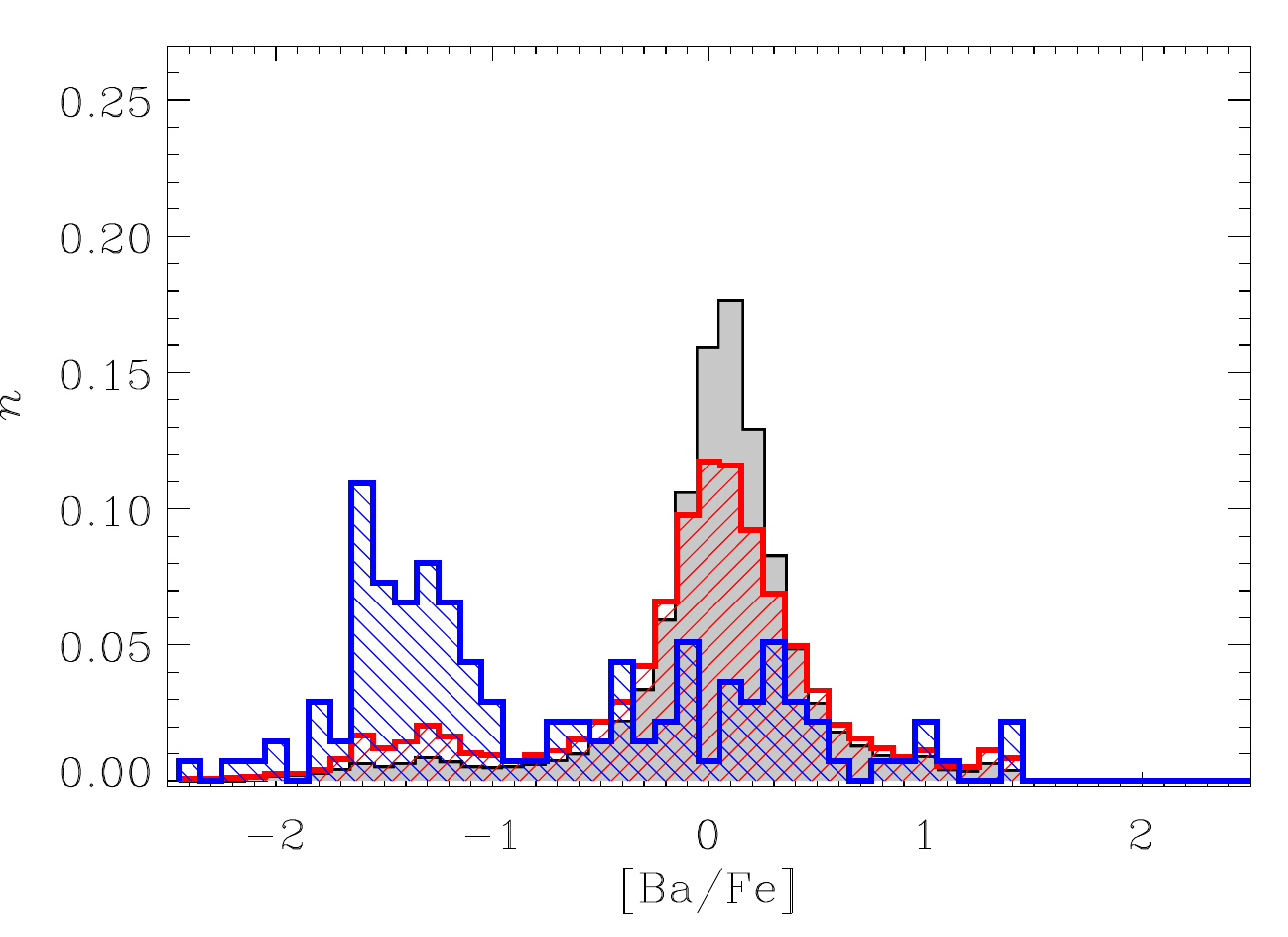}

 \caption{Normalized distribution functions of the various element ratios
 for our  fiducidal halo sample, which assumes an age threshold for halo stars of  $\tau=1$ Gyr (black line with grey shade). We also show results for the very old stars, formed until 100 Myr after the formation of the first star ($\tau=0.24$ Gyr, blue), and for an intermediate value of $\tau=0.5$ Gyr (red).  Note that the various panels have different $x-$ranges in order to highlight the features of the distributions for different elements.}
\label{fig:hists_ME_halo}
\end{figure*}

\subsection{Fingerprints of formation history on abundances and scatter}\label{sec:stellar_age_dependence}

  As discussed above, the chemical properties of the stars encode information on the enrichment history; in the case of the halo stars the very old populations are particularly sensitive to the early stages of enrichment that follows the beginning of the star formation (SF) activity. In our cosmological simulation ME, the enrichment history is much more complex than the one obtained in the isolated case (Section~\ref{sec:isolated}), as stars that form the present-day stellar halo  were formed in  different systems, each of them with its own SF and enrichment properties. In particular,  the onset of SF in  the main progenitor and in the smaller satellites can occur at different times, and the SF rates can have different levels depending on the mass of the system. In our simulation, the very old stars of today's halo were formed mainly ex-situ and, furthermore, the ex-situ stars have a higher fraction of very old stars compared to the in-situ component, as can be seen from Fig.~\ref{fig:hist_age_insitu_exsitu} where we show the formation time distributions of the ex-situ and in-situ populations (normalized separately for clarity, as ex-situ stars are highly dominant). 

  The very old stars have in fact a significant impact on the element ratios, producing particular features that are not seen for the rest of the stellar halo population. Furthermore, such impact depends strongly on the chemical element considered and the characteristics of their yields and corresponding mass/metallicity dependencies.  Fig.~\ref{fig:hists_ME_halo} shows the various abundance ratios  for the halo stars (i.e. our fiducidal sample, assuming a formation time threshold of $\tau=1~$Gyr, black lines), as well as for the very old stars with $\tau=240$  Myr (i.e., 100 Myr after the formation of the first star, blue) and for an intermediate age threshold of $\tau=0.5~$Gyr (red). In the case of the $\alpha$-elements, restricting the sample to older populations produces a shift to higher abundances, because enrichment with these elements is faster compared to iron: even though both the $\alpha-$elements and iron are produced in SNII, 
  the iron yields are comparatively lower for metal-poor stars. For $\tau=240$ Myr, the interstellar medium did not yet have time to enrich sufficiently so that the most massive stars produce significant amounts of  $\alpha$-elements compared to iron.

  In contrast, for the neutron-capture elements Sr and Ba we find that the very old stars have lower abundance ratios compared to the fiducidal halo sample, with  larger scatter. For Eu, the abundances are similar, but the scatter is also larger when we consider the very old stars. In this case, stars formed from non/low contaminated gas produce both iron and the neutron-capture elements in the same progenitors, but as the interstellar medium is contaminated,  enrichment with iron occurs faster compared to the neutron-capture elements. For this reason, once the very early stages of star formation have completed, the interstellar medium becomes more homogeneous, producing narrower abundance ratio distributions.
In the case of the scatter, considering the very old stars produces wider distributions for all elements, because the interstellar medium is less homogeneous compared to later times.

The case of [Ba/Fe] is particularly important, as we have already seen that the [Ba/Fe] versus [Fe/H] distribution showed the presence of a stellar group with  [Ba/Fe]$<-1$, different from the overall population. From Fig.~\ref{fig:hists_ME_halo} it is clear that it is the very old stars that contribute to this group, and these are also the stars that populate the lowermost [Sr/Fe] tail.

\subsection{The [Sr/Ba] ratio}\label{sec:srba}

\begin{figure*}
  \centering
  \includegraphics[width=7cm]{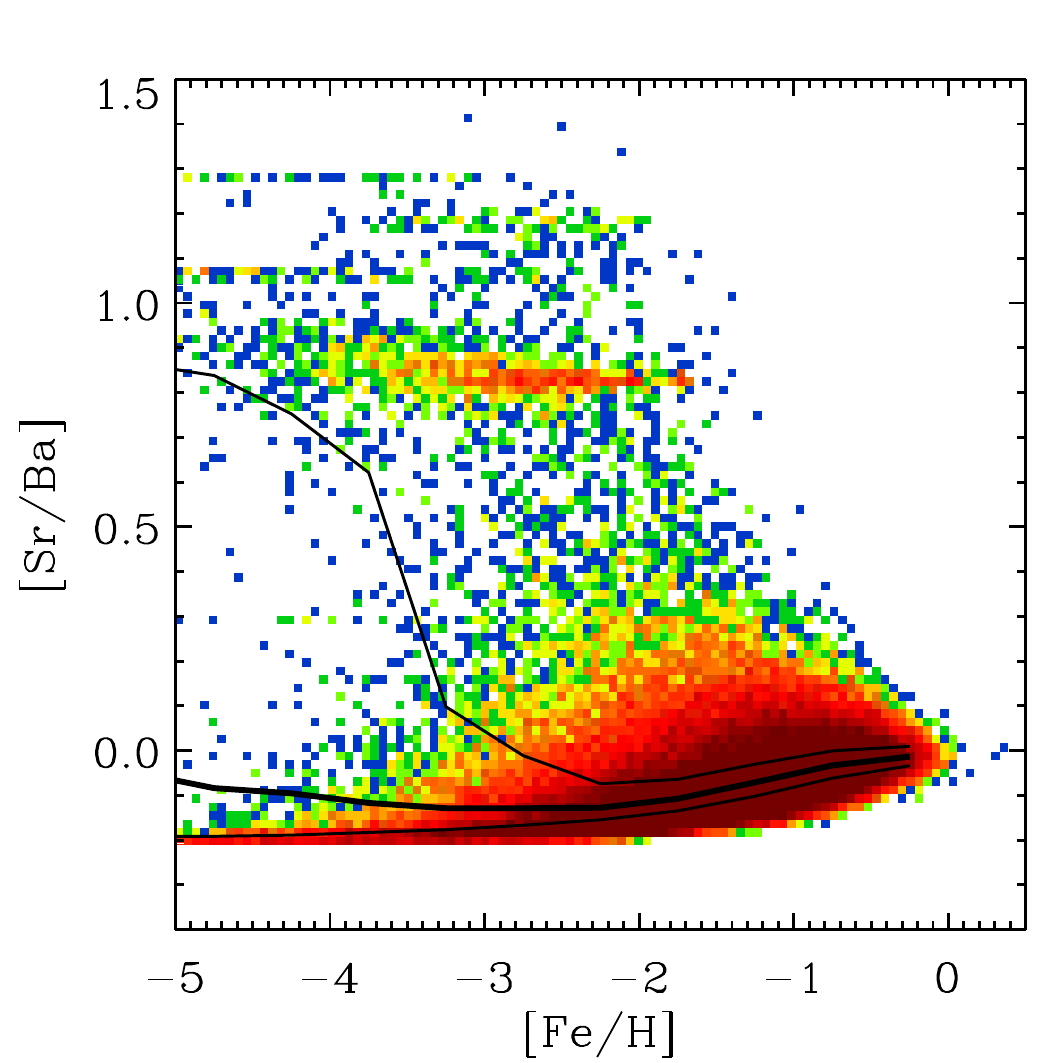}\hspace{-1.2cm}\includegraphics[width=7cm]{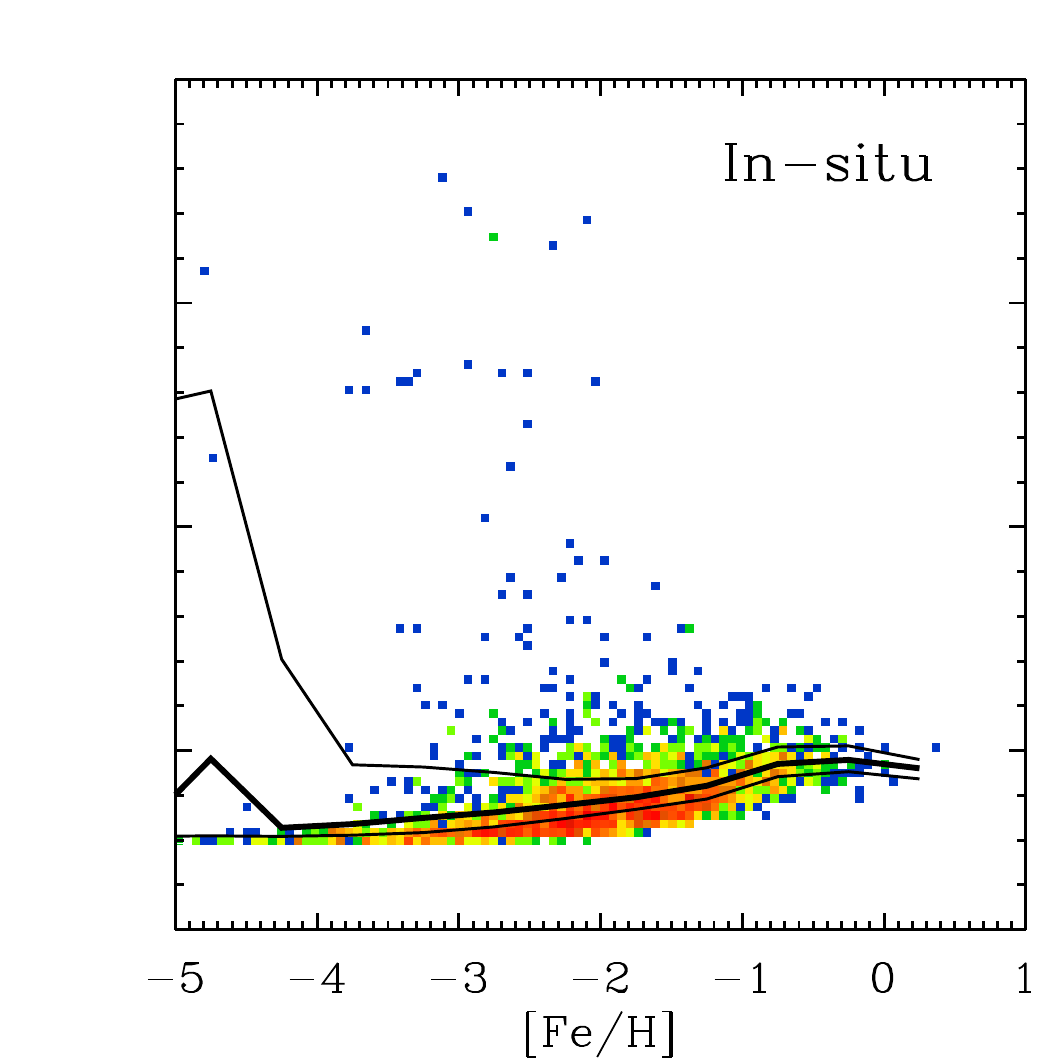}

\caption{[Sr/Ba] abundance as a function of [Fe/H] for all halo stars (left-hand panel) and restricting the sample to the in-situ population (left-hand panel). The solid thick lines indicate the median of the distributions, and the thin lines show the  corresponding  $70$ and $30$  percentiles. The colour scale is also shown. 
}
\label{fig:sr_over_ba_all_and_insitu}
\end{figure*}

The [Sr/Ba] ratio is  an important chemical ratio,  as it is sensitive to the origin of the heavy elements. Observationally, not only the [Sr/Fe] and [Ba/Fe] abundances show a dispersion in our Galactic halo, but also their ratio [Sr/Ba] is not constant \citep{Francois07}. These observations confirm the need for a second neutron-capture process which synthesised differently the light elements (such as Sr) and the heavy ones (as Ba), in addition to the r-process.  Early contribution from fast rotating massive stars to s-process elements has been proposed \citep{Pignatari08,Chiappini11,Frischknecht12,Chiappini13}, as normal s-process enrichment by AGBs operates on long time-scales ($>300$ Myr, see \citealt{Cescutti06}); moreover, at low metallicity,  [Sr/Ba]$<0$ is expected from AGB nucleosynthesis
(\citealt{Cristallo09}, see also \citealt{Frebel18}). 

These ideas were  incorporated in stochastic inhomeogenous chemical evolution models which were  able to predict a scatter in the [Sr/Ba] ratio \citep{Cescutti13,Cescutti14}. These models however, not being in the cosmological framework, do not take into account additional scatter caused by the addition of ex-situ stars resulting from mergers. These smaller galaxies would have their own star formation histories which, in principle, could add more scatter into the more simplified picture of the stochastic inhomogenous chemical evolution models.  This is in fact what we see in our cosmological simulations.

The left-hand panel of Fig.~\ref{fig:sr_over_ba_all_and_insitu} shows the [Sr/Ba] abundance ratio for halo stars, as a function of the [Fe/H] abundance. 
Halo stars  have diverse [Sr/Ba] ratios, in the range $-0.2$ to   $1.3$ dex; 
stars with  [Fe/H]$\gtrsim -2$ have near solar abundances, but  low metallicity stars appear both at sub-solar values and at [Sr/Ba]$\sim 0.8$, although at a lower rate.

It is worth noting
  that the simulation predicts a high scatter for [Sr/Ba], due to  the  combined pollution via r- and s-processes  in fast rotating massive stars, our implementation of the differential enrichment of stars with different masses, and the cosmological evolution.

\begin{figure}
  \centering
\includegraphics[width=7cm]{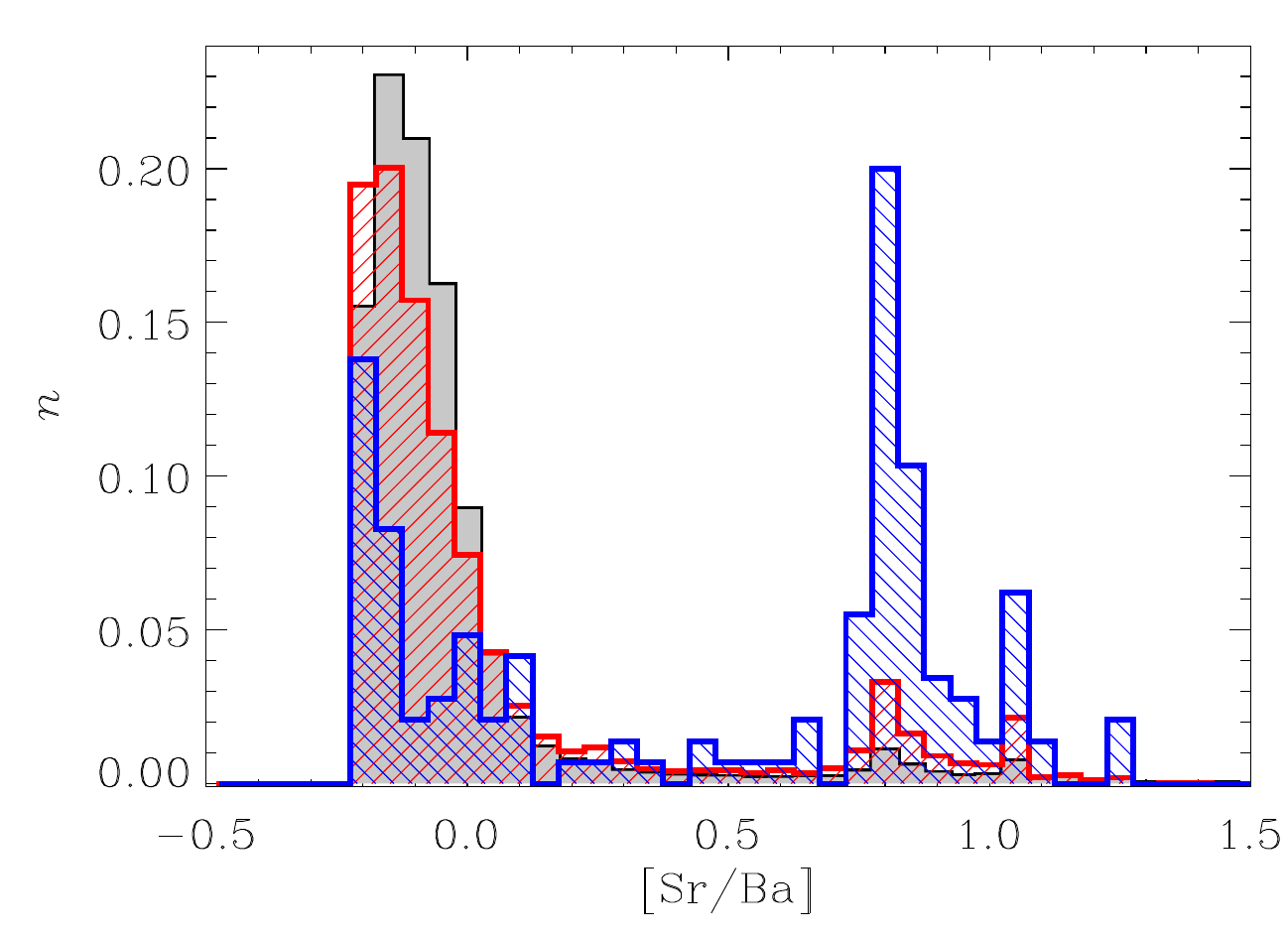}

\caption{Normalized distribution functions of [Sr/Ba] in simulation ME, for our
  fiducidal halo sample, which assumes an age threshold for halo stars of  $\tau=1$ Gyr (black line with grey shade). We also show results for the very old stars, formed until 100 Myr after the formation of the first star ($\tau=0.24$Gyr, red), and for an intermediate value of $\tau=0.5$ Gyr (red). 
  }
\label{fig:sr_ba_halo_and_oldest}
\end{figure}

An important feature of this figure, which follows the characteristics of the [Ba/Fe] and [Sr/Fe] abundances seen above, is the stellar group with  [Sr/Ba]$\sim 0.8$, different from the overall population. Such abundance ratios are only seen for the ex-situ population, and dissapear if we only consider the in-situ stars, as can be observed from the right-hand panel of Fig.~\ref{fig:sr_over_ba_all_and_insitu}\footnote{Note that there are also few in-situ stars with high [Sr/Ba] ratios, although this might be  an artifact of the algorithm to identify subhaloes in the simulation, as it is not able to properly follow the very small systems.}. 
However, note that the key ingredient for forming the high [Sr/Ba] stars is the time-scale, because it is only the very old stars that produced such high ratios. This can be  seen from  Figure~\ref{fig:sr_ba_halo_and_oldest}, where we show the [Sr/Ba] distributions for our fiducidal halo sample with $\tau=1$ Gyr (black), as well as for $\tau=0.5$ Gyr (red) and for the very old stars, with $\tau=0.24$ Gyr (blue). According to our simulation,  the existence of the high [Sr/Ba] stars is a result of the enrichment time-scales for rotating massive stars compared to those assumed for the r-process, to the ages of accreted stars and to the slower enrichment occuring in the small systems that contribute to the formation of the stellar halo, compared to that of the main progenitor.

\section{Discussion}

  The work presented here is a first implementation of the enrichment of various neutron-capture  in cosmological simulations, and it might be relevant to compare, even at this early stage,  to make a comparison with observational data.
Note that a proper comparison with observations requires an extensive study on the probable observational biases and the selection of the most comparable stellar sample of the simulation, and this will be done in future work. 
 In this Section, we make a  broad comparison with observations, and we also compare our results with those from stochastic chemical evolution models which use the same chemical yields and stellar ages. Note that the scatter predicted in the simulation is more important than the exact levels obtained, as the abundance ratios can be easily changed by rescaling the yields, unlike the scatter which depends on the properties of their age/metallicity dependencies.

\subsection{Simulations compared to observations}

In this Section we compare our results for the abundance ratios and scatter of the different elements with the
observations of 
\cite{Roederer14}  and \cite{Yong13}  (only carbon-normal stars), which provide data on the abundances of metal-poor stars in our Galaxy. 
Note that the comparison between simulations and observations  is complex, and it needs to be taken with caution for various reasons. 
On one side, the observational samples are biased to low metallicity stars, and
are complete only for the low metallicity range (approximately at [Fe/H] abundances lower than $-2.5$ or $-3$). Second, these observations are for stars in the solar vicinity, unlike the simulation data which include all stars in the simulated stellar halo. Also, the number of stars in both observational samples is relatively small and so there is no good statistics. Finally, it is worth noting that  the two observational samples are not always compatible with each other,  most notably in terms of the Si and Mg abundance ratios. 
Despite these difficulties, the two datasets are useful as they include the abundances
 of all or
most of the chemical elements studied here.

\begin{figure*}
  \centering

\includegraphics[width=5.5cm]{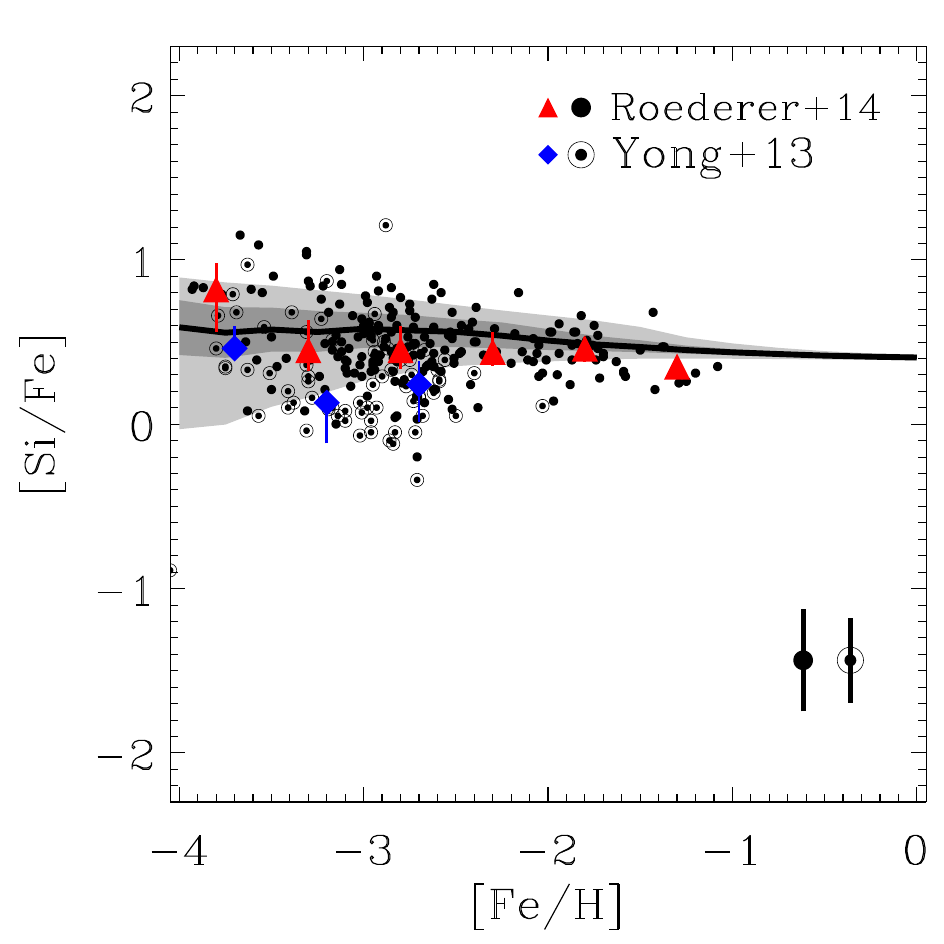}\hspace{0.5cm}\includegraphics[width=5.5cm]{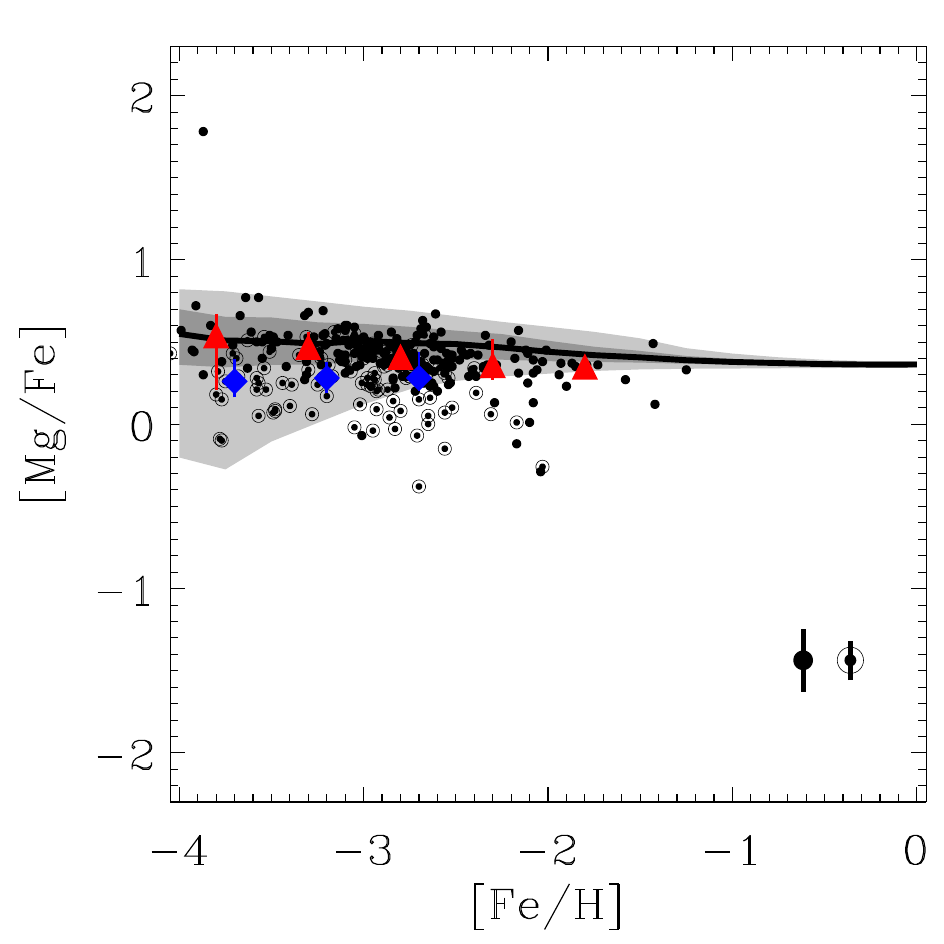}\hspace{0.5cm}\includegraphics[width=5.5cm]{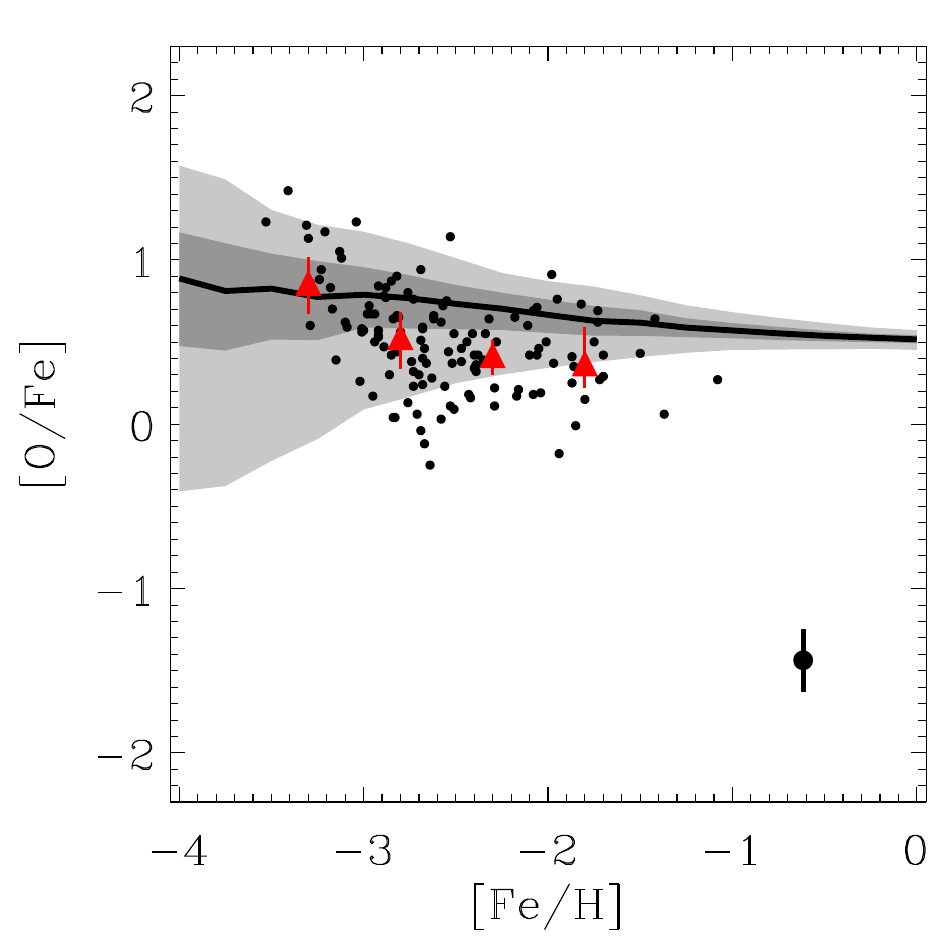}

\includegraphics[width=5.5cm]{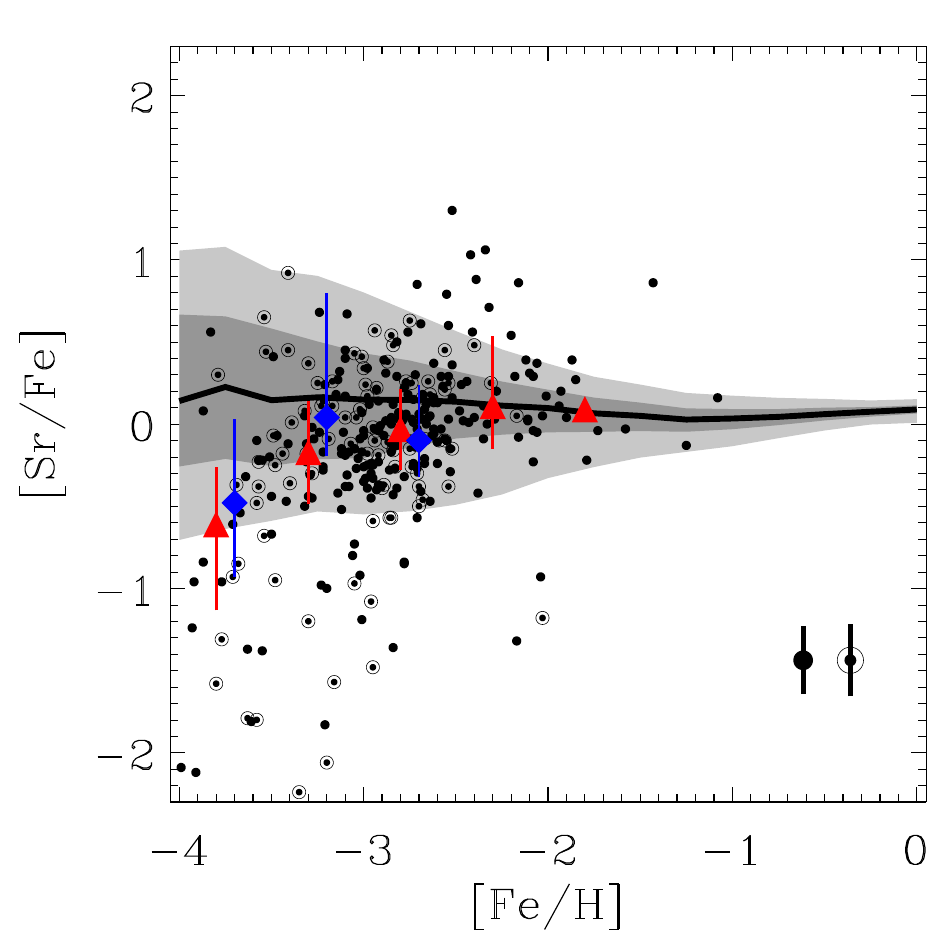}\hspace{0.5cm}\includegraphics[width=5.5cm]{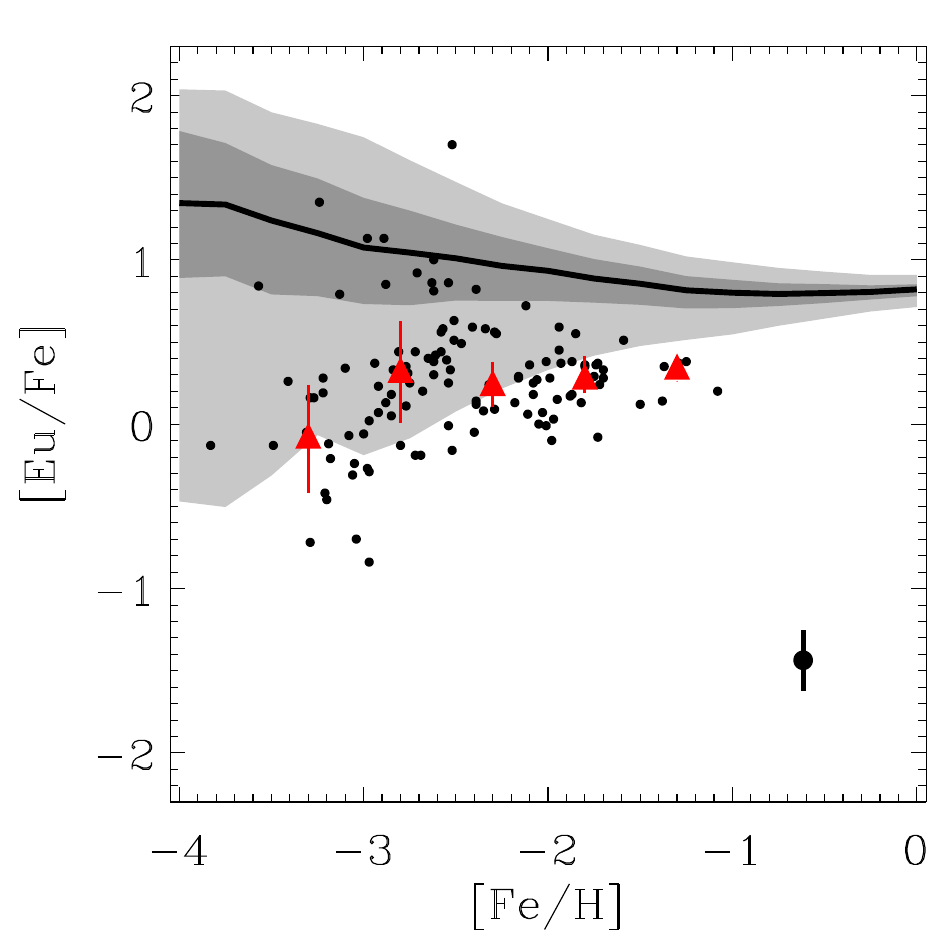}\hspace{0.5cm}\includegraphics[width=5.5cm]{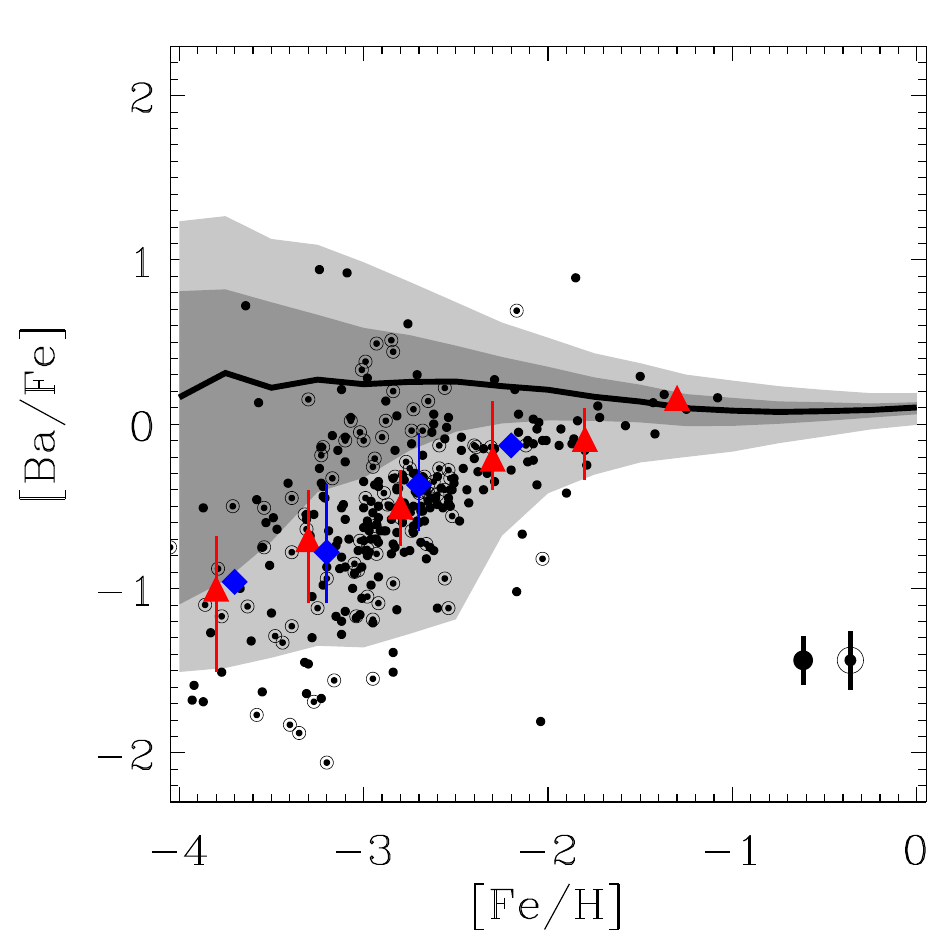}

\caption{ Median (black lines) and $90$, $70$, $30$ and $10$ percentiles (shaded regions) for the different element distributions in our halo stars in simulation ME.
  We also include the observations of  Roederer et al. 2014 (solid black circles) and Yong et al. 2013 (encircled black symbols), as well as the median and corresponding standard deviations in 0.5 dex wide bins (red triangles and blue diamonds, respectively).
}
\label{fig:mean_median_sigma}
\end{figure*}

Fig.~\ref{fig:mean_median_sigma} shows the predicted median  values for the different elements, as a function of [Fe/H] (black lines), together with the  $70/30$ and $90/10$ percentiles (dark and light shaded areas, respectively). The observations of \cite{Roederer14} (solid black circles) and \cite{Yong13} (enricled symbols) are also included, as well as the median and corresponding standard deviations in 0.5 dex wide bins (red triangles and blue diamonds, respectively), as to include at least 5 observational points per bin. 
We find that the predicted abundances for the $\alpha-$elements are consistent with the observations, particularly in the case of the \cite{Roederer14} data set. On the other hand, the observations of  \cite{Yong13} for [Si/Fe] and [Mg/Fe] are in general below the predictions and also below the \cite{Roederer14} observations.
In terms of scatter, we find a good agreement for [Si/Fe] and [O/Fe], but the simulation predicts a higher scatter compared to the observations for the [Mg/Fe] ratio; the disagreement is more important for the low-metallicity regime. As explained above, the predictions for the scatter of the  different $\alpha-$elements can vary due to the particular dependencies of the yields with the stellar mass and metallicity.

In the case of the neutron-capture elements, we find that the predicted ranges of [Sr/Fe] are consistent with the observational data
  but  the [Ba/Fe]  and [Eu/Fe] predicted values lie above the observations.
Similarly, the scatter of [Sr/Fe] in the simulation is consistent with the data, but the scatter found for the other n-capture elements is higher in the simulation -- this effect is more significant for lower metallicities. As explained above, Eu is almost completely produced by r-processes in contrast to the other n-capture elements, and so we expect different levels of scatter.

The discrepancies between the predicted and observed distributions for the neutron capture-elements 
can be interpreted as a signature that the real r-process
contribution is more complex than what is assumed in this work. In fact, the rate of r-process production is fixed (from the narrow mass range of $8-10$M$_\odot$)
and we do not consider any delay apart from the life-time of the considered stars. Moreover, our nucleosynthesis for the r-process component is determined in
\cite{Cescutti13} for Ba, and Eu is simply scaled based on the solar r-process residual that is the the same as the r-process ratio observed in r-process
rich stars \citep{Sneden08}. With these assumptions, our simulation overpredicts the [Eu/Fe] ratios compared to the data. It is worth mentioning that a similar outcome is observed in the \cite{Cescutti14} chemical evolution model which uses a similar approach.
A possible solution would be to assume that the [Eu/Ba] ratio in the r-process is lower or not constant.

\begin{figure*}
  \centering
  \includegraphics[width=7.5cm]{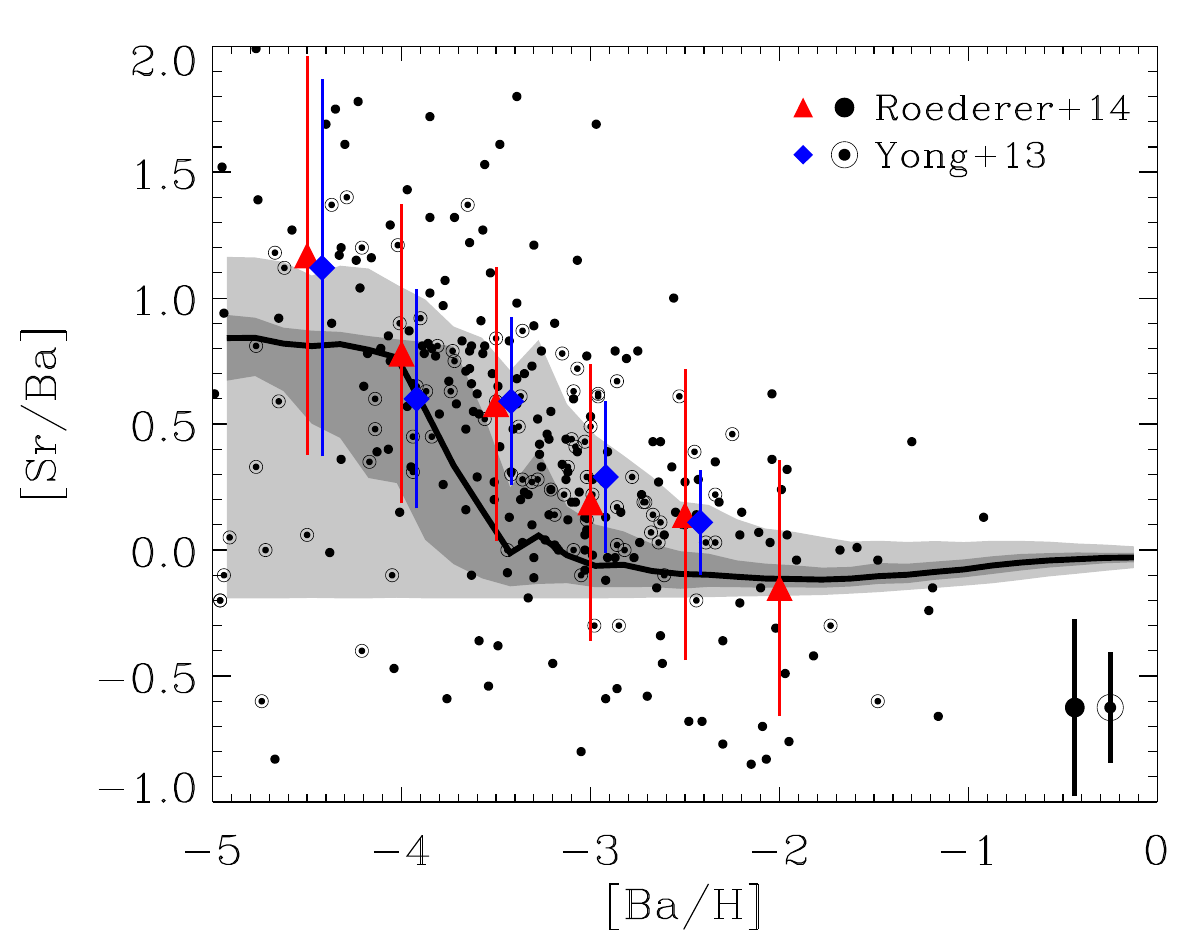}\hspace{0.5cm}\includegraphics[width=7.5cm]{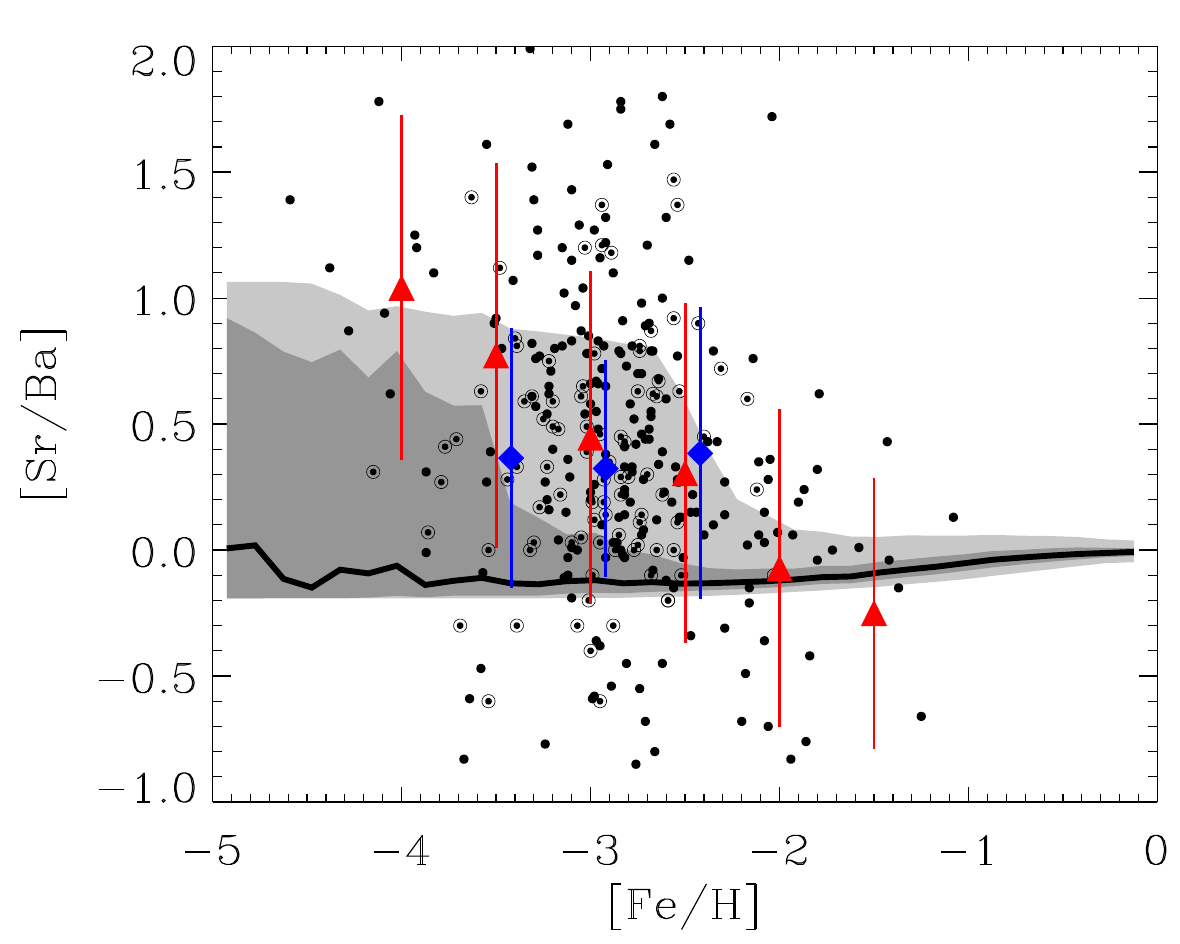}

  \caption{
     Median (black lines) and $90$, $70$, $30$ and $10$ percentiles (shaded regions) for the  [Sr/Ba] ratio, as a function of [Ba/H] (left-pand panel) and [Fe/H] (right-hand panel) in our halo stars in simulation ME.
  We also include the observations of  Roederer et al. 2014 (solid black circles) and Yong et al. 2013 (encircled black symbols), as well as the median and corresponding standard deviations in 0.5 dex wide bins (red triangles and blue diamonds, respectively).
  }
\label{fig:sr_ba_vs_feh_and_bah}
\end{figure*}

 Figure~\ref{fig:sr_ba_vs_feh_and_bah}
compares the predicted distribution of [Sr/Ba] with observations. This figure shows [Sr/Ba] as a function of [Ba/H] (left-hand panel), which allows to better describe the two typical [Sr/Ba] ranges seen in Fig.~\ref{fig:sr_ba_vs_gab} which overlap when we plot [Sr/Ba] as a function of [Fe/H] (right-hand panel). For the simulations, we show the median abundances of [Sr/Ba] for halo stars at $z=0$ (black lines), together with the 90, 70, 30 and 10 percentile levels (shaded areas).  The observations of \cite{Roederer14} (solid black circles) and \cite{Yong13} (enricled black symbols) are also included as data points, as well as the median and corresponding standard deviations in 0.5 dex wide bins (red triangles and blue diamonds, respectively).
 The simulation and observations of [Sr/Ba] are consistent in general terms, and while the simulation median is systematically lower than the observational result, they both have a similar trend with the  [Ba/H] abundance. Furthermore, the deviation around the mean for the simulation and the observations is similar, 
  but  in the simulation there are no stars with [Sr/Ba]$< -0.2$, a limit which is determined by the stellar yields adopted, particularly for the r-process.
 This is because for the early
stages of chemical enrichment, the only relevant sources are the r-process events and the s-process products from rotating massive stars. The [Sr/Ba] ratio is fixed at $-0.2$ for the r-process events and, without rotating massive stars, we would obtain a fixed ratio for all  stars in the simulation, as shown in \cite{Cescutti13}. The yields from massive stars at this metallicity can vary  in the range $0<$[Sr/Ba]$<$2 according to \cite{Frischknecht16}. Therefore, the final result is the mixing of these
two sources which stay within these boundaries. Also, note that the absolute yields of r-process events are higher compared to those of rotating massive stars,
therefore the former tend to dominate the final composition of stars in our model.

   As we discussed above, the ex-situ stars  formed very early on play a fundamental role in the scatter observed
   in the [Sr/Ba] ratio. Observationally, the contribution of satellites to the Milky Way stellar halo could be tested at least in two different ways.
 First, by studying the fractions of high and low [Sr/Ba] stars in present dSphs, and the possible dependence on galactic masses, e.g.  whether the lightest galaxies are those with the highest [Sr/Ba] ratios.
     Previous works  suggest that most classical and ultra faint dwarf galaxies have high [Sr/Ba] ratios (\citealt{Venn04, Francois16}) although Reticulum II is a small galaxy which got an  early enrichment of r-process and shows a low [Sr/Ba] \citep{Ji16}. Combining observations with simulations which provide a theoretical framework to understand how low and high [Sr/Ba] systems are formed and how they might contribute to the stellar halo of larger systems after accretion will likely help to better understand and interpret observational data. 
   The second opportunity would
be to check whether 
stars now in the Galactic halo with a high [Sr/Ba] show 
 different characteristics from those with low or intermediate [Sr/Ba] ratio, 
which would point to their origin as debris of satellites merged to our halo
 (e.g. \citealt{Helmi18},  see also \citealt{Aguado21,Limberg21,Farouqi22,Myeong22}). On this subject, \cite{Roederer18}
have found an indication that r-process rich stars could actually  
have belonged to small systems.

\subsection{Simulations compared to inhomogeneous chemical  evolution models}

\begin{figure*}
  \centering
\vspace{0.5cm}  {{\hspace{-0.5cm} Cescutti et al. (2013)  \hspace{3.5cm} Ex-situ \hspace{4.5cm} In-situ }}
{\includegraphics[width=12cm]{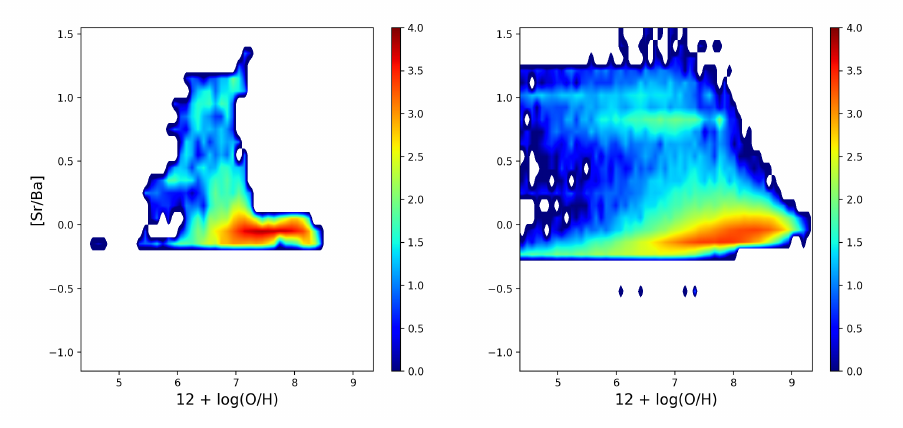}\hspace{-1cm}\includegraphics[width=6cm]{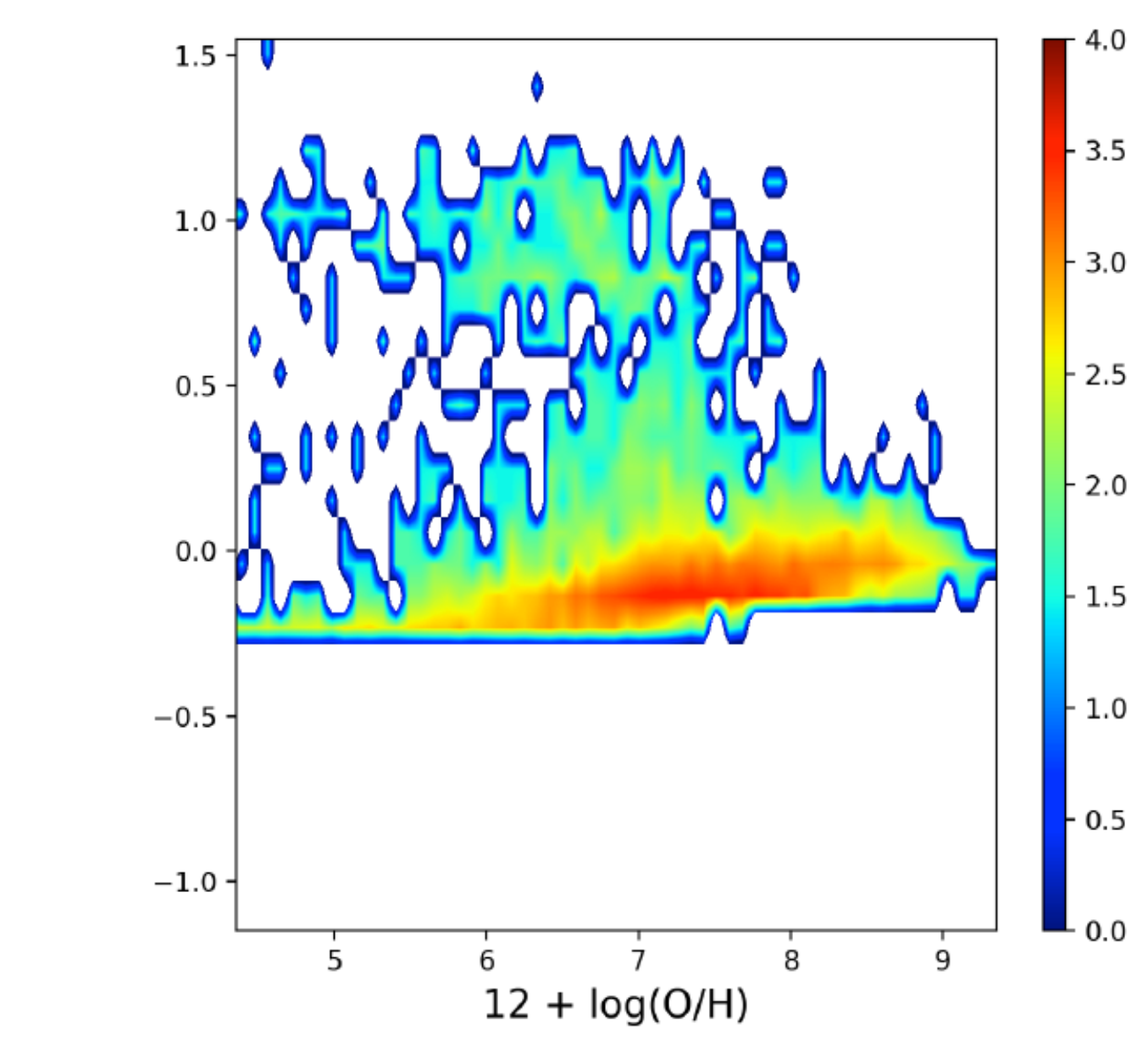}}

\caption{Distribution of [Sr/Ba] as a function of oxygen abundance for the halo stars in the \citealt{Cescutti13} model (left-hand panel), and the ex-situ (middle-hand panel) and in-situ (right-hand panel) components of the stellar halo in our cosmological simulation. }

\label{fig:sr_ba_vs_gab}
\end{figure*}

 In this Section we compare the results of our simulation and the inhomogeneous chemical evolution models of \cite{Cescutti13}, focusing on the [Sr/Ba] abundance ratio.
 In Fig.~\ref{fig:sr_ba_vs_gab} we show the [Sr/Ba] vs oxygen abundance distributions obtained with the \cite{Cescutti13} model (left-hand panel) 
 and with the cosmological simulations, for ex-situ (middle-hand panel) and in-situ (right-hand panel) halo stars separately.
The simulation, particularly in the case of the ex-situ stars which are the dominant component of the stellar halo, predicts a broader range of oxygen abundances. At the metal-rich end,  the stochastic models do not extend to very high oxygen abundances because the present-day metallicity distribution of the halo at the solar vicinity is used as a constraint. However, it is now known that large fraction of the stars between [Fe/H] = $-$1 and below are actually part of debris of  galaxies accreted onto the Milky Way in the past (\citealt{Helmi20} and references therein), and therefore are an ex-situ population, which is not included in the model. 
On the other hand, the low-metallicity tail in  the \cite{Cescutti13} model is affected by the low dilution of the contribution of a single SNII by the ISM, which is a free parameter.  
Given that the chemical evolution models are a much more simplified approach compared to the simulations, it is encouraging that results for the [Sr/Ba] abundances and scatter are similar, particularly in the case of the ex-situ stars.

Note that, consistently with the results obtained with the cosmological simulation (Section~\ref{sec:srba}),
in the stochastic model the high-[Sr/Ba] population
is very old, formed before the r-process events start to enrich the ISM. 
The reason is that in the  stochastic volume the mixing is instantaneous and the typically abundant 
r-process enrichment erases the chemical signature of the rotating massive stars pollution. 
 We expect this outcome to be correct even if we consider a different r-process event,
as it is the case for the stochastic approach (as illustrated in \citealt{Cescutti14} for magneto-rotational driven SNe and in  
\citealt{Cescutti15} for neutron star mergers), although the exact definition of the age range expected for these stars
can vary depending on the nucleosynthesis assumptions. This stresses the importance
to obtain a high-resolution in age estimations in the very early phases of the galaxy assembly.
The first steps in this direction were recently shown by \cite{Montalban21}.

\section{Conclusions}\label{conclu}

In this paper we present a novel implementation of chemical enrichment
in hydrodynamical, cosmological simulations, which considers
the production of $\alpha-$ 
and  neutron-capture elements, and
their distribution into the interstellar medium during SNII explosions. 
The novel aspect of our work is twofold: first, we use
chemical yields which consider
the impact of the rotation of massive stars at
low metallicity \citep{Cescutti14,Cescutti15} and, second, we consider
 the differential enrichment due to
the different life-times of stars of different mass 
within the massive star range. 
In other words, we do not use IMF weighted yields for core collapse SNe.
This means that 
the star particles of the simulations, which represent a mix of
individual stars with given masses,   {\it explode}
in several episodes rather than at a single time, as usually done
in this type of numerical studies. 
This might produce small variations in the overall chemical properties of galaxies; however, it has   a critical role for the oldest/most metal-poor  stars which
are severely affected by the early enrichment stages, a point which we demonstrated here.

The main aim of the work was  to investigate whether our model
can reproduce the different 
scatter levels for the abundance ratios of the 
$\alpha-$ and neutron-capture elements, as suggested by observations
of old stars in our Galaxy.
We focused on 
the abundance ratios and scatter of
 the neutron-capture elements Ba, Eu and Sr,
and the $\alpha-$elements O, Si and Mg,  relative
to the Fe abundance, and compared the predictions of simulations that
assumed either one or multiple explosions per star particle -- i.e.
the ``single explosion'' (SE) and  ``multiple
explosions'' (ME) models.
 All simulations of this work assumed, for the ME model, a number
of 5 explosions per SNII event. With this choice, we can properly describe
the enrichment of the ISM avoiding large numerical overcosts.

Our implementation is grafted onto the {\sc gadget3} code \citep{Springel08}
with the sub-grid modules of \cite{S05,S06} for 
cooling, star formation, chemical enrichment and feedback from supernova (type II and Ia) explosions. 
As the current work focuses  on the chemical abundances of the old stellar population, which are mainly  determined 
by the products of SNII explosions, we have switched off the module
for SNIa, as well as the \citealt{Poulhazan18} extensions for AGB stars, in
all simulations presented here.

In order to validate and test our implementation, we run a number
of idealized simulations of isolated galaxies, as well as the cosmological
formation of a Milky Way-mass galaxy, and compared the
chemical properties of the resulting systems for the SE and ME models.
We found that the chemical abundance ratios obtained for all chemical elements
are different for the SE and ME models:
while in the former the abundance ratios are restricted to a narrow
range given by the yields tables, the latter are able to produce
much more diverse abundance ratios  and higher scatter levels, in better agreement with observational
results. 
The ME models produce a  faster enrichment of the interstellar medium, because
the first {\it explosion} event associated to star particles occurs earlier compared
to the SE models, which affects the cooling, star formation and feedback levels at the very early times.

After passing the various numerical tests performed to our implementation, we focused
on the chemical properties of 
the {\it stellar halo} formed in our cosmological simulation ME.
Our main findings can be summarized as follows:

\begin{itemize}

\item 
The scatter found for the abundance ratios of   neutron-capture elements
(relative to iron) of the  halo stars
is larger compared to that of $\alpha$-elements. 
We find that the $\alpha$-elements
have a typical scatter of the order of  $\lesssim 0.35$ dex (even at low metallicity),
while for all neutron-capture elements studied here the scatter levels are  systematically higher, with values $\gtrsim 0.3$ and up to $1$ dex for [Fe/H]$<-2.5$. For both types of elements, the scatter increases for
 decreasing
metallicity, although this trend is much more pronounced for the neutron-capture
elements.

\item  The scatter levels of the various elements are affected by the cosmological evolution: in our simulation,
  most stars that form the present-day stellar halo ($\sim 80\%$) formed ex-situ and later accreted onto the main progenitor.
  The oldest stars formed in the satellites present particular features in the abundance ratios, most notably in the case
  of n-capture elements, that are fingerprints of the ISM levels of enrichment at very early times.

\item  Our model also predicts a high scatter for the [Sr/Ba] ratio in halo stars,
  and a non-negligible fraction of  stars with [Sr/Ba]$\sim 0.8$.  The high [Sr/Ba] ratios are driven by the pollution via the fast-rotating stars at very early times.

\end{itemize}

In summary,
{\it the main reasons for obtaining different levels of scatter  for the neutron-capture and $\alpha-$elements are the dependence of the stellar yields on the properties of the massive stars and the description of the differential enrichment during the early phases. }
 The early enrichment of the ISM leaves an imprint in the abundances of the
 very old stars, in terms of the abundances and scatter of particular elements.
 This opens the  possibility to reconstruct the enrichment level  of galaxies at early epochs from observational data of the old stars, 
information that is otherwise inaccessible.

The success of our implementation in reproducing the scatter
levels of the different elements  allows to study, in a much more realistic
manner than possible before, the chemical properties of all old stellar
components in galaxies, most notably the bulge, which is the subject
of the next paper of this series. In combination with 
the effects of SNIa and AGB stars, these models have the potential
to provide detailed chemical abundances for all components in galaxies,
which can be used to put additional constrains on the sub-grid physics
describing star formation and feedback, and that are necessary to
provide more predictive power to simulations in the $\Lambda$CDM cosmology.

\section*{Acknowledgments}
We thank the anonymous referee whose suggestions helped improve this paper. CS gratefully acknowledges the Gauss Centre for Supercomputing e.V. (www.gauss-centre.eu) for funding this project by providing computing time on the GCS Supercomputer SuperMUC at Leibniz Supercomputing Centre (www.lrz.de) through project pr49zo, the  Leibniz Gemeinschaft for funding this
project through grant SAW-2012-AIP-5 129, and the TUPAC supercomputer of CONICET in Argentina, where we have run some of the simulations.
CC acknowledges  support  from  DFG  Grant  CH1188/2-1. GC and CC  acknowledge partial support from ChETEC COST Action (CA16117), supported by COST (European Cooperation in Science and Technology) and from the European Union (ChETEC-INFRA, project no. 101008324).

\section*{Data Availability}

The data underlying this article will be shared on reasonable request to the corresponding author.

\bibliographystyle{mnras}
\bibliography{biblio}

\begin{thebibliography}{}
\makeatletter
\relax
\def\mn@urlcharsother{\let\do\@makeother \do\$\do\&\do\#\do\^\do\_\do\%\do\~}
\def\mn@doi{\begingroup\mn@urlcharsother \@ifnextchar [ {\mn@doi@}
  {\mn@doi@[]}}
\def\mn@doi@[#1]#2{\def\@tempa{#1}\ifx\@tempa\@empty \href
  {http://dx.doi.org/#2} {doi:#2}\else \href {http://dx.doi.org/#2} {#1}\fi
  \endgroup}
\def\mn@eprint#1#2{\mn@eprint@#1:#2::\@nil}
\def\mn@eprint@arXiv#1{\href {http://arxiv.org/abs/#1} {{\tt arXiv:#1}}}
\def\mn@eprint@dblp#1{\href {http://dblp.uni-trier.de/rec/bibtex/#1.xml}
  {dblp:#1}}
\def\mn@eprint@#1:#2:#3:#4\@nil{\def\@tempa {#1}\def\@tempb {#2}\def\@tempc
  {#3}\ifx \@tempc \@empty \let \@tempc \@tempb \let \@tempb \@tempa \fi \ifx
  \@tempb \@empty \def\@tempb {arXiv}\fi \@ifundefined
  {mn@eprint@\@tempb}{\@tempb:\@tempc}{\expandafter \expandafter \csname
  mn@eprint@\@tempb\endcsname \expandafter{\@tempc}}}

\bibitem[\protect\citeauthoryear{{Aguado} et~al.,}{{Aguado}
  et~al.}{2021}]{Aguado21}
{Aguado} D.~S.,  et~al., 2021, \mn@doi [\apjl] {10.3847/2041-8213/abdbb8},
  \href {https://ui.adsabs.harvard.edu/abs/2021ApJ...908L...8A} {908, L8}

\bibitem[\protect\citeauthoryear{{Argast}, {Samland}, {Thielemann}  \&
  {Qian}}{{Argast} et~al.}{2004}]{Argast04}
{Argast} D.,  {Samland} M.,  {Thielemann} F.~K.,   {Qian} Y.~Z.,  2004, \mn@doi
  [\aap] {10.1051/0004-6361:20034265}, \href
  {https://ui.adsabs.harvard.edu/abs/2004A&A...416..997A} {416, 997}

\bibitem[\protect\citeauthoryear{{Belokurov}, {Deason}, {Koposov}, {Catelan},
  {Erkal}, {Drake}  \& {Evans}}{{Belokurov} et~al.}{2018}]{Belokurov18}
{Belokurov} V.,  {Deason} A.~J.,  {Koposov} S.~E.,  {Catelan} M.,  {Erkal} D.,
  {Drake} A.~J.,   {Evans} N.~W.,  2018, \mn@doi [\mnras]
  {10.1093/mnras/sty615}, \href
  {https://ui.adsabs.harvard.edu/abs/2018MNRAS.477.1472B} {477, 1472}

\bibitem[\protect\citeauthoryear{{Belokurov}, {Deason}, {Erkal}, {Koposov},
  {Carballo-Bello}, {Smith}, {Jethwa}  \& {Navarrete}}{{Belokurov}
  et~al.}{2019}]{Belokurov19}
{Belokurov} V.,  {Deason} A.~J.,  {Erkal} D.,  {Koposov} S.~E.,
  {Carballo-Bello} J.~A.,  {Smith} M.~C.,  {Jethwa} P.,   {Navarrete} C.,
  2019, \mn@doi [\mnras] {10.1093/mnrasl/slz101}, \href
  {https://ui.adsabs.harvard.edu/abs/2019MNRAS.488L..47B} {488, L47}

\bibitem[\protect\citeauthoryear{{Brauer}, {Ji}, {Drout}  \& {Frebel}}{{Brauer}
  et~al.}{2020}]{Brauer20}
{Brauer} K.,  {Ji} A.~P.,  {Drout} M.~R.,   {Frebel} A.,  2020, arXiv e-prints,
  \href {https://ui.adsabs.harvard.edu/abs/2020arXiv201015837B} {p.
  arXiv:2010.15837}

\bibitem[\protect\citeauthoryear{{Busso}, {Gallino}  \& {Wasserburg}}{{Busso}
  et~al.}{1999}]{Busso99}
{Busso} M.,  {Gallino} R.,   {Wasserburg} G.~J.,  1999, \mn@doi [\araa]
  {10.1146/annurev.astro.37.1.239}, \href
  {https://ui.adsabs.harvard.edu/abs/1999ARA&A..37..239B} {37, 239}

\bibitem[\protect\citeauthoryear{{Caughlan} \& {Fowler}}{{Caughlan} \&
  {Fowler}}{1988}]{Caughlan88}
{Caughlan} G.~R.,  {Fowler} W.~A.,  1988, \mn@doi [Atomic Data and Nuclear Data
  Tables] {10.1016/0092-640X(88)90009-5}, \href
  {http://adsabs.harvard.edu/abs/1988ADNDT..40..283C} {40, 283}

\bibitem[\protect\citeauthoryear{{Cayrel} et~al.,}{{Cayrel}
  et~al.}{2004}]{Cayrel04}
{Cayrel} R.,  et~al., 2004, \mn@doi [\aap] {10.1051/0004-6361:20034074}, \href
  {http://adsabs.harvard.edu/abs/2004A%26A...416.1117C} {416, 1117}

\bibitem[\protect\citeauthoryear{{Cescutti}}{{Cescutti}}{2008}]{Cescutti08}
{Cescutti} G.,  2008, \mn@doi [\aap] {10.1051/0004-6361:20078571}, \href
  {http://adsabs.harvard.edu/abs/2008A%26A...481..691C} {481, 691}

\bibitem[\protect\citeauthoryear{{Cescutti} \& {Chiappini}}{{Cescutti} \&
  {Chiappini}}{2010}]{Cescutti10}
{Cescutti} G.,  {Chiappini} C.,  2010, \mn@doi [\aap]
  {10.1051/0004-6361/201014086}, \href
  {http://adsabs.harvard.edu/abs/2010A%26A...515A.102C} {515, A102}

\bibitem[\protect\citeauthoryear{{Cescutti} \& {Chiappini}}{{Cescutti} \&
  {Chiappini}}{2014}]{Cescutti14}
{Cescutti} G.,  {Chiappini} C.,  2014, \mn@doi [\aap]
  {10.1051/0004-6361/201423432}, \href
  {https://ui.adsabs.harvard.edu/abs/2014A&A...565A..51C} {565, A51}

\bibitem[\protect\citeauthoryear{{Cescutti}, {Fran{\c{c}}ois}, {Matteucci},
  {Cayrel}  \& {Spite}}{{Cescutti} et~al.}{2006}]{Cescutti06}
{Cescutti} G.,  {Fran{\c{c}}ois} P.,  {Matteucci} F.,  {Cayrel} R.,   {Spite}
  M.,  2006, \mn@doi [\aap] {10.1051/0004-6361:20053622}, \href
  {https://ui.adsabs.harvard.edu/abs/2006A&A...448..557C} {448, 557}

\bibitem[\protect\citeauthoryear{{Cescutti}, {Chiappini}, {Hirschi}, {Meynet}
  \& {Frischknecht}}{{Cescutti} et~al.}{2013}]{Cescutti13}
{Cescutti} G.,  {Chiappini} C.,  {Hirschi} R.,  {Meynet} G.,   {Frischknecht}
  U.,  2013, \mn@doi [\aap] {10.1051/0004-6361/201220809}, \href
  {http://adsabs.harvard.edu/abs/2013A%26A...553A..51C} {553, A51}

\bibitem[\protect\citeauthoryear{{Cescutti}, {Romano}, {Matteucci}, {Chiappini}
   \& {Hirschi}}{{Cescutti} et~al.}{2015}]{Cescutti15}
{Cescutti} G.,  {Romano} D.,  {Matteucci} F.,  {Chiappini} C.,   {Hirschi} R.,
  2015, \mn@doi [\aap] {10.1051/0004-6361/201525698}, \href
  {https://ui.adsabs.harvard.edu/abs/2015A&A...577A.139C} {577, A139}

\bibitem[\protect\citeauthoryear{{Chiappini}}{{Chiappini}}{2013}]{Chiappini13}
{Chiappini} C.,  2013, \mn@doi [Astronomische Nachrichten]
  {10.1002/asna.201311902}, \href
  {https://ui.adsabs.harvard.edu/abs/2013AN....334..595C} {334, 595}

\bibitem[\protect\citeauthoryear{{Chiappini}, {Hirschi}, {Meynet},
  {Ekstr{\"o}m}, {Maeder}  \& {Matteucci}}{{Chiappini}
  et~al.}{2006}]{Chiappini06}
{Chiappini} C.,  {Hirschi} R.,  {Meynet} G.,  {Ekstr{\"o}m} S.,  {Maeder} A.,
  {Matteucci} F.,  2006, \mn@doi [\aap] {10.1051/0004-6361:20064866}, \href
  {http://adsabs.harvard.edu/abs/2006A%26A...449L..27C} {449, L27}

\bibitem[\protect\citeauthoryear{{Chiappini}, {Ekstr{\"o}m}, {Meynet},
  {Hirschi}, {Maeder}  \& {Charbonnel}}{{Chiappini} et~al.}{2008}]{Chiappini08}
{Chiappini} C.,  {Ekstr{\"o}m} S.,  {Meynet} G.,  {Hirschi} R.,  {Maeder} A.,
  {Charbonnel} C.,  2008, \mn@doi [\aap] {10.1051/0004-6361:20078698}, \href
  {http://adsabs.harvard.edu/abs/2008A%26A...479L...9C} {479, L9}

\bibitem[\protect\citeauthoryear{{Chiappini}, {Frischknecht}, {Meynet},
  {Hirschi}, {Barbuy}, {Pignatari}, {Decressin}  \& {Maeder}}{{Chiappini}
  et~al.}{2011}]{Chiappini11}
{Chiappini} C.,  {Frischknecht} U.,  {Meynet} G.,  {Hirschi} R.,  {Barbuy} B.,
  {Pignatari} M.,  {Decressin} T.,   {Maeder} A.,  2011, \mn@doi [\nat]
  {10.1038/nature10000}, \href
  {https://ui.adsabs.harvard.edu/abs/2011Natur.472..454C} {472, 454}

\bibitem[\protect\citeauthoryear{{Choplin}, {Hirschi}, {Meynet}, {Ekstr{\"o}m},
  {Chiappini}  \& {Laird}}{{Choplin} et~al.}{2018}]{Choplin18}
{Choplin} A.,  {Hirschi} R.,  {Meynet} G.,  {Ekstr{\"o}m} S.,  {Chiappini} C.,
   {Laird} A.,  2018, \mn@doi [\aap] {10.1051/0004-6361/201833283}, \href
  {https://ui.adsabs.harvard.edu/abs/2018A&A...618A.133C} {618, A133}

\bibitem[\protect\citeauthoryear{{Cowan}, {Sneden}, {Lawler}, {Aprahamian},
  {Wiescher}, {Langanke}, {Mart{\'\i}nez-Pinedo}  \& {Thielemann}}{{Cowan}
  et~al.}{2021}]{Cowan21}
{Cowan} J.~J.,  {Sneden} C.,  {Lawler} J.~E.,  {Aprahamian} A.,  {Wiescher} M.,
   {Langanke} K.,  {Mart{\'\i}nez-Pinedo} G.,   {Thielemann} F.-K.,  2021,
  \mn@doi [Reviews of Modern Physics] {10.1103/RevModPhys.93.015002}, \href
  {https://ui.adsabs.harvard.edu/abs/2021RvMP...93a5002C} {93, 015002}

\bibitem[\protect\citeauthoryear{{Cristallo}, {Straniero}, {Gallino},
  {Piersanti}, {Dom{\'\i}nguez}  \& {Lederer}}{{Cristallo}
  et~al.}{2009}]{Cristallo09}
{Cristallo} S.,  {Straniero} O.,  {Gallino} R.,  {Piersanti} L.,
  {Dom{\'\i}nguez} I.,   {Lederer} M.~T.,  2009, \mn@doi [\apj]
  {10.1088/0004-637X/696/1/797}, \href
  {https://ui.adsabs.harvard.edu/abs/2009ApJ...696..797C} {696, 797}

\bibitem[\protect\citeauthoryear{{Farouqi}, {Thielemann}, {Rosswog}  \&
  {Kratz}}{{Farouqi} et~al.}{2022}]{Farouqi22}
{Farouqi} K.,  {Thielemann} F.~K.,  {Rosswog} S.,   {Kratz} K.~L.,  2022,
  \mn@doi [\aap] {10.1051/0004-6361/202141038}, \href
  {https://ui.adsabs.harvard.edu/abs/2022A&A...663A..70F} {663, A70}

\bibitem[\protect\citeauthoryear{{Fran{\c c}ois}, {Matteucci}, {Cayrel},
  {Spite}, {Spite}  \& {Chiappini}}{{Fran{\c c}ois} et~al.}{2004}]{Francois04}
{Fran{\c c}ois} P.,  {Matteucci} F.,  {Cayrel} R.,  {Spite} M.,  {Spite} F.,
  {Chiappini} C.,  2004, \mn@doi [\aap] {10.1051/0004-6361:20034140}, \href
  {http://adsabs.harvard.edu/abs/2004A%26A...421..613F} {421, 613}

\bibitem[\protect\citeauthoryear{{Fran{\c{c}}ois} et~al.,}{{Fran{\c{c}}ois}
  et~al.}{2007}]{Francois07}
{Fran{\c{c}}ois} P.,  et~al., 2007, \mn@doi [\aap]
  {10.1051/0004-6361:20077706}, \href
  {https://ui.adsabs.harvard.edu/abs/2007A&A...476..935F} {476, 935}

\bibitem[\protect\citeauthoryear{{Fran{\c{c}}ois}, {Monaco}, {Bonifacio}, {Moni
  Bidin}, {Geisler}  \& {Sbordone}}{{Fran{\c{c}}ois} et~al.}{2016}]{Francois16}
{Fran{\c{c}}ois} P.,  {Monaco} L.,  {Bonifacio} P.,  {Moni Bidin} C.,
  {Geisler} D.,   {Sbordone} L.,  2016, \mn@doi [\aap]
  {10.1051/0004-6361/201527181}, \href
  {https://ui.adsabs.harvard.edu/abs/2016A&A...588A...7F} {588, A7}

\bibitem[\protect\citeauthoryear{{Frebel}}{{Frebel}}{2018}]{Frebel18}
{Frebel} A.,  2018, \mn@doi [Annual Review of Nuclear and Particle Science]
  {10.1146/annurev-nucl-101917-021141}, \href
  {https://ui.adsabs.harvard.edu/abs/2018ARNPS..68..237F} {68, 237}

\bibitem[\protect\citeauthoryear{{Frischknecht}, {Hirschi}  \&
  {Thielemann}}{{Frischknecht} et~al.}{2012}]{Frischknecht12}
{Frischknecht} U.,  {Hirschi} R.,   {Thielemann} F.~K.,  2012, \mn@doi [\aap]
  {10.1051/0004-6361/201117794}, \href
  {https://ui.adsabs.harvard.edu/abs/2012A&A...538L...2F} {538, L2}

\bibitem[\protect\citeauthoryear{{Frischknecht} et~al.,}{{Frischknecht}
  et~al.}{2016a}]{Frisch16}
{Frischknecht} U.,  et~al., 2016a, \mn@doi [\mnras] {10.1093/mnras/stv2723},
  \href {http://adsabs.harvard.edu/abs/2016MNRAS.456.1803F} {456, 1803}

\bibitem[\protect\citeauthoryear{{Frischknecht} et~al.,}{{Frischknecht}
  et~al.}{2016b}]{Frischknecht16}
{Frischknecht} U.,  et~al., 2016b, \mn@doi [\mnras] {10.1093/mnras/stv2723},
  \href {https://ui.adsabs.harvard.edu/abs/2016MNRAS.456.1803F} {456, 1803}

\bibitem[\protect\citeauthoryear{{Gaia Collaboration, Brown} et~al.,}{{Gaia
  Collaboration, Brown} et~al.}{2018}]{Brown18}
{Gaia Collaboration, Brown} A.~G.~A.,  et~al., 2018, \mn@doi [\aap]
  {10.1051/0004-6361/201833051}, \href
  {https://ui.adsabs.harvard.edu/abs/2018A&A...616A...1G} {616, A1}

\bibitem[\protect\citeauthoryear{{Gaia Collaboration, Brown} et~al.,}{{Gaia
  Collaboration, Brown} et~al.}{2021}]{Brown21}
{Gaia Collaboration, Brown} A.~G.~A.,  et~al., 2021, \mn@doi [\aap]
  {10.1051/0004-6361/202039657e}, \href
  {https://ui.adsabs.harvard.edu/abs/2021A&A...650C...3G} {650, C3}

\bibitem[\protect\citeauthoryear{{Grevesse}, {Asplund}, {Sauval}  \&
  {Scott}}{{Grevesse} et~al.}{2010}]{Grevesse10}
{Grevesse} N.,  {Asplund} M.,  {Sauval} A.~J.,   {Scott} P.,  2010, \mn@doi
  [\apss] {10.1007/s10509-010-0288-z}, \href
  {http://adsabs.harvard.edu/abs/2010Ap%26SS.328..179G} {328, 179}

\bibitem[\protect\citeauthoryear{{Gudin} et~al.,}{{Gudin}
  et~al.}{2021}]{Gudin21}
{Gudin} D.,  et~al., 2021, \mn@doi [\apj] {10.3847/1538-4357/abd7ed}, \href
  {https://ui.adsabs.harvard.edu/abs/2021ApJ...908...79G} {908, 79}

\bibitem[\protect\citeauthoryear{{Guidi}, {Scannapieco}, {Walcher}  \&
  {Gallazzi}}{{Guidi} et~al.}{2016}]{Guidi16}
{Guidi} G.,  {Scannapieco} C.,  {Walcher} J.,   {Gallazzi} A.,  2016, \mn@doi
  [\mnras] {10.1093/mnras/stw1790}, \href
  {http://adsabs.harvard.edu/abs/2016MNRAS.462.2046G} {462, 2046}

\bibitem[\protect\citeauthoryear{{Gull}, {Frebel}, {Hinojosa}, {Roederer}, {Ji}
   \& {Brauer}}{{Gull} et~al.}{2021}]{Gull21}
{Gull} M.,  {Frebel} A.,  {Hinojosa} K.,  {Roederer} I.~U.,  {Ji} A.~P.,
  {Brauer} K.,  2021, \mn@doi [\apj] {10.3847/1538-4357/abea1a}, \href
  {https://ui.adsabs.harvard.edu/abs/2021ApJ...912...52G} {912, 52}

\bibitem[\protect\citeauthoryear{{Haynes} \& {Kobayashi}}{{Haynes} \&
  {Kobayashi}}{2019}]{Haynes19}
{Haynes} C.~J.,  {Kobayashi} C.,  2019, \mn@doi [\mnras]
  {10.1093/mnras/sty3389}, \href
  {https://ui.adsabs.harvard.edu/abs/2019MNRAS.483.5123H} {483, 5123}

\bibitem[\protect\citeauthoryear{{Helmi}}{{Helmi}}{2020}]{Helmi20}
{Helmi} A.,  2020, \mn@doi [\araa] {10.1146/annurev-astro-032620-021917}, \href
  {https://ui.adsabs.harvard.edu/abs/2020ARA&A..58..205H} {58, 205}

\bibitem[\protect\citeauthoryear{{Helmi}, {Babusiaux}, {Koppelman}, {Massari},
  {Veljanoski}  \& {Brown}}{{Helmi} et~al.}{2018}]{Helmi18}
{Helmi} A.,  {Babusiaux} C.,  {Koppelman} H.~H.,  {Massari} D.,  {Veljanoski}
  J.,   {Brown} A. G.~A.,  2018, \mn@doi [\nat] {10.1038/s41586-018-0625-x},
  \href {https://ui.adsabs.harvard.edu/abs/2018Natur.563...85H} {563, 85}

\bibitem[\protect\citeauthoryear{{Hirschi}}{{Hirschi}}{2007}]{Hirschi07}
{Hirschi} R.,  2007, \mn@doi [\aap] {10.1051/0004-6361:20065356}, \href
  {http://adsabs.harvard.edu/abs/2007A%26A...461..571H} {461, 571}

\bibitem[\protect\citeauthoryear{{Honda}, {Aoki}, {Kajino}, {Ando}, {Beers},
  {Izumiura}, {Sadakane}  \& {Takada-Hidai}}{{Honda} et~al.}{2004}]{Honda04}
{Honda} S.,  {Aoki} W.,  {Kajino} T.,  {Ando} H.,  {Beers} T.~C.,  {Izumiura}
  H.,  {Sadakane} K.,   {Takada-Hidai} M.,  2004, \mn@doi [\apj]
  {10.1086/383406}, \href
  {https://ui.adsabs.harvard.edu/abs/2004ApJ...607..474H} {607, 474}

\bibitem[\protect\citeauthoryear{{Ji}, {Frebel}, {Simon}  \& {Geha}}{{Ji}
  et~al.}{2016}]{Ji16}
{Ji} A.~P.,  {Frebel} A.,  {Simon} J.~D.,   {Geha} M.,  2016, \mn@doi [\apj]
  {10.3847/0004-637X/817/1/41}, \href
  {https://ui.adsabs.harvard.edu/abs/2016ApJ...817...41J} {817, 41}

\bibitem[\protect\citeauthoryear{{Kennicutt}}{{Kennicutt}}{1998}]{Kennicutt98}
{Kennicutt} Jr. R.~C.,  1998, \mn@doi [\araa] {10.1146/annurev.astro.36.1.189},
  \href {http://adsabs.harvard.edu/abs/1998ARA%26A..36..189K} {36, 189}

\bibitem[\protect\citeauthoryear{{Koppelman}, {Helmi}, {Massari}, {Roelenga}
  \& {Bastian}}{{Koppelman} et~al.}{2019}]{Koppelman19}
{Koppelman} H.~H.,  {Helmi} A.,  {Massari} D.,  {Roelenga} S.,   {Bastian} U.,
  2019, \mn@doi [\aap] {10.1051/0004-6361/201834769}, \href
  {https://ui.adsabs.harvard.edu/abs/2019A&A...625A...5K} {625, A5}

\bibitem[\protect\citeauthoryear{{Limberg} et~al.,}{{Limberg}
  et~al.}{2021}]{Limberg21}
{Limberg} G.,  et~al., 2021, \mn@doi [\apjl] {10.3847/2041-8213/ac0056}, \href
  {https://ui.adsabs.harvard.edu/abs/2021ApJ...913L..28L} {913, L28}

\bibitem[\protect\citeauthoryear{{Limongi} \& {Chieffi}}{{Limongi} \&
  {Chieffi}}{2018}]{Limongi18}
{Limongi} M.,  {Chieffi} A.,  2018, \mn@doi [\apjs] {10.3847/1538-4365/aacb24},
  \href {https://ui.adsabs.harvard.edu/abs/2018ApJS..237...13L} {237, 13}

\bibitem[\protect\citeauthoryear{{Maeder} \& {Meynet}}{{Maeder} \&
  {Meynet}}{1989}]{Maeder89}
{Maeder} A.,  {Meynet} G.,  1989, \aap, \href
  {https://ui.adsabs.harvard.edu/abs/1989A&A...210..155M} {210, 155}

\bibitem[\protect\citeauthoryear{{Majewski} et~al.,}{{Majewski}
  et~al.}{2017}]{Majewski17}
{Majewski} S.~R.,  et~al., 2017, \mn@doi [\aj] {10.3847/1538-3881/aa784d},
  \href {https://ui.adsabs.harvard.edu/abs/2017AJ....154...94M} {154, 94}

\bibitem[\protect\citeauthoryear{{McWilliam}}{{McWilliam}}{1998}]{McWilliam98}
{McWilliam} A.,  1998, \mn@doi [\aj] {10.1086/300289}, \href
  {https://ui.adsabs.harvard.edu/abs/1998AJ....115.1640M} {115, 1640}

\bibitem[\protect\citeauthoryear{{Meynet} \& {Maeder}}{{Meynet} \&
  {Maeder}}{2002}]{Meynet02}
{Meynet} G.,  {Maeder} A.,  2002, \mn@doi [\aap] {10.1051/0004-6361:20020755},
  \href {http://adsabs.harvard.edu/abs/2002A%26A...390..561M} {390, 561}

\bibitem[\protect\citeauthoryear{{Miglio} et~al.,}{{Miglio}
  et~al.}{2021}]{Miglio21}
{Miglio} A.,  et~al., 2021, \mn@doi [\aap] {10.1051/0004-6361/202038307}, \href
  {https://ui.adsabs.harvard.edu/abs/2021A&A...645A..85M} {645, A85}

\bibitem[\protect\citeauthoryear{{Montalb{\'a}n} et~al.,}{{Montalb{\'a}n}
  et~al.}{2021}]{Montalban21}
{Montalb{\'a}n} J.,  et~al., 2021, \mn@doi [Nature Astronomy]
  {10.1038/s41550-021-01347-7}, \href
  {https://ui.adsabs.harvard.edu/abs/2021NatAs.tmp...90M} {}

\bibitem[\protect\citeauthoryear{{Montes} et~al.,}{{Montes}
  et~al.}{2007}]{Montes07}
{Montes} F.,  et~al., 2007, \mn@doi [\apj] {10.1086/523084}, \href
  {https://ui.adsabs.harvard.edu/abs/2007ApJ...671.1685M} {671, 1685}

\bibitem[\protect\citeauthoryear{{Myeong}, {Vasiliev}, {Iorio}, {Evans}  \&
  {Belokurov}}{{Myeong} et~al.}{2019}]{Myeong19}
{Myeong} G.~C.,  {Vasiliev} E.,  {Iorio} G.,  {Evans} N.~W.,   {Belokurov} V.,
  2019, \mn@doi [\mnras] {10.1093/mnras/stz1770}, \href
  {https://ui.adsabs.harvard.edu/abs/2019MNRAS.488.1235M} {488, 1235}

\bibitem[\protect\citeauthoryear{{Myeong}, {Belokurov}, {Aguado}, {Evans},
  {Caldwell}  \& {Bradley}}{{Myeong} et~al.}{2022}]{Myeong22}
{Myeong} G.~C.,  {Belokurov} V.,  {Aguado} D.~S.,  {Evans} N.~W.,  {Caldwell}
  N.,   {Bradley} J.,  2022, arXiv e-prints, \href
  {https://ui.adsabs.harvard.edu/abs/2022arXiv220607744M} {p. arXiv:2206.07744}

\bibitem[\protect\citeauthoryear{{Naidu}, {Conroy}, {Bonaca}, {Johnson},
  {Ting}, {Caldwell}, {Zaritsky}  \& {Cargile}}{{Naidu} et~al.}{2020}]{Naidu20}
{Naidu} R.~P.,  {Conroy} C.,  {Bonaca} A.,  {Johnson} B.~D.,  {Ting} Y.-S.,
  {Caldwell} N.,  {Zaritsky} D.,   {Cargile} P.~A.,  2020, \mn@doi [\apj]
  {10.3847/1538-4357/abaef4}, \href
  {https://ui.adsabs.harvard.edu/abs/2020ApJ...901...48N} {901, 48}

\bibitem[\protect\citeauthoryear{{Naiman} et~al.,}{{Naiman}
  et~al.}{2018}]{Naiman18}
{Naiman} J.~P.,  et~al., 2018, \mn@doi [\mnras] {10.1093/mnras/sty618}, \href
  {https://ui.adsabs.harvard.edu/abs/2018MNRAS.477.1206N} {477, 1206}

\bibitem[\protect\citeauthoryear{{Navarro} \& {White}}{{Navarro} \&
  {White}}{1993}]{Navarro93}
{Navarro} J.~F.,  {White} S.~D.~M.,  1993, \mn@doi [\mnras]
  {10.1093/mnras/265.2.271}, \href
  {http://adsabs.harvard.edu/abs/1993MNRAS.265..271N} {265, 271}

\bibitem[\protect\citeauthoryear{{Nuza}, {Parisi}, {Scannapieco}, {Richter},
  {Gottl{\"o}ber}  \& {Steinmetz}}{{Nuza} et~al.}{2014}]{Nuza14}
{Nuza} S.~E.,  {Parisi} F.,  {Scannapieco} C.,  {Richter} P.,  {Gottl{\"o}ber}
  S.,   {Steinmetz} M.,  2014, \mn@doi [\mnras] {10.1093/mnras/stu643}, \href
  {http://adsabs.harvard.edu/abs/2014MNRAS.441.2593N} {441, 2593}

\bibitem[\protect\citeauthoryear{{Pignatari}, {Gallino}, {Meynet}, {Hirschi},
  {Herwig}  \& {Wiescher}}{{Pignatari} et~al.}{2008}]{Pignatari08}
{Pignatari} M.,  {Gallino} R.,  {Meynet} G.,  {Hirschi} R.,  {Herwig} F.,
  {Wiescher} M.,  2008, \mn@doi [\apjl] {10.1086/593350}, \href
  {https://ui.adsabs.harvard.edu/abs/2008ApJ...687L..95P} {687, L95}

\bibitem[\protect\citeauthoryear{{Poulhazan}, {Scannapieco}  \&
  {Creasey}}{{Poulhazan} et~al.}{2018}]{Poulhazan18}
{Poulhazan} P.-A.,  {Scannapieco} C.,   {Creasey} P.,  2018, \mn@doi [\mnras]
  {10.1093/mnras/sty2080}, \href
  {http://adsabs.harvard.edu/abs/2018MNRAS.480.4817P} {480, 4817}

\bibitem[\protect\citeauthoryear{{Prantzos}, {Abia}, {Limongi}, {Chieffi}  \&
  {Cristallo}}{{Prantzos} et~al.}{2018}]{Prantzos18}
{Prantzos} N.,  {Abia} C.,  {Limongi} M.,  {Chieffi} A.,   {Cristallo} S.,
  2018, \mn@doi [\mnras] {10.1093/mnras/sty316}, \href
  {https://ui.adsabs.harvard.edu/abs/2018MNRAS.476.3432P} {476, 3432}

\bibitem[\protect\citeauthoryear{{Qian} \& {Wasserburg}}{{Qian} \&
  {Wasserburg}}{2000}]{Qian00}
{Qian} Y.~Z.,  {Wasserburg} G.~J.,  2000, \mn@doi [\physrep]
  {10.1016/S0370-1573(00)00017-X}, \href
  {https://ui.adsabs.harvard.edu/abs/2000PhR...333...77Q} {333, 77}

\bibitem[\protect\citeauthoryear{{Re Fiorentin}, {Spagna}, {Lattanzi}  \&
  {Cignoni}}{{Re Fiorentin} et~al.}{2021}]{ReFiorentin21}
{Re Fiorentin} P.,  {Spagna} A.,  {Lattanzi} M.~G.,   {Cignoni} M.,  2021,
  \mn@doi [\apjl] {10.3847/2041-8213/abd53d}, \href
  {https://ui.adsabs.harvard.edu/abs/2021ApJ...907L..16R} {907, L16}

\bibitem[\protect\citeauthoryear{{Roederer}, {Preston}, {Thompson}, {Shectman},
  {Sneden}, {Burley}  \& {Kelson}}{{Roederer} et~al.}{2014}]{Roederer14}
{Roederer} I.~U.,  {Preston} G.~W.,  {Thompson} I.~B.,  {Shectman} S.~A.,
  {Sneden} C.,  {Burley} G.~S.,   {Kelson} D.~D.,  2014, \mn@doi [\aj]
  {10.1088/0004-6256/147/6/136}, \href
  {http://adsabs.harvard.edu/abs/2014AJ....147..136R} {147, 136}

\bibitem[\protect\citeauthoryear{{Roederer}, {Hattori}  \&
  {Valluri}}{{Roederer} et~al.}{2018}]{Roederer18}
{Roederer} I.~U.,  {Hattori} K.,   {Valluri} M.,  2018, \mn@doi [\aj]
  {10.3847/1538-3881/aadd9c}, \href
  {https://ui.adsabs.harvard.edu/abs/2018AJ....156..179R} {156, 179}

\bibitem[\protect\citeauthoryear{{Ryan}, {Norris}  \& {Beers}}{{Ryan}
  et~al.}{1996}]{Ryan96}
{Ryan} S.~G.,  {Norris} J.~E.,   {Beers} T.~C.,  1996, \mn@doi [\apj]
  {10.1086/177967}, \href
  {https://ui.adsabs.harvard.edu/abs/1996ApJ...471..254R} {471, 254}

\bibitem[\protect\citeauthoryear{{Sawala}, {Scannapieco}, {Maio}  \&
  {White}}{{Sawala} et~al.}{2010}]{Sawala10}
{Sawala} T.,  {Scannapieco} C.,  {Maio} U.,   {White} S.,  2010, \mn@doi
  [\mnras] {10.1111/j.1365-2966.2009.16035.x}, \href
  {http://adsabs.harvard.edu/abs/2010MNRAS.402.1599S} {402, 1599}

\bibitem[\protect\citeauthoryear{{Sawala}, {Guo}, {Scannapieco}, {Jenkins}  \&
  {White}}{{Sawala} et~al.}{2011}]{Sawala11}
{Sawala} T.,  {Guo} Q.,  {Scannapieco} C.,  {Jenkins} A.,   {White} S.,  2011,
  \mn@doi [\mnras] {10.1111/j.1365-2966.2010.18163.x}, \href
  {http://adsabs.harvard.edu/abs/2011MNRAS.413..659S} {413, 659}

\bibitem[\protect\citeauthoryear{{Sawala}, {Scannapieco}  \& {White}}{{Sawala}
  et~al.}{2012}]{Sawala12}
{Sawala} T.,  {Scannapieco} C.,   {White} S.,  2012, \mn@doi [\mnras]
  {10.1111/j.1365-2966.2011.20181.x}, \href
  {http://adsabs.harvard.edu/abs/2012MNRAS.420.1714S} {420, 1714}

\bibitem[\protect\citeauthoryear{{Scannapieco}, {Tissera}, {White}  \&
  {Springel}}{{Scannapieco} et~al.}{2005}]{S05}
{Scannapieco} C.,  {Tissera} P.~B.,  {White} S.~D.~M.,   {Springel} V.,  2005,
  \mn@doi [\mnras] {10.1111/j.1365-2966.2005.09574.x}, \href
  {http://adsabs.harvard.edu/abs/2005MNRAS.364..552S} {364, 552}

\bibitem[\protect\citeauthoryear{{Scannapieco}, {Tissera}, {White}  \&
  {Springel}}{{Scannapieco} et~al.}{2006}]{S06}
{Scannapieco} C.,  {Tissera} P.~B.,  {White} S.~D.~M.,   {Springel} V.,  2006,
  \mn@doi [MNRAS] {10.1111/j.1365-2966.2006.10785.x}, \href
  {http://adsabs.harvard.edu/abs/2006MNRAS.371.1125S} {371, 1125}

\bibitem[\protect\citeauthoryear{{Scannapieco}, {Tissera}, {White}  \&
  {Springel}}{{Scannapieco} et~al.}{2008}]{S08}
{Scannapieco} C.,  {Tissera} P.~B.,  {White} S.~D.~M.,   {Springel} V.,  2008,
  \mn@doi [MNRAS] {10.1111/j.1365-2966.2008.13678.x}, \href
  {http://adsabs.harvard.edu/abs/2008MNRAS.389.1137S} {389, 1137}

\bibitem[\protect\citeauthoryear{{Scannapieco}, {White}, {Springel}  \&
  {Tissera}}{{Scannapieco} et~al.}{2009}]{S09}
{Scannapieco} C.,  {White} S.~D.~M.,  {Springel} V.,   {Tissera} P.~B.,  2009,
  \mn@doi [MNRAS] {10.1111/j.1365-2966.2009.14764.x}, \href
  {http://adsabs.harvard.edu/abs/2009MNRAS.396..696S} {396, 696 (S09)}

\bibitem[\protect\citeauthoryear{{Scannapieco}, {Gadotti}, {Jonsson}  \&
  {White}}{{Scannapieco} et~al.}{2010}]{S10}
{Scannapieco} C.,  {Gadotti} D.~A.,  {Jonsson} P.,   {White} S.~D.~M.,  2010,
  \mn@doi [MNRAS] {10.1111/j.1745-3933.2010.00900.x}, \href
  {http://adsabs.harvard.edu/abs/2010MNRAS.407L..41S} {407, L41}

\bibitem[\protect\citeauthoryear{{Scannapieco}, {White}, {Springel}  \&
  {Tissera}}{{Scannapieco} et~al.}{2011}]{S11}
{Scannapieco} C.,  {White} S.~D.~M.,  {Springel} V.,   {Tissera} P.~B.,  2011,
  \mn@doi [MNRAS] {10.1111/j.1365-2966.2011.19027.x}, \href
  {http://adsabs.harvard.edu/abs/2011MNRAS.417..154S} {417, 154 }

\bibitem[\protect\citeauthoryear{{Scannapieco} et~al.,}{{Scannapieco}
  et~al.}{2012}]{S12}
{Scannapieco} C.,  et~al., 2012, \mn@doi [\mnras]
  {10.1111/j.1365-2966.2012.20993.x}, \href
  {http://adsabs.harvard.edu/abs/2012MNRAS.423.1726S} {423, 1726}

\bibitem[\protect\citeauthoryear{{Shen}, {Cooke}, {Ramirez-Ruiz}, {Madau},
  {Mayer}  \& {Guedes}}{{Shen} et~al.}{2015}]{Shen15}
{Shen} S.,  {Cooke} R.~J.,  {Ramirez-Ruiz} E.,  {Madau} P.,  {Mayer} L.,
  {Guedes} J.,  2015, \mn@doi [\apj] {10.1088/0004-637X/807/2/115}, \href
  {https://ui.adsabs.harvard.edu/abs/2015ApJ...807..115S} {807, 115}

\bibitem[\protect\citeauthoryear{{Sneden}, {Cowan}  \& {Gallino}}{{Sneden}
  et~al.}{2008}]{Sneden08}
{Sneden} C.,  {Cowan} J.~J.,   {Gallino} R.,  2008, \mn@doi [\araa]
  {10.1146/annurev.astro.46.060407.145207}, \href
  {http://adsabs.harvard.edu/abs/2008ARA%26A..46..241S} {46, 241}

\bibitem[\protect\citeauthoryear{{Springel} \& {Hernquist}}{{Springel} \&
  {Hernquist}}{2003}]{Springel03}
{Springel} V.,  {Hernquist} L.,  2003, \mn@doi [\mnras]
  {10.1046/j.1365-8711.2003.06206.x}, \href
  {http://adsabs.harvard.edu/abs/2003MNRAS.339..289S} {339, 289}

\bibitem[\protect\citeauthoryear{{Springel} et~al.,}{{Springel}
  et~al.}{2008}]{Springel08}
{Springel} V.,  et~al., 2008, \mn@doi [MNRAS]
  {10.1111/j.1365-2966.2008.14066.x}, \href
  {http://adsabs.harvard.edu/abs/2008MNRAS.391.1685S} {391, 1685}

\bibitem[\protect\citeauthoryear{{Sutherland} \& {Dopita}}{{Sutherland} \&
  {Dopita}}{1993}]{SD93}
{Sutherland} R.~S.,  {Dopita} M.~A.,  1993, \mn@doi [\apjs] {10.1086/191823},
  \href {http://adsabs.harvard.edu/abs/1993ApJS...88..253S} {88, 253}

\bibitem[\protect\citeauthoryear{{Tissera}, {Scannapieco}, {Beers}  \&
  {Carollo}}{{Tissera} et~al.}{2013}]{Tissera13}
{Tissera} P.~B.,  {Scannapieco} C.,  {Beers} T.~C.,   {Carollo} D.,  2013,
  \mn@doi [\mnras] {10.1093/mnras/stt691}, \href
  {http://adsabs.harvard.edu/abs/2013MNRAS.432.3391T} {432, 3391}

\bibitem[\protect\citeauthoryear{{Tissera}, {Beers}, {Carollo}  \&
  {Scannapieco}}{{Tissera} et~al.}{2014}]{Tissera14}
{Tissera} P.~B.,  {Beers} T.~C.,  {Carollo} D.,   {Scannapieco} C.,  2014,
  \mn@doi [\mnras] {10.1093/mnras/stu181}, \href
  {http://adsabs.harvard.edu/abs/2014MNRAS.439.3128T} {439, 3128}

\bibitem[\protect\citeauthoryear{{Travaglio}, {Gallino}, {Arnone}, {Cowan},
  {Jordan}  \& {Sneden}}{{Travaglio} et~al.}{2004}]{Travaglio04}
{Travaglio} C.,  {Gallino} R.,  {Arnone} E.,  {Cowan} J.,  {Jordan} F.,
  {Sneden} C.,  2004, \mn@doi [\apj] {10.1086/380507}, \href
  {https://ui.adsabs.harvard.edu/abs/2004ApJ...601..864T} {601, 864}

\bibitem[\protect\citeauthoryear{{Valentini} et~al.,}{{Valentini}
  et~al.}{2019}]{Valentini19}
{Valentini} M.,  et~al., 2019, \mn@doi [\aap] {10.1051/0004-6361/201834081},
  \href {https://ui.adsabs.harvard.edu/abs/2019A&A...627A.173V} {627, A173}

\bibitem[\protect\citeauthoryear{{Venn}, {Irwin}, {Shetrone}, {Tout}, {Hill}
  \& {Tolstoy}}{{Venn} et~al.}{2004}]{Venn04}
{Venn} K.~A.,  {Irwin} M.,  {Shetrone} M.~D.,  {Tout} C.~A.,  {Hill} V.,
  {Tolstoy} E.,  2004, \mn@doi [\aj] {10.1086/422734}, \href
  {https://ui.adsabs.harvard.edu/abs/2004AJ....128.1177V} {128, 1177}

\bibitem[\protect\citeauthoryear{{Wanajo}, {Kajino}, {Mathews}  \&
  {Otsuki}}{{Wanajo} et~al.}{2001}]{Wanajo01}
{Wanajo} S.,  {Kajino} T.,  {Mathews} G.~J.,   {Otsuki} K.,  2001, \mn@doi
  [\apj] {10.1086/321339}, \href
  {https://ui.adsabs.harvard.edu/abs/2001ApJ...554..578W} {554, 578}

\bibitem[\protect\citeauthoryear{{Wehmeyer}, {Pignatari}  \&
  {Thielemann}}{{Wehmeyer} et~al.}{2015}]{Wehmeyer15}
{Wehmeyer} B.,  {Pignatari} M.,   {Thielemann} F.~K.,  2015, \mn@doi [\mnras]
  {10.1093/mnras/stv1352}, \href
  {https://ui.adsabs.harvard.edu/abs/2015MNRAS.452.1970W} {452, 1970}

\bibitem[\protect\citeauthoryear{{Wehmeyer}, {Fr{\"o}hlich}, {C{\^o}t{\'e}},
  {Pignatari}  \& {Thielemann}}{{Wehmeyer} et~al.}{2019}]{Wehmeyer19}
{Wehmeyer} B.,  {Fr{\"o}hlich} C.,  {C{\^o}t{\'e}} B.,  {Pignatari} M.,
  {Thielemann} F.~K.,  2019, \mn@doi [\mnras] {10.1093/mnras/stz1310}, \href
  {https://ui.adsabs.harvard.edu/abs/2019MNRAS.487.1745W} {487, 1745}

\bibitem[\protect\citeauthoryear{{Woosley} \& {Weaver}}{{Woosley} \&
  {Weaver}}{1995}]{WW95}
{Woosley} S.~E.,  {Weaver} T.~A.,  1995, \mn@doi [\apjs] {10.1086/192237},
  \href {http://adsabs.harvard.edu/abs/1995ApJS..101..181W} {101, 181}

\bibitem[\protect\citeauthoryear{{Yong} et~al.,}{{Yong} et~al.}{2013}]{Yong13}
{Yong} D.,  et~al., 2013, \mn@doi [\apj] {10.1088/0004-637X/762/1/26}, \href
  {http://adsabs.harvard.edu/abs/2013ApJ...762...26Y} {762, 26}

\bibitem[\protect\citeauthoryear{{van de Voort}, {Quataert}, {Hopkins},
  {Kere{\v{s}}}  \& {Faucher-Gigu{\`e}re}}{{van de Voort}
  et~al.}{2015}]{VandeVoort15}
{van de Voort} F.,  {Quataert} E.,  {Hopkins} P.~F.,  {Kere{\v{s}}} D.,
  {Faucher-Gigu{\`e}re} C.-A.,  2015, \mn@doi [\mnras] {10.1093/mnras/stu2404},
  \href {https://ui.adsabs.harvard.edu/abs/2015MNRAS.447..140V} {447, 140}

\bibitem[\protect\citeauthoryear{{van de Voort}, {Pakmor}, {Grand}, {Springel},
  {G{\'o}mez}  \& {Marinacci}}{{van de Voort} et~al.}{2020}]{VandeVoort20}
{van de Voort} F.,  {Pakmor} R.,  {Grand} R. J.~J.,  {Springel} V.,
  {G{\'o}mez} F.~A.,   {Marinacci} F.,  2020, \mn@doi [\mnras]
  {10.1093/mnras/staa754}, \href
  {https://ui.adsabs.harvard.edu/abs/2020MNRAS.494.4867V} {494, 4867}

\makeatother
\end{thebibliography}

\appendix

\section[]{Additional tests of the code}

\subsection{The effects of resolution}\label{sec:app_resolution}

In this Section we test the effects of resolution on our results, by
comparing simulations using the idealized initial conditions of 
Section~\ref{sec:isolated} and varying the
number of particles, $N$ (note that $N$ refers here to the number
of particles in one dimension, i.e. the total number of particles
of the simulation is $N_{\rm tot}\sim N^3$). In particular,
we compare the runs with $N=64$ of Section~\ref{sec:isolated} (i.e. SE64 and ME64) 
with identical tests using $N=32$ and $N=128$
Following our naming convention used in this work, we refer to the
simulations as SE/ME to refer to the single/multiple events per SN explosion,
followed by the number  $N$ denoting the number of particles.
The mass resolution of the simulations with $N=32, 64$ and $128$ is,
respectively, 
 $600$, $70$ and $9\times 10^{4}$ M$_\odot$ for
the gas particles and 
$500$, $60$ and $8\times 10^{5}$ M$_\odot$ for dark matter particles,
and we have used the same gravitational softenings in all runs.

\begin{figure*}
  \centering
  \includegraphics[height=4.7cm]{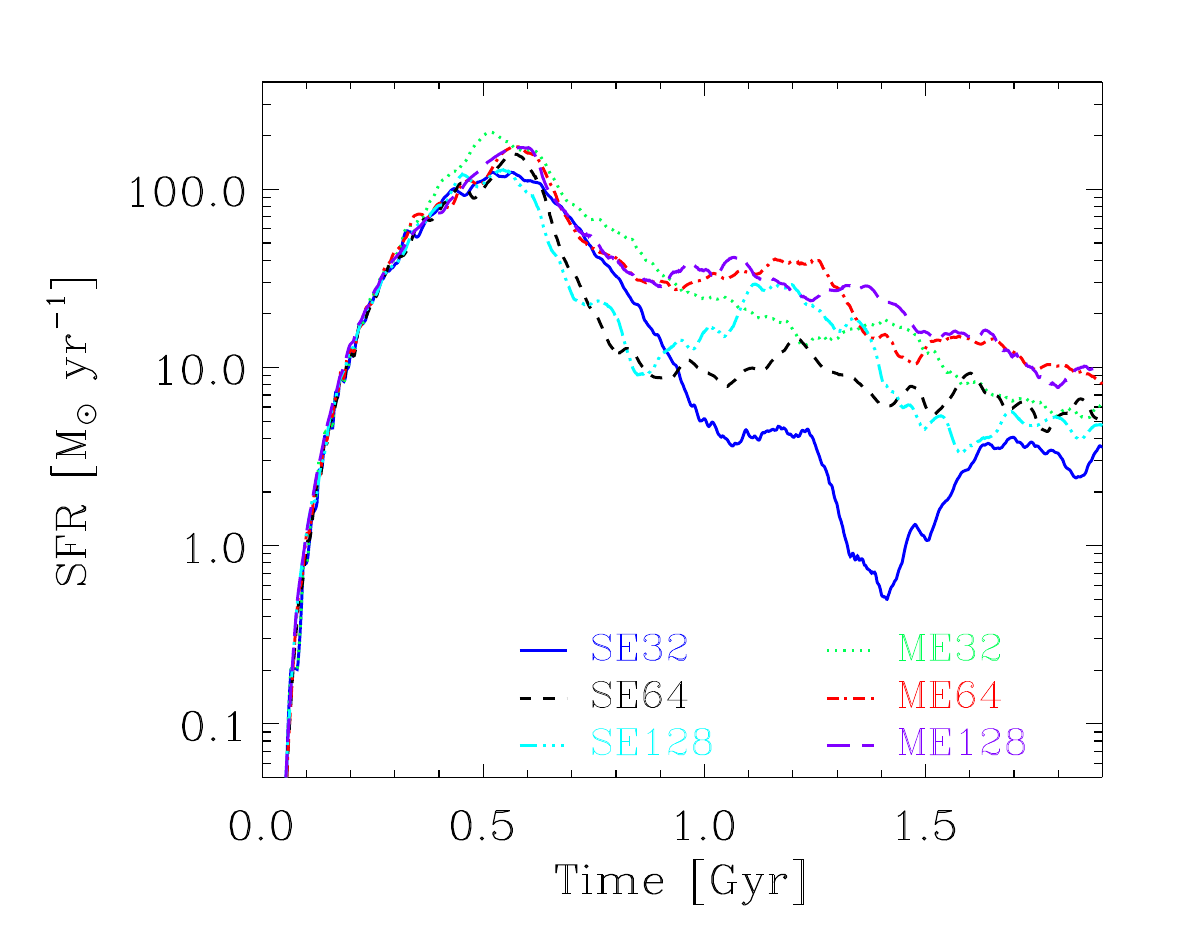}\includegraphics[height=4.7cm]{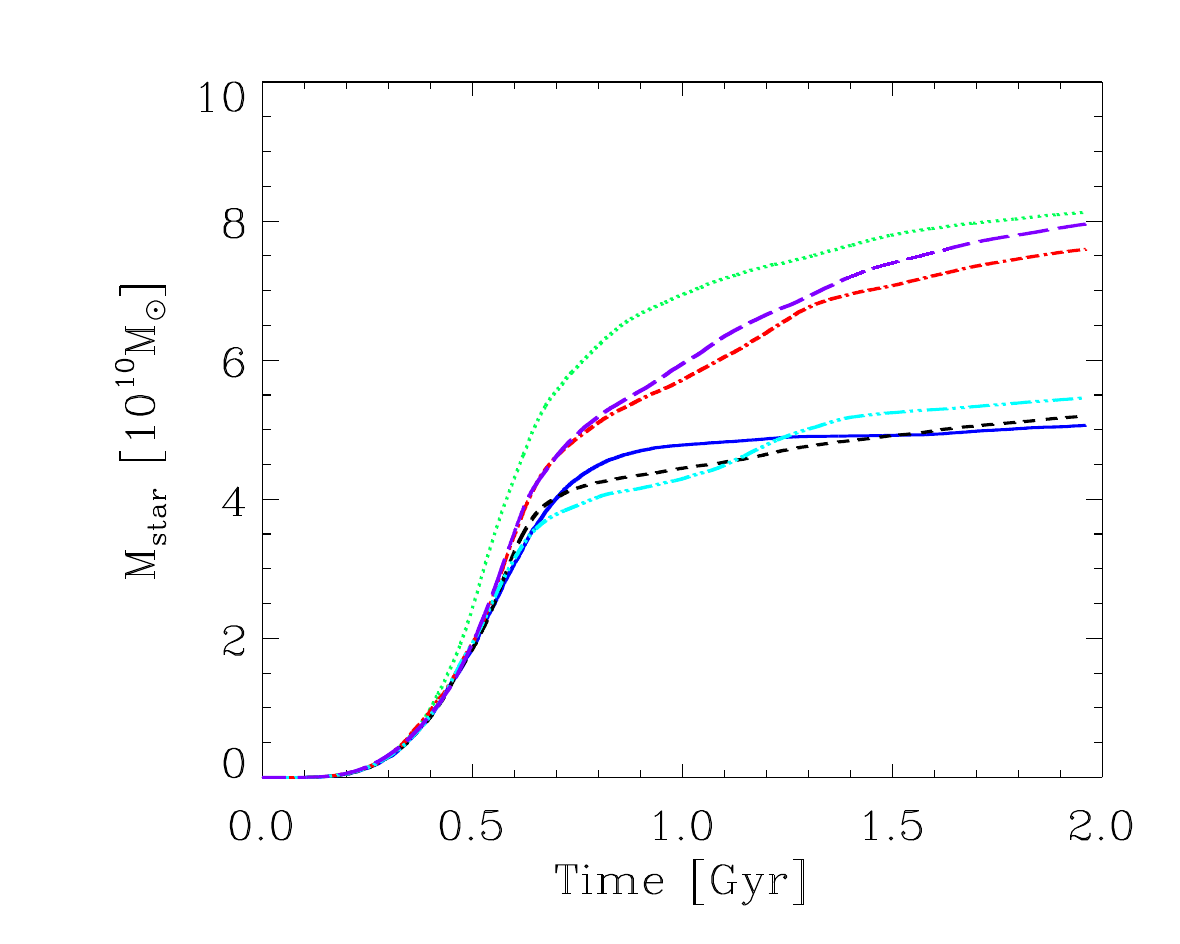}\includegraphics[height=4.7cm]{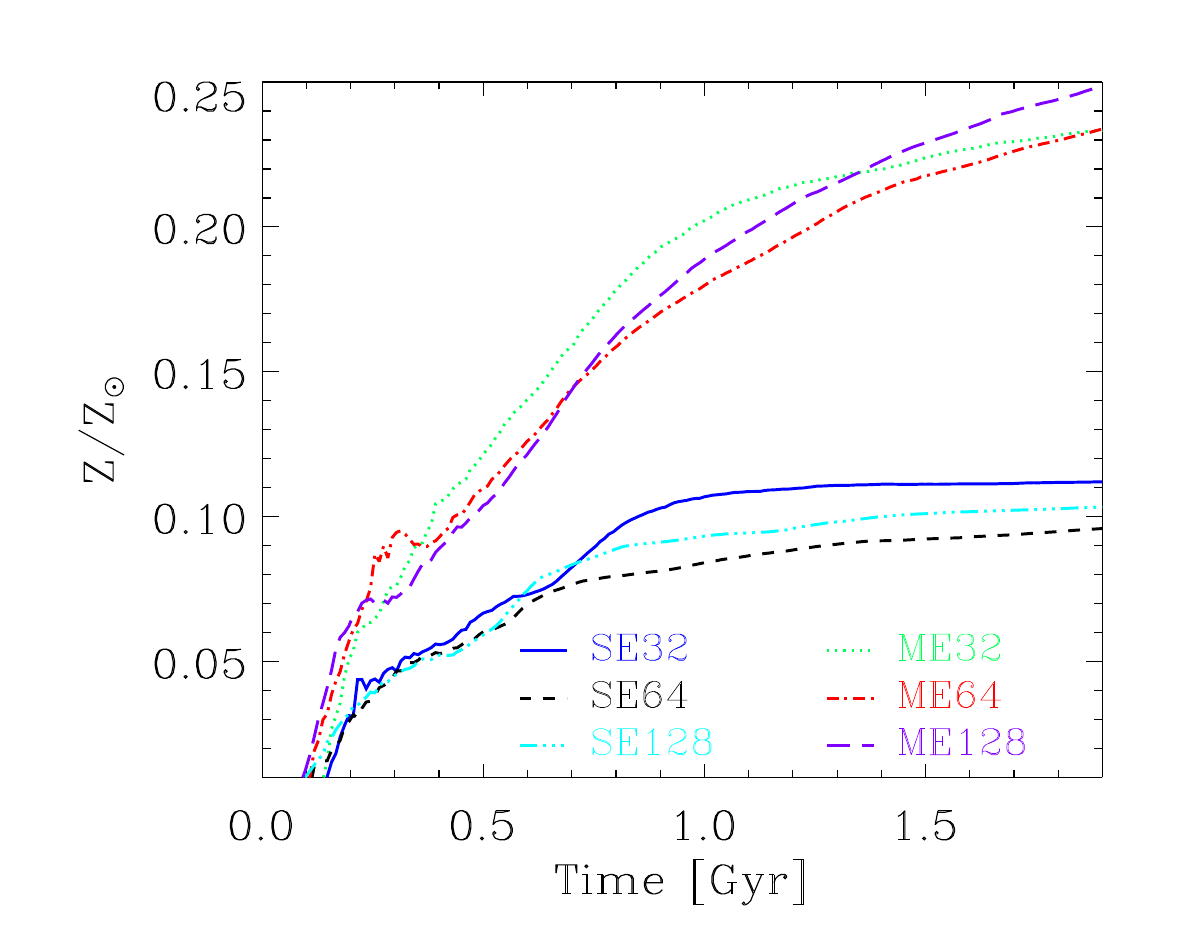}
\caption{Star formation rates  and  integrated stellar mass and metallicity as a function of time  for our simulations with single/multiple explosion times per SN event and various resolutions. }
\label{fig:SFR_isolated_resolution}
\end{figure*}

The left- and middle-hand panels of 
Fig.~\ref{fig:SFR_isolated_resolution} compare the
SFR and cumulative stellar mass as a function of time for our various tests.
The runs with
$N=64$ and $128$ show very good convergence, particularly at the early
times when the star formation activity is at its maximum.
In contrast, the lowest resolution runs show somewhat stronger differences.
In the SE case, the star formation rate at late times is significantly lower than their
counterparts in the runs with higher resolution, although this is
not significant in terms of the total stellar mass formed. For ME32,
the differences are more important, and appear right at the peak of
star formation, producing a noticeable difference, with respect to
ME64 and ME128, in the evolution of the stellar mass formed. We note
that simulations using multiple events per SN explosion are expected to
induce more important differences because the changes in chemical abundance
that occur right after the first SN explosions
translate into changes in the cooling and the subsequent star formation activity.  
The faster enrichment in the ME simulations can be observed from the
right-hand panel
of Fig.~\ref{fig:SFR_isolated_resolution} where we show the evolution of the
total stellar metallicity of our tests with different resolution.
After 1 Gyr of evolution, and regardless of resolution, the ME runs
have have significantly higher metallicities compared to the SE tests.

\begin{figure*}
  \centering
\includegraphics[width=4.5cm]{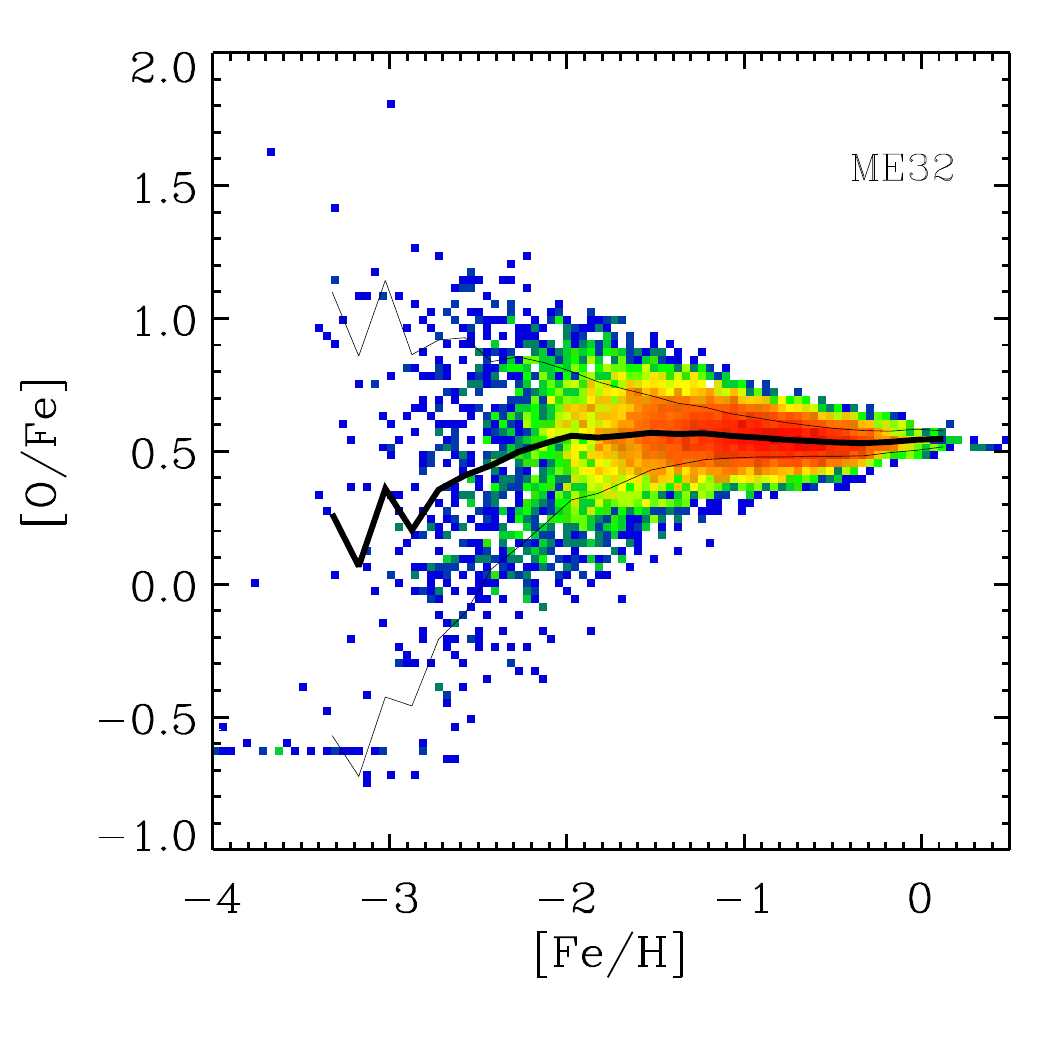}\includegraphics[width=4.5cm]{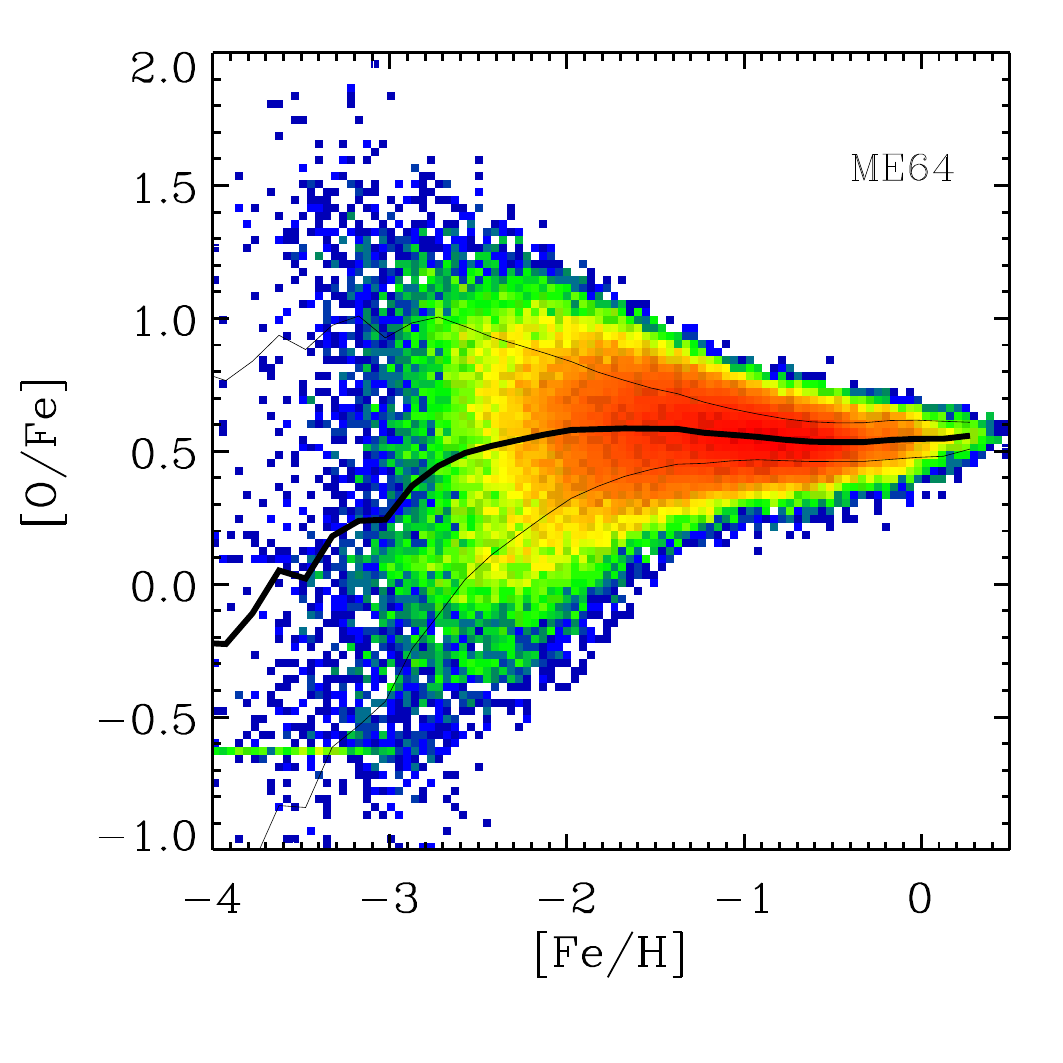}\includegraphics[width=4.5cm]{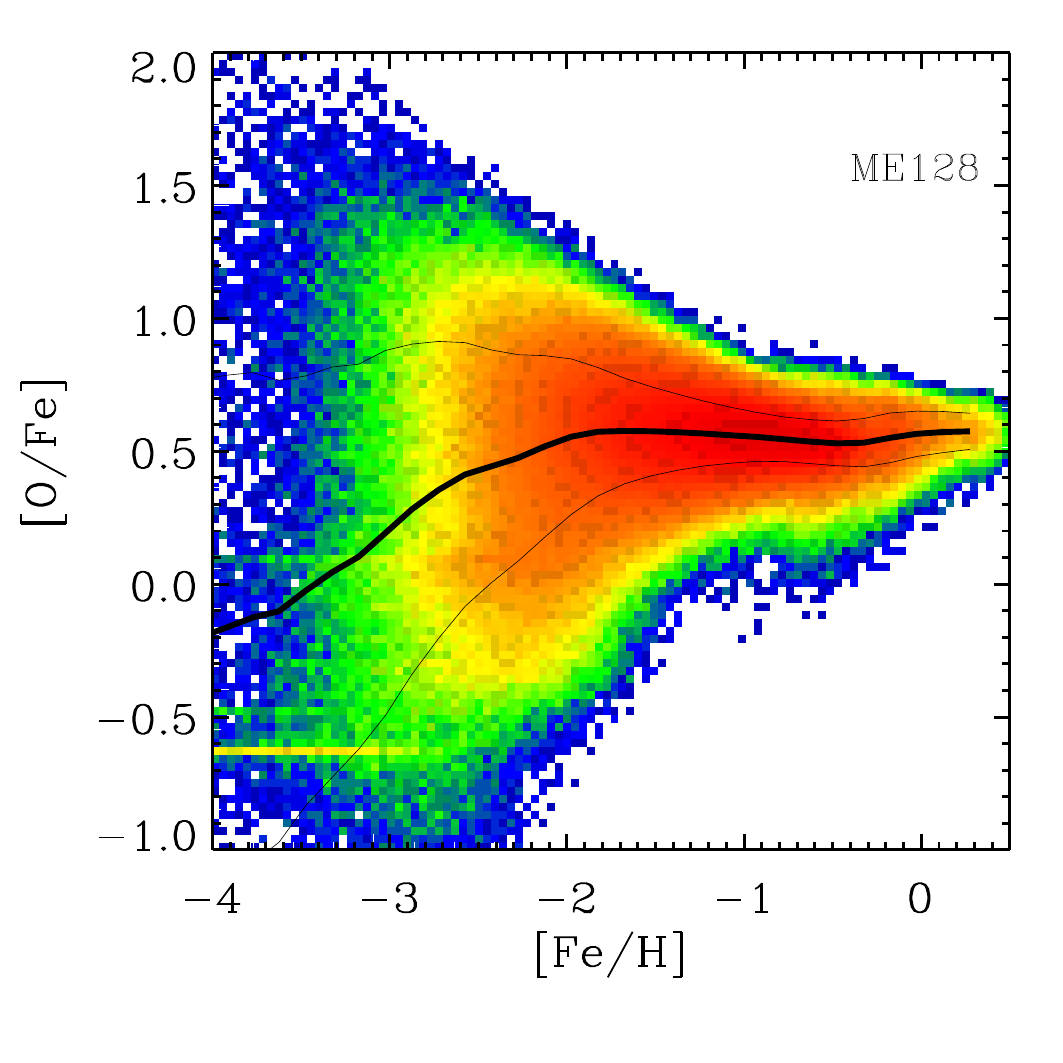}\includegraphics[width=4.5cm]{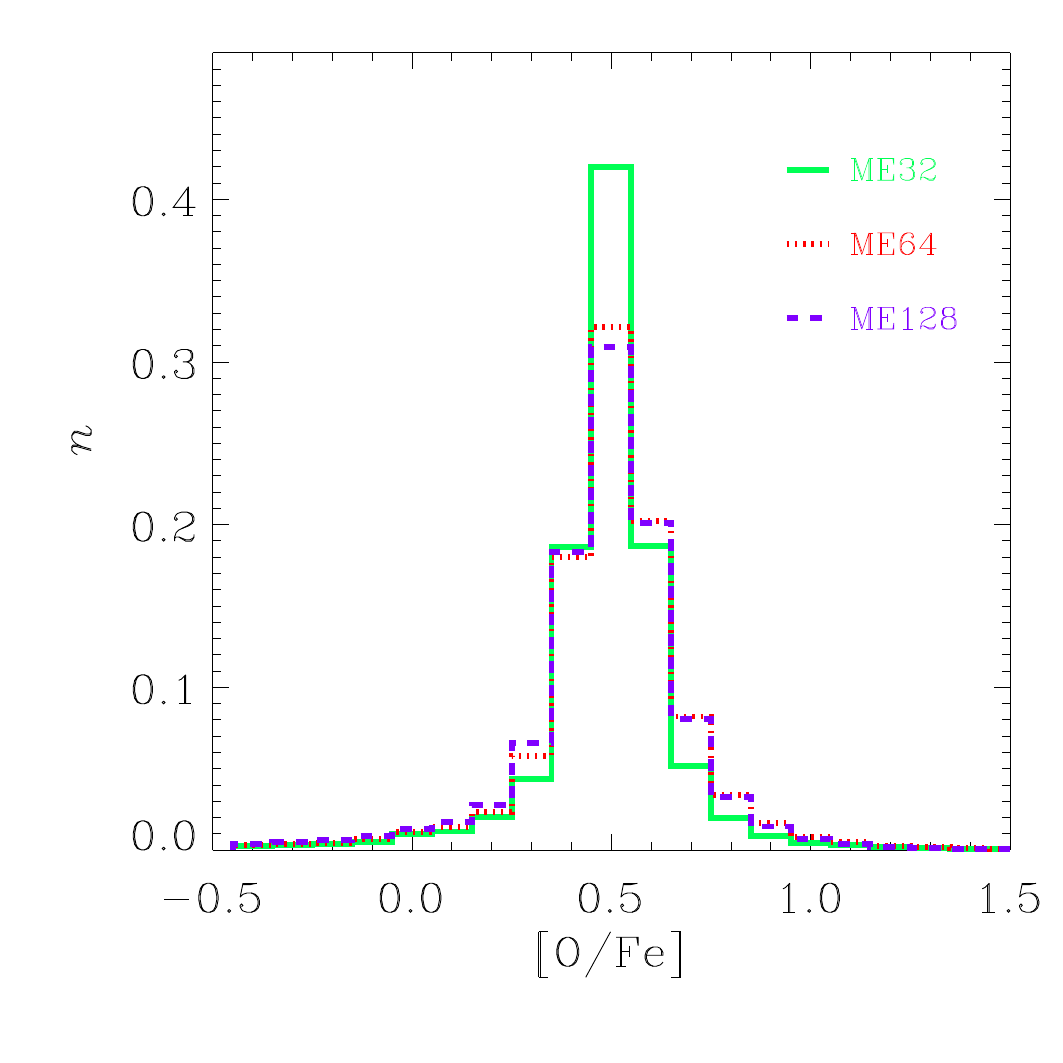}

\includegraphics[width=4.5cm]{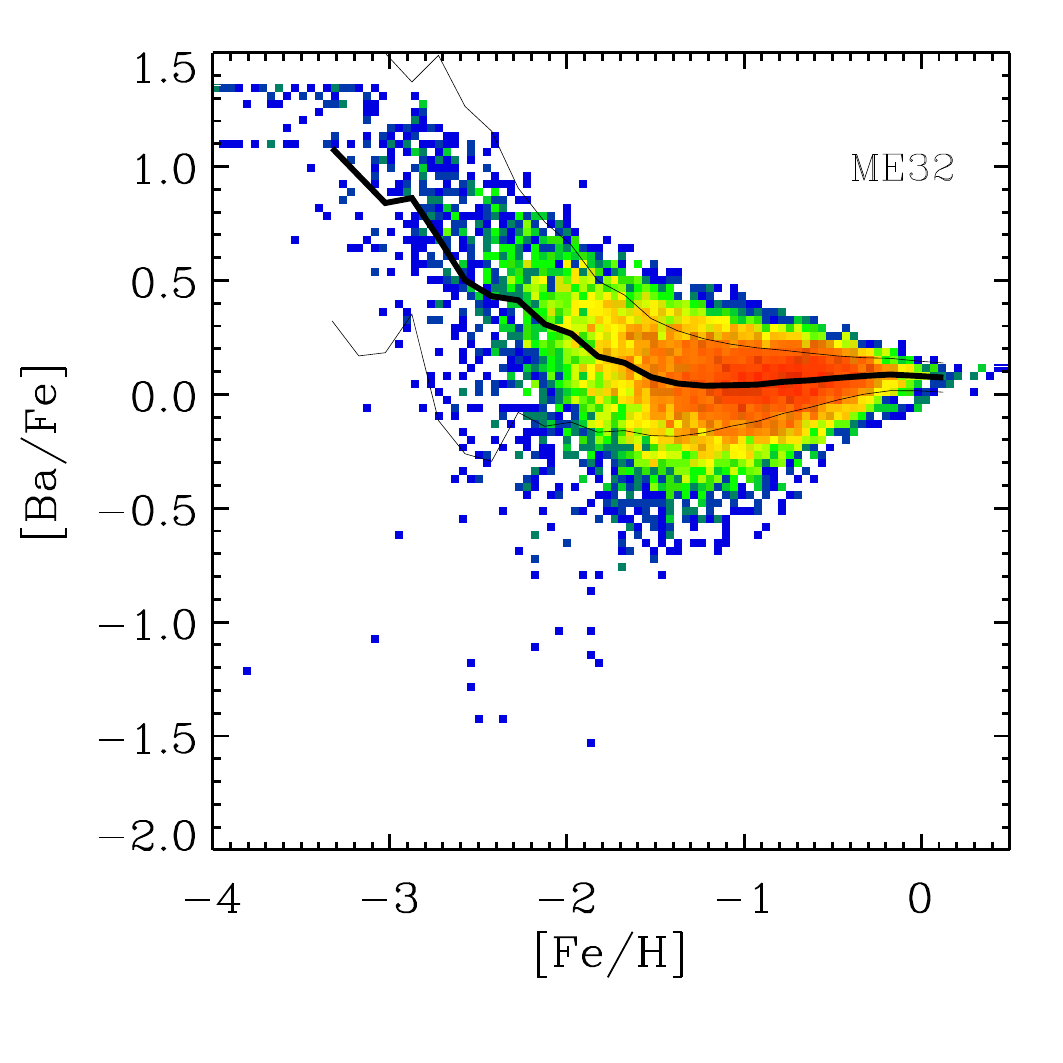}\includegraphics[width=4.5cm]{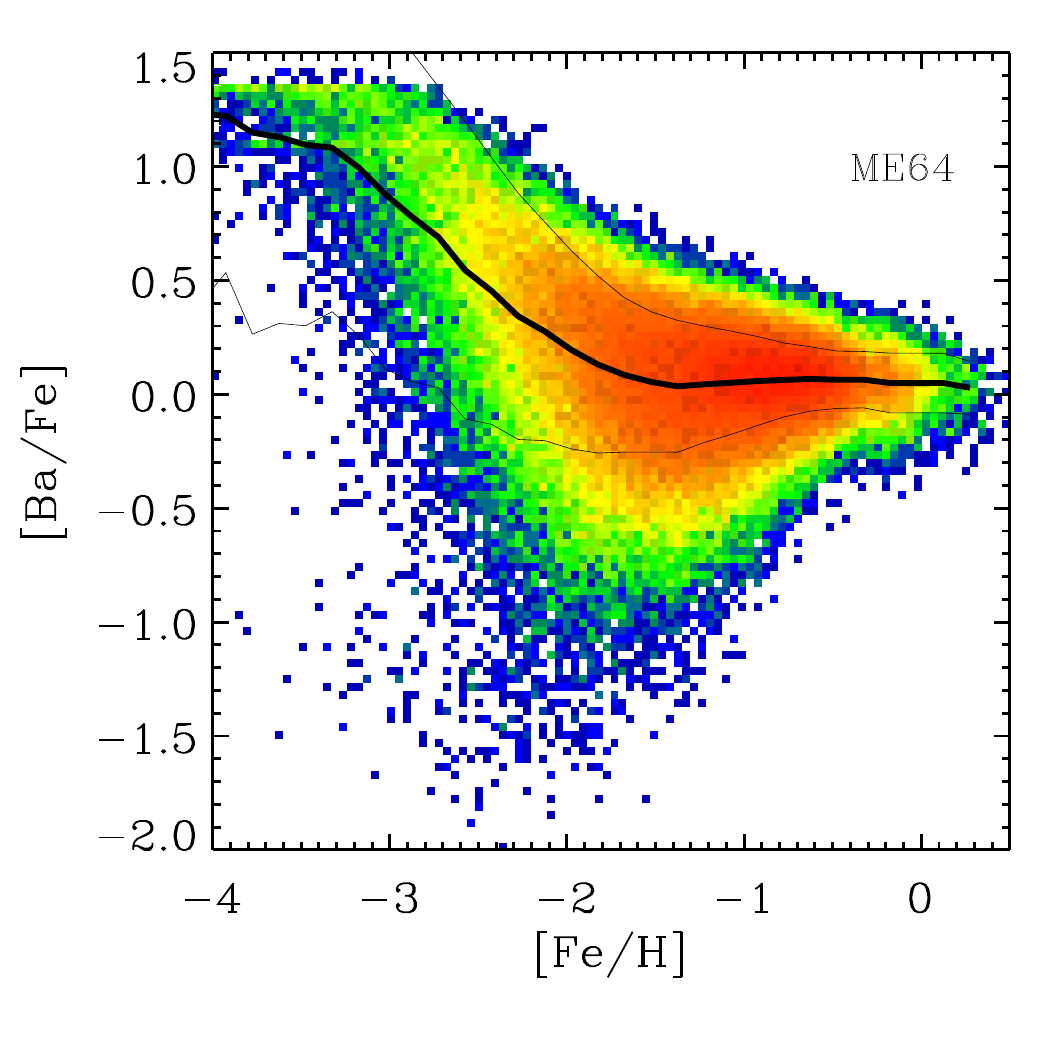}\includegraphics[width=4.5cm]{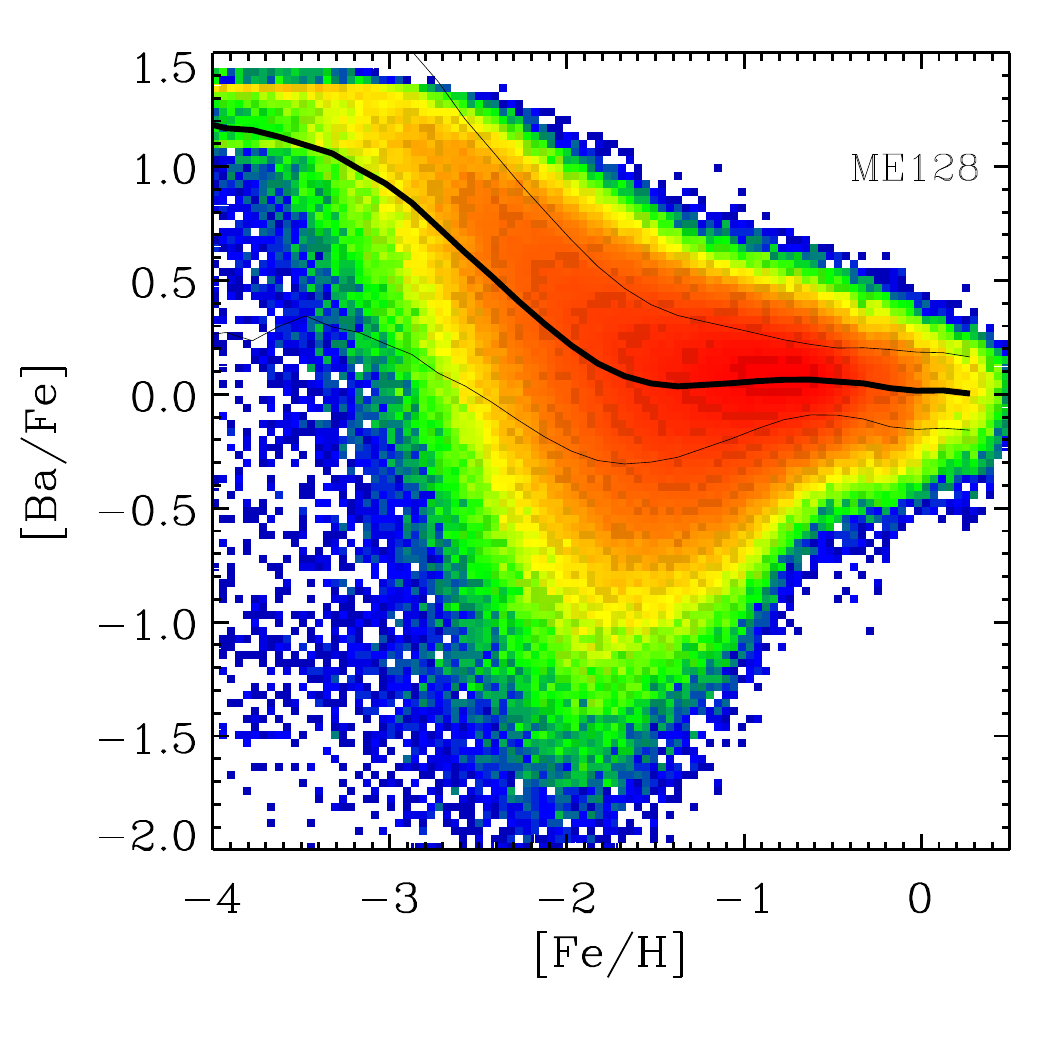}\includegraphics[width=4.5cm]{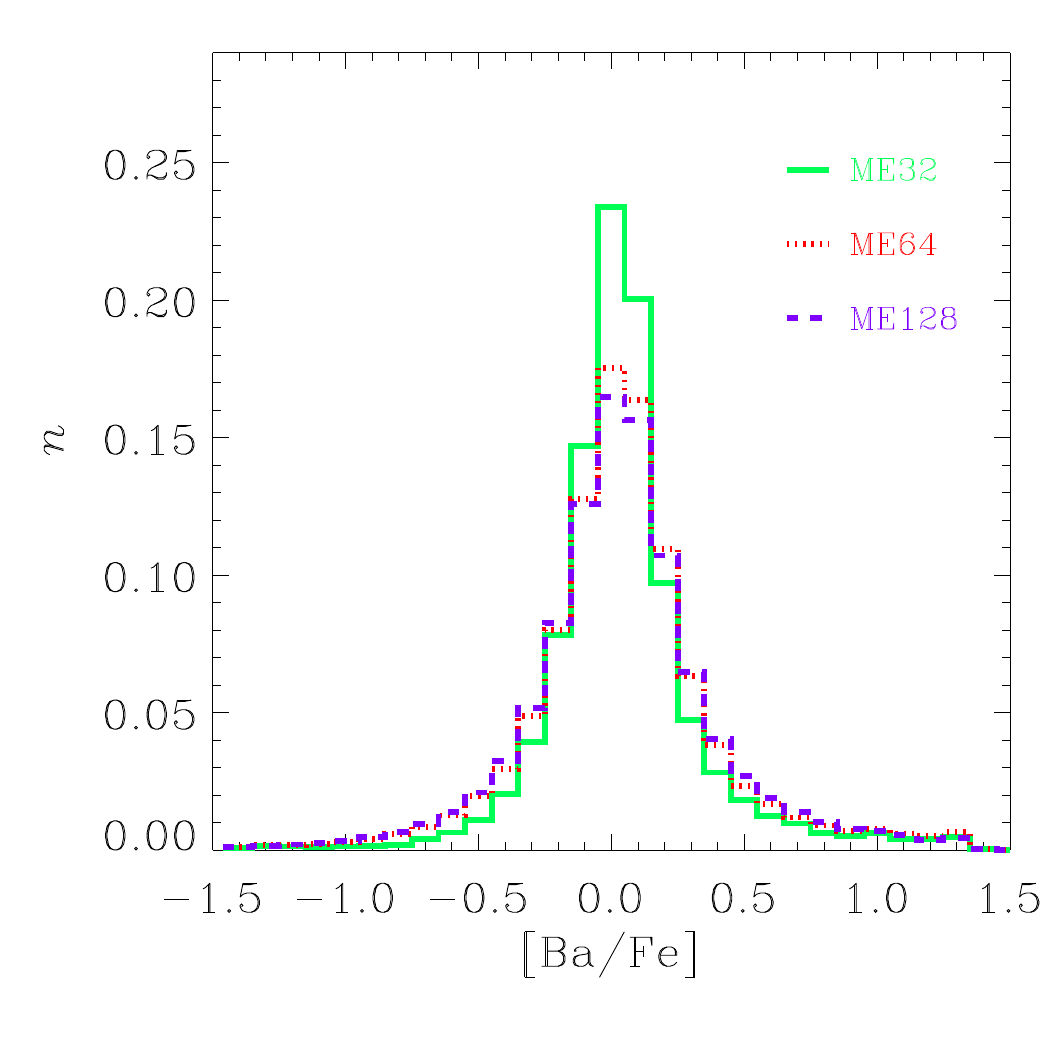}

\caption{The distribution of [O/Fe] and [Ba/Fe]  versus [Fe/H] for our
simulations assuming multiple events per SN explosion and different number of
particles, after 1 Gyr of evolution. 
The colour scale is normalized to the
total number of star particles in each simulation, and we also show
the corresponding median (thick lines) and $\pm\sigma$ contours (thin lines).
The right-hand panel shows a quantitative comparison of the three distributions.}
\label{fig:element_ratios_isolated_resolution}
\end{figure*}

We have also studied the effects of resolution on the predicted
element ratios after  1 Gyr of evolution. As an example, we show in
  Fig.~\ref{fig:element_ratios_isolated_resolution}
 the results of the ME32, ME64 and ME128 runs, for [O/Fe] and [Ba/Fe]. 
We find very similar results in all cases, indicating that resolution
effects, particularly for $N\geq 64$, are unimportant. Note the
very good agreement for the mean/dispersion values, as clearly
seen from the right-hand panel of this figure, despite the expected
differences in terms of higher resolutions being able to better sample
stars at the extremes of the distributions. 
We conclude that our models are robust against resolution
provided $64^3$ or more particles are used.

\subsection[]{Dependence on number of SPH neighbours}\label{sec:app_ngb}

We discuss in this Section the effects of varying the number of SPH neighbours,
which determine the spatial extent over which metals are distributed after
SN explosion events. 
This could have an effect on the predicted distributions of chemical abundances
through the effects on the  mixing of elements in the interstellar medium. 
In order to investigate this we have run additional tests, both for
the SE and ME models, where we increased the number of
neighbours from the fiducidal value of $N_{\rm ngb}=40$ (used in the SE64 and ME64 
simulations of Section~\ref{sec:isolated}) to $64/128$. 

\begin{figure*}
  \centering
  \includegraphics[height=4.7cm]{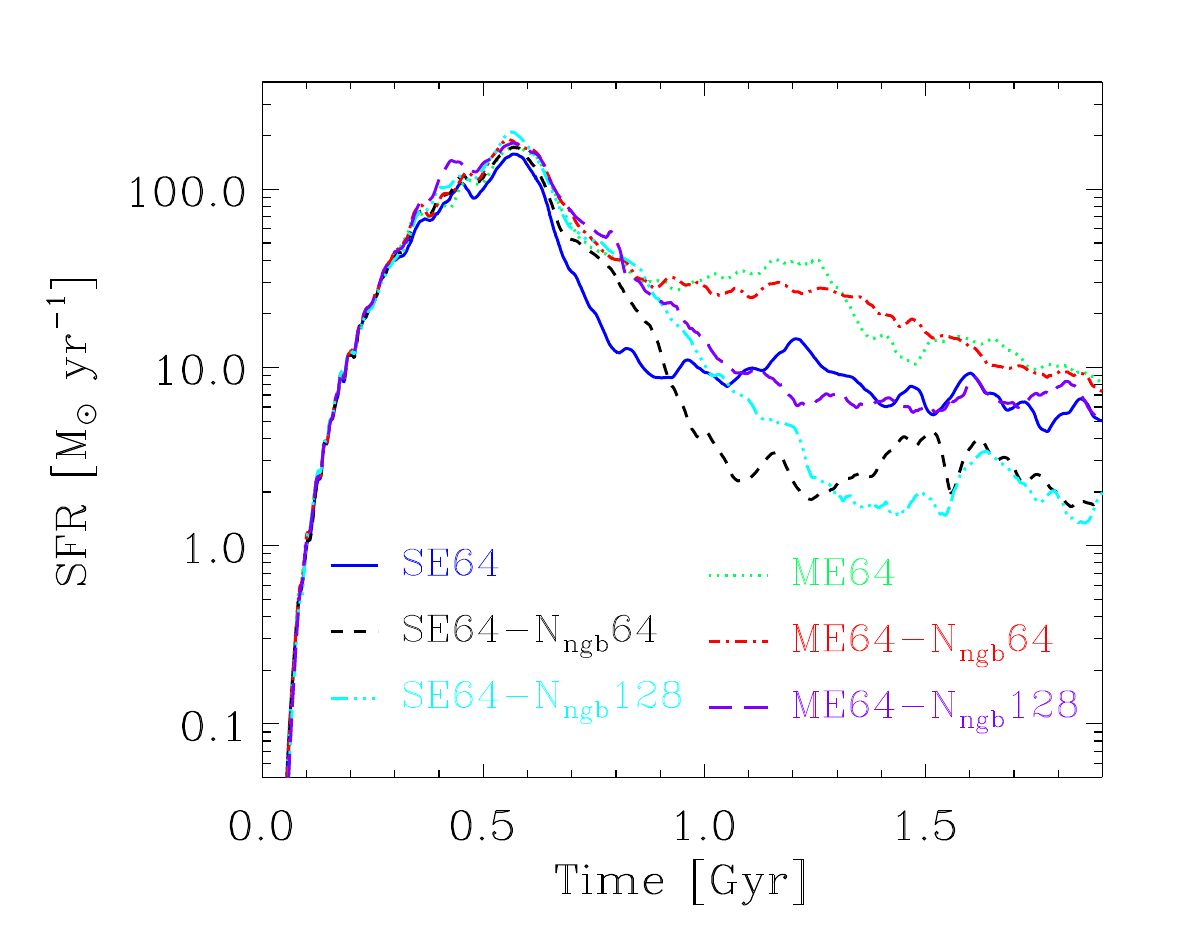}\includegraphics[height=4.7cm]{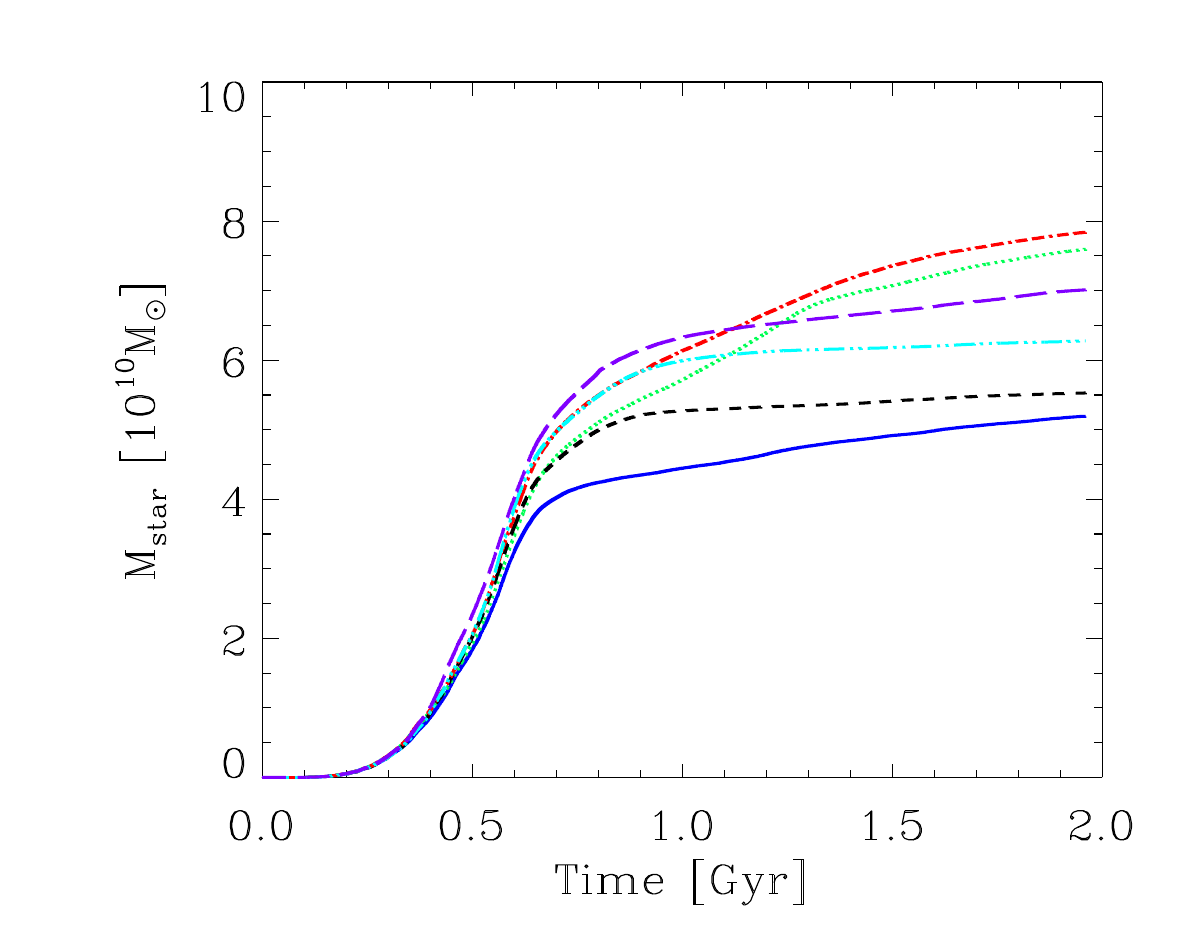}\includegraphics[height=4.7cm]{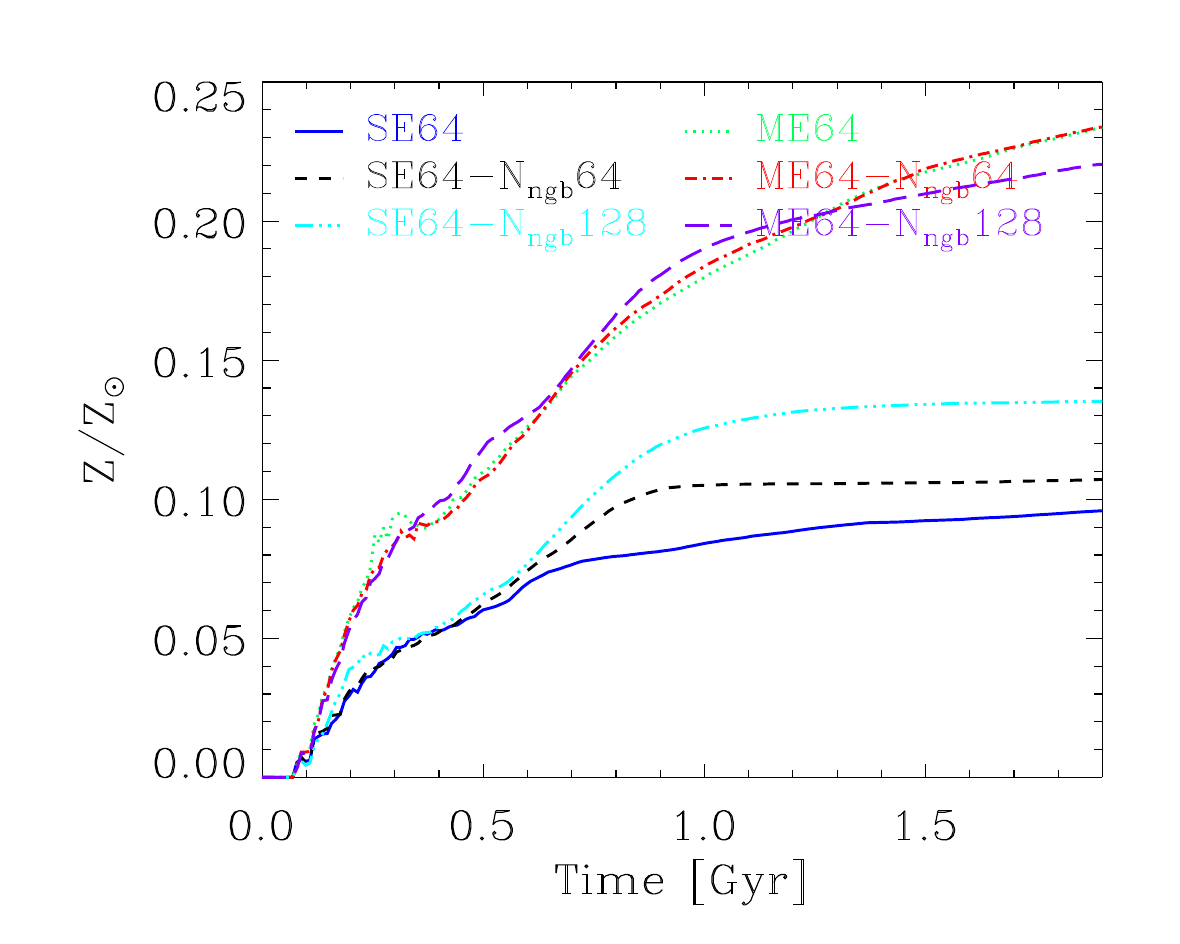}

\caption{Star formation rates  and integrated stellar mass and stellar metallicity  as a function of time for our simulations with single/multiple explosion times per SN event and various choices for the number of SPH neighbours. 
}
\label{fig:SFR_isolated_NgbNumber}
\end{figure*}

Fig.~\ref{fig:SFR_isolated_NgbNumber}  compares
the SFR and the evolution of the stellar mass and metallicity
for these tests.
We find very good convergence of the tests until the peak of the star formation
rate, when differences become more significant.
In particular, an increase in $N_{\rm ngb}$ translates into an enhancement
of star formation activity, which is produced because the chemical elements
that are distributed after SN explosions spread over comparatively larger
regions. Note that this effect is most clear right after the SF peak, as
later on the effects of varying the number of SPH neighbours
becomes non trivial: on one hand, the polluted regions are larger
enhancing the cooling efficiency and star formation activity, while,
on the other, higher SFRs produce larger amounts
of feedback that reduce subsequent star formation levels.

Following the behaviour of the SFRs, the total stellar mass formed
is in general larger for the simulations with larger $N_{\rm ngb}$, with
more significant differences in the case of the SE runs.
A similar behaviour is detected in the case of the stellar metallicities, where
differences are more important for the SE runs and not significant
for the ME tests. Variations in the stellar metallicity follow
the variations in the stellar masses, such that runs with larger $N_{\rm ngb}$
are systematically more enriched.

\begin{figure*}
  \centering

  \includegraphics[width=4.5cm]{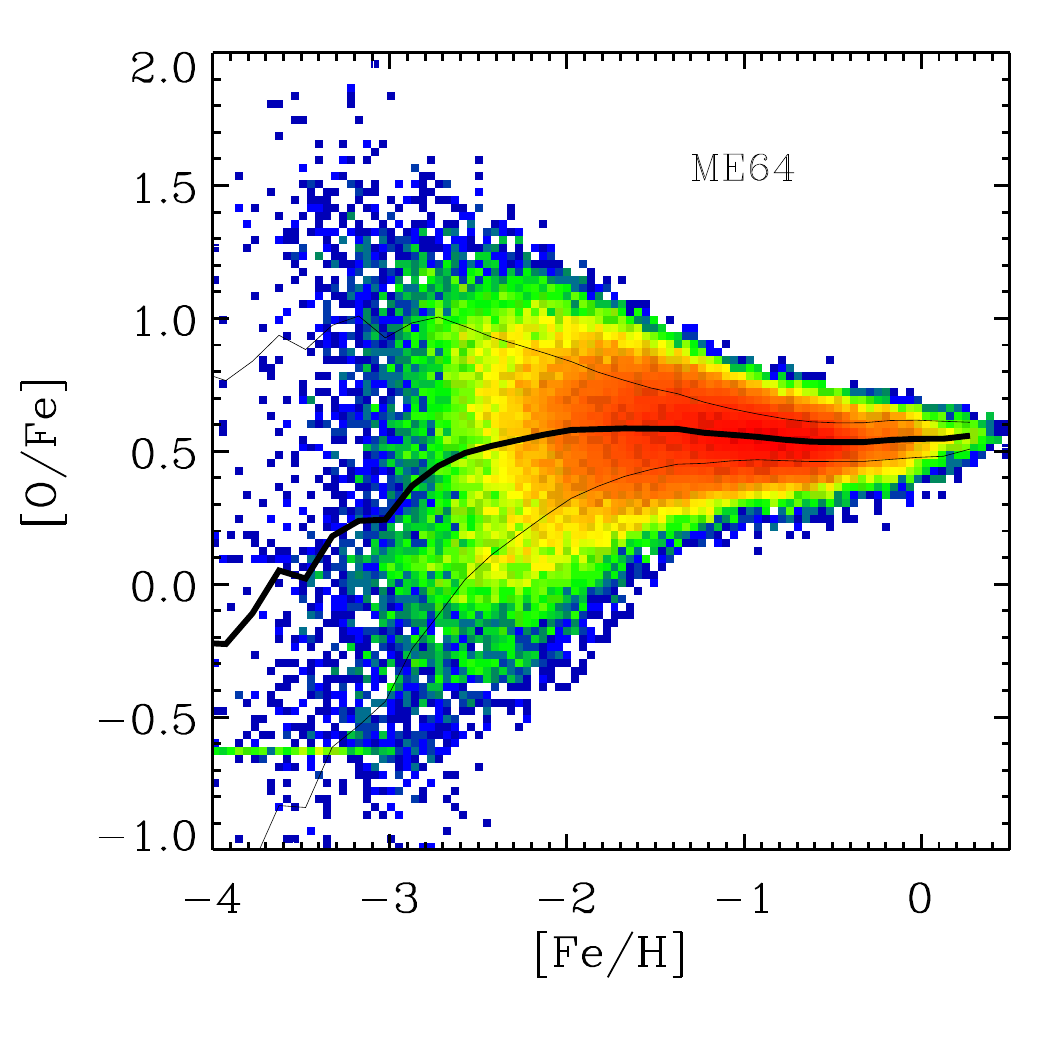}\includegraphics[width=4.5cm]{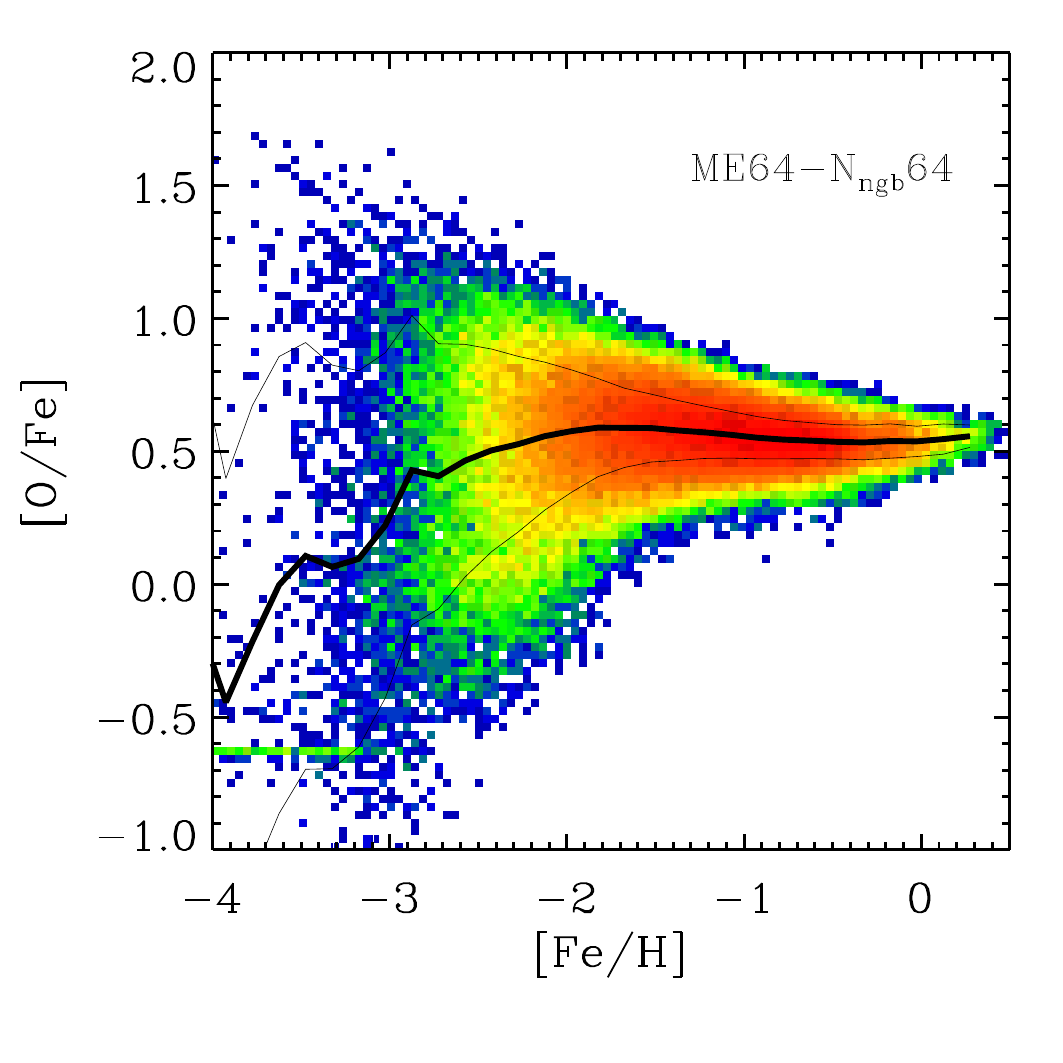}\includegraphics[width=4.5cm]{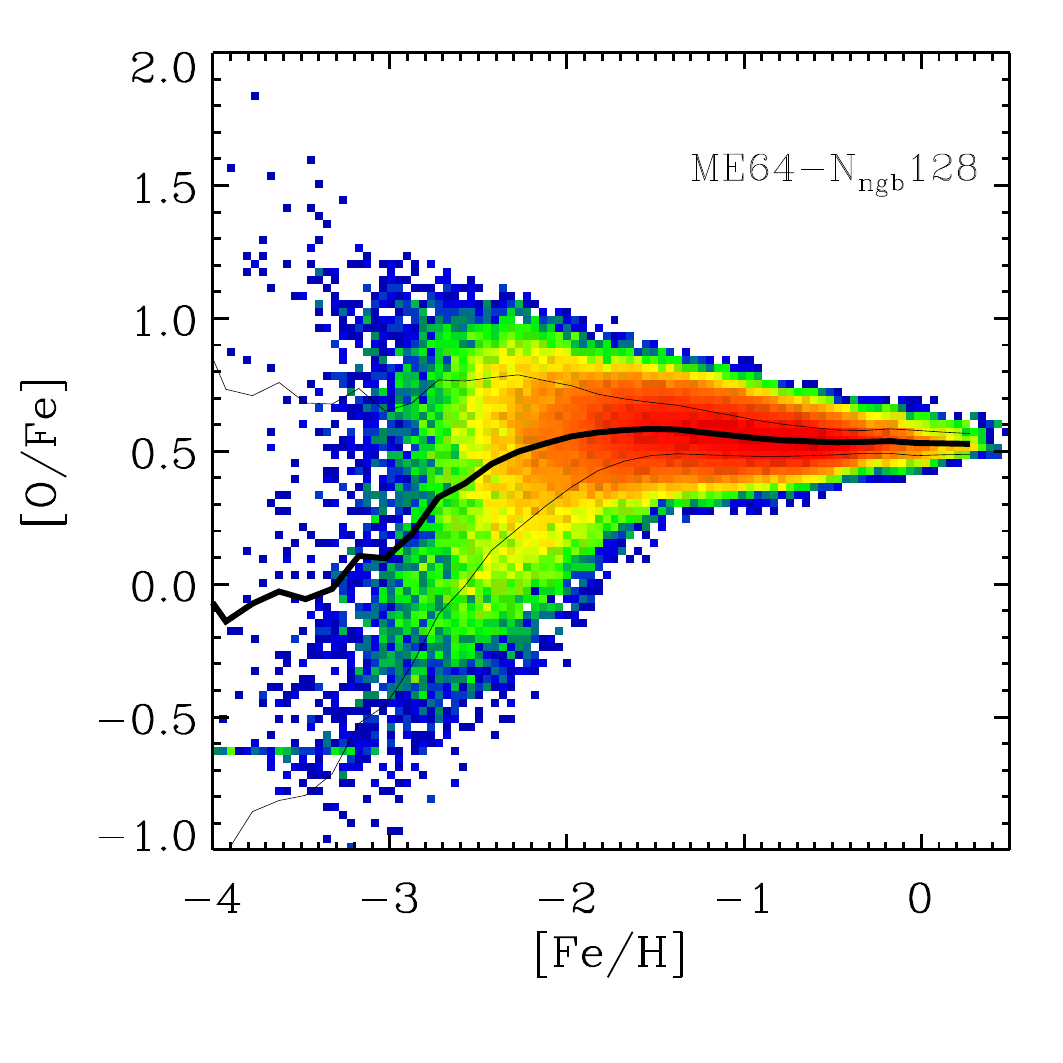}\includegraphics[width=4.5cm]{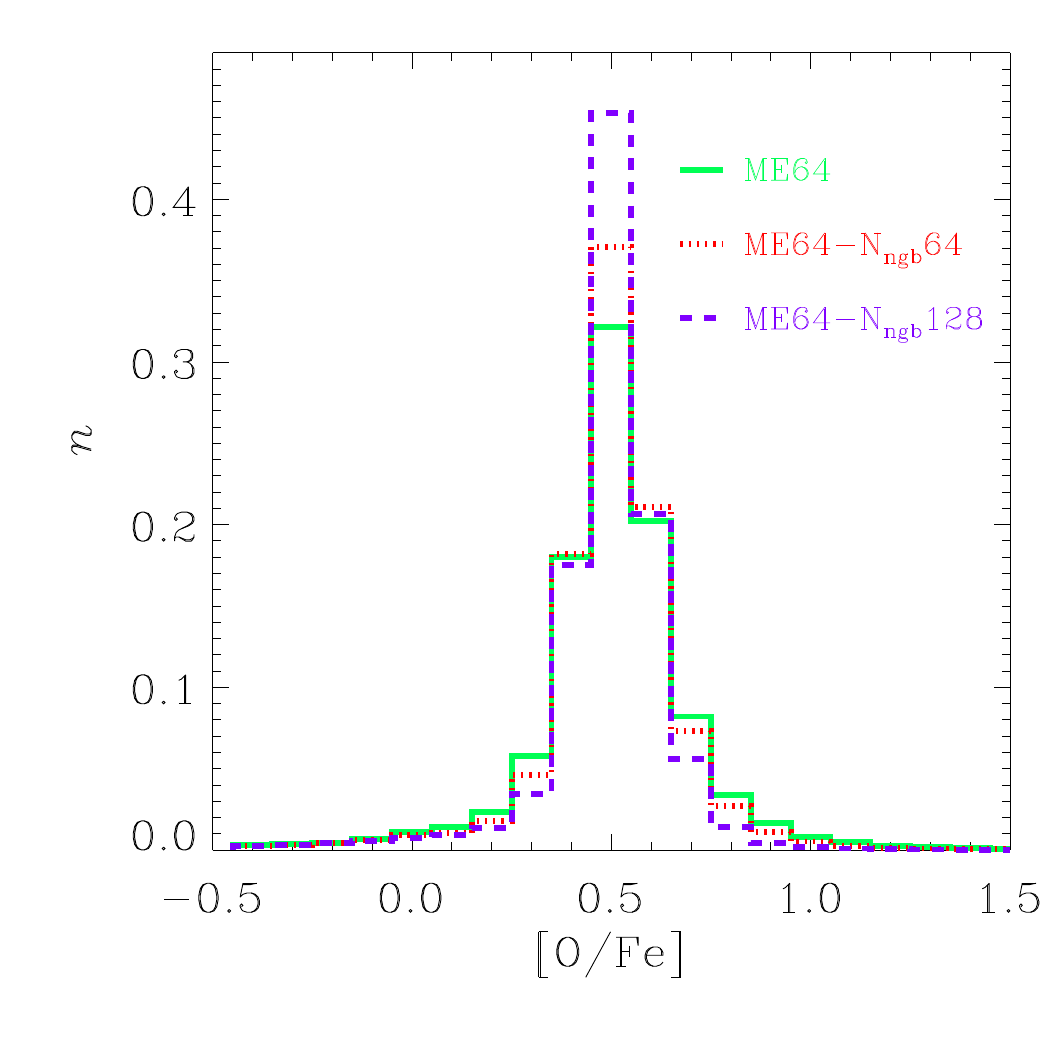}

  \includegraphics[width=4.5cm]{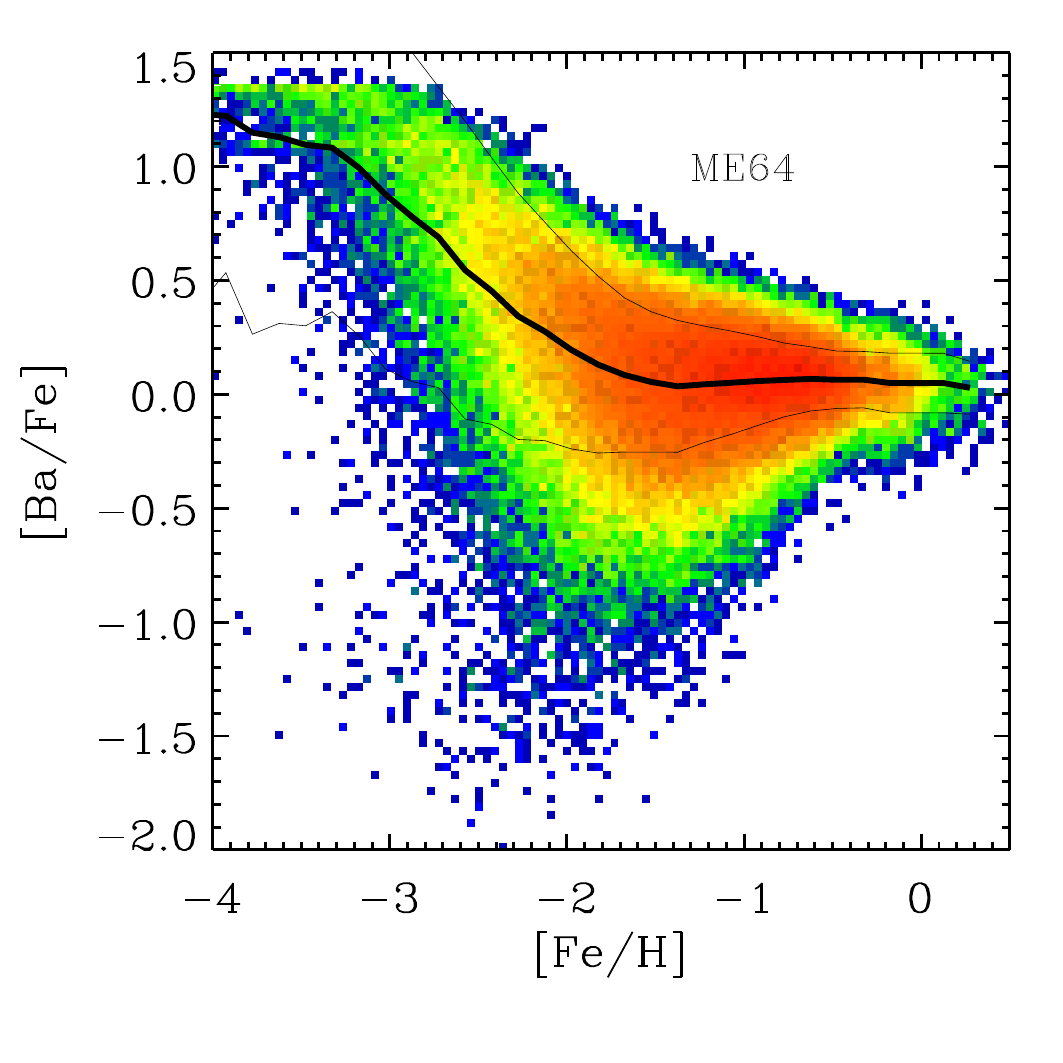}\includegraphics[width=4.5cm]{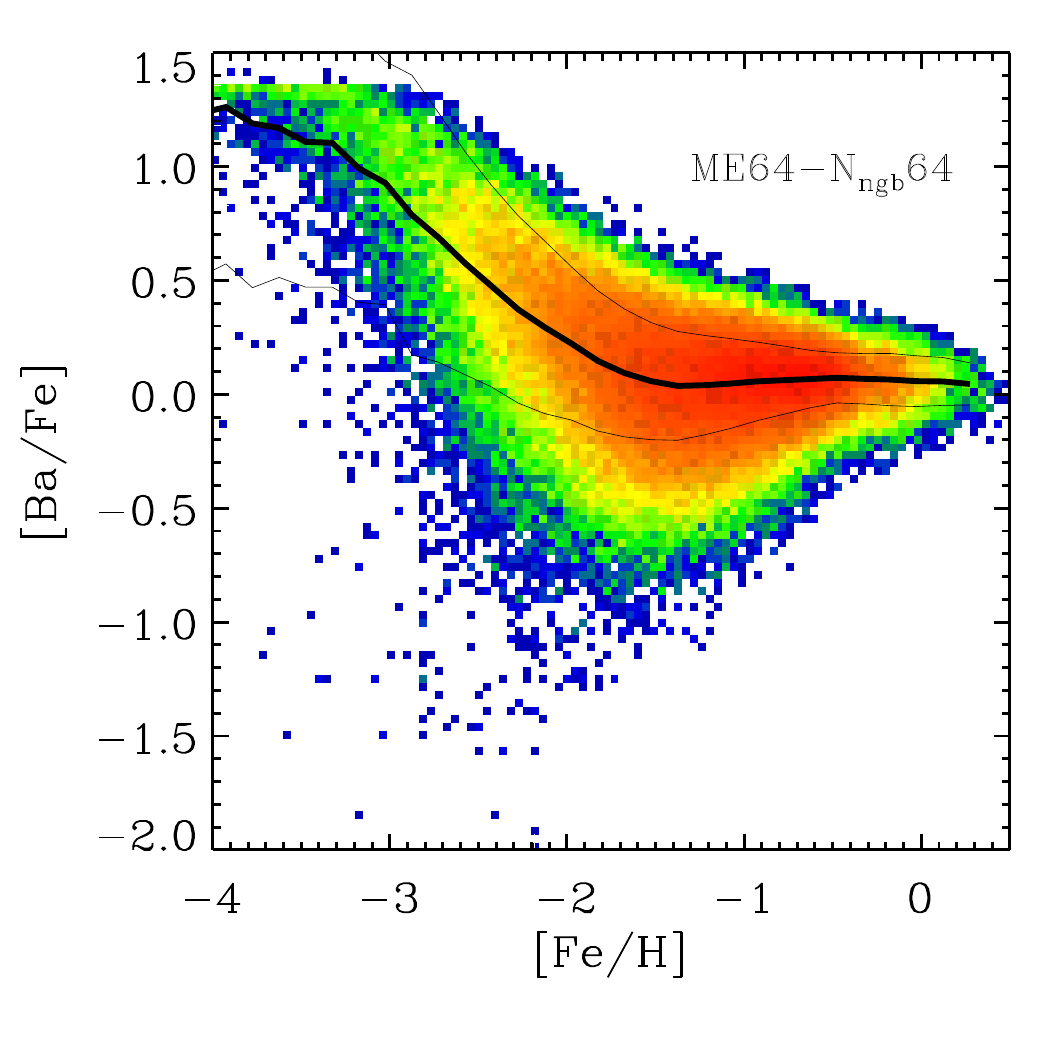}\includegraphics[width=4.5cm]{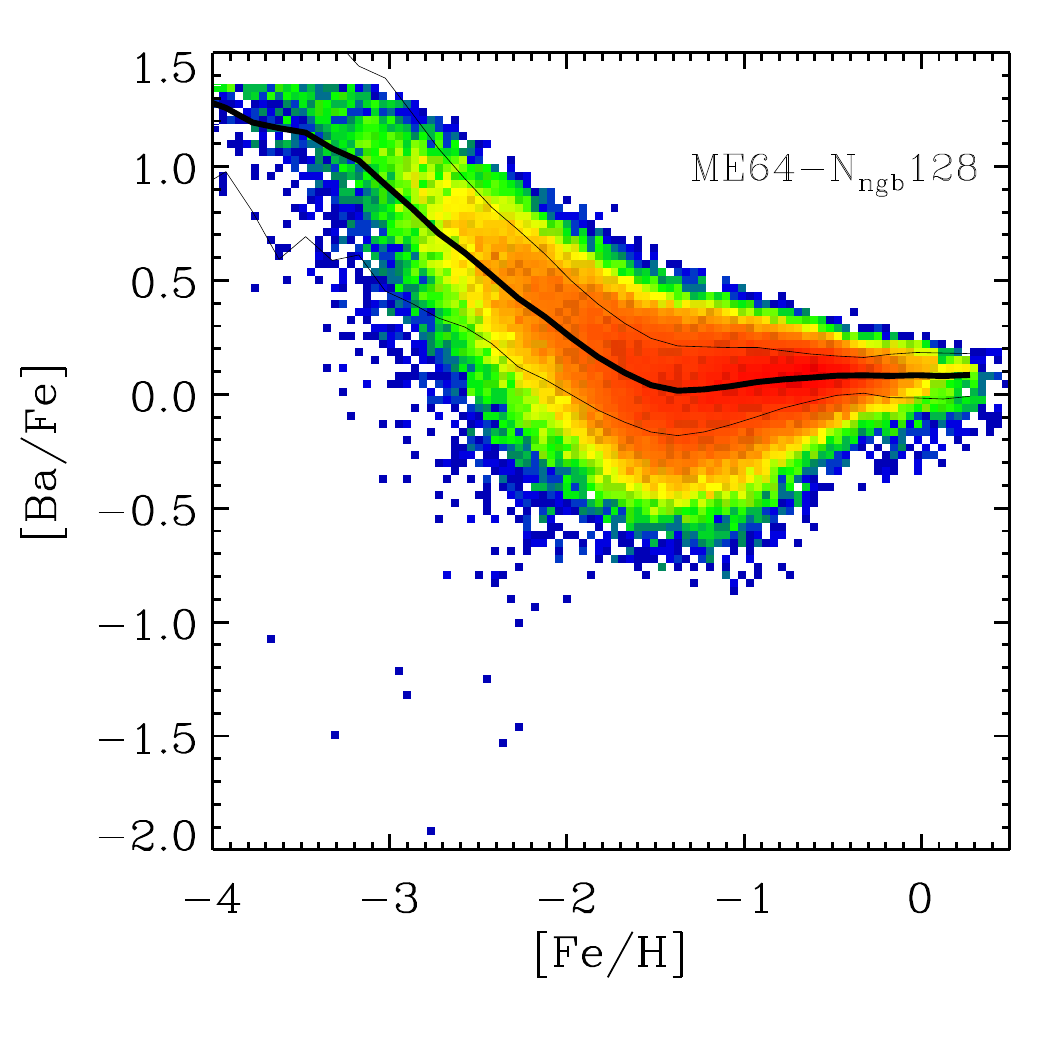}\includegraphics[width=4.5cm]{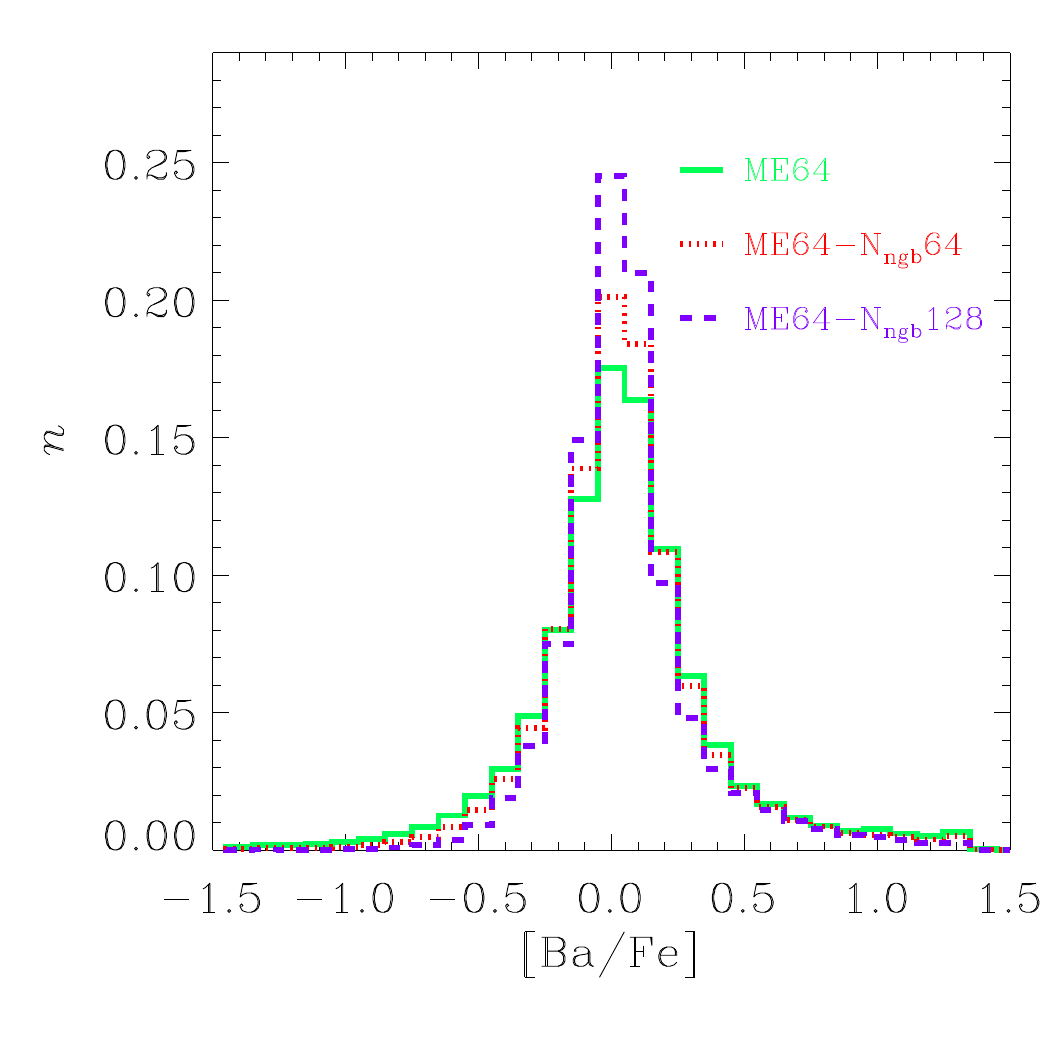}

\caption{The distribution of [O/Fe] and [Ba/Fe] versus [Fe/H] for our
simulations assuming multiple events per SN explosion and different number of
SPH neighbours, after 1 Gyr of evolution. 
The colour scale is normalized to the
total number of star particles in each simulation, and we also show
the corresponding median (thick lines) and $\pm\sigma$ contours (thin lines).
The right-hand panel shows a quantitative comparison of the three distributions.}
\label{fig:element_ratios_isolated_Ngb}
\end{figure*}

Finally, we show in
  Fig.~\ref{fig:element_ratios_isolated_Ngb}
 the results of runs with different number of neighbours, for [O/Fe] and [Ba/Fe]
 after 1 Gyr of evolution, and their comparison. 
We find very similar results for both element ratios, indicating that our results
are robust against changes in  $N_{\rm ngb}$. Note the excellent
agreement of the mean values and also of the scatter, evidenced in the right-hand panel of this figure.
We conclude that the results of our model
are robust against changes in the number  of SPH neighbours used in the
simulations.

\section{Cosmological simulations}\label{app:cosmo}

  In this Appendix we show that   
  the effects of differential enrichment on the  chemical properties of the old stars, 
  discussed in Section~\ref{sec:isolated}, are reproduced in cosmological simulations.
  For this, we compare
  the results of  cosmological  simulations SE and ME  (see Table~\ref{table:simulations}),
  which assume a single or multiple explosions per SNII, respectively.
  As explained in Section~\ref{sec:halo}, these correspond to cosmological simulations of a Milky Way-mass galaxy. 
  In order to show that differential enrichment affects the properties of old stars, we 
  focus here on stars formed during the first Gyr of evolution (which also allows
  comparison with our idealized simulations), even though the cosmological simulations have been run up to $z=0$.

\begin{figure}
  \centering
\includegraphics[width=8cm]{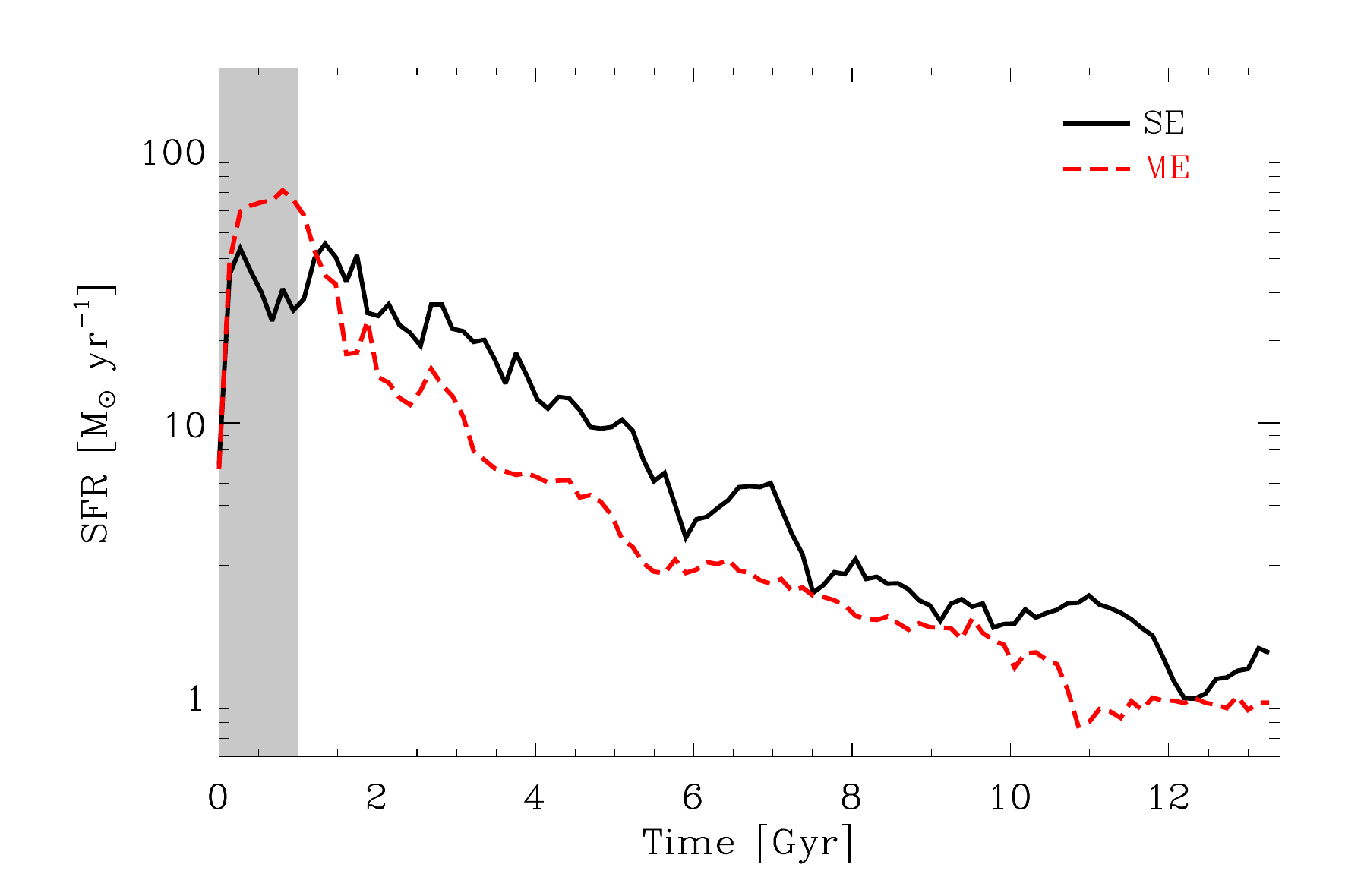}

\includegraphics[width=8cm]{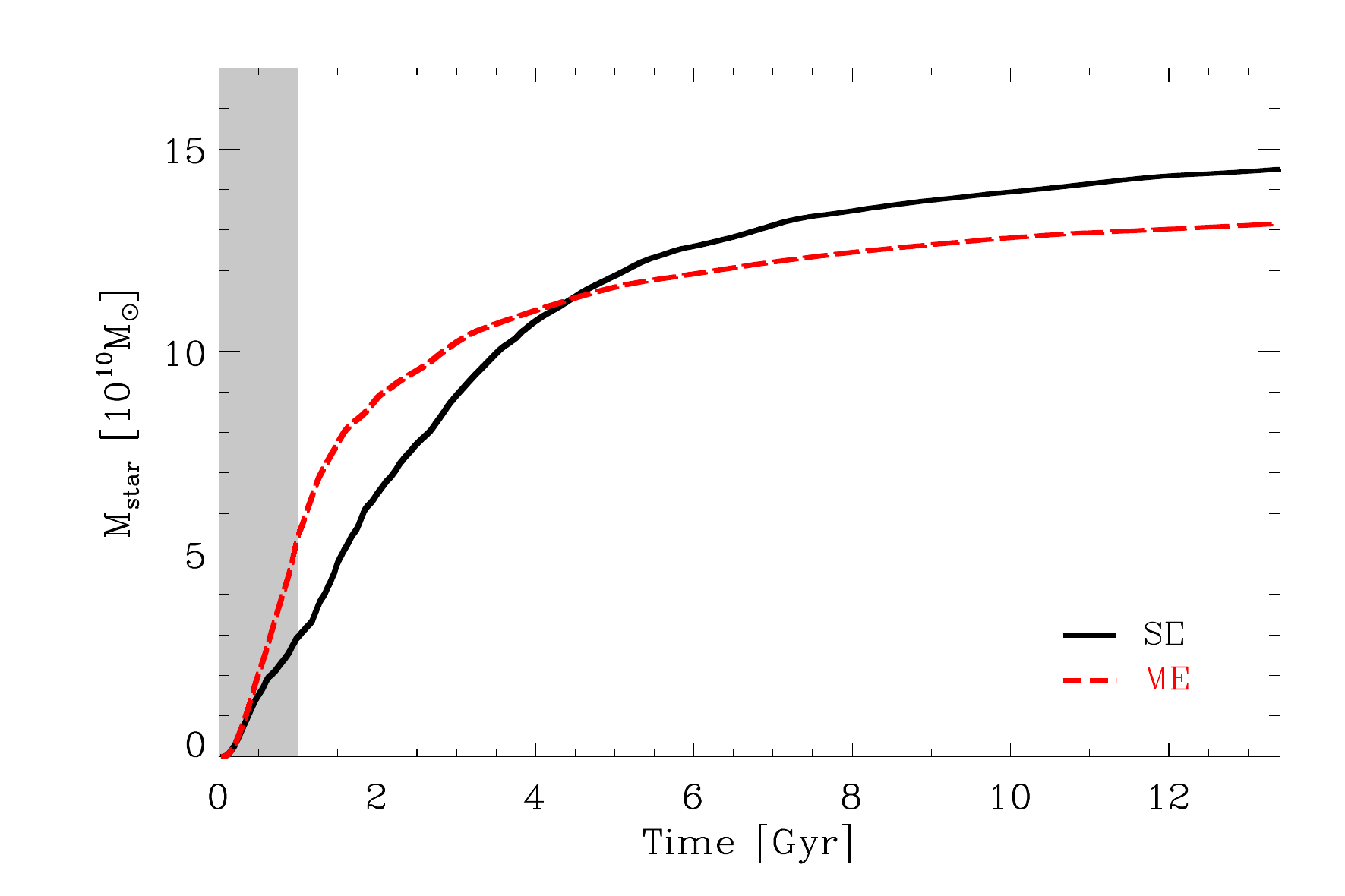}

\caption{Evolution of the star formation rate (upper panel) and integrated stellar mass  (lower panel) for
our cosmological simulations SE and ME, which assume single and multiple events per SN explosion, respectively. The shaded area highlights the first 1Gyr of evolution, where we focus our study.}
\label{fig:sfr_cosmo}
\end{figure}

Fig.~\ref{fig:sfr_cosmo}
  shows the SFRs and evolution of the stellar mass in simulations SE and ME.
  Similarly to our findings of the idealized simulations,  ME has a higher SFR compared to ME during the first Gyr of
  evolution,  due to the faster enrichment of the ISM when multiple events per
  SNII are assumed and the enhanced cooling rates.
  Later on, however, the behaviour of the SFRs inverts, as the amount of feedback
in ME is larger than in SE, producing a stronger reduction of the star formation
activity after the first starburst.
As a result, at the end of the simulation the galaxy formed in ME has lower stellar
mass.

\begin{figure*}
  \centering

\includegraphics[width=7.5cm]{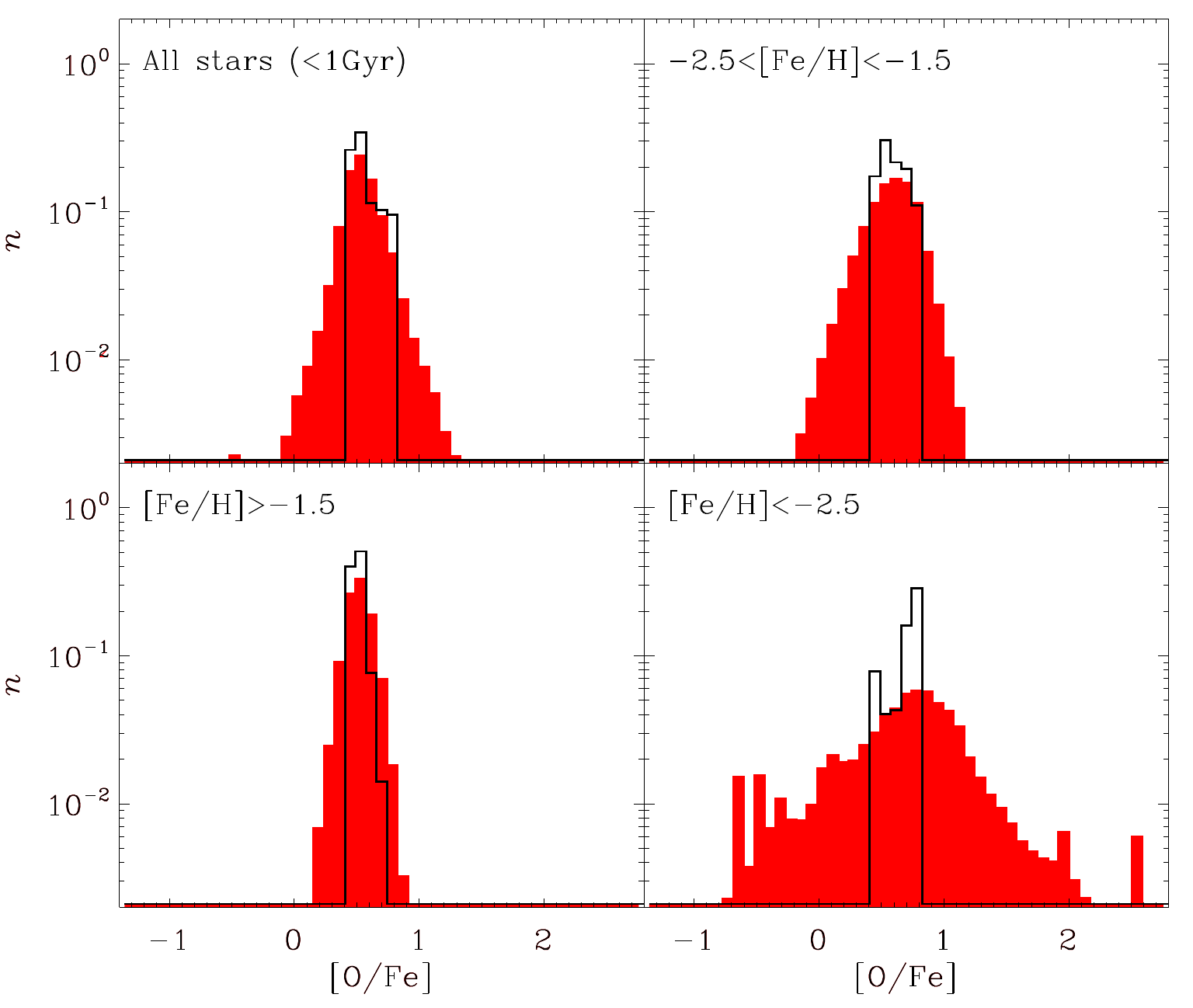}\includegraphics[width=7.5cm]{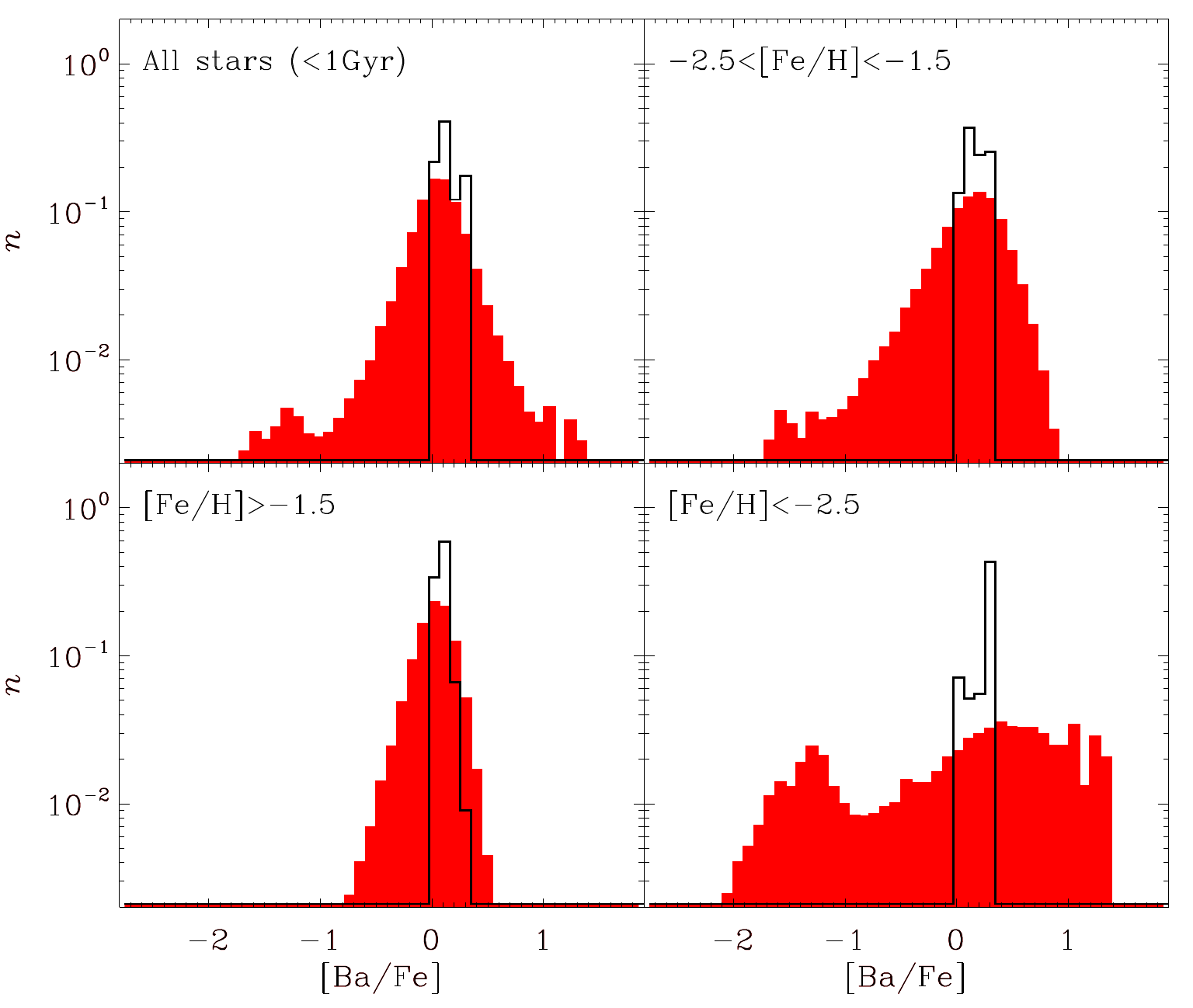}

\vspace{0.1cm}
\includegraphics[width=7.5cm]{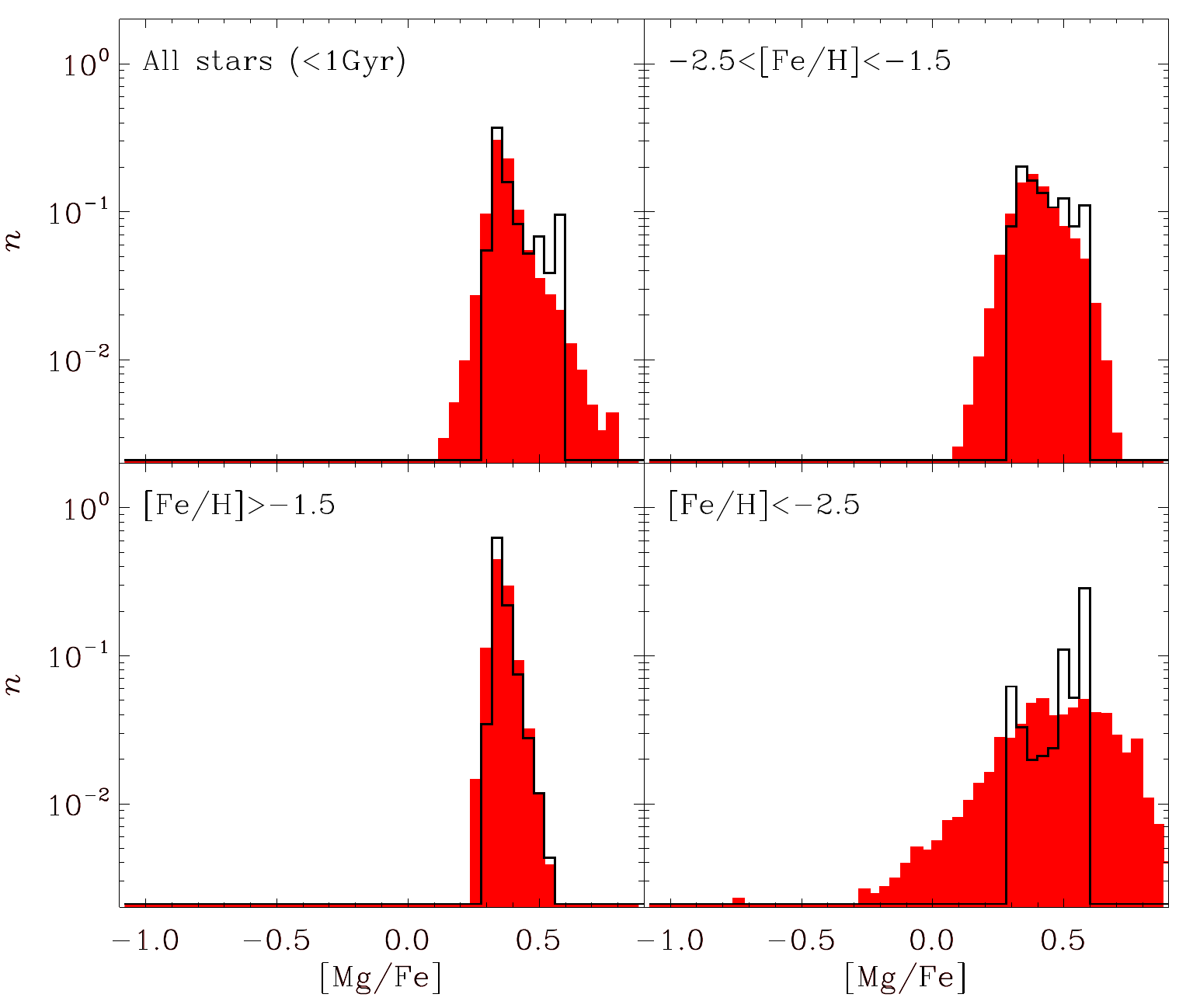}\includegraphics[width=7.5cm]{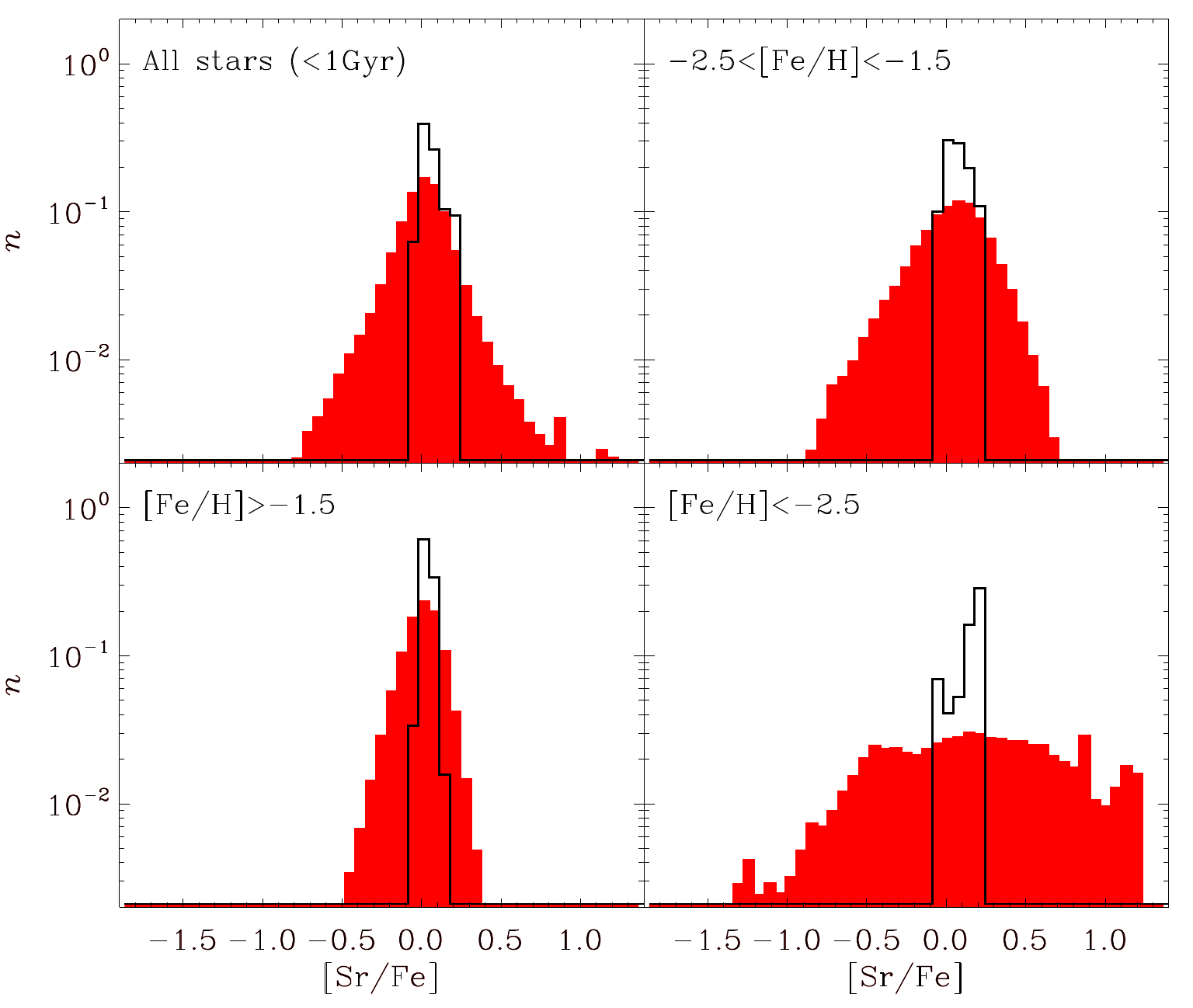}

\vspace{0.1cm}
\includegraphics[width=7.5cm]{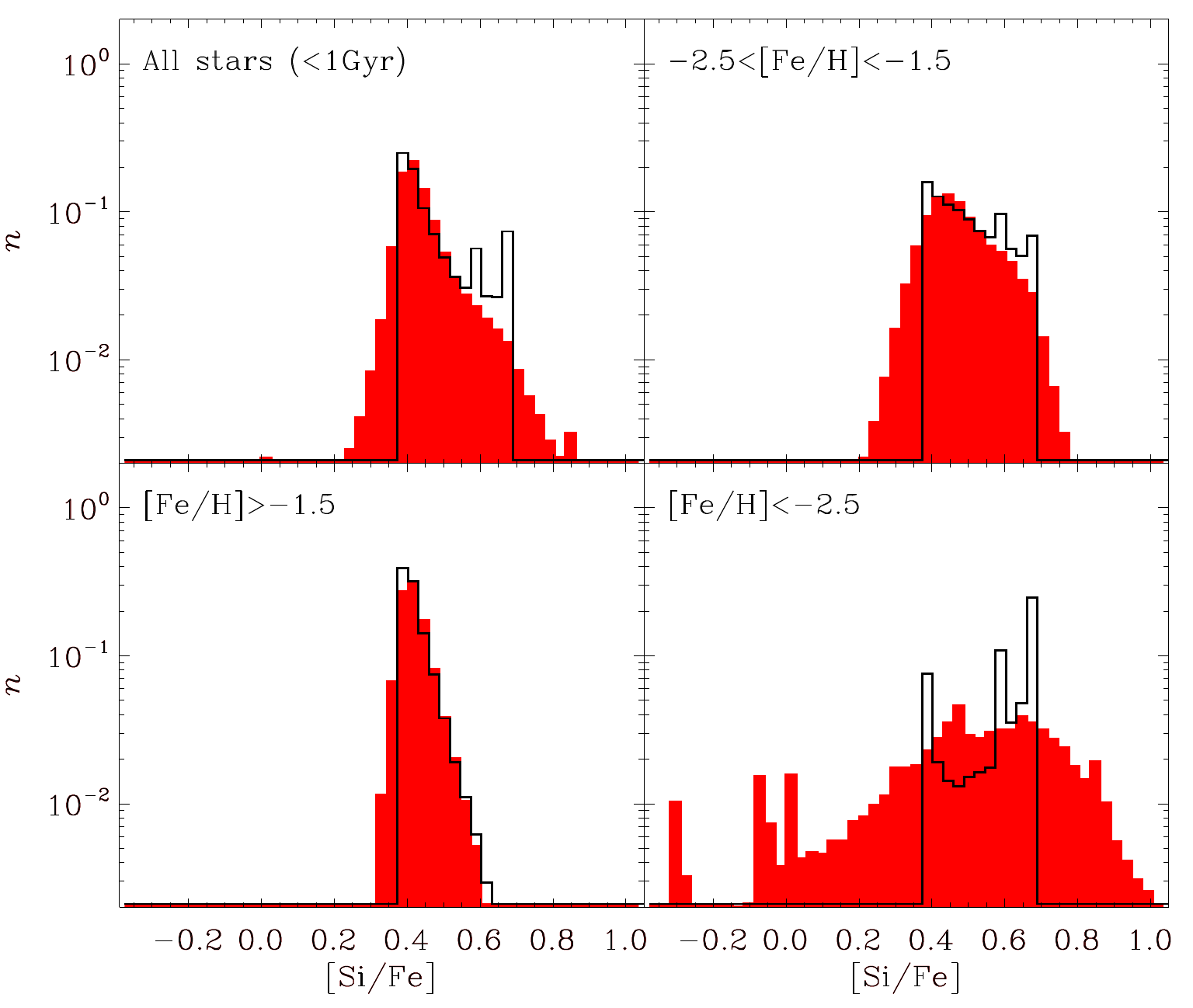}\includegraphics[width=7.5cm]{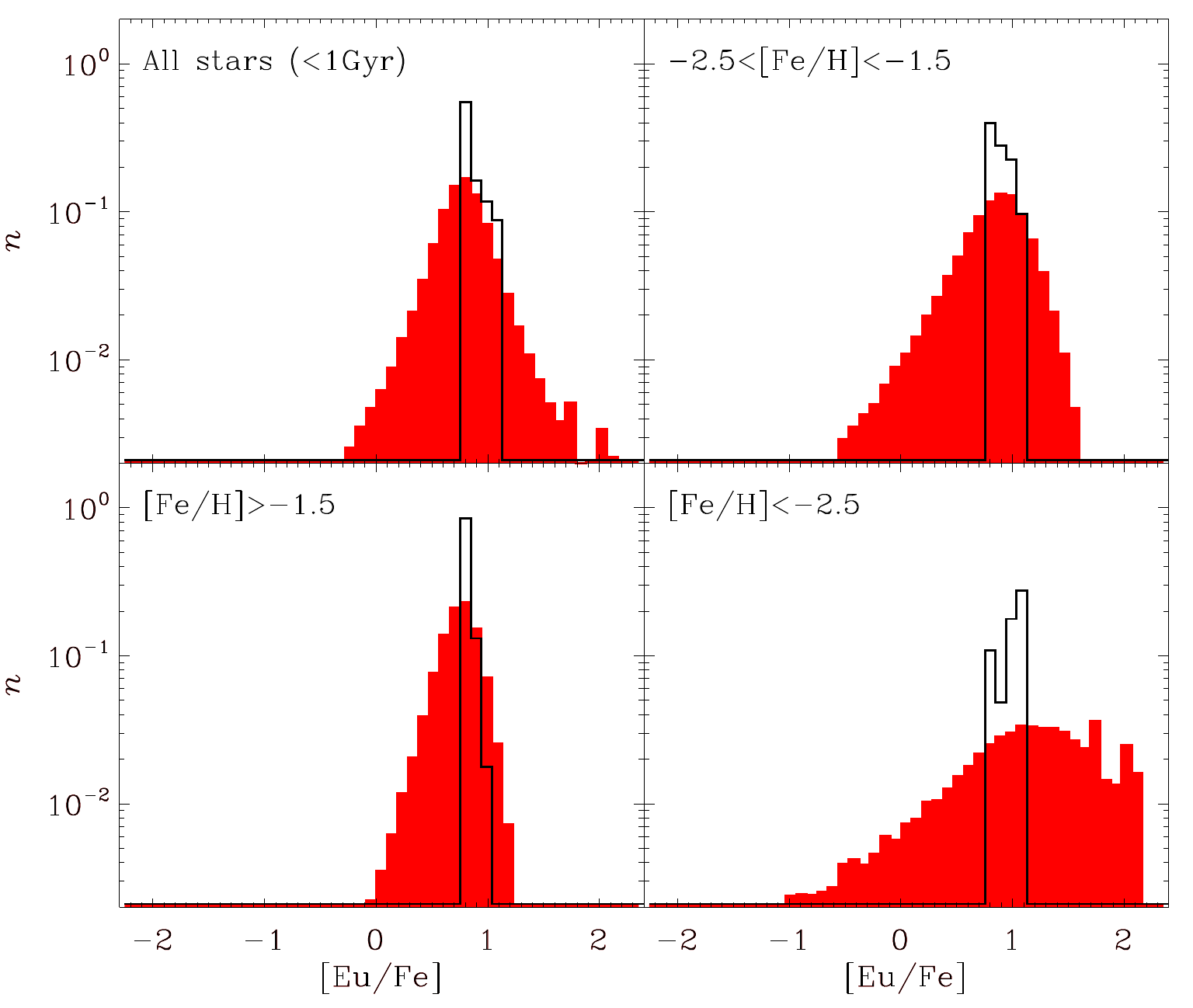}

\caption{Distribution functions of various stellar abundance ratios, for our cosmological simulations SE (solid lines) and ME (shaded areas), at $z=0$, considering only the old stars (formed during the first Gyr of evolution). The different panels show distributions for four ranges in [Fe/H].}
\label{fig:el_ratios_fehbins}
\end{figure*}

In Fig.~\ref{fig:el_ratios_fehbins} we compare the  distributions
of the various element ratios in SE and ME for stars formed during the 1 Gyr of evolution. We show results for all these
stars and, because differences are expected to be more important for the very metal-poor stars, we show the same
distributions restricting the sample according to the [Fe/H] abundance.
The differences between SE and ME are dramatic, and confirm that the proper treatment of the
early chemical enrichment significantly affects the scatter of the various element ratios. ME predicts
 much broader element ratio distributions compared
 to SE, and this effect is sronger as we move to more metal-poor populations. These results are consistent with our findings
 using the idealized simulations. It is worth noting that SE is
 unable to produce broad element ratio distributions, even though the cosmological setting is taken into account. This means
 that it is the differential enrichment what induces the scatter in the distributions, despite the fact that the cosmological
 evolution could add even more scatter. In particular, in the case of the stellar halo we expect a high contribution of
 ex-situ stars adding to this scatter (see Section~\ref{sec:high_srba}.

\bsp

\label{lastpage}

\end{document}